\documentstyle[twocolumn,aps]{revtex}
\begin{document}
\title{
A Survey of the Van Hove Scenario for High-T$_c$ Superconductivity \\
With Special Emphasis on Pseudogaps and Striped Phases}

\author{R.S. Markiewicz} 

\address{Physics Department and Barnett Institute, 
Northeastern University,
Boston MA 02115}
\maketitle
\twocolumn[
\widetext
\centerline{\it In Memoriam:}
\centerline{\it Carson Jeffries}
\centerline{\it Leo Falicov}

\begin{abstract}
The Van Hove singularity (VHS) provides a paradigm for the study of the
role of peaks in the density of states (dos) on electronic properties.  More 
importantly, it appears to play a major role in the physics of the high-T$_c$
superconductors, particularly since recent photoemission studies have found that
the VHS is close to the Fermi level in most of the high-T$_c$ cuprates near the
composition of optimum T$_c$.  This paper offers a comprehensive survey of the
VHS model, describing both theoretical properties and experimental evidence for
the picture.  Special topics discussed include a survey of the Fermi surfaces of
the cuprates and related compounds, and an analysis of the reliability of the 
slave boson approach to correlation effects.  
\par
While many properties of the cuprates can be qualitatively understood by a 
simple rigid-band-filling model, this is inadequate for more quantitative 
results, since correlation effects tend to pin the Fermi level near
the VHS over an extended doping range, and can lead to a nanoscale phase 
separation.  Furthermore, the peaks in the dos lead to competition from other
instabilities, both magnetic and structural (related to charge density waves).  
A novel form of {\it dynamic} structural instability, involving dynamic
VHS-Jahn-Teller effects has been predicted.  Scattered through the literature,
there is considerable experimental evidence for both nanoscale phase separation
of holes, and for local, possibly dynamic, structural disorder.  This review
attempts to gather these results into a comprehensive database, to sort the
results, and to see how they fit into the Van Hove scenario.  Recent experiments
on underdoped cuprates are found to provide a strong confirmation that the 
pseudogap is driven by a splitting of the VHS degeneracy.

\end{abstract}

\pacs{PACS numbers~:~~71.27.+a, ~71.38.+i, ~74.20.Mn  }

]
\narrowtext
\tableofcontents

\section{Introduction}

\subsection{What is a Van Hove Singularity?}

In 1953, Leon Van Hove\cite{vH} demonstrated the crucial role played by topology
in the band structure of either electrons or phonons.  He showed that any
nonanalytic behavior is caused by a change in the topology of the
bands -- changes now known as Van Hove singularities (VHS's)\cite{VHJr}.
The simplest such changes are associated with the opening or closing of
new sheets of the energy bands ($M_0$ and $M_3$ points in a three-dimensional
(3D) band structure).  But there are additional VHS's interior to the band,
associated with changes of the signs of one of the principal curvatures of the
band ($M_1$ and $M_2$ points in 3D).  These interior VHS's produce peaks in the 
density of states (dos) of the bands.  As such, these VHS's are expected to play
an important role in the electronic properties of these materials.
\par
In materials of lower spatial dimension, the role of some VHS's is enhanced,
because the dos $N(E)$ can actually diverge at a VHS.  In one-dimensional (1D) 
materials, the divergence is power law, $N(E)\propto (\Delta E)^{-1/2}$, where
$\Delta E=E-E_{VHS}$ is the distance in energy from the VHS.  However, the
importance of this VHS is somewhat limited, because the VHS falls at a
threshold at which a new band opens or closes, so the number of free carriers is
very small.  On the other hand, in a two-dimensional (2D) material, the relevant
VHS falls near the center of the band, where the number of free carriers is
maximal.  At this VHS, the dos diverges logarithmically $N(E)\propto ln(B/2
\Delta E)$, where $B$ is the bandwidth.  The topology of this VHS is that of a 
{\it saddle point}, associated with a crossover from electron-like to hole-like 
conduction, Figure~\ref{fig:1}.  This figure will be discussed in more detail in
Section IV.A, but the logarithmic divergence of the dos is clear from 
Fig.~\ref{fig:1}c.  The electron-to-hole crossover can be seen by looking at the
Fermi surfaces themselves, Fig.~\ref{fig:1}b.  [The dashed lines are the
generic behavior expected for the cuprates; the solid lines represent an 
anomalously strong singularity -- perfect nesting -- which corresponds to
special choices of the band parameters.]  For an underdoped sample (surface
tagged with a filled square) the Fermi surface is an electron-like surface
closed around the $\Gamma$-point of the Brillouin zone; for overdoping 
(circle-tagged curves) the Fermi surface is hole-like, closed about the 
$S$-point.
\par
At a saddle point, the two principal curvatures
have opposite signs; in electronic terms, the effective mass is electron-like in
one direction, and hole-like in the other (see the $X$-point in 
Fig.~\ref{fig:1}a).  In general, if an energy band has one saddle point, there 
will be other, symmetry related saddle points at the same energy.  In addition 
to the divergence of the dos, there can also be divergences of the (spin and/or 
charge) susceptibilities at wave vectors $Q$ coupling two of these saddle 
points.
\par
Hence, the anomalies of this VHS are of three sorts: associated with divergences
in the dos $N(E)$, divergent susceptibilities $\chi (Q,\omega )$, and with the
electron-hole crossover which implies that both electron-like and hole-like
carriers are simultaneously present at the VHS.
This paper is intended to serve as an introduction to the `natural history' of
this 2D saddle point VHS, pointing out some of the consequences of these three,
competing anomalies.  One of these consequences may be high-temperature 
superconductivity.

\subsection{The Van Hove Scenario}

In broadest terms, the `Van Hove scenario' seeks to answer the question, how is
the physics of an interacting electron liquid modified by the presence of a
(saddle point) Van Hove singularity near the Fermi level?  In more detail, this 
may be stated as follows. (1) The saddle point Van Hove singularity (VHS) of a 
two dimensional (2D) metal is the simplest model with a peak in the density of
states (dos), and as such can act as a paradigm for analyzing the role of dos 
structure in the physics of Fermi liquids. (2) In the 2D limit, the dos 
actually diverges, suggesting that the Fermi liquid is unstable as $T\rightarrow
0$, and hence that the underlying physics is as rich as that of 1D metals.
(3) A 2D VHS is present in the immediate vicinity of the Fermi level in the 
high-T$_c$ cuprates, and it is necessary to understand the role of this VHS in 
order to develop a correct picture of the physics of these new materials.  (4) 
The VHS can explain a number of normal-state anomalies in the cuprates, 
including the linear-in-T and $\omega$ electron-electron scattering rate and a 
pronounced crossover in properties from the underdoped to the overdoped regime. 
(5) The most exciting possibility 
(but only a possibility at this point) is that the VHS itself can provide a new 
pairing mechanism which drives the transition to high-T$_c$ superconductivity.
The following paragraphs are intended to clarify these points. 

The reasoning behind point (1) is straightforward.
For years prior to the discovery of the cuprates, an empirical rule
stated that the best way to find high-T$_c$ superconductors is to look for a
material with a peak in the dos, which could drive both structural (CDW) and
superconducting instabilities.  The highest T$_c$'s occur just on the edge of a
structural instability, and T$_c$ can be appreciably raised by inhibiting the
structural transition.  In light of this, the 2D saddle point VHS should be 
looked on as {\it the} generic dos peak. Indeed, all 3D dos peaks are associated
with 3D VHS's, and the largest dos peaks are generally produced by
closely-spaced $M_1-M_2$ VHS's\cite{Muel}, which can be generated from the 2D
saddle point by adding a small 3D coupling!
\par
Point (2): A peak in the electronic dos signals that the electrons can behave 
collectively, leading to a tendency for magnetic, or structural, or
superconducting instability.  The 2D saddle point VHS has a (logarithmically)
diverging dos, signalling an absolute instability, which must be very
carefully dealt with.  This was clearly brought out in the early work of
Dzyaloshinskii\cite{Dzy} and Schulz\cite{Sch}: the 2D VHS is a natural 
generalization of the one-dimensional electron gas\cite{Soly}, with competing 
spin and charge density wave (S/CDW), and s- and d-wave superconducting 
instabilities, and a correspondingly 
complicated renormalization group (RG) derived phase diagram.
\par
While the 1D VHS shows a stronger divergence, it falls near the top or bottom of
the electronic band, where there are too few carriers for collective effects to
be important.  The most interesting physics in a 1D metal is generally found 
near the middle of the electronic band.  In 2D, the saddle point VHS provides
the natural generalization of this regime of competing instabilities.  Since a
one-band model of the VHS can be constructed, with the same generic features,
it should be possible to find essentially exact solutions describing this
complex interplay of instabilities, and to much better understand under which
conditions superconductivity can become the dominant instability.
\par
Point (3) is most clearly demonstrated by angle-resolved photoemission 
experiments, which reveal the presence of a VHS close to the Fermi level in
optimally doped Bi$_2$Sr$_2$Ca$_{n-1}$Cu$_n$O$_x$ (Bi-22[n-1]n) 
(n=2)\cite{PE1}, YBa$_2$Cu$_3$O$_{7-\delta}$ (YBCO)\cite{PE2,PE3a}, and 
YBa$_2$Cu$_4$O$_8$\cite{PE3}.  However, Bi-2201\cite{PE4} offers
evidence that the relation between the VHS and superconductivity is not simple: 
despite having a VHS close to the Fermi level, T$_c$ is low, $\approx 10K$.
A similar result has also been found in Sr$_2$RuO$_4$, with $T_c=0.93K
$\cite{SRO}. Could this have something to do with competing interactions?  
Curiously, Bi-2201 has a squarer Fermi surface than Bi-2212, yet lower T$_c$.  
This is the same relation as found in comparing La$_{2-x}$Sr$_x$CuO$_4$ (LSCO) 
to YBCO, and explained in terms of pseudogaps.  But more of that later.
\par
Given that a VHS is present quite close to the Fermi level, then it will have a
very profound effect on {\it any theory} of high temperature superconductivity.
This point must be stressed.  It is often naively assumed that there is a 
straightforward competition between, e.g., nearly antiferromagnetic models and
VHS models of superconductivity, with one excluding the other.  This is by no
means the case.  The nearly antiferromagnetic models involve a peak in the
susceptibility near $Q=(\pi /a,\pi /a)$, the VHS nesting vector, and hence are
sensitive to the proximity of a VHS to the Fermi level.  
\par
But if this is granted, a whole new set of questions arises.  {\it How} does one
model the VHS?  Isn't it smeared out by correlation effects, disorder, and
interlayer coupling (not to mention the competing structural and magnetic 
instabilities)?  Just how smeared out?  Is there still a peak (albeit
finite) in the dos?  {\it The systematic analysis of these, and related issues
constitutes the scientific program of the Van Hove scenario.}  A partial
answer to these questions, based on our present state of understanding, is one
goal of the current review.
\par
Point (4): Many researchers have noted that, whereas the superconductivity of 
the cuprates is relatively conventional, except for the high T$_c$ values and
the (probably) d-wave symmetry of the gap, the normal
state properties are highly anomalous, and a key to understanding the cuprates
is to understand these anomalous normal state properties.  In this respect, the
fact that the normal state anomalies follow rather naturally from the proximity
of the VHS is very significant.
\par
Point (5): It is 
entirely conceivable that the role of the VHS in superconductivity is
indirect, with the density of states peak simply enhancing some underlying
conventional (electron-phonon) or unconventional (spin) pairing mechanism.
Nevertheless, there is a possibility that the VHS leads to
a fundamentally new kind of pairing mechanism.  At least two possibilities have 
been suggested.  First, the VHS can lead to an {\it electronic} pairing
mechanism, in that it is associated with the simultaneous appearence of
electron-like and hole-like Fermi surface sections.  Hence, there will be 
an attractive component to the Coulomb interaction, which leads to either
reduced Coulomb repulsion or to a possible net attraction.  Secondly, there
can be an enhanced {\it electron-phonon} coupling, via low energy phononic
excitations, associated with a dynamic Jahn-Teller effect.  (A similar mechanism
has been postulated for buckyballs.)

\subsection{The Purpose of this Review}

There is no universally accepted model of the physics associated with a VHS.
The purpose of the present review is therefore twofold: first, to survey the
experimental situation, summarizing various strands of evidence which point to
a role of the VHS in the physics of the high-T$_c$ cuprates; and secondly, to
summarize the various theoretical approaches which have been applied to the
VHS model.  In trying to make this review useful tutorially, I have found it
necessary to incorporate new calculations of some properties of the VHS's.
\par
There are a number of earlier reviews in this 
field\cite{Fried,RMrev,NewCom,Recon}, but most were 
written before the angle-resolved photoemission data became available.
While the VHS model seems capable of providing a unified picture of many
seemingly disparate experimental observations on the cuprates, there remain a
number of loose ends, puzzles, etc., which at present cannot be neatly fit into
the VHS scheme.  By including a discussion of these puzzling features, it is
hoped that they will invite the reader to think more deeply about the VHS
scenario, and perhaps to derive yet another test of the theory.
\par
It goes without saying that no model of high-T$_c$ superconductivity has yet
attained a high level of acceptance.  For almost every experimental observation
explained by the VHS scenario, there are a number of alternative 
interpretations.  While it is difficult to do justice to all these
alternative viewpoints in this review, I will at least try to
briefly note some competing theories of particular phenomena, such as the
pseudogap and nanoscale phase separation.
For discussions of various scenarios and the related spectroscopic evidence, 
see the recent series of conference proceedings, Refs. 
\cite{JCPS1,JCPS2,JCPS3,JCPS4}.
\par
A number of features of this review will have an interest beyond the immediate
Van Hove scenario.  Thus, there is a detailed collection of the Fermi surfaces 
of the cuprates (Section IV), including parameters for tight binding models
(Tables I-IV); a discussion of the accuracy of slave boson calculations (Section
VII); a catalog of experiments observing anomalous short-range order (Section
IX) and pseudogaps (Section IX.A); and a discussion of the evidence for (a 
possibly nanoscale) phase separation in the cuprates, both in underdoped and 
overdoped materials (Section XI).  
\par
It is particularly important to provide a database of experimental evidence for
the most anomalous non-superconducting features of the cuprates: (a) nanoscale
phase separation and (b) strong electron-phonon coupling effects and local
structural disorder.  These data
are widely scattered through\-out the literature.  Taken in isolation, the 
evidence of any one article could easily be overlooked, but taken collectively, 
the data present a coherent picture of highly anomalous behavior.
\par
At this early stage, I feel that it is important to provide a general overview,
summarizing all of the data even if its relation to the VHS scenario is not 
clear -- particularly in the case of nanoscale phase separation.  Nevertheless,
the picture that emerges seems to be largely explicable in terms of the VHS,
complicated by commensurability effects.

\subsection{Abbreviations}
\par
It is convenient to summarize the various abbreviations used in this review.
{\bf Materials:} Y\-Ba$_2$\-Cu$_3$\-O$_{7-\delta}$ = YBCO, YBa$_2$Cu$_4$O$_8$ = 
Y-124, YBa$_2$Cu$_{3.5}$O$_{7.5-\delta}$ = Y-123.5, Bi$_2$Sr$_2$Ca$_{n-1}$Cu$_n
$O$_x$ = Bi-22[n-1]n, Tl$_m$Ba$_2$Ca$_{n-1}$Cu$_n$O$_x$ = Tl-m2[n-1]n (m=1,2), 
La$_2$CuO$_{4+\delta}$ = LCO, La$_{2-x}$A$_x$CuO$_4$ = LACO (where A can stand 
for either S=Sr, B=Ba, or C=Ca), and Nd$_{2-x}$Ce$_x$CuO$_4$ = NCCO. In 
addition, the nickelates,
which have the same formula as LCO, LSCO, with Ni replacing Cu, are abbreviated
LNO or LSNO.  Finally, to refer to a specific oxygen stoichiometry, I will
either quote $\delta$ or write the stoichiometry explicitly in the abbreviation,
as, e.g., La$_{2-x}$Sr$_x$NiO$_{4.12}$ = LSNO$_{4.12}$.  The bismuthates will be
abbreviated as BPBO = BaPb$_{1-x}$Bi$_x$O$_3$, BKBO = Ba$_{1-x}$K$_x$BiO$_3$.
\par
{\bf Structures:}  The various phase transitions in LBCO, LSCO, etc., will be
abbreviated as high-temperature tetragonal = HTT (space group {\it I4/mmm}); 
low-temp\-er\-ature orthorhombic = LTO ({\it Bmab}, in the conventional, but
non-standard notation); low-temperature tetragonal = LTT ({\it P4$_2$/ncm}); and
the {\it Pccn} phase, which will just be called 
that, rather than LTO2 or some equivalent.  Generic structures such as 
charge-density wave (CDW), spin-density wave (SDW), antiferromagnet (AFM), and
antiferromagnetic insulator (AFI) are also used.  In drawing the Brillouin
zones and energy dispersions, I will generally simplify the structure to a
2D tetragonal (or orthorhombic) cell, defining the special points
$S=(\pi /a,\pi /a)$ and $\bar S=S/2$.  (In some figures taken from earlier
publications, these points are called $M$ and $\bar M$, respectively.)
\par
{\bf Doping:} For the hole doping, I will use the symbol $P$, where, e.g., $P=
x$ in LSCO and $P\simeq 2\delta$ in LNO.  The relation between $P$ and $\delta$
in YBCO and LCO is more involved, and is discussed in Subsection VI.B.  For
convenience, and where no confusion is likely to arise, I will often quote $x$
or $\delta$ instead of $P$.

\subsection{Partial Synonyms}

\par
In the recent literature, a number of papers have discussed calculations which
can in principle apply to a larger class of phenomena than VHS theories -- and
hence have been given a new name.  Nevertheless, in practice the phenomena are 
often found to be associated with a VHS.  Thus, {\bf flat bands}\cite{flatb} can
in principle arise in any band structure, but in practice the interesting bands 
are those which are flat in the immediate vicinity of a VHS, thereby enhancing 
the effects of the dos divergence.  
\par
Again, a {\bf hot spot} is a point on a Fermi surface associated with 
anomalously strong scattering, which may therefore be relevant to transport
anomalies in the cuprates.  When Hlubina and Rice\cite{HlR} introduced this 
term, they had two examples in mind -- the VHS itself and the anomalous 
scattering associated with the peak in spin susceptibility postulated in a 
number of nearly antiferromagnetic theories.  These latter need not in principle
be associated with the VHS, although, as discussed below, the VHS is the most
likely candidate for self-consistently generating the susceptibility peak.
In a recent calculation of these spin-related hot spots, Stojkovic and 
Pines\cite{Pin} showed that the hot spots arose at those points heavily shaded 
in Fig.~\ref{fig:a1}. Clearly, the hot spots are just those points on the Fermi
surface closest to the VHS.  Two points of interest: there are strong effects 
even though the Fermi level is not exactly at the VHS, and these VHS hot spots 
can explain\cite{Pin} the anomalous T-dependence of the 
Hall effect in these materials.

\section{Instability and Fermi Surface Topology}

The connection between structural instability and the topology of the Fermi 
surface was first made when Hume-Rothery noted that, in a series of metallic
alloys based on Cu, structural phase transitions are found to occur at fixed
values of the conduction electron concentration\cite{HumR}.  Jones\cite{Jon} 
suggested an interpretation of these Hume-Rothery alloys, assuming a nearly free
electron gas with spherical Fermi surface.  At the critical concentrations, the 
size of the Fermi surface approximately matches that of the Brillouin zone 
(actually, of the Jones zone), so that, by introducing small gaps associated 
with the superstructure, large sections of Fermi surface can be gapped, 
lowering the electronic energy.  Despite the great progress in band structure 
theory in the intervening years, this fundamental insight still seems to be 
correct\cite{Haf}, even though detailed calculations have proven 
difficult\cite{M&M}.
\par
In a one-dimensional (1D) metal, the analogous situation leads to a charge 
density wave (CDW) accompanied by a Peierls distortion. Since all real materials
have some interchain or interlayer coupling, no metal is truly 1D.
Quasi-1D character is described in terms of Fermi surface nesting.  This 
{\it conventional nesting} must be carefully distinguished from {\it VHS 
nesting}.  In conventional nesting, the single particle dos $N(E)$ of a material
is featureless, so to get a peak in the joint density of states (jdos),
\begin{equation}
J(\vec Q)=\int d^3k N(E_{\vec k})N(E_{\vec k+\vec Q}),
\end{equation}
it is necessary that two parts of the Fermi surface run parallel over a 
considerable distance, separated by a common vector $\vec Q$ -- i.e., that the 
two parts of the Fermi surface `nest'.  In contrast, near a VHS the dos $N(E)$
already has a strong peak, so if $\vec Q$ joins two VHS's, a large peak in
$J$ is assured.  Under ordinary circumstances, this peak generally overwhelms
the peak due to conventional nesting.  For example, in Fig.~\ref{fig:2}, the 
jdos calculated by Pickett, et al.\cite{Pick1} for LSCO, peaks associated with 
both conventional nesting ($Q_1$, $Q_2$) and VHS nesting ($Q_0$) are clearly
evident (see inset $a$).  However, the dominant feature is clearly that
associated with VHS nesting.
\par
The VHS nesting can be seen in the (spin or charge) susceptibility
\begin{equation}
\chi (\vec q,\omega )=-\sum_{\vec k}{f(E_{\vec k})-f(E_{\vec k+\vec q})\over 
E_{\vec k}-E_{\vec k+\vec q}-\hbar\omega -i\delta}.
\label{eq:02}
\end{equation}
At the VHS, both the intra-VHS ($\vec q=0$) and the inter-VHS ($\vec q=\vec Q_0
=(\pi /a,\pi /a)$) susceptibilities have a logarithmic (or $ln^2$) divergence. 
In the absence of electron-phonon coupling, the RPA dielectric constant
becomes $\epsilon_e(\vec q,\omega )=1+V_c(q)\chi (\vec q,\omega ),$
with $V_c(q)=4\pi e^2/q^2$.  Near the VHS, $\epsilon_e>>1$, so the screened 
Coulomb interaction becomes 
\begin{equation}
\tilde V_c(\vec q,\omega )={V_c(q)\over\epsilon_e(\vec q,\omega )}\simeq{1\over
\chi (\vec q,\omega )}.
\label{eq:03}
\end{equation}
Thus, at a VHS, the screening is perfect, $\tilde V_c=0$, both at $\vec q=0$ and
at $\vec q=\vec Q_0$.  This is the standard form for $\tilde V_c$, and will be 
adequate for qualitative calculations.  However, it must be kept in mind that 
very near the VHS, the polarizability has significant corrections due to 
electron-hole attraction (ladder diagram summation)\cite{RM4}.
\par
If the electrons are coupled to a phonon of bare frequency $\omega_l$, the 
phonon frequency is renormalized to $\tilde\omega =\omega_lR^{1/2}$, with
\begin{equation}
R=1-\bar V_c\chi (\vec Q,\omega ),
\label{eq:0}
\end{equation}
with $\bar V_c=V_c/ P_{e3}(Q)$ and $P_{e3}(Q)\simeq 0.15$, a local field
correction, is a complicated function of the dielectric parameters\cite{RM5}.  
Assuming 
\begin{equation}
\chi (\vec Q,\omega\simeq 0)={1\over 2B}ln({1.13B\over T}),
\end{equation}
there will be a structural instability, $R=0$, at
\begin{equation}
T_{CDW}=1.13Be^{-1/\lambda_1},
\label{eq:00}
\end{equation}
\begin{equation}
\lambda_1={\bar V_c\over 2B}.
\end{equation}
This result must be corrected for strong fluctuations in two 
dimensions\cite{RM5}, but it qualitatively shows the role of the VHS in
promoting structural phase transitions.

\section{Van Hove Scenario Before High T$_c$}

\subsection{Generic VHS's -- Mostly 3D}

Van Hove\cite{vH} showed that the dos for both electrons and phonons is 
dominated by singularities associated with changes in the topology of the 
constant energy surfaces in momentum space (in particular, the electronic Fermi 
surface).  For present purposes, the most important VHS is the saddle-point VHS 
found in the interior of a 2D energy band.  This is associated with a crossover 
between electron-like and hole-like Fermi surface sections.  In a strictly 2D 
limit, the dos has a logarithmic divergence at such a point\cite{vH}; Van Hove
actually was calculating the dos for phonons, but since the results are purely
topological, the same dos is found for electronic bands.  In the presence of a 
weak interlayer dispersion, this VHS splits into a closely spaced pair of M1 and
M2 VHS's -- i.e., the logarithmic spike is replaced by a peak with a flat top --
but the integrated dos under the peak is unchanged\cite{SHF,RM6,3dVHS}. 
Figure~\ref{fig:a2} illustrates this development, showing the dos of an 
individual layer (Fig.~\ref{fig:a2}a), of four coupled layers 
(Fig.~\ref{fig:a2}b), and of an infinite number of coupled layers 
(Fig.~\ref{fig:a2}c).  Even in 3D, the resulting dos peak can be quite sharp.
Thus, the insert to Fig.~\ref{fig:a2}c shows the calculated dos for 
Ni\cite{Muel}.
\par
As part of a general analysis of Fermi surfaces in metals, Lifshitz\cite{Lif} 
initiated a study of how the properties of a material are altered by varying
the topology of the Fermi surface.  He showed that in general, if the material
properties could be varied in a quasi-continuous fashion (via pressure or 
doping), the change of the Fermi surface topology would produce a weak (order
2.5) phase transition, with consequences for the thermodynamic properties of the
materials.  In addition to the order 2.5 phase transition exactly {\it at} the
VHS, Lifshitz showed that there could also be a {\it first order} phase
transition, with volume discontinuity (compressibility $\kappa <0$) in the
immediate vicinity of the VHS.
Lifshitz' analysis was restricted to the VHS's of three-dimensional (3D)
metals, and the extension to the 2D saddle point is given in Appendix A.
Considerable research on this order 2.5 phase transition was carried out, 
particularly in Russia, and this has been the subject of extensive
reviews\cite{Varl,BKPV,BGGRV}.  However, the transition was typically 
studied in 3D materials, where the VHS leads only to a slope 
discontinuity in the dos, and the stronger discontinuity associated with the 2D 
VHS was not studied.  There was some theoretical work analyzing how a VHS would 
affect superconductivity\cite{VHSsc}, but again, in 3D materials.  
\par
From the BCS formula for the superconducting $T_c$, it is clear that enhancing
the dos should enhance superconductivity.  This simple idea was applied to the
`old' high-T$_c$ superconductors, in particular the A15 compounds.  The A15
structure is very suggestive of three interpenetrating chains, so the dos peaks 
were initially interpreted in terms of one-dimensional (1D) VHS's, with even 
stronger (square-root) singularities\cite{Lab}.  Full 3D band structure
calculations find that the Fermi level is at a peak in the dos\cite{A15}, but  
this peak cannot be interpreted in terms of 1D features. Gorkov suggested that 
the physics is dominated by a 3D VHS, near the [111]-directions of the Brillouin
zone\cite{Gor}.  Bilbro and McMillan\cite{BM}
developed Gorkov's model and showed that it leads to increased T$_c$'s and could
explain the strong competition between superconductivity and structural 
instability observed in the A15's.  Weger\cite{Weg} has stated that ``the normal
state properties [of the A15's] ... are dominated by the Van Hove 
scenario\cite{WegG}, [but] are entirely different from those of the cuprates.'' 
The result of this review will be a qualified agreement with this statement:
to the extent that the VHS can be treated perturbatively, the striking
differences in physical properties suggest that the VHS model applied to the
A15's does not work in the cuprates.  Instead, the review will attempt to
demonstrate that the differences between the A15's and the cuprates are due to 
the reduced dimensionality of the latter, which greatly enhances the role of the
VHS's.  For present purposes, the important feature is that the VHS's play an 
important role in the A15's. 

In searching for higher temperature superconductivity in the A15's, the dual
role played by a competing structural instability was first recognized. 
The instability is often associated with phonon mode softening, and these soft,
anharmonic phonons can contribute to a further enhancement of the 
electron-phonon coupling; after the structural phase transition, the
remaining phonons are much stiffer, with weaker residual electron-phonon
coupling, and hence poorer superconducting properties.  Hence, the lore in
high-T$_c$ research is that the best place to search for superconductivity is in
the immediate vicinity of a structural instability.  For instance, T$_c$ can be
enhanced in A15 compounds by preparing quench-condensed sputtered films, in
which the structural transition is arrested by disorder\cite{splat}.
\par
Many papers published in the period prior to the discovery of the cuprate
superconductors concentrated on the structural instabilities.  Thus, 
Dagens\cite{Dag} calculated the phonon softening associated with a 3D VHS --
the examples he gave were from superconductors Nb and (doped)
Pb.  These ideas were used\cite{Pick0} to explain the structural instability
found in La under pressure.  

\subsection{2D Saddle-Point VHS}

\par
Starting in $\approx 1970$, theoretical interest was drawn to the unusual
properties of 2D saddle point VHS's.  Thus, Roth, et al.\cite{RZK} showed that
the VHS would lead to a generalized Kohn anomaly, with a cusp in the magnetic
susceptibility.  Rice\cite{Ric} noted that near a VHS, there is a {\it breakdown
of analyticity} in expansions of the free energy.  He showed that this could 
lead to a transition to an antiferromagnetic phase.  Rice and Scott\cite{RiSc} 
introduced the concept of VHS 
nesting, in which the jdos has a peak at a $Q$-vector joining two VHS's, Fig.
~\ref{fig:2}. Balseiro and Falicov\cite{BFal} explored the competition between 
superconductivity and charge density waves (CDW's) near a 2D VHS, using a model
which is now a standard one-band model for the cuprates.  While their
calculations were numerical, they clearly displayed the enhanced instability
associated with the 2D VHS: a superconducting or structural instability always
occured for any arbitrarily small electron-phonon coupling.  Scalapino and
coworkers\cite{GSST,HiSc} showed the result analytically, both for the 
superconducting and for the density wave transitions.

\par
Within a year, high-T$_c$ superconductivity had been discovered in the cuprates,
and band structure calculations had shown that a VHS is close to the Fermi
level at optimum doping.  A number of papers applied the VHS model to these
materials, starting with Labb\'e and Bok\cite{LB} and Hasegawa and 
Fukuyama\cite{HaF}.  Dzyaloshinskii\cite{Dzy} and
Schulz\cite{Sch} showed that VHS theory was formally very similar to the g-ology
theory of 1D metals, and applied a renormalization group (RG) theory to study 
the competition between CDW's and s-wave superconductivity, {\it and} between 
spin-density waves (SDW's) and d-wave superconductivity!  This RG method will be
discussed in Section XIV, below. 
\par
However, when early band structure calculations failed to find a VHS near the
Fermi level in YBCO, most interest shifted to other models.  (In fact, these
early calculations failed to properly account for chain-plane charge transfer,
and more recent calculations find the VHS close to the doping of optimum T$_c$, 
in excellent agreement with photoemission measurements.)

\subsection{Extended  Saddle-Points}

Photoemission experiments have found evidence for a {\it potentially new kind of
VHS}, initially in YBCO and Y-124\cite{PE2,PE3}, but now in most cuprates.  For 
an ordinary saddle point, the curvature of the energy bands has the opposite
sign in two orthogonal directions. The magnitudes of these curvatures (inverse
effective masses) can take any two independent values.  In an `extended saddle
point', one of these masses is infinite -- that is, the corresponding
curvature lacks a term in $k^2$, and so generically starts out as $k^4$.  Thus,
in one direction, the saddle point is extremely flat.  This makes the energy
dispersion quasi-1D, so the dos diverges more strongly, with a square-root
rather than a logarithmic divergence.  Such behavior can greatly amplify the
effects of a VHS, leading to enhanced T$_c$'s\cite{Ab2}.
\par
However, it is important to recognize that the extended VHS does not follow
automatically from Van Hove's theorem -- it is not necessary that an extended
VHS appear in an energy band.  Indeed, LDA calculations have been unable to
produce any structure in the cuprates which leads to a power-law dos
divergence, and one of the puzzles of VHS theory is to understand where the
extended VHS comes from.  Extended VHS's are discussed further in Section V.D.4.

\section{Fermi Surfaces of the Cuprates}

\subsection{Generic Features of a VHS}

Before presenting the detailed Fermi surfaces of the individual cuprates, it is
convenient to begin with a generic Fermi surface for (the antibonding band of) 
the CuO$_2$ planes.
These planes are generally believed to provide the dominant contribution to 
superconductivity.  Figure~\ref{fig:1} shows the generic shape of this band, 
with Fig.~\ref{fig:1}a showing the dispersion relations, Fig.~\ref{fig:1}b a 
series of Fermi surfaces, Fig.~\ref{fig:1}c the dos, and Fig~\ref{fig:1}d the
integrated number of electrons per unit cell.  The saddle point VHS 
is indicated in the figures; it is the point at which the Fermi surface first 
touches the Brillouin zone boundary.  
\par
Figure~\ref{fig:1} illustrates two different possibilities for the VHS.  If
the Fermi surface is {\it square} at half filling, the VHS nesting is 
supplemented by a conventional nesting, the Fermi surface is perfectly nested,
and the susceptibility divergence is stronger.  This is illustrated by the set
of Fermi surfaces drawn in solid lines in Fig.~\ref{fig:1}.  The dashed lines
illustrate the more usual situation: the Fermi surface is curved at the VHS, so
conventional nesting is weak.  This greatly reduces the possibility of a CDW or
SDW instability, while retaining a strong superconducting instability.
\par
By definition, a VHS occurs at a point where the Fermi surface topology changes.
The saddle-point VHS is characterized by an electron-hole crossover.  Thus, just
below the VHS in Fig.~\ref{fig:1}a, the Fermi surface is a 
closed, electron-like Fermi surface centered at the Brillouin zone center 
($\Gamma$ point); above it, the Fermi surface is again closed, but now it is a 
hole-like Fermi surface centered at the zone corner ($S$ point).  Exactly at the
VHS, the two orbits merge, at the $X$ and $Y$ points of the zone.  The Fermi 
surface is simultaneously electron-like and hole-like.  
\par
This ambiguity and degeneracy are characteristic features of the VHS.  The
fact that there are two degenerate VHS's at $X$ and $Y$ will give rise to a 
novel form of band Jahn-Teller effect.  The electron-hole crossover leads to 
complications in the magnetotransport properties -- Hall effect and
magnetooscillatory phenomena -- as will be discussed below.  

\subsection{LDA and Experimental Fermi surfaces}

A striking result which has come out of the studies of high-T$_c$
superconductivity is to learn just how good the LDA (local density 
approximation) bandstructures can be, for such complicated materials, as long as
correlation effects are not too strong.  Of course, there is Luttinger's
theorem, which exercises a powerful constraint on the result: the {\it area}
of the Fermi surface must enclose the correct number of holes, independently
of any correlation effects.  Nevertheless, there is still considerable
flexibility: the Fermi surface of YBCO consists of four (interacting) sections, 
associated with carriers on several different layers, and there have been
experimental reports that all four sections are close to the calculated
shapes and sizes.  For present purposes, the most important result is that
band structure calculations generally find that the Fermi level is close to the
VHS at optimum doping, not only for LSCO\cite{Xu}, YBCO, and Bi-2212, but also
for the new Hg compounds\cite{NoFree} and the borocarbides\cite{boro}.  
\par
In principle, the Fermi surfaces can be found by a number of probes, such as
de Haas - van Alphen (dHvA) measurements and positron annihilation studies.
However, these probes have had restricted utility in the cuprates.  The dHvA
oscillations are most sensitive to the smallest Fermi surface pockets, and so
far have only found some minority pockets and chain related structure in 
YBCO\cite{dHvA,dHvA1} and Tl-2201\cite{dHvA1}.  
Positron annihilation tends to be sensitive to specific electronic states, 
and so far the technique has only been successful in identifying unambiuously 
the chain Fermi surface in YBCO\cite{Ch0,Ch1,Ch2}.
High-resolution Compton scattering will likely provide new 
opportunities with regard to Fermiology related problems, as the third 
generation light sources come on line\cite{Cosc}.
\par
By far the most direct experimental data on
band dispersion and Fermi surface shape in the cuprates has come from 
angle-resolved photoemission studies.  These studies have found that the Fermi 
level is very close to a VHS at the compositions of optimum T$_c$, both in 
YBCO\cite{PE2} and in Bi$_2$Sr$_2$CaCu$_2$O$_8$ 
(Bi-2212)\cite{PE1}.  In contrast, the Fermi surface in the electron-doped 
cuprates, with much lower $T_c$'s, is much further from the VHS.  So far, no
photoemission results are available for the lanthanates or the Tl or Hg
compounds, due to the difficulty in obtaining high enough quality samples.
Shen and Dessau recently reviewed photoemission studies on the cuprates, with 
related results on a variety of Mott insulators\cite{PE0}.
\par
Figures~\ref{fig:4}-\ref{fig:8} illustrate the variety of Fermi surfaces found 
in the cuprates.  Figure~\ref{fig:4} shows the {\it experimentally derived 
dispersion} of the CuO$_2$
antibonding band(s) for a series of cuprates\cite{PE0}, showing a remarkable 
similarity to one another, and to the generic form of Fig.~\ref{fig:1}a.  
Figures~\ref{fig:5}-\ref{fig:8} illustrate the Fermi surfaces of individual
cuprates in more detail.  These are taken from both LDA calculations and 
photoemission experiments, and are representative of the optimally doped 
compounds.  In all cases studied, the VHS is very close to the Fermi level -- 
always within a phonon energy.  In the following subsections, the various 
materials will be discussed individually.

\subsubsection{Bi-2212}

Bi-2212 is perhaps the best studied material.  Photoemission measurements have
been carried out by a number of groups\cite{Ols1,PE1,Aeb,One1,One3}, and there 
is broad agreement on a number of features of the energy dispersion and Fermi 
surfaces.  Figure~\ref{fig:5} shows a representative collection of the Fermi 
surfaces reported by different groups. (Figure~\ref{fig:5}a is included for 
historical purposes, but Figs.~\ref{fig:5}b and c are believed to be more
representative.)  In particular, there is a large Fermi surface associated with 
the CuO$_2$ planes, with a VHS very close to the Fermi level. By studying the 
transverse dispersion, Ma, et al.\cite{One3} have confirmed that this is an 
extended saddle-point VHS.  Several groups have measured
the superconducting gap directly from the shift of the photoemission edge below 
the Fermi level\cite{Pgap1,Pgap2,Pgap3,Pgap4,Pgap5,Pgap6,Pgap10,Pgrev}.  
It is generally agreed to be highly anisotropic, with a maximum gap in the $\bar
M$ direction, along the Cu-O-Cu direction (i.e., close to the VHS), and nearly 
zero gap at 45$^o$ from this direction\cite{Pgap2,Pgap4}. The gap and its 
symmetry will be discussed in Section VI.D.10.  There are also strong
indications of a shift of spectral weight below the Fermi level, for $T<T_c
$\cite{Pdip,One1}.  A very similar gap is found in Bi-2223\cite{Pgap7}.
\par
In addition to the CuO$_2$ plane bands, a number of particular features in the
band structure have been predicted or identified.  First, the LDA calculations 
predict that a BiO pocket should intersect the Fermi level near $\bar M$, 
thereby coupling strongly to the VHS.  This is not seen experimentally, except
possibly in O-overdoped materials\cite{Ols1,One2}. Note that this is one place 
where inverse photoemission data would be quite welcome. This discrepancy 
probably arises because the real atomic structure of Bi-2212 is more complex 
than that assumed in the LDA calculations, due to superlattice structure
associated with the BiO planes.  There is both a commensurate $\sqrt{2}\times
\sqrt{2}$ orthorhombic supercell, and an incommensurate modulation along
the orthorhombic $b$-axis.  Singh and Pickett\cite{SiPi} have shown that in the
presence of the orthorhombic supercell the BiO pockets are smaller, but should
still be present.  No one has yet calculated the effect of the incommensurate 
modulations on the pockets, but it is plausible to suppose that they might push 
the BiO pocket above the Fermi level, thereby explaining the residual 
discrepancy.  However, there remains a problem: the BiO should form an electron
pocket, and hence if it is shifted off of the Fermi level, it should reappear in
{\it underdoped} material (assuming a rigid band filling).  Yet curiously 
enough, the underdoped material has very low c-axis conductivity, which is 
greatly enhanced in the overdoped material, suggesting that the latter is more 
3D (conducting BiO layer).  One possibility is that this is a charging effect
-- when too many holes are added to the CuO$_2$ planes, additional holes are
repelled, and must populate other planes.  This could also be a VHS effect, 
since the CuO$_2$ plane dos drops rapidly in the overdoped regime (see, for 
example, Ref. \cite{rgpin1}).
\par
The incommensurate modulation leads to additional Fermi surfaces produced by 
zone-folding along the b-direction, which are observed 
experimentally\cite{Pgap3,Pgap4,Pg7,Pg6,Aeb2,Pgap9}, Fig.~\ref{fig:5}c.  
A number of detailed features differ between the various experimental groups.
This may be partly due to differences in oxygen stoichiometry, and a curious
change of the normal-state wave function symmetry has been reported between
optimally-doped and overdoped materials\cite{One2}.  
Aebi, et al.\cite{Aeb} have applied a novel modification of the photoemission
technique to provide a detailed map of the full Fermi surface.  They interpret
their data as two sets of Fermi surfaces, the original surfaces, as found in an
LDA calculation, plus weak structure due to a set of zone folded `ghost' Fermi 
surfaces, associated with incipient superlattice formation, Fig.~\ref{fig:5}b.  
Bansil and Lindroos\cite{BaLi} have pointed out the importance of first 
principle photo-intensity computations for a proper understanding of how Fermi 
surface features manifest themselves in an angle-scanned photoemission 
experiment of the sort reported by Aebi et al.
The ghost Fermi surfaces will be discussed in more detail below.
\par
Photoemission experiments on underdoped Bi-2212 have provided striking evidence
for pseudogap formation, Section IX.A.2.

\subsubsection{YBCO, YBa$_2$Cu$_4$O$_8$}

\paragraph{\bf Extended VHS near $Y$-Point.}
Photoemission studies of the Fermi surfaces of YBCO have also been carried out
by a number of groups\cite{PEY1,PEY2,PE2}, including studies of the variation of
Fermi surfaces with O stoichiometry\cite{Liu,Liu2} and Zn and Co 
substitution\cite{PEY3}.  While there is good agreement between
experiment and LDA calculations at optimal doping, the experiments find much 
less change with doping than expected, as long as the material is 
superconducting; as soon as YBCO becomes insulating, the Fermi surface features 
abruptly disappear (between O$_{6.35}$ and O$_{6.4}$\cite{Liu,Liu2}).  These
results will be discussed in more detail in Section VII.D.3, where they will be
compared to slave boson calculations.  In this section, I will concentrate on
optimally doped samples.
\par 
In YBCO and YBa$_2$Cu$_4$O$_8$ (Y-124), the overall agreement between 
theory\cite{Pick2,OKA,Mass,YMFK,sen}, Fig.~\ref{fig:a5}, and 
experiment\cite{PE3a,PE3,PEY2,Liu}, Fig.~\ref{fig:b5} is quite good.  Both find 
a VHS at optimum doping near the $Y$-point of the Brillouin zone.  While the 
theoretical calculations all find a {\it bifurcated} VHS, the experiments find 
an {\it extended} VHS very close to the Fermi level: within 6meV (resolution 
limited) in YBCO, and $\approx 19$meV below the Fermi level in Y-124, 
Fig.~\ref{fig:c5}.  These two modified VHS's, particularly the extended VHS, can
be important in enhancing the role of the VHS; they will be discussed further in
Section V.D.4.
\par
While 19meV is close enough to the Fermi energy to play an important role (it is
within a phonon energy for most of the O modes), this is actually rather far for
many VHS models of superconductivity to apply.  
Loram, et al.\cite{Gp3} have suggested that the feature identified as the VHS
is actually associated with a {\it pseudogap} -- the separation $\Delta
E=E_F-E_{VHS}$ is quite close to the pseudogap energy.  This is
a very exciting possibility, and would give a strong confirmation of the VHS
theory of pseudogap formation, e.g., Figs.~\ref{fig:c22}d and e below.  A 
thorough discussion of this possibility is presented in Section IX.A.2, below.

\paragraph{\bf VHS Near $X$-Point?}
Due to the orthorhombic distortion in YBCO, the $X$ and $Y$ point VHS's need
not be degenerate in energy.  Unfortunately, the issue of this possible VHS 
splitting is not well
understood.  The region near the $X$-point is very poorly characterized, both 
experimentally (most high resolution experimental results have been presented 
only near the $Y$-point) and theoretically, with considerable 
disagreement between different calculations, as to whether the $X$ point VHS is 
bifurcated or merely broadened by coupling to the chains, Fig.~\ref{fig:a5}.  
Experimentally, Veal, et al.\cite{Veal2} state that this peak is absent along 
the $X$ axis (perhaps above the Fermi level), while Campuzano and 
Gofron\cite{CampG} state that a feature is present, but broadened, perhaps by 
chain-plane coupling.  
\par
Recently, Schabel, et al.\cite{Scha} have reported results on an untwinned
YBCO$_{6.95}$ sample near the X-point.  They find a Fermi level crossing 
similar to that in Bi-2212, but with complications from a strong chain-derived
feature.  By adjusting the polarization to minimize this chain feature, they
found a plane-derived branch with significant dispersion, $\approx 0.5eV$, 
although a second, flatter band might also be present. This dispersion may be 
responsible for the absence of a well-defined quasiparticle peak along the 
$X-S$ line of the Brillouin zone.  More importantly, this is the first YBCO 
sample for which an (anisotropic) gap has been observed, with the largest gap 
along $X-S$.

\paragraph{\bf Complications from the Chains}

An additional complication in YBCO is the chain layer, which can also be
superconducting and can enhance three-dimensional coupling.
The chain band crosses the plane band near the $X$ point, which has led to some
confusion as to whether the band nearest the $X$-point is more plane-like or 
chain-like.  Actually, there is no plane-chain coupling in the $k_z=0$ plane of 
the Brillouin zone, so the bands simply cross at that point\cite{OKA}.  For
arbitrary $k_z$ values, there is a weak hybridization in the immediate vicinity
of the point at which they cross, leading to a band anticrossing effect.  
However, away from this crossing, the 1D band dispersion is predominantly 
associated with chain electrons, on both sides of the crossing, as illustrated 
in Fig.~\ref{fig:a5}.  Hence, the band nearest the 
$X$-point is predominantly plane-like, and there should be a VHS similar to that
at the $Y$-point.  
\par
The plane-chain coupling leads to a larger c-axis dispersion, which 
could smear out the VHS singularity near $X$. While the LDA calculations do find
significant dispersion near $X$, the calculations tend to overestimate
dispersion, particularly for interlayer coupling, so it would be important 
to have experimental data.  
\par
Chain ordering is also the cause of the orthorhombicity in YBCO, Y-124, which
could split the degeneracy of the two VHS's.  For instance, LDA 
calculations\cite{OKA2} in YBCO find both $Y$ and $X$ point VHS's 
below the Fermi level, but the $Y$-point VHS is predicted to be 180-200meV(!) 
below, while the $X$-point is only 0-20meV below (the spreads in values are due 
to c-axis dispersion).  

\subsubsection{NCCO}

Photoemission studies\cite{PEn,PEn2} of Nd$_{2-x}$Ce$_x$CuO$_4$, an 
electron-type superconductor, find a {\it hole-like} and roughly circular Fermi 
surface, Fig.~\ref{fig:6}. It is $\approx 200meV$ 
from the nearest VHS, and the authors have suggested that this may be why $T_c$ 
is so low.  The apparent simplicity of the Fermi surface is deceptive, since the
transport properties imply that the majority carriers are electron-like.  This
puzzling material will be discussed further in Section XV.  The remarkable
similarity of the Fermi surfaces of the various cuprates, is well illustrated by
Fig.~\ref{fig:a6}\cite{PE0}, which compares the Fermi surface of NCCO with those
of the Bi compounds.

\subsubsection{LSCO}

While there are as yet no photoemission studies on this compound, band structure
calculations find the Fermi level coincides with a VHS at the doping of optimum 
$T_c, Fig.~\ref{fig:7}$\cite{Xu}.  Positron annihilation studies of the doping 
dependence of the Fermi surface are consistent with the LDA results, at least
near optimal doping\cite{PASt}.

\subsubsection{Hg-22(n-1)n}

Again, photoemission studies have not yet been successful on these materials.
However, Novikov and Freeman have carried out extensive calculations of the band
structures, as a function of both stage number $n$\cite{NoFree} and pressure 
$p$\cite{NoF2}, comparing them to the infinity phase compounds, Sr$_{1-x}$Ca$_x
$CuO$_2$\cite{NoF3}.  In all cases, they find that the Fermi level is near one
or more VHS's (there are $n$ VHS's in a stage $n$ Hg compound), and that the
calculated doping (away from half filling) which places the first VHS at the
Fermi level corresponds well with the experimental doping of optimum T$_c$. The 
Fermi surface has the same familiar form at the VHS, Fig.~\ref{fig:8}. (It would
be useful to repeat this analysis for the Tl compounds.)
\par
Pressure causes the apical O's to move closer to the CuO$_2$ planes\cite{NoF2}, 
which reduces the Fermi surface curvature, moving the VHS closer to half 
filling.
(This is explained in a tight-binding model in Section V.C.2.)  Now the closer 
the VHS is to half filling, the stronger are the VHS dos and susceptibility
divergences.  Unless a structural instability intervenes, this should lead to a
higher value of T$_c$.  Hence, the LDA calculations provide an explanation of
why pressure can lead to such a large enhancement of T$_c$ (up to an
unprecedented 164K in the 3-Cu-layer compound) and, moreover, in a 
way which cannot be mimicked by simple charge transfer.  It remains to be seen
whether this insight can be used to synthesize a chemical substitution which 
would stabilize these highest T$_c$ compounds at atmospheric pressure.

\subsubsection{Sr$_2$CuO$_2$Cl$_2$}

Many of the cuprate superconductors undergo a Mott-Hubbard transition to an
antiferromagnetic insulating state, when the hole concentration is reduced to
exactly half filling.  So far it has not proven possible to measure 
angle-resolved photoemission from any of these samples.  However, Wells, et
al.\cite{Well} have measured photoemission from the related insulating cuprate,
Sr$_2$CuO$_2$Cl$_2$ (SCOC), Fig.~\ref{fig:a8}a.  Attempts to describe the
results via calculations of a single hole in an AFM background (tt$^{\prime}$J 
or 3-band models) have found difficulty in reproducing the measured energy 
dispersion, Fig.~\ref{fig:a8}b\cite{LMan,DagN,Naz1}.  A good fit is found if the
system is instead assumed to be in a flux phase\cite{Lau1,WeL}, 
Fig.~\ref{fig:a8}a\cite{RM11}. This is discussed further in Section VII.B.
\par
Some compounds related to SCOC become super\-con\-ducting when doped, including 
Sr$_2$CuO$_2$F$_{2+\delta}$ (T$_c$\-=\-46K)\cite{Tc1} and (Ca$_{1-x}$Na$_x$)$_2
$CuO$_2$Cl$_2$ (T$_c$=26K)\cite{Tc2}.  Novikov and coworkers\cite{NoF4} have 
calculated the LDA band structure, and find a prominent VHS upon doping.

\subsubsection{BKBO/BPBO}

A future research area for the VHS is the exploration of the role of the VHS in
3D materials.  Whereas peaks are often found in the dos of 3D metals, they are 
actually associated with a pair of VHS: the $M_1$ and $M_2$ VHS's form the
onset and terminus of a plateau in the dos, and a peak in the dos is really a
narrow plateau.  Formally, such a peak can be formed from a 2D saddle point
VHS, with a small dispersion in the third dimension.  The question is, whether
this is an adequate representation of the dos peaks in real materials.  The
data in this and the following subsections are meant as preliminary hints only,
that the Fermi surfaces of some high-T$_c$ materials are suggestive of such a
picture.
\par
Figures~\ref{fig:c8} and \ref{fig:AB} show the calculated energy dispersion,
dos, and Fermi surface in BaBiO$_3$\cite{MaHa,Hama,ABan}.  Figure~\ref{fig:AB}b
shows a series of cross sections through the Fermi surface, as the Fermi level 
crosses the VHS.  The resemblance to the cuprates, e.g., Fig.~\ref{fig:7} and
Fig.~\ref{fig:4}, is striking, as is the fact that the Fermi level is predicted 
to nearly coincide with the $X$-point VHS. However, this is a fully 3D band 
structure, for the high temperature simple cubic phase, and the Fermi surface,
Fig.~\ref{fig:AB}a, is fully three-dimensional, with the VHS's 
much less in evidence.
\par
In the real material, there is a structural phase transition, presumed to be 
related to a CDW, which produces a $\approx 1.9eV$ gap at the Fermi level. Angle
resolved photoemission\cite{BBOx1} and positron annihilation\cite{BBOx2} studies
have reported Fermi surfaces for the doped, metallic phases which are 
consistent with the calculated surfaces.

\subsubsection{Fullerenes}
\par
Finally, Figure~\ref{fig:d8} shows a `disorder-averaged' Fermi surface for the
fullerene superconductors, M$_3$C$_{60} (M=K,Rb,Cs)$\cite{ME2}.  For perfectly 
ordered buckyballs, the Fermi surfaces depend sensitively on the relative 
orientations of adjacent balls.  However, there is strong orientational 
(merohedral)
disorder in these materials, which broadens structure in the dos.  By averaging
over the disorder, a residual, but smeared Fermi surface is found, which bears a
striking resemblance to that of the cuprates.  In fact, Mele and Erwin\cite{ME2}
have introduced a disorder-averaged virtual-crystal Hamiltonian model for the
bands.  They find that the bands reduce to three nearly independent 2D bands,
which interact by level repulsion near the points where they intersect (the 
three bands differ in the choice of dispersing axes, $x,y$ $\rightarrow$ $x,y$,
$x,z$, or $y,z$).  The
dispersion of the 2D bands is described by Eq.~\ref{eq:4a}, below, with $t_1=0
$, the simplest band structure used to describe the cuprates.

\section{Tight-Binding Fermi Surface Models}

\subsection{1 Band and 3 Band Models}
\par
The LDA or experimental (photoemission) band dispersion can be reproduced by a 
tight-binding model
of the Cu-O$_2$ plane involving three orbitals, Cu $d_{x^2-y^2}$ and one $p$ 
orbital on each of the planar O's (the $p$ orbital with a lobe pointing
towards the Cu).  This is the standard three-band model.  The Hamiltonian is 
\begin{eqnarray}
H=\sum_j\bigl(\Delta d^{\dagger}_jd_j
+\sum_{\hat\delta}t_{CuO}[d^{\dagger}_jp_{j+\hat\delta}+(c.c.)] \nonumber \\
+\sum_{\hat\delta^{\prime}}t_{OO}[p^{\dagger}_{j+\hat\delta}p_{j+\hat\delta^
{\prime}}+(c.c.)] 
+Un_{j\uparrow}n_{j\downarrow}\bigr),
\label{eq:1}
\end{eqnarray}
where the band parameters are $\Delta$, the splitting between the Cu and O 
energy levels, $t_{CuO}$, the Cu-O hopping parameter, and $t_{OO}$, the O-O 
hopping parameter.  Furthermore,  
$d^{\dagger}$ ($p^{\dagger}$) is a creation operator for holes on Cu (O),
j is summed over lattice sites, $\hat\delta$ over nearest neighbors, 
$\hat\delta^{\prime}$ over next-nearest (O-O) neighbors, and c.c.
stands for complex conjugate.  Energies are measured from the center of the
O bands, and $U$ is the on-site Coulomb repulsion.  
In a slave boson calculation, $U$ produces correlations which renormalize the 
one-electron band parameters.  Thus, a simple model is to assume that the 
one-electron parameters are properly renormalized, and to otherwise neglect $U$.
In this case, one need only solve the resulting one-electron Hamiltonian.
The energy bands are the zeroes of 
\begin{equation}
(E-\Delta)(E^2-u^2)-4Et^2(s_x^2+s_y^2)-8ut^2s_xs_y=0
\label{eq:2a}
\end{equation}
with $t=t_{CuO}$, $s_i=sin(k_ia/2)$, $c_i=cos(k_ia/2)$, $i=x,y$,
and $u=4t_{OO}s_xs_y$.  The VHS of the antibonding band is at
\begin{equation}
E_{VHS}={\Delta\over 2}+\sqrt{({\Delta\over 2})^2+4t^2}.
\label{eq:2b}
\end{equation}
For the special case $t_{OO}=0$, the dispersion simplifies.  One
O-band decouples, at $E=0$, the other two bands have dispersion
\begin{equation}
E={\Delta\over 2}\pm\sqrt{({\Delta\over 2})^2+4t^2(s_x^2+s_y^2)}
\label{eq:3}
\end{equation}
and the Fermi surface is exactly square at half filling.
\par
The parameter $t_{OO}$ is important in shifting the VHS away from half filling.
The sign is important: $t_{OO}>0$ shifts the VHS in the direction of larger
hole doping, as found in the cuprates.
\par
In the cuprates, two of these bands are filled, and one, the antibonding band,
is approximately half filled.  For most purposes, it is sufficient to consider 
only this one band.  In this case, it is often simpler in calculations to 
employ a one band model,
\begin{equation}
E^{(1)}=-2t_0(\bar c_x+\bar c_y)-4t_1\bar c_x\bar c_y,
\label{eq:4a}
\end{equation}
where $\bar c_i=cosk_ia$, $t_0$ is a nearest-neighbor hopping energy, and $t_1$ 
a next-nearest-neighbor hopping energy, and
\begin{equation}
E_{VHS}^{(1)}=4t_1.
\label{eq:4b}
\end{equation}
This model is convenient when only the shape of the Fermi surface matters.  
Also, it forms the basis of the tJ and tt$^{\prime}$J models, which include
magnetic correlations but simplify the problem by neglecting the oxygens. The 
parameters of the one and three band models may be interconverted as follows.  
Let $\tau =t_1/t_0$ and $\hat u=4t_{OO}/E_{VHS}$.  Then both models will have 
exactly the same Fermi surface at the VHS whenever
\begin{equation}
\tau ={-y\over 2(1+y)},
\label{eq:5a}
\end{equation}
with $y=\hat u+\hat u^2/2$.  The parameter $t_0$ can then be chosen to yield
approximately the same bandwidth as in the 3-band model.  There is considerable
flexibility in this choice.  One possibility, adopted here, is to make both
models have the same curvature near the VHS [specifically, the same value of
$(\partial s_x^2/\partial E)|_{s_y=0}$].  This yields
\begin{equation}
t_0(1+2\tau)={t^2\over\sqrt{\Delta^2+16t^2}}\equiv{\hbar^2\over 8m_+^*a^2}.
\label{eq:5b}
\end{equation}
\par
While the one-band model can exactly reproduce the shape of the 3-band model
Fermi surface, the overall dispersion is very different in the two models, so
for fitting the angle-resolved photoemission, it is preferable to use the
3-band model.  The difference can be seen by comparing Eqs. \ref{eq:2b} and
\ref{eq:4b}: $E_{VHS}$ is independent of $t_{OO}$ in the 3-band model, but is
proportional to $t_1$ in the one-band case.  A large (negative) $t_1$ pushes the
VHS close to the bottom of the band, so to fit the dispersion below the VHS
(seen in photoemission), a large value of $t_0$ must be assumed, which
exaggerates the dispersion above the VHS.

\subsection{Parabolic Bands}

For analytic calculations in the vicinity of a VHS, it is convenient to
introduce a further simplification by expanding the above dispersion relations 
near the VHS, and retaining only quadratic terms.  This yields
\begin{equation}
E-E_{VHS}={\hbar^2\over 2}({k_y^2\over m_-^*}-{k_x^2\over m_+^*}),
\label{eq:6a}
\end{equation}
with 
\begin{equation}
{\hbar^2\over 8m_{\pm}^*a^2}=t_0(1\pm 2\tau ).
\label{eq:6b}
\end{equation}
The second VHS has $k_x$ and $k_y$ interchanged.
\par
This can be further simplified to
\begin{equation}
E-E_{VHS}={2\hbar^2k_x^{\prime}k_y^{\prime}\over m^*},
\label{eq:7a}
\end{equation}
with $k_y^{\prime}=k_y\sqrt{m^*/m_-^*}$, $k_x^{\prime}=k_x\sqrt{m^*/m_+^*}$,
and $m^*=\sqrt{m_-^*m_+^*}$, or
\begin{equation}
{\hbar^2\over 8m^*a^2}=t_0(\sqrt{1-4\tau^2}).
\label{eq:7b}
\end{equation}
Note that the scaling of the $k^{\prime}$'s would involve the opposite effective
masses for the second VHS, so Eq.~\ref{eq:7a} cannot be used to 
describe inter-VHS scattering. However, it affords a simple approximation for 
calculating intra-VHS properties.

\subsection{Parameters for the 3 Band Model}

\subsubsection{LSCO: Tight Binding Parameters}

It is widely believed that the essential physics of the cuprates should be
contained within the three band model of the CuO$_2$ planes.  This belief is
far more general that the VHS scenario.  In Anderson's model\cite{AnCh}, for 
example, the CuO$_2$ plane bands must be supplemented by an interlayer coupling
contribution, but the three-band model still plays an important role.  It
is thus essential to know the `correct' values to assign to the parameters.
The best procedure seems to be to derive parameter values from the LDA
band calculations.  
\par
However, extreme care must be taken in extracting the parameters, and it is
by no means clear whether the best values have yet been found, even for the 
simplest case of LSCO.  In this section, I will discuss some of the problems,
and provide a Table of recommended values for LSCO, with suggestions of how
these values might change in the other cuprates.
\par
The simplest procedure is to directly fit the LDA-derived Fermi surfaces with a
tight-binding model.  Unfortunately, the LDA calculations underestimate the role
of correlation effects, particularly near half filling, where LDA calculations 
do not reproduce the transition to the insulating state.  For optimal doping, 
correlation problems are less severe, and a simple tight-binding fit might be 
adequate.  Table I lists tight-binding fits\cite{Xu,YFree,OKA2} to the Fermi 
surfaces for a number of cuprates in the {\it one-band model}, Eq.~\ref{eq:4a}. 
Andersen, et al.\cite{OKA2} also include values for more distant neighbor 
hopping.  Table II lists the corresponding values for the {\it three-band 
model}, Eq.~\ref{eq:1}.  Three sets of values are listed in Table II. Those for
LSCO and YBCO are derived from the values in Table I, using 
Eqs.~\ref{eq:5a} and \ref{eq:5b}. This inversion is not unique; however, it
is found that $\Delta$ is small, so $\Delta =0$ is assumed.  The values listed 
for Bi-2212 are the renormalized (via slave boson calculation) parameters which
reproduce the photoemission-derived band dispersion of Bi-2212. 
\par
In the one band model, it is important to know how the doping $x_c$, at which  
the Fermi level coincides with the VHS, varies with the parameter $\tau$. 
This relation is derived in Appendix B.
\par
Many physical properties are given by integrals over the band structure.  For
example, $\mu$SR results are typically calibrated in terms of $\lambda^{-2}
\propto n/m$, where $\lambda$ is the penetration depth, and the factor $n/m$ is
derived from a parabolic-band model.  What is actually measured is the plasma
frequency, which cannot simply represent $n/m$, since it must vanish for a
full or empty band.  It can be written, for a 2D band, as 
\begin{equation}
\omega_{pl,x}^2=({e\over\pi\hbar})^2{2c\over\pi}\int d^2k{\partial^2E\over
\partial k_x^2}.
\label{eq:7c}
\end{equation}
For the one-band model, this is illustrated in Fig.~\ref{fig:b8}.  Note that 
$\omega_{pl}$ has a peak near half filling, but not exactly at the VHS.  This
could help explain the anomalous doping dependence of T$_c$ in overdoped
cuprates (the `boomerang' effect), Section VI.D.5.

\subsubsection{LSCO: Correlation Corrections}

\par
More care must be applied if one wishes to include correlation effects in some
approximation beyond the LDA.  That is because the LDA-derived parameters
already include correlation effects in an average way. These correlation
effects must first be eliminated to produce bare parameters.  A striking
example is the transition to a charge-transfer insulator at half filling.  This
transition occurs when $\Delta /t_{CuO}$ exceeds a critical value.  In LDA
calculations, correlations reduce the value of $\Delta /t_{CuO}$ so much that a 
correlation calculation, using these starting values, will find a metallic state
at half filling. Hence, it is essential to first find the bare, unrenormalized 
three-band parameters before undertaking a correlation calculation.
For LSCO, several groups have provided estimates of these bare parameters, Table
III, using LDA or cluster calculations\cite{Hyb,GraM,Esk}.  Earlier calculations
are summarized in Ref. \cite{MAM}.
\par
However, even these parameters require further modification: they
have been determined as part of a larger parameter set.  When this larger
parameter set is further reduced to a three band model, the remaining parameters
have to serve double duty, and the best choice of parameters will be modified 
from the values found in the larger parameter set.  Two examples will be given.
\par
First, at half filling, a hole transferred from a Cu to an O will almost always
find that there are already holes on each of the surrounding Cu's.  Hence,
it will have an energy $\Delta+2V$, where $V$ is the nearest neighbor Coulomb
repulsion\cite{BARA}.  Away from half filling, this is modified to\cite{Rai}
\begin{equation}
\Delta_{eff}=\Delta+2V(1-x-4r_0^2) ,
\label{eq:8}
\end{equation}
where the term in $r_0$, the mean field amplitude of the slave boson, can play a
role in phase separation, as will be discussed further below.  Neglecting this 
term for now, the constant term $2V(1-x)$ provides a (doping-dependent) 
renormalization for $\Delta$, for which the $x=0$ value is listed in Table III.
In a three-band model, $\Delta_{eff}$ will provide the best single parameter
approximation of the energy difference between adding a hole to a Cu or to an O.
\par
As a second example, it is essential in the VHS model to correctly predict the
doping at which the Fermi level coincides with the VHS; within the model, this
is also the doping of optimum $T_c$.  Now, in the full LDA calculation, this
doping is fixed by a rich interplay of many band parameters\cite{LZAM}, whereas 
in the three-band model, it is entirely fixed by a single parameter, $t_{OO}$.
Hence, in a strictly three-band model, the value chosen for $t_{OO}$ does not
coincide with the LDA-derived value of this parameter, but must be taken as an
effective parameter which correctly reproduces the location of the VHS.
It appears that this value will be considerably smaller
than the LDA value.  Thus, Aligia\cite{AAA} has shown that the Fermi surface
curvature is controlled by two opposing effects, the direct O-O hopping and the 
Cu $d_{x^2-y^2}-d_{z^2}$ mixing term, which acts to reduce the curvature.  This
problem has recently been studied in detail\cite{FeJR}.
\par
In Ref. \cite{RMXA}, I attempted to derive an effective value of $t_{OO}$ from 
a five-band model of the CuO$_2$ planes, including the Cu $d_{z^2}$ and apical O
orbitals\cite{FGD}.  I found a reduction by more than a factor of two, but due
to uncertainties in the parameter values, I do not believe that the final 
results are quantitatively correct.  Hence, I suggest that at present, the
best way to choose the value for $t_{OO}$ is empirical: to adjust the value of
$t_{OO}$ to reproduce the LDA-value of Fermi surface curvature at optimum
doping.  This doping is chosen since correlation effects are smaller away from
half filling, and the LDA calculations seem to do a good job of reproducing
the Fermi surfaces.  Clearly, a full derivation of the appropriate $t_{OO}$
is a desideratum, but at present, this seems to be the best available choice.
The resulting corrected parameter values are listed in Table IV.

\subsubsection{Other Cuprates}

Unfortunately, such `bare' parameters are available only for LSCO, whereas it
is known that the correlated parameters differ significantly for different
cuprates.  
\par
For other superconductors, there is considerably less information.  Only for
YBCO are estimates available\cite{sen,OKA2}, Table III, and these have not been 
corrected for correlation effects.  In analyzing
other superconductors, two factors enter: first, the values of the three
primary parameters may be different, and secondly, additional parameters may
be important.  These issues will be addressed consecutively.  First, since
the dominant physics lies within the CuO$_2$ planes, the parameter values would 
be expected to be fairly similar.  However, small changes can be important.
For example, $t_{OO}$ seems to be larger in YBCO than in LSCO.  This means
that the VHS is shifted further away from half filling in the former compound.
Correspondingly, the optimum doping is about 0.15 holes/Cu in LSCO and 0.25
holes/(planar) Cu in YBCO.  In both cases, the optimum doping brings the Fermi 
level into registry with the VHS.  (In a recent reanalysis of the 
data\cite{Tall1}, it has been suggested that the optimum doping is the same for 
all cuprates, but this result remains controversial.  It is further discussed in
Section VI.D.5.)  A second example: the value of $\Delta$ in
LSCO is only a little larger than the critical value $\Delta_c$ needed to drive 
the transition to a charge transfer insulator at half filling.  In the Bi- and 
Tl-based cuprates, it is much harder to find evidence of this insulating phase.
Could this be because for these compounds $\Delta<\Delta_c$?
\par
Physically, the parameter $t_{CuO}$ should depend sensitively on the 
Cu-O separation, $d_{CuO}$, but otherwise may not be sensitive to the
chemical environment off of the planes.  On the other hand, $\Delta$ depends
on the Madelung energy at the Cu and O sites, and hence can differ in
different compounds.  Similarly, since $t_{OO}$ is modified by additional,
off-planar orbitals (particularly the apical O's, which influence the Cu
$d_{z^2}$'s), it can also vary from compound to compound.  Ideally, the same
detailed procedure should be followed to extract the corrected
parameters for all these compounds from the LDA calculations, but in the short 
term, the simplest procedure is to do a simple tight-binding fit to the LDA
data, and then use this to `adjust' the corrected values found for LSCO.
\par
An additional problem is that, for other compounds, parameters beyond the 
three-band model can be important. A number of examples will be provided in the 
following subsections.

\subsection{Beyond the 3 Band Model}

\subsubsection{Several CuO$_2$ Layers per Unit Cell}

\par
The cuprates behave two-dimensionally because they are layered compounds, and 
one of the layers is nearly insulating, so that hopping across that layer is 
extremely improbable.  These insulating layers
break the material up into cells, such that there can still be strong
interlayer coupling within a cell, even though there is very little coupling
between cells.  While the intercell coupling eliminates the dos divergence at
a VHS, intracell coupling does not: it only splits the VHS into several 
components, each of which diverges logarithmically.  In the Bi-compounds, this
intracell hopping is so small that it can approximately be neglected.  In YBCO,
it is quite large, leading to two well-resolved Fermi surfaces, corresponding
to symmetric and antisymmetric combinations of holes on the two planes.  
Liechtenstein, et al.\cite{LGAM} have shown that correlation effects can
reduce the intercell splitting by nearly an order of magnitude.  These
two Fermi surfaces appear to have been resolved in photoemission studies 
(Fig.~\ref{fig:b5}).  Interband coupling can be accounted for by adding a
contribution due to hopping along the c-axis.  However, the angular dependence 
of this hopping term is in dispute.  To describe the calculated and measured
dispersion in YBCO, I have simply added or subtracted a constant term in the 
energy dispersion:
\begin{equation}
E^{(1)}=-2t_0(\bar c_x+\bar c_y)-4t_1\bar c_x\bar c_y\pm t_z,
\label{eq:9}
\end{equation}
where $t_z$ is given in Table I.  In contrast, Chakravarty, et al.\cite{AnCh} 
have proposed that the $t_z$ term is proportional to $(c_x-c_y)^2$, which would
make the two bands degenerate along the diagonals of the Brillouin zone, in 
contrast to experiments on YBCO, Fig.~\ref{fig:b5}b, although it may hold for 
Bi-2212.  This issue is discussed in Ref.  \cite{OKA2}.  For present purposes,
the precise shape of the Fermi surface is not crucial.  What matters is that, by
Including a $t_c$ term, it is found that the 
antisymmetric Fermi surface is considerably larger than the symmetric
combination.  Indeed, virtually all of the doped holes go into the 
antisymmetric band, leaving the symmetric band half filled.  In optimally doped 
YBCO, it is the antisymmetric band whose Fermi surface coincides with the VHS.

\subsubsection{Intercell Coupling and c-axis Resistivity}

\par
Since the intercell coupling cuts off the VHS divergence, it can lead to an
important restriction on VHS theory.  However, the superconducting transition
depends only on the {\it integrated dos}, within $\sim 4k_BT_c$ of the Fermi
level.  Thus, if the splitting due to interlayer coupling, $\sim 4t_z$ is
smaller than this, the broadening of the dos peak will have no effect.  While
the calculated values of $t_z$ are already small, there is evidence that
correlation effects reduce its importance further, and a strong intracell 
confinement has been predicted by the spinon-holon theory\cite{WHA,AnCh}.  This 
same factor is responsible for coherent interlayer conductivity, $\sigma_c$.  
Yu and Freeman\cite{YF} have found that the calculated values for $\sigma_c$ in 
YBCO exceed the experimental values by over an order of magnitude, and suggest 
that polaronic effects may be responsible for the difference.  Correlation
effects may be more important for interlayer coupling -- they can reduce
intracell interlayer hopping by nearly an order of magnitude\cite{LGAM}.  From 
measurements of the optical conductivity of LSCO, Tamasaku, et al.\cite{Tim3} 
find that the effective number of electrons contributing to $\sigma_c$ is orders
of magnitude smaller than that expected from the theoretical plasma frequency.
Indeed, it has been possible to understand the c-axis conductivity of most
cuprates below T$_c$ as due solely to Josephson coupling between the 
superconducting layers\cite{Tim2,Tim6}. Tallon, et al.\cite{Tall11} were able to
experimentally vary the interlayer coupling in YBCO by simultaneously doping Ca 
for Y and removing O, to disrupt the chains while staying at optimal plane 
doping.  They found that $T_c$ actually decreases slightly as the interlayer 
coupling decreases.  This result only means that even for maximal chain 
coupling, the interlayer hopping is too small to reduce $T_c$ (the residual
decrease in $T_c$ is presumably due to a crossover to a 2D Kosterlitz-Thouless
type superconductivity).

\subsubsection{Coupling to Other Layers}
\par
The Fermi surface at the VHS illustrated in Fig.~\ref{fig:1} is the form most 
commonly found for the cuprates, but it is not the only possible form.  When two
Fermi surface sections approach closely in energy, they can hybridize, giving 
rise to new Fermi surfaces of strikingly different shape.  For instance, LDA 
calculations of Bi-2212 generally find a Bi pocket just beginning to open near 
the $X$($Y$) points of Fig.~\ref{fig:1} (in the orthorhombic unit cell of 
Bi-2212, this is referred to as the $\bar M$ point).  Since this point coincides
with the CuO$_2$ VHS's, the interaction is very strong.  The resulting Fermi
surface shapes have been discussed\cite{RM6}.  The VHS's are not eliminated,
but are split and shifted in energy.  (Van Hove proved that there must
always be at least one saddle point VHS for any 2D electronic band.)

While the LDA calculations find BiO hole pockets crossing the Fermi level, 
they are not seen experimentally in optimally doped samples, as discussed in
Section IV.B.1.  However, they should be present just above the Fermi level, and
hence will modify the dispersion. Also, doping away from the VHS may push them 
through the Fermi level.  

In addition to distorting the shape of the Fermi surface, these additional
layers, if metallic, can also be rendered superconducting via proximity
effects.  This has been proposed for both the chain layer in YBCO and the
Tl layers in the 2-Tl-layer Tl-cuprates\cite{Tl}.  In YBCO, there is evidence
that the chains are superconducting (at optimum doping), but do not greatly
modify the value of T$_c$\cite{Chsc}.  There is an interesting 
complication in YBCO: the chains are very sensitive to oxygen stoichiometry, 
since the O vacancies lie in the chains, and O vacancy disorder quickly disrupts
chain conductivity.  This strong disorder effect can lead to {\it gapless
superconductivity} on the chains, with anomalies in many of the
superconducting properties\cite{Klei}.

\subsubsection{Extended and Bifurcated VHS's}

\par
Just as the BiO pocket can push the VHS's away from the Brillouin zone corners
in Bi-2212, interlayer coupling in YBCO is predicted to lower the band energy
at the $X$ (and $Y$) points, leading to a camelback band structure.  The energy
has a local maximum along the $\Gamma -X$ ($\Gamma -Y$) line, and the change in 
topology as the energy moves through this maximum leads to a pair of VHS's, 
degenerate in energy, Fig.~\ref{fig:b5}c.  Andersen, et al.\cite{sen,OKA2} have 
provided a detailed calculation, explaining the origin of these {\it bifurcated 
VHS's}. The bifurcation comes about through a small admixture of the three bands
(plus Cu $4s$) with a group of c-axis directed $\pi$-orbitals, including the 
planar O $p_z$ orbitals.  This mixing becomes allowed due to the dimpling (or 
tilting) of the CuO$_2$ planes.
\par
Experiments on YBCO and Y-124 have not found these bifurcated VHS's, but
instead the potentially more interesting {\it extended VHS's}\cite{Ab2}:
compare the theoretical and experimental band dispersions found in Fig.~
\ref{fig:b5}c. In an extended VHS, the saddle point extends over a finite length
of Fermi surface, enhancing the associated dos singularity.  There does seem to 
be a connection between the two types of VHS, in that the
length of the extended VHS is approximately the separation between the two
bifurcated VHS's.  However, the bifurcated VHS's are not equivalent to the 
extended VHS's: at a bifurcated VHS, the dos always has a logarithmic 
divergence, even at the point of bifurcation, and indeed it is difficult, in 
looking only at the dos, to tell when the bifurcation has occured.  Correlation
effects can reduce the dispersion, leading to a stronger logarithmic
divergence -- this is particularly true for interlayer correlations\cite{LGAM}.
However, within slave boson theory, correlations will not change the
divergence to the power law form, which has been postulated for an extended 
VHS\cite{Ab2}.
\par
An interesting question has to do with (inter-VHS) nesting at an extended
saddle point.  For the bifurcated VHS's there is very strong inter-VHS
scattering\cite{OKA}.  Since the VHS's are shifted off of the $X$ and $Y$ points
of the Brillouin zone, the inter-VHS scattering is very far from $Q_0=(\pi /a,
\pi /a)$ -- for example, see the arrow in Fig.~\ref{fig:b5}c.  For an extended 
VHS, the $Y$-point VHS is stretched along $\Gamma -Y-\Gamma$, but still centered
at $Y$; the $X$-point VHS would be stretched at right angles to this.  The 
resulting inter-VHS scattering should be strong over an extended region, {\it 
centered at $Q_0$}.  This difference between bifurcated and extended VHS's 
should be reflected in the spin susceptibility $\chi$: whereas inter-VHS
scattering from bifurcated VHS's should lead to well-defined incommensurate 
peaks in $\chi$, extended VHS's would lead to a broad peak in the spin 
susceptibility at $Q_0$, the AFM point.  Recent experimental evidence on this
point will be discussed in Section VI.E.3 (Fig. \ref{fig:15}b).
\par
It would be useful to generalize the three band model, to incorporate bifurcated
saddle points, and then study how these are modified by correlation effects.
Andersen, et al.\cite{sen} have attempted to produce a set of tight-binding
parameters which do just this.  Unfortunately, the effects of correlations have 
not been removed from the parameters, so, just as in LSCO, these parameters 
cannot be used in slave boson calculations.  Thus, when the tight-binding 
parameters are reduced to a three band model (Table III), the Cu-O splitting is 
found to be $\Delta =3.0eV$. This is better than the situation in LSCO, where
neglect of correlations led to estimates of $\Delta\approx 1.4eV$\cite{nocor}, 
but
$\Delta /t_{CuO}\simeq 1.9$ is still too small to lead to an insulating state at
half filling -- the critical value of this ratio is 3.353, Eq.~\ref{eq:16a}.  
Since the balance of parameters is so delicate, simply increasing
the value of $\Delta$ is found to move the camelback away from the Fermi 
surface.  For more extensive parameter sets, which reproduce the bifurcated
VHS's, the original references\cite{sen,OKA2} should be consulted.  (Note 
further that when the model is reduced to three band or one band form, 
additional interactions arise which are not included in Eqs.~\ref{eq:1} and
\ref{eq:4a}.)
\par
The microscopic origin of the extended VHS's remains unclear, although there
are a number of theoretical possibilities.  One can generate extremely flat 
bands by renormalizing the band dispersion near a VHS by, e.g., strong 
electron-phonon coupling\cite{RMXB}, as discussed in Section VIII.D (see 
Fig.~\ref{fig:18}) or in the presence of a (magnetic) pseudogap\cite{DagN}.  
However, in both these examples, the actual divergence remains logarithmic,
and the extended VHS arises from band narrowing.  If one actually wants the dos 
to change from a logarithmic to a square-root divergence, this can be done only 
by making the dispersion quasi-1D -- e.g., formally, by making the Cu-O hopping 
parameter anisotropic, $t_x<<t_y$.  These two situations can be readily 
distinguished: as the bands become more one-dimensional, the VHS moves closer to
a band extremum -- either the top or bottom of the band.  Note that this is 
approximately what happens when a gap is opened at the VHS, Fig.~\ref{fig:c22}. 
The careful study of the dispersion near the VHS in Y-124, where the VHS is 
19meV below the Fermi level, would appear to place severe limitations on any 
quasi-1D model of the extended VHS.

\subsubsection{Chain Ordering}  

\par
In YBCO, the CuO chains are also conducting, and must be included in any 
quantitative model.  In addition to proximity-effect superconductivity, 
discussed above, chains can play a number of roles.
First, the chain-ordering transition breaks the symmetry of the $X$ and $Y$
points, and can split the degeneracy of the two VHS's.  This can affect the
doping dependence of properties, structural instabilities, and the magnetic
susceptibility at $(\pi ,\pi )$, important for antiferromagnetism.
Unfortunately, the structure near the $X$ point is one of the features in
which different LDA calculations are most in disagreement, and more 
photoemission studies should be made near the $X$ point. Since the chains 
conduct only parallel to their lengths, along the b-direction, they greatly 
enhance the in-plane anisotropy of the normal-state transport properties.

\subsubsection{Ghost Fermi Surfaces}  

\par
There is considerable experimental and theoretical evidence for the existence of
`pseudogaps' (Section IX.A) -- an incipient gapping
of (part of) the Fermi surfaces of these materials.  This may be caused by
magnetic effects (incipient antiferromagnetism) or by structural effects.  In
either case, the gap formation could be associated with an increase in the
size of the in-plane unit cell.  If so, this would lead to the appearence of 
additional bands in the Brillouin zone, by a process of zone folding.  
The gaps are not fully formed, but some residual trace is expected to appear
in the photoemission, in the form of `ghost bands' with anomalously
large broadening.  Such bands are seen in Bi-2212
in photoemission, where they were attributed to short range antiferromagnetic 
correlations\cite{Aeb}, Fig.~\ref{fig:5}b.  However, there is no evidence for 
long-range antiferromagnetic order in Bi-2212, except by substituting Y for 
Ca\cite{BiAF}. It seems more probable that these ghost surfaces are associated 
with the known orthorhombic distortion of the Bi-2212 structure\cite{SiPi}: the 
pseudogap transition temperature is generally found to be close to the 
superconducting $T_c$ in optimally doped 90K materials, whereas the ghost Fermi 
surfaces persist up to room temperature in Bi-2212.  In other cuprates, the 
ghost Fermi surfaces may be too smeared out to observe.

\subsubsection{Spin-Orbit Coupling}

\par
Finally, spin-orbit coupling can play an important role in modifying the shape 
of Fermi surfaces.  Thus, in LSCO, the LTO phase transition leads to a doubling
of the unit cell, hence to a reduced Brillouin zone, with consequent zone 
folding.  However, because the unit cell contains a glide plane, the energy 
bands can cross without opening a gap on the Brillouin zone face.  (Because
of this, the presence of the orthorhombic supercell can often be neglected, as 
in Fig.~\ref{fig:1}, above.)  However, spin-orbit scattering leads to a coupling
between these bands, opening gaps and strongly modifying the topology of the 
Fermi surface\cite{RMXA}.  Since spin-orbit coupling can split the VHS 
degeneracy, it was originally thought that it might play an important role in 
structural phase transitions, but the energy involved seems to be too small.
Nevertheless, this coupling will have a significant
influence on the exact Fermi surface shape, and hence on low-energy transport 
properties, such as Hall effect and magnetooscillatory phenomena.  Spin-orbit
coupling also plays an important role in the magnetic properties, opening a
gap in the spin-wave spectrum and contributing to the 3D coupling which gives
rise to a finite Curie temperature for the antiferromagnetic transition.

\section{Properties of a Simple VHS Theory}

\subsection{Divergences in dos and Susceptibility}

While the dos and the charge and spin susceptibilities of the VHS model must
generally be calculated numerically, some analytical results are possible in
the one band and parabolic models.  In general, the carrier density is given
by the area of the Fermi surface,
\begin{equation}
n={2n_{Cu}\over\pi^2}\int_0^{\pi}\phi_xd\phi_y,
\label{eq:27}
\end{equation}
where $\phi_i=k_ia$, $i=x,y$, and $n_{Cu}$ is the density of the planar Cu 
atoms.  The dos is then
\begin{equation}
N(E)={2n_{Cu}\over\pi^2}\int_0^{\pi}{\partial\phi_x\over\partial E}d\phi_y.
\label{eq:28}
\end{equation}
For the one-band model, in the special case $t_1=0$, this can be evaluated
exactly
\begin{eqnarray}
N(E)={n_{Cu}\over\pi^2t_0}\int_0^{\pi}{d\phi_y\over\bar s_x}
={n_{Cu}\over\pi^2t_0}K\Bigl(\sqrt{1-({E\over 4t_0})^2}\Bigr) \nonumber \\
\simeq{n_{Cu}\over\pi^2t_0}ln({16t_0\over E}).
\label{eq:29}
\end{eqnarray}
\par
The dos can also be evaluated exactly in the parabolic model, 
\begin{equation}
N(E)=2{m^*a^2\over 2\hbar^2k_0^2}ln({2\hbar^2k_0^2\over m^*a^2E})
\label{eq:30}
\end{equation}
where $k_0$ is a momentum cutoff and the first factor of two on the right-hand 
side comes from the fact that there are two VHS's.  Using the definition of 
$m^*$, Eq.~\ref{eq:7b}, the prefactor of the parabolic model will agree with
that of the one band model, $\tau =0$, if $k_0=\pi /2\sqrt{2}$.
\par
The bare static susceptibility, $\chi_0(\vec q)=\chi(\vec q,\omega =0)$, Eq.~
\ref{eq:02}, diverges at both $\vec q=0$, where it is equal to the dos, and
at $\vec q=\vec Q_0$, when the VHS is at the Fermi level.  This latter
divergence is due to inter-VHS scattering, and is sensitive to the shape of
the Fermi level.  In particular, it has a $ln^2$ divergence when the Fermi
surface is square (perfect nesting), which is cut off to $ln$ when the
Fermi surface is curved (e.g., $t_{OO}\ne 0$ in the three band 
model)\cite{RM5,Patt}.  The origin of this cutoff can be understood by 
linearizing the Fermi surface about the VHS, as in Fig.~\ref{fig:a9}.
Near the $X$-point VHS, $\phi_x,\phi_y$ are defined as $k_xa=\pi -2\phi_x$,
$k_ya=2\phi_y$, and the denominator of $\chi (\vec Q_0)$ becomes $E_{\vec k+\vec
Q_0}-E_{\vec k}=2t(\phi_x^2-\phi_y^2)$, so $\chi$ is proportional to
\begin{equation}
\Gamma (A)=\int_0^{\phi_m}d\phi_x\int_0^{A\phi_x}d\phi_y[{1\over\phi_x^2-\phi_y
^2}],
\label{eq:a30}
\end{equation}
\begin{equation}
\Gamma (A)=\cases{{1\over 2}ln({1+A\over 1-A})ln({\phi_m\over\phi_c})&if $A<1$
\cr
             {1\over 2}ln^2({\phi_m\over\phi_c})&if $A=1$}
\label{eq:b30}
\end{equation}
with $\phi_m\approx\pi /4$ a large angle cutoff, $\phi_c\approx max(T,\omega )$ 
a small angle cutoff, $A=tan\alpha$, and $\alpha$ defined in Fig.~\ref{fig:a9}.

\subsection{Estimating Hole Density}

Let $P$ designate the excess hole doping per CuO$_2$ plane of a cuprate away 
from half filling.  Experimentally, it can be measured by iodometric titration.
In many cuprates, $P$ can be estimated from the chemical valence.  Thus, in
LSCO and LBCO, $P=x$, the number of dopant Sr's or Ba's.  When the doping is
related to added O's, $P$ can be estimated from O content, measured, e.g., by
thermogravimetric analysis (TGA), but the relation between $P$ and hole doping 
must be known.  Thus, for YBCO, there is evidence that, when O's are first
added to the chains ($y\equiv 1-\delta\simeq 0$), the holes remain localized on
the chains, up to a critical doping $y_0$.  Tokura, et al.\cite{Tok} proposed
that $y_0=0.5$, and $P=0.25(y-y_0)$ for larger dopings.  However, there is
evidence from studying modifications to the N\'eel temperature, that plane 
doping starts at $y_0\simeq 0.2$\cite{plans,R-M2}.  More recently, Tallon, et
al.\cite{Tall1} have proposed $P=0.187-0.21\delta$, for $\delta\le 0.55$, 
followed by a more rapid drop for larger $\delta$, and $P\approx 0$ for $\delta
\ge 0.7$.
\par
In La$_2$CuO$_{4+\delta}$, there is considerable uncertainty about the relation
between $P$ and $\delta$ (note that in La$_2$CuO$_{4+\delta}$, adding oxygen 
increases $\delta$, whereas adding oxygen decreases $\delta$ in YBCO).
Grenier and coworkers\cite{GrenX} found a large discrepancy between TGA and 
iodometric results, which they could correlate with excess O adsorbed on 
nonbonding sites from the electrolytic solution.  Correcting for the adsorbed
fraction, they found that the two measurements would be reconciled if each O 
contributes two holes to the CuO$_2$ plane, or $P=2\delta$.  On the other hand, 
Radaelli, et al.\cite{Rad2} found better agreement with the relation $P=\delta$,
which would suggest that the interstitial O's have a partially covalent bonding
with the apical O's, forming the peroxide $O_2^{2-}$.  In more recent work,
Grenier\cite{Gren3} suggested that he may have overestimated $\delta$.  Chou, et
al.\cite{Chou} find an intermediate result, that the susceptibility of La$_2
$CuO$_{4+\delta}$ matches that of optimally doped LSCO ($x=0.15$) for $\delta 
=0.11$.  More recently, Johnston\cite{DCJ} found a crossover
\begin{equation}
P=\cases{\delta&if $\delta\le\delta_0$;\cr
         \delta_0+2(\delta -\delta_0)&if $\delta >\delta_0$,\cr}
\label{eq:30b}
\end{equation}
with $\delta_0\simeq 0.06$, close to the upper limit of the two-phase region.
In this paper, I will use this last result, but generally quote the doping in
terms of $\delta$ (note however that Grenier's group
usually measures $P$ directly by iodometric titration, and converts that into an
effective $\delta$ by the relation $\delta =P/2$).  Even lower values of hole
doping -- $P\approx 1.1\delta$ at $\delta =0.12$ -- see Ref. \cite{Qui} and
references therein.

\subsection{Normal State}

\par
There is considerable evidence that the normal state of the cuprates is highly
anomalous, particularly in the immediate vicinity of optimum doping.  This has 
led to the notion of a marginal Fermi liquid\cite{MFL,MFL2}, where the 
renormalization factor $Z$ vanishes weakly (logarithmically) at the Fermi 
energy, so that quasiparticles are not quite well defined.  In this section, it 
will be shown that most of the anomalous normal-state features can be accounted 
for in the context of the VHS model.
\par
In calculating the properties of the VHS model, a number of complications must
be taken into consideration.  First, in studying the doping dependence, it is
not correct to assume a tight-binding model, since correlation effects pin the
Fermi level close to the VHS over an extended doping range, and the dos may
actually increase with underdoping.
\par
More importantly, it must be recognized that the Van Hove scenario can actually
be subdivided into two classes, the {\it basic} and the {\it generalized}.  In 
the {\it basic} scenario, one analyzes the role of a peak in the dos on 
superconductivity and the normal state properties, taking correlation effects 
into account, but neglecting any complications from structural instabilities or
phase separation.  In the {\it generalized} scenario, one notes that there are
competing instabilities in the vicinity of a VHS, and that these instabilities
can profoundly modify physical properties in the vicinity of a VHS.  For 
instance, when a pseudogap opens up, the splitting of the VHS peak will drive 
the large dos away from the Fermi level.
\par
In this section, the predictions of the {\it basic} scenario will be worked out
and compared to experiment.  The conclusion of this exercise is that while
there are some tantalizing correspondences, there are enough differences of 
detail to suggest that the basic scenario is too simplified to completely
describe the cuprates.  In particular, recent heat capacity and photoemission 
studies show that the pseudogap strongly modifies the properties of the 
underdoped cuprates in ways that are inconsistent with the basic scenario, but
that have actually been predicted with the generalized scenario.
\par
In subsequent sections, the complications of the {\it generalized} scenario will
be discussed. These are of two types: (a) strong electron-phonon coupling, which
leads to pseudogaps; and (b) a tendency to (nanoscale) phase separation, which
produces striped phases.  In principle, these are two independent phenomena,
and it is possible that some materials may display one without the other 
(although, within the VHS scenario, the strong electron-phonon coupling
stabilizes phase separation).  However, in the cuprates, both phenomena appear
to coexist, and the greatest challenge lies in understanding how both can be
simultaneously present.

\subsubsection{Thermodynamics}

\paragraph{\bf Peak in $\Delta C$.}
The peak in the dos associated with the VHS should be directly observed in
a number of measurements, such as heat capacity, compressibility, and magnetic 
susceptibility.  These and other thermodynamic properties of a VHS are 
calculated in Appendix A.  Here, the simple VHS model has difficulty in
describing the data, since these are dominated by effects of pseudogap formation
and/or phase separation.  For example, in studying the heat capacity, some
groups found it more convenient to analyze the jump in specific heat, $\Delta 
C$, associated with the superconducting transition\cite{NewC,Carb3}, due to the 
difficulty of separating out the phonon contribution to the heat capacity.  
Theoretically, this should be a measure of the thermally averaged dos:
\begin{equation}
\Delta C(\delta )=A^2B^2k_B^2T_c<N(\delta )>_{T_c}(1+\lambda_{ep}),
\label{eq:10}
\end{equation}
with $A^2=-d[\Delta (T)/\Delta (0)]^2/d(T/T_c)|_{T=T_c}$, $B=\Delta (0)/T_c$,
and
\begin{equation}
<N(\delta )>_{T_c}=\int_{-\infty}^{\infty}N(\epsilon )(-{\partial f\over
\partial\epsilon})d\epsilon .
\end{equation}
This thermal averaging smears out the logarithmic divergence, but still leaves a
distinct peak (see Eq.~\ref{eq:A9d}).  For a weak coupling treatment of the 
VHS\cite{NewC} (using the modified parabolic band, Eq.~\ref{eq:7a}), $A=1.74$ 
retains its BCS value, while $B$ varies from 1.84 at the VHS ($\delta \simeq 0$)
to the BCS value, 1.76, as $\delta$ increases.  The last term in Eq.~\ref{eq:10}
is included to account for an effective mass enhancement associated with 
electron-phonon coupling, and must also be calculated in the VHS 
model\cite{NewC}.  Figure~\ref{fig:10}a shows the measured\cite{C1,C2,C3,C4}
$\Delta C$ as a function of oxygen deficiency $\delta$ in YBCO, compared with
both a weak coupling VHS theory (solid line) and a strong coupling version 
(dashed line).  The samples were quenched from high temperatures to minimize
problems of phase separation.  A clear peak is observed at the optimum value of 
$T_c$, which is actually {\it sharper} than the weak coupling VHS peak.  While 
a strong coupling calculation gives better agreement, this might be evidence for
an extended VHS\cite{SarD}.  It is also possible that
$\Delta C$ contains a chain contribution, which is very sensitive to oxygen
deficiency.  Similar peaks in $\Delta C$ have been found in LSCO\cite{C5}, 
Y$_{1-x}$Ca$_x$Sr$_2$Cu$_2$Tl$_{0.5}$Pb$_{0.5}$O$_7$\cite{C7}, and overdoped 
Tl-2201\cite{Nied2,C6}.  Note that the analysis of the jump $\Delta C$ is
nontrivial, due to strong fluctuations, which cause the heat capacity to
weakly diverge at the transition\cite{C4,Cfluct}.  In fact, a reanalysis of the
YBCO data\cite{Breit} finds that when the fluctuations are subtracted, the jump
in heat capacity $\Delta C$ continues to increase, significantly (by $\approx 45
\%$) in the overdoped regime.  The authors suggest that this effect, unique to
YBCO, may be associated with chain ordering.
\par
However, the situation is somewhat more complicated.   Figure~\ref{fig:10}b 
shows that there are {\it two} contributions to $\gamma$\cite{Kuma}: that 
associated with the jump $\Delta C$ at T$_c$ (diamonds) plus a residual 
contribution, $\gamma T$ present in the normal state (open and closed circles 
are two different measures of this quantity).  This is suggestive of phase 
separation away from the optimum doping, with the two contributions 
representative of the two phases.  The sum of the two contributions (solid line)
has a broad peak near the optimum doping, consistent with a broadened
VHS peak (for $x\le 0.15$ -- the sharp falloff at lower $x$ is more likely
associated with the pseudogap). Hence, the strong doping dependence of $\Delta
C$, Fig.~\ref{fig:10}a, may be more indicative of the doping dependence of the
superconducting fraction, due perhaps to phase separation, rather than the 
doping dependence of the dos.  
\par
\paragraph{\bf Peak in Heat Capacity, Susceptibility.}
This conclusion is reinforced by additional studies of the full electronic
contribution to the heat capacity\cite{Gp3} and the susceptibility 
$\chi$\cite{suscx,susc1,Gp3}, which tend to resemble the solid line in 
Fig.~\ref{fig:10}b, dominated by a pseudogap in the underdoped regime.
At $\vec q =0$, a Pauli-like (T-independent) susceptibility
is found only near the doping of optimum T$_c$\cite{KLA} in YBCO; at lower
dopings in YBCO, and all dopings in LSCO, a stronger T-dependence is found,
which has been attributed to pseudogap effects.  Thus, Loram, et al.\cite{Lor2} 
show that in YBCO, the temperature and doping dependence of $\chi$ is consistent
with that of $C_v$, and both are strongly influenced by pseudogap formation.  
\par
The influence of the pseudogap on $C$ and $\chi$ will be discussed in Section
IX.A.2.  Here, I wish to address the question of whether, if the pseudogap
could be eliminated, there would be any residual evidence for the VHS.  First,
the opening of the pseudogap is quite similar to what is expected for VHS
nesting, accompanied by the splitting of the VHS degeneracy.  Hence, it might be
thought that the dos derived from $\chi$ and $C$ should resemble that of a
split VHS, which in turn would resemble a single VHS, shifted off of the Fermi
energy.  Loram, et al.\cite{Gp3} have considered that possibility, and conclude
that a gap function produces a considerably better fit to the data: for a
shifted VHS, there would still be a substantial dos at the Fermi level, whereas
the data clearly reveal a (d-wave) gap.
\par
Hence, the remaining question is whether a VHS can be seen to underlie the
pseudogap.  In LSCO, this appears to be the case\cite{Gp6}.  As the hole doping 
is increased, the pseudogap collapses, revealing a peak in the dos, Fig. 
\ref{fig:b10}, open squares, which gradually crosses over to a flat dos at 
higher doping, $x\simeq 0.45$.  Surprisingly, the dos peak is maximal in the 
overdoped regime, $x\approx 0.24-0.27$ (rather close to the doping of optimum 
T$_c$ for {\it YBCO}, as determined by $\mu$SR).  This is in fact consistent 
with the theory discussed below.  
It is predicted that (1) there is a finite pseudogap at optimal doping, which
disappears in the overdoped regime; and (2) in this overdoped regime, the 
material is in a state of nanoscale phase separation, with one phase at the
VHS, and the other a normal metal, with presumably a flat dos.  Thus, the model
predicts that there would be a pseudogap at optimum doping, which closes to
reveal a VHS dos peak in overdoped materials, as observed.  Consistent with
this, as doping is further increased, the dos peak does not appear to shift off 
of the Fermi level, but only to decrease in intensity, relative to a flat 
background.  The susceptibility peak at $x=0.27$ can be well fitted to a 
slightly broadened VHS, Fig.~\ref{fig:b10}, solid line, where the broadening is 
described by an interlayer hopping, $t_z=4meV$, Section V.D.2.
\par
In YBCO, there is a similar pseudogap collapse with doping, but in the most
overdoped sample, $\delta =0.03$, the dos revealed by the susceptibility or the
heat capacity\cite{Gp3} appears flat, and not peaked.  One part of the problem
is that it is not possible to overdope YBCO nearly as much as LSCO.  However, 
the pseudogap collapse occurs much closer to optimal doping in YBCO, so one 
would have expected a hint of a dos peak. Perhaps the main problem is that there
are other bands present in YBCO -- the symmetric plane band and the chain band.
In particular, the chain band anticrosses the antisymmetric plane band, with
zero gap in the $k_z=0$ plane.  Hence, there will be interband contributions to
the dos, which should increase with frequency, or temperature, starting almost
from $T=0$ (not to mention the symmetric plane band VHS's).  Evidence for this 
explanation comes from Y NMR\cite{Allo}, which gives a measure of the local 
susceptibility at the Y site, and hence weighs the CuO$_2$ plane bands more 
strongly.  In optimally doped material, the ${}^{89}Y$ NMR shift does indeed
have a peak at $T=0$, open circles in Fig.~\ref{fig:b10} (here, the Knight shift
has been transformed into a susceptibility by assuming that the peak value 
coincides with the susceptibility of Ref. \cite{Gp3}).  There are two simple 
ways to explain the dos lineshape: in the dashed line of Fig.~\ref{fig:b10}, the
data are fit to a broadened VHS, $t_z=20meV$; while the filled circles of 
Fig.~\ref{fig:b10} are the susceptibility data after subtracting a constant 
contribution $\chi_b=1.32\times 10^{-4}emu/mole$, assumed to represent the
contribution of the other bands, in particular the bonding plane band.  It can
be seen that the contribution of the antibonding band (filled circles) is
very similar to the susceptibility of LSCO, modified by a larger combined 
superconducting and residual pseudogap.  Since both bands would be expected to 
make a comparable contribution to $\chi$, and since $\chi$ for YBCO is about
twice as large as for LSCO, the latter interpretation is to be preferred.
Thus a clear VHS is present in both
materials, when the pseudogap closes, while the pseudogap itself can be
interpreted as a splitting of the VHS degeneracy, Section IX.A.2.
\par
If, on the other hand, it is assumed that the VHS and pseudogap are two
unrelated phenomena, and that the VHS accidentally crosses the Fermi level in
the midst of the pseudogap opening, one would expect to see two peaks in the
susceptibility, one of which follows the VHS, the other associated with the
pseudogap.  Two such peaks have not been observed.
\par
The susceptibility also has a strong peak at or near $\vec q=\vec Q_0$, just as 
predicted for a VHS.  In the literature, this is called an antiferromagnetic 
peak, since $\vec Q_0$ is the wavevector of the AFM superlattice.  It is 
discussed further in Section VI.E.

\subsubsection{Transport Properties}

\paragraph{\bf Linear-in-$\omega$ and $T$ scattering rate ($\tau^{-1}$).}
\par
In a normal Fermi liquid, in any dimension, electron-electron scattering
should lead to a resistivity $\rho$ proportional to the square of the excitation
energy away from the Fermi energy -- hence, $\rho\propto max(T^2,\omega^2)$.  
Instead, it was noted that $\rho\propto T$ in optimally doped cuprates, crossing
over to $T^2$ in overdoped materials, and to insulating behavior $\propto T^{-1
}$ on the underdoped side\cite{Tsu}.  Similarly, it was found that $\rho\propto
\omega$ in optimally doped material.
\par
Lee and Read\cite{LR} first noted that the dos divergence near a VHS should lead
to an anomalous scattering rate.  They showed that the electron-electron
scattering rate $\tau^{-1}_{ee}\propto T$ when the Fermi level falls at a VHS.  
For a Fermi liquid, linear in $T$ is equivalent to linear in $\omega$; more
precisely\cite{RMopt},
\begin{equation}
\tau^{-1}_{ee}\propto\sqrt{\hbar^2\omega^2+\pi^2k_B^2T^2},
\label{eq:rho}
\end{equation}
where the factor $\pi^2$ is empirical (but c.f. Eq. 4.13 of Ref. \cite{pi}).
\par
A number of groups have reanalyzed these calculations, and found significant
modifications of the results.  There are a number of subtleties, and the final
results are not yet clear.  The first problem is that the Lee-Read calculation
involves inter-VHS scattering, due to a divergence of the inter-VHS
susceptibility, $Im\chi (\vec Q_0,\omega )\propto ln(\omega )$.  Pattnaik, et 
al.\cite{Patt} have shown that this divergence is cut off when $t_{OO}\ne 0$,
and the Fermi surface is not square at half filling, Eqs.~\ref{eq:a30} and 
\ref{eq:b30}, due to weaker nesting (compare Fig.~\ref{fig:a9}a
and \ref{fig:a9}b).  They and Gopalan, et 
al.\cite{Gop} showed that Eq. \ref{eq:rho} still holds, but it is dominated by
{\it intra}-VHS scattering.  The idea is as follows: the scattering rate is
calculated from the imaginary part of the self energy, $\tau^{-1}_{ee}\propto Im
\Sigma$, which in turn is given by an integral (over $q_x,q_y$) of $Im\chi (\vec
q,\omega )$.  For a parabolic band, $Im\chi\propto\omega$, so the integral gives
$Im\Sigma\propto\omega^2$.  At a VHS, however, $Im\chi\propto sign(\omega )$ for
small $q$ ($\hbar\omega\ge |E_{\vec q}|$) so $Im\Sigma\propto\omega$, giving 
$\tau^{-1}_{ee}\propto max(\omega ,T)$.  A linear-in-T
and $\omega$ scattering rate was also thought to follow from an 
extended VHS\cite{Ab2}, but this claim has been withdrawn\cite{Ab4}.  
Figure~\ref{fig:a10} shows the calculated scattering lifetime as a function of 
$T$ or $\omega$ -- good agreement is found with the photoemission line 
broadening\cite{Ols1,PEY2} and infrared reflectivity\cite{Schl}, even to the 
magnitude of the scattering.  It should be cautioned that the optical data
extend into the frequency range of the `mid-infrared peak', and sorting out this
contribution could modify the observed frequency dependence.  However, a
linear-in-$\omega^{-1}$ optical conductivity is found in Bi-2212 and along
the a-axis in YBCO, where there is little sign of the mid-infrared 
peak\cite{CooRz}; moreover, in LSCO the free carrier component has $\sigma
\propto\omega^{-1}$ even after the midinfrared feature is subtracted 
out\cite{Gao}.

\paragraph{\bf Resistivity: theory}
\par
Thus, the VHS model predicts a linear-in-$T$ and $\omega$ scattering rate, 
$\tau^{-1}_{ee}$, in good agreement with experiment.  However, experiments also 
find that the {\it resistivity} is linear in T and $\omega$.  For 
electron-electron scattering it is non-trivial to calculate the resistivity,
both because of the Fermi surface averages and because of the importance of
including Umklapp scattering.  While the average scattering rate will be
dominated by the strong scattering at a VHS, the resistivity can be dominated
by those carriers with small scattering, hence far from the hot spots near the
VHS.  Taking these factors into account, Hlubina and Rice\cite{HlR} find a $T^2$
resistivity near a VHS.  (Similar modifications also
arise in the calculation of the thermopower\cite{HlR2}).  The reason for this
discrepancy is not clear at present.  There are a number of additional factors
which could be included in the calculation.  For instance, at a VHS, any
scattering at all will lead to a vanishing mean-free path, $l=v_F\tau$, since
$v_F=0$, which suggests that multiple scattering corrections will be large.  In
particular, since both holes and electrons are simultaneously present at a VHS, 
there should 
be strong excitonic effects\cite{RM4}.  Equivalently, Phillips\cite{JC1} has
pointed out that, if the VHS is assumed to cause marginal Fermi liquid behavior,
then the renormalization factor $Z\rightarrow 0$ at the Fermi level.  Since 
$\rho\propto ZN(E_F)$, a self-consistent calculation of $\rho$ is required.
Newns, et al.\cite{NewHR} have also noted that the Boltzmann equation 
calculation of the resistivity involves weighted scattering probabilities, and
the weighing functions $\Phi_{\vec k}$ are not well known.  In particular, 
they may themselves have a logarithmic divergence at a VHS, which would strongly
modify the resulting T-dependence of $\rho$.  A more recent 
calculation\cite{MaKr}, including correlation effects, finds that $\rho\propto 
T^2$ very close to $T=0$, but $\rho\propto T$ for an extended range of 
temperatures above $\sim 0.01B$, with $B$ the electronic bandwidth. [However, it
is not clear if Umklapp scattering was correctly accounted for in this 
calculation.]

\paragraph{\bf Resistivity: experiment}

Experimentally, a universal crossover is found in the doping dependence of the
resistivity, which is tied to the doping at optimum $T_c$, $P^*$.  For optimally
doped material, the in-plane resistivity is linear in T down to T$_c$, except 
for superconducting fluctuations\cite{Tsuei}, while for overdoped material, 
$\rho\propto T^{\alpha}$, with $\alpha\rightarrow 2$ (Fermi liquid behavior) in
strongly overdoped material.  For underdoped material, $\rho$ is sublinear in T
($\alpha <1$), ultimately developing a semiconducting behavior ($\alpha <0$), 
and increasing with decreasing T.
Batlogg, et al.\cite{Batl} have found a stronger result at optimum doping: not
only do $\rho(T)$ and the scattering rate $\Gamma^*(\omega )$ increase linearly
with T or $\omega$, but the slopes $d\rho /dT$ and $d\Gamma^*/d\omega$ are 
essentially the same for all families of cuprates studied (LSCO, YBCO, Bi, Tl),
even though the T$_c$ differ.  Moreover, for an overdoped sample of LSCO, $x=
0.34$, both $\rho (T)$ and $\Gamma^*(\omega )$ were found to be superlinear
functions of their arguments, with essentially the same exponent, $\alpha\approx
1.5-1.6$.
\par
In a rigid band picture there is an approximate symmetry between over and 
underdoping which is not present in the experimental data.  While the crossover 
from optimal doping to overdoping is consistent with a rigid band model, the 
anomalous transport in the underdoped material is associated with correlation
effects and proximity to an insulating phase.  Even so, the small Hall density
is hard to reconcile with the large Fermi surface.  However, both resistivity 
and Hall effect can be naturally understood in terms of a percolation picture, 
if there is a (nanoscale) phase separation between the VHS phase and a magnetic
insulator. (See Fig. 2 of Ref. \cite{Halla} and Fig. 9 of Ref. \cite{Hallb}a). 
Indeed, this observation was one of the initial motivations for postulating the
existence of such a phase separation\cite{Halla,Hallb}.

\paragraph{\bf Zn and Ni Impurities}
\par
A long-standing puzzle in the cuprates has been the specificity of doping
effects at in-plane Cu sites -- in particular, the fact that non-magnetic Zn
depresses $T_c$ so much more strongly than magnetic Ni, even though both
enhance the resistivity by comparable amounts.  Fehrenbacher\cite{Fehr2} has
been able to explain these results by combining d-wave pairing (see below) with
proximity to a VHS.  The d-wave symmetry means that potential scatterers (such 
as Zn) will strongly suppress superconductivity, similar to magnetic scatterers 
in an s-wave superconductor.  The VHS plays several roles: it breaks 
electron-hole symmetry, so Zn can reduce T$_c$ more than Ni if the two atoms
have impurity potentials of opposite signs; it can cause Zn to act as a resonant
scatterer, even for not-too-large impurity potentials; and it can explain the
varying role of Zn doping as a function of hole concentration, in terms of
proximity to the VHS.

\paragraph{\bf Thermopower}
\par
The thermoelectric power (TEP) $S$ is also found to display a universal behavior
in most of the cuprates, Fig.~\ref{fig:11}a\cite{TEP,TEP1}: the TEP has a {\it
negative slope}, as a function of temperature.  Doping leaves the T-dependence
nearly unchanged, adding a constant (P-dependent) contribution to S.  
This leads to a {\it universal curve} of TEP {\it at
room temperature} vs. doping, Fig.~\ref{fig:11}b, with $S(290K)$ $>0$ for
underdoped, =0 for optimal doping, and $<0$ for overdoped material.  However, 
more complicated behavior is found in YBCO, Fig.~\ref{fig:11}c\cite{TEPY}, 
presumably due to the chains, which form a nearly 1D electron-like band.  Thus,
the negative chain contribution changes the sign of $S$ near O$_7$, where the 
chains are nearly perfect, but when oxygen is removed from the chains, the 
nearly-1D chain electron conduction is rapidly disrupted, leaving a plane 
contribution to $S$ which resembles the thermopower of the other cuprates.  
[LSCO with $x\ge 0.12$ is also an exception\cite{ZGx,Tal10}.]
\par
Newns, et al.\cite{NewS} have calculated how the thermopower should vary near a 
VHS, assuming rigid band filling.  They find that near a VHS, $S\propto (E_{VHS}
-E_F)/T$ -- that is, $S$ changes sign as a function of doping and vanishes at 
the doping of highest $T_c$, Fig.~\ref{fig:11}d.  This behavior is closer to the
anomalous behavior found in YBCO than to the universal behavior of the other 
cuprates\cite{PRC1}.  However, if a small, doping-independent term linear in $T$
is subtracted from the experimental normal-state thermopowers, the doping
dependence of $S(T)$ is in reasonable agreement with the Newns, et al.
model\cite{PRC2}.  Moreover, the model successfully reproduces the trend of 
$S(290K)$, Fig.~\ref{fig:11}b.  One residual problem is that the VHS should be
associated with zero thermopower, while the optimum T$_c$ is shifted to 
positive thermopower.  Thus, in Fig.~\ref{fig:11}a, by subtracting curve b from
all the curves for Bi-2212 (curve `$T_c<4K$' for Tl-2201), the curves will
closely resemble the theoretical curves of Fig.~\ref{fig:11}d.  However, the
curves of optimum $T_c$ (curve d for Bi-2212, 84K for Tl-2201) now have a
positive thermopower, with the corrected $S(290K)\approx 8\mu V/K$ for both.
This is presumably because Newns, et al.\cite{NewS} employed a simplified
model for the VHS, Eq. \ref{eq:7a}, which has electron-hole symmetry, whereas
in the actual bands the VHS is shifted to the hole doped side of half filling.
Additional discussion of these calculations is found in Refs. \cite{GoodNewns}.
[Alternative interpretations of the thermopower
involve strong electron-phonon coupling\cite{Trod} or (in-plane) Jahn-Teller 
polarons\cite{ZBG}, similar to those discussed in Section VIII.D, below.]   The
calculations of Newns, et al.\cite{NewS} have been questioned by 
Hlubina\cite{Hl2}, who finds an effect of the opposite sign.

\subsubsection{Marginal Fermi Liquid Behavior}

Noting the anomalous linear-in-T and $\omega$ behavior of the resistivity in
optimally-doped cuprates, Varma and coworkers\cite{MFL} introduced the 
phenomenological model of a {\it marginal Fermi liquid} (MFL): the cuprates 
can be described by a Fermi-liquid like theory, but the quasiparticle 
renormalization constant $Z$ vanishes logarithmically at the Fermi level.  This
follows from postulating the existence of excitations which produce an anomalous
contribution to both the charge and spin polarizabilities:
\begin{equation}
Im\tilde P(\vec q,\omega )\propto\cases{-N(0){\omega\over T}&for $|\omega |<T$,
\cr
                                    -N(0)sgn(\omega )&for $|\omega |>T$.}
\label{eq:10c}
\end{equation}
Exchange of these fluctuations leads to a self-energy of the form 
\begin{equation}
\Sigma (\vec q,\omega )\simeq g^2N(0)^2[\omega ln(z/\omega_c)-i(\pi /2)z],
\label{eq:10d}
\end{equation}
with $N(0)$ the dos, $g$ a coupling constant, $\omega_c$ an ultraviolet cutoff,
and $z=max(|\omega |,T)$.
\par
Given this form of $\Sigma$, a number of anomalous features of the cuprates can
be explained\cite{MFL}: $\tau^{-1}\propto Im\Sigma$, explaining $\rho\propto z$,
as well as the broadening of the photoemission peaks with $\omega$.
The model can also describe the $\omega$ and $T$ independent background seen in
Raman scattering out to $\sim 0.5eV$, the mid-infrared peak in the optical 
conductivity (although this must be modified\cite{MFL2} in light of the fact
that this peak is at least in part due to the chains\cite{opt2}), and several 
other features.
\par
In searching for a microscopic origin for the anomalous self energy, two early
candidates were a conventional Fermi surface nesting\cite{VR} and the 
VHS\cite{RM4}.  The VHS model leads to both the self energy correction and the
anomalous polarizability, the latter due to excitonic interactions of the holes 
and electrons simultaneously present at the VHS.  While Viroszek and 
Ruvalds\cite{VR} initially chose a model wherein the nesting was well 
distinguished from a VHS, a more recent calculation\cite{VR2} finds the 
strongest effects near a square Fermi surface at half filling -- i.e., at a VHS.
Moreover, calculations\cite{Gala} for the Hubbard model also find $Z
\rightarrow 0$ only near half filling (i.e., at the VHS).  Even this may be an
artefact of the calculation technique: in the tJ model, $Z$ is found to be 
finite even for a single doped hole\cite{ElDa}.
\par
An important difference between the original MFL theory and the VHS and 
conventional nesting models is that the MFL model assumes a $\vec q$-independent
self energy, while nesting leads to a strong $\vec q$ dependence.  
However, in order to describe the dynamical structure factor\cite{Mon}, a $\vec 
q$ dependent extension of the MFL theory has been introduced\cite{LZAM,AFMFL}. 

\subsubsection{de Haas - van Alphen Effect}

At the saddle point VHS, the Fermi surface crosses over from an electron-like
Fermi surface centered on the $\Gamma$ point to a hole-like Fermi surface,
centered on the Brillouin zone corner $S$.   This crossover leads to
anomalous transport and thermodynamic properties in a magnetic field.
In a strong magnetic field, an electron follows a cyclotron orbit, along a
path around the Fermi surface, in a plane perpendicular to the applied field.
As the Fermi level crosses the VHS, the cyclotron orbit switches between two
branches, Fig.~\ref{fig:12}.  When the Fermi level is exactly at the VHS, there
will be a finite probability of switching onto either branch.  This orbital
switching is analogous to magnetic breakdown\cite{magb}, but is actually a
distinct physical effect.  It is one of the `classical' (pre-high-T$_c$) VHS
effects, and its anomalous properties have been calculated by a number of
different techniques\cite{magsw}.  
\par
The electron orbit need not be localized in k-space, but can spread out just as
in an open orbit.  Hence, the magnetoresistance saturates, and, if scattering is
weak, there can be a novel, metallic magnetic band structure, similar to the 
`quantum coherent structure' proposed by Pippard\cite{magb}.  Unfortunately, 
dHvA studies have so far only detected some small Fermi surface 
pockets\cite{dHvA,dHvA1}, and not the larger plane Fermi surfaces where these 
effects are expected.
\par
The presence of switching orbits at the VHS can also explain the Hall effect 
anomalies in the cuprate superconductors: whereas photoemission finds evidence 
for large (`Luttinger') Fermi surfaces, the measured Hall density only 
corresponds to the excess holes beyond half filling of the band.  A low-field 
Hall effect calculation\cite{Hallb} suggests that the switching orbit, coupled 
with strong correlation effects, may be able to account for this anomaly.

\subsubsection{Excitons}

At a switching orbit, the carriers can equally follow either electron-like or 
hole-like orbits, leading to unusual excitonic effects\cite{RM4,RMex}. A carrier
localized near one VHS has a positive mass -- i.e., behaves like an electron -- 
for certain directions in $k$-space, while in other directions
it has a negative, hole-like mass.  For a carrier at the other VHS, most of
these $k$-directions are reversed.  Hence, if these two carriers move together 
in the same direction, they will usually feel a {\it Coulomb attraction}.  This 
may actually lead to an excitonic binding of the pair\cite{RMex}.  (I should
caution that, while I think the concept of excitonic attraction introduced in
this paper is correct, the calculations are wrong, due to an incorrect
definition of the excitonic binding energy.)  It should be recalled that the 
CDW/SDW's can be interpreted as forms of excitonic 
instability, wherein electron-like and hole-like Fermi surface sections nest
into one another (bind into excitons)\cite{RM4}.  Recently, Varma\cite{Varm2}
has explored the consequences of an excitonic instability, without relating it 
to a particular nesting instability.  He suggests that it might give rise to an
orbital antiferromagnetic-like instability (Appendix D).

This {\it attractive} electron-hole interaction means that the net Coulomb
interaction is anomalously small near the VHS.  As will be demonstrated below, 
this can lead to an {\it electronic} pairing mechanism.

\subsection{Superconducting State}

\subsubsection{Weak Coupling Theory}

The initial interest in the VHS arose because of the possibility that its
large dos could substantially enhance the superconducting transition
temperature.  Indeed, the logarithmic divergence of the dos gives rise to
a modified form of the BCS result for $T_c$\cite{HiSc,LB,Dzy,NewT}.  In a weak
coupling model, the BCS formula for $T_c$ is
\begin{equation}
1=V\int_0^{\hbar\omega_c}{N(E)\over E}tanh({E\over 2k_BT_c})
\end{equation}
For the one band model (Eq.~\ref{eq:29}),
\begin{equation}
N(E)={8\over \pi^2B}log|{2B\over E-E_{VHS}}|,
\end{equation}
where $B=8t_0$ is the electronic bandwidth, and $N$ is normalized per Cu.  
Approximating $tanh(x)\simeq min (x,1)$, the transition temperature becomes
\begin{equation}
k_BT_c=eBe^{-1/\sqrt{\tilde\lambda}},
\label{eq:11}
\end{equation}
with
\begin{equation}
{1\over\tilde\lambda}={2\over\lambda_0}+ln^2({2B\over\hbar\omega_c})-1,
\end{equation}
$\lambda_0=N_0V$, and $N_0=8/\pi^2B$.  This approximation is qualitatively 
correct, but overestimates $T_c$ by $\approx 15\%$\cite{BCS}.  If $2/\lambda_0>>
ln^2(2B/\hbar\omega_c)$, Eq.~\ref{eq:11} becomes
\begin{equation}
k_BT_c=eBe^{-\sqrt{2/\lambda_0}}.
\label{eq:12}
\end{equation}
At a VHS, T$_c$ is enhanced from the BCS value in two ways.  First, the
exponent depends only on the square root of $\tilde\lambda$, and secondly, the
prefactor is an electronic energy, rather than the Debye frequency.  
However, the first factor is important only when $\lambda$ is small, and in this
case, $T_c$ is low.  As $\lambda$ increases, the presence of a VHS leads to less
of an increase in $T_c$.  This can be approximately seen from Eq.~\ref{eq:11}.  
For $\lambda_0=2$, this predicts $k_BT_c=e\hbar\omega_c/2$.
Now with the same approximation of the tanh function, the BCS equation for a
material with a constant dos, $N_0$, would find
\begin{equation}
k_BT_c={e\hbar\omega_c\over 2}e^{-1/\lambda_0},
\end{equation}
so for this value $\lambda_0=2$, the VHS enhancement is only a factor of $\sqrt{
e}=1.65$.
\par
At the same time that T$_c$ is enhanced by a VHS, the Coulomb repulsion is 
reduced.  Bok\cite{Bok2} has calculated the reduced Coulomb repulsion $V_C^*$ 
due to a band with a VHS of width $2D$ plus a flat band of width $2W$:
\begin{equation}
V_C^*={V_C\over 1+V_C[{n_1\over 2}(ln{D\over\hbar\omega_0})^2+n_0ln{W\over\hbar
\omega_0}]},
\end{equation}
where $n_0$ ($n_1ln|D/E|$) is the dos in the flat (VHS) band.  When $n_1=0$,
this becomes the usual Anderson-Morel result\cite{AM}; when $n_0=0$, this shows
the enhanced, $ln^2$, reduction of $V_C$ by a VHS.  However, Bok feels that a
substantial flat band contribution is necessary to really suppress $\mu^*$. 
\par
While the weak coupling VHS model predicts a $T_c$ enhancement and a reduced
isotope effect (see below), there are quantitative discrepancies between its
predictions and experiment for the magnitude of the isotope effect and of the
BCS universal ratios.  For instance, the ratio, $2\Delta (0)/k_BT_c$ is only 
slightly enhanced from its BCS value, 3.53\cite{BCS,Goic}, whereas experiments 
generally find a considerable enhancement, although there is considerable spread
in the values quoted.  Thus, in photoemission, a ratio of $\sim 6-8$ is
typically found for the maximum gap value, whereas smaller values $\sim 4-5$
are often found for Fermi surface averages (e.g., heat capacity finds a ratio, 
$\approx 4.8$\cite{Lor2}).  Likewise, the 
ratio $\Delta C(T_c)/\gamma T_c$ is predicted to be only weakly enhanced from 
the BCS value, 1.43\cite{Goic} (here, $\Delta C(T_c)$ is the jump in specific 
heat at $T_c$, while $C_N=\gamma T_c$ is the normal state heat capacity at $T_c
$).  Experimentally, the ratio is close to 5\cite{Goic,DelC,Phil}, and has an 
anomalously strong temperature dependence in the superconducting state.  Note 
however that the heat capacity data are for YBCO, whereas smaller values of this
ratio are found in other cuprates; hence part of the enhancement may be chain 
related (although the mechanism is unclear).  Larger values for these BCS ratios
can be produced by strong coupling (see below) or gap anisotropy\cite{Ab1}, or
by extended VHS's.

\subsubsection{Strong Coupling Theory}

\par
Strong coupling calculations\cite{RM7,RaNo} find that the increase in $T_c$ due 
to a VHS is slightly larger (a factor of $\sim 2-3$ for $\lambda\sim 1$) than 
predicted by weak coupling theory; for $\lambda >>1$, the VHS produces no 
enhancement.  A similar result is found for an extended VHS\cite{Zey}.  Since 
$\lambda$ appears to be close to 1, the VHS could make the difference between a 
40K superconductor and a 90K superconductor.  Perhaps more importantly, the VHS 
could be responsible for the large values of $\lambda$, as will be discussed 
below (Section VIII.C.3,4).  Mahan\cite{GDM} found a related result:
he found that essentially the only effect the dos has on the Eliashberg 
equations is through $\lambda\propto N(E_F)$ -- but $\lambda$ has a logarithmic
divergence at the VHS, which he suggests may be cut off by additional factors,
such as screening. Marsiglio\cite{Mars1} showed that the divergence is not real,
but only analyzed the problem at half filling, where the superconductivity is
unstable with respect to CDW order.  Mahan's result has been confirmed by
Cappelluti and Pietronero\cite{CapaPie}, who further find that the peak leads
to a breakdown in Migdal's theorem, and that inclusion of nonadiabatic effects
can lead to further enhancement of T$_c$.  This breakdown of Migdal's theorem is
intimately connected to nanoscale phase separation: at a VHS, the 
compressibility, which is proportional to the dos (Appendix A), diverges, 
leading to a divergence in the vertex function (Section VIII.A).  In this case,
the Migdal self energy should be replaced by a nesting enhanced version, as in
Eq. \ref{eq:rev1}.
\par
Straightforward generalization of the VHS model to strong coupling also has
difficulty in reproducing the large values of the BCS ratios.  However, the
strong quasiparticle scattering discussed above acts as a strong and 
temperature-dependent pairbreaker.  In particular, pairbreaking can 
significantly inhibit the superconducting onset near $T_c$, but has little
effect on the low-T values of the gap.  Hence, by including pairbreaking 
effects, the VHS model can generate larger values of $2\Delta/k_BT_c\approx 6-8
$, close to experimental observations\cite{RM7}.  

\subsubsection{VHS Pairing Mechanism}

The presence of both electrons and holes at the VHS leads to an attractive
component to the Coulomb interaction.  While the net Coulomb interaction is
repulsive, it is anomalously small, and its Fourier transform, $V(\vec q,
\omega)=e^2/q^2\epsilon (\vec q,\omega )$ has a sharp minimum at $q=\omega =0$,
Fig.~\ref{fig:14}b (with a strong $\vec q$ dependence, Fig.~\ref{fig:14}c).  
Newns, et al.\cite{NewP} noted the similarity between this 
situation and the usual electron-phonon interaction, Fig.~\ref{fig:14}a. In this
latter case, the potential is always repulsive, since $V_c>V_{ep}$, but 
superconductivity is possible since $V_c$ is renormalized to
\begin{equation}
V_c^*={V_c\over 1+N(0)V_cln(E_f/\hbar\omega_c)};
\end{equation}
the BCS formula for $T_c$ involves $\lambda -\mu^*$, with $\lambda =N(0)V_{ep}$,
$\mu^*=N(0)V_c^*$.
\par
In just the same way, the VHS notch in $V_c$ leads to a purely electronic 
contribution to the pairing mechanism.  Neglecting any phonon coupling, the
superconducting transition temperature is approximately of the form of 
Eq.~\ref{eq:11}, with the substitutions $\hbar\omega_c\rightarrow E^*$, the 
width of the notch, and
\begin{equation}
\lambda_0\rightarrow N_0\bigl(V_e-{V_c\over 1+N_0V_cln^2(2B/E^*)/4}\bigr),
\end{equation}
with $V_e$ the depth of the notch.  Newns, et al.\cite{NewP} showed that 
combining this electronic mechanism with a phonon-mediated pairing could lead to
reasonable estimates of $T_c$.  
\par
On the other hand, Mahan\cite{Mah2} has recently
shown that in a one-band, tight binding model, the effective potential due to
{\it purely electron-electron interactions} is always repulsive, and
superconductivity is only possible if the gap function changes sign (as in a
d-wave superconductor).  For electron-phonon coupling, attraction is possible,
but ``the interaction is most attractive for electrons on scattering to opposite
sides of the Fermi surface, and it is repulsive for scattering by small wave 
vectors.''  In incorporating this behavior into a model calculation, he was able
to find nonzero gap functions only near a 2D VHS\cite{Mah3}.  A combined
phonon-electron coupling mechanism remains possible -- at least in the weak
sense that the Coulomb repulsion is anomalously small near a VHS.
\par
Liu and Levin\cite{LiLe} have shown that local field corrections to the Coulomb
interaction become very important near a VHS, and can lead to an attractive
pairing at near neighbor sites due to the ensuing Friedel oscillations 
(Kohn-Luttinger mechanism\cite{KL}).  The resulting gap has d-wave symmetry; in 
this model, there is a competition with electron-phonon coupling, which favors 
an s-wave gap.

\subsubsection{Isotope Effect}

Experimentally, the isotope effect is found to be extremely small at the doping
of optimum T$_c$, and this was initially taken as evidence against the
conventional electron-phonon coupling mechanism.  However, in the weak coupling
limit, the isotope effect vanishes at a VHS, even for electron-phonon 
coupling\cite{LB,NewT}.  This result can be seen from Eq.~\ref{eq:12}.  Since 
$T_c$ is independent of $\omega_c$, the isotope effect is identically zero.  
However, this equation is only valid when $\lambda_0<<1$, which is not 
consistent with a large transition temperature.  Using the fuller expression, 
Eq.~\ref{eq:11}, the isotopic mass exponent becomes\cite{NewT}
\begin{equation}
\alpha\equiv-{\partial lnT_c\over\partial lnM}={1\over 2}{ln(2B/\hbar\omega_c)
\over ln(eB/k_BT_c)},
\end{equation}
with $M$ the isotope mass, and it is assumed that $\omega_c\propto M^{-1/2}$.
While no longer vanishing, this is still significantly reduced below the BCS
value, $\alpha =0.5$.  Moreover, as a function of doping, it is found that
$\alpha$ has a minimum at the VHS, Fig.~\ref{fig:a14}, as found 
experimentally\cite{Fran}.
\par
However, it is difficult to make $\alpha$ at the VHS as small as that found in 
optimally doped YBCO\cite{RM7,Carb2}.  Thus, for the parameters assumed in Ref. 
\cite{NewT}, $\alpha_{VHS}\approx 0.2$.  If the parameters are adjusted to give 
a small $\alpha$ at optimal doping, then the variation of $\alpha$ with doping 
is found to be small.  The extended VHS also has difficulty in fitting the 
isotope effect\cite{Ab3}.  Carbotte\cite{Carb} has suggested that strong 
magnetic pairbreaking in the underdoped material can enhance the doping 
dependence, and by incorporating this effect, a good fit to experiment can be 
obtained\cite{RM7}. Strong anharmonic effects can also produce an anomalously 
small $\alpha$\cite{Cres,Kres}.  Figure~\ref{fig:a14} illustrates various 
calculations for $\alpha$ in the vicinity of a VHS, comparing them to 
experimental data\cite{isot1,isot2,isot3}. 
\par
In weak coupling theory, $\alpha$ is found to diverge when $E_F-E_{VHS}=\hbar
\omega_0$\cite{NewT,Goic,Kish}.  This divergence is associated with the
sharp cutoff in BCS theory, and should be greatly reduced if a smooth cutoff is
assumed.  It is not observed in strong coupling calculations\cite{RM7,NewP}.
It was initially compared to the highly anomalous isotope effect found in LSCO 
and LBCO in the vicinity of the structural transition to the LTT 
phase\cite{Craw}.  This is more probably understood as a consequence of the 
structural instability\cite{Plakx,Chern}.
\par
An interesting finding is that the isotope effect in YBCO is negative for Cu, 
but positive for O, even though both have a similar doping dependence (very
small at the VHS).  Franck\cite{Franck} has pointed out that this sign change 
can be understood if there is structure in the dos (such as a VHS), and the
gap energy falls between the Cu and O phonon frequencies.
\par
Franck\cite{Franck} has found that the Cu isotope effect in YBCO is small not 
only at optimum doping, but at a specific underdoping corresponding to the 60K 
plateau of $T_c$.  He suggests that this might be evidence that a second VHS 
underlies the 60K plateau.
\par
Some studies have attempted to determine which O's were playing the main role
in the isotope effect. A recent study\cite{iso1} found that the dominant isotope
effect in YBCO is associated with the planar O's, and concluded: ``This suggests
that theories of electron pairing in high-temperature
superconductors have to consider a phononic contribution, in which the planar 
tilting or buckling modes in the CuO$_2$ layers play an essential role.''

\subsubsection{Doping Dependence of $T_c$}

\par
The universal doping dependence of T$_c$ was originally revealed in the `Uemura 
plot'\cite{Uem1,Uem4}.  Uemura and coworkers showed that $n$, or more generally 
$n_s/m\propto\lambda^{-2}$ could be extracted from $\mu$SR measurements, where 
$n_s$ is the superconducting pair density (measured at low temperatures) and 
$\lambda$ is the London penetration depth.  Furthermore, a plot of T$_c$ vs $n_
s/m$ displays universal properties, Fig.~\ref{fig:c14}a, particularly in the 
underdoped regime, where the relation is linear, and all cuprates have 
essentially the same slope.  
\par
The VHS model provides a natural explanation for this universal observation of
an optimum doping (i.e., when the Fermi level coincides with the VHS), with
T$_c$ falling off for either under- or overdoping.  Figure~\ref{fig:b14}a shows 
that the overall trend of the data can be readily reproduced\cite{MG}.  (In this
early work, the phase separation was taken into account approximately.  The
double peak for YBCO is suggestive of what might happen for a split VHS peak; 
the arrows represent the approximate dopings at which the $Y$ and $X$ point
VHS's of the antibonding hole band cross the Fermi level in the LDA
calculations\cite{YMFK}.)
\par
More recently, it has been suggested that there is a {\it universal} relation 
between $T_c$ and hole doping $n$\cite{Univ}, Fig.~\ref{fig:b14}b.  While a 
flat-topped T$_c$(P) curve has been suggested\cite{uni0}, most groups find a 
parabolic form\cite{Univ,Uni1}.  It has further been proposed\cite{Tall1}
that, when plotted as $T_c(P)/T_{c,max}$ vs $P$, the curve is universal for all 
the cuprates!  That is, for all the cuprates, $T_c$ maximizes when $P=0.16$:
\begin{equation}
{T_c\over T_{c,max}}=1-({P-0.16\over 0.11})^2,
\label{eq:12b}
\end{equation}
Fig.~\ref{fig:b14}b.  It is not clear how to reconcile this with the Uemura 
plot.  It would seem more plausible that $m^*$ is the same for all cuprates, in 
which case, the Uemura plot suggests that the optimum doping is different in
different cuprates.
\par
If it is further assumed that the main dependence of $T_c$ on pressure $p$ or 
isotopic mass $M$ comes in through the dependence of $T_c$ on $n$, then there 
should also be universal trends of the isotope effect coefficient $\alpha$ and 
the pressure coefficient $\beta =dlnT_c/dp$ as functions of $T_c$, with {\it 
both coefficients vanishing at the optimum T$_c$}!  Such universal trends are 
indeed observed\cite{Univ}; the isotope effect has already been discussed, while
the pressure dependence will be discussed further below.
\par
The overdoped regime also has a universal, but unexpected behavior, 
Fig.~\ref{fig:c14}b\cite{Uem2,Nied2}.  Here, the anomaly is in 
$n_s$: instead of growing with doping, it reaches a maximum value, and then 
decreases with additional doping.  This is actually expected near a VHS, as 
illustrated in Fig.~\ref{fig:b8}, and replotted as Fig.~\ref{fig:c14}c.
While the general shapes are in agreement, the experimental points at 
which $n_s\rightarrow 0$ are anomalous.  In the simple theory, they should lie
at the top and bottom of the antibonding band.  Instead, they are separated by a
much smaller doping.  On the underdoped side, this is well understood: this is 
the correlation-induced Mott-Hubbard insulator at half filling.  The point at 
high doping is somewhat of a surprise, since when $n_s\rightarrow 0$, the 
material is a good metal.  The suggestion that it is related to a phase 
separation\cite{Uem3} is consistent with much data discussed in Section XI.
Alternatively, it has been attributed to pairbreaking\cite{Nied}.  Strong
pairbreaking effects in the overdoped regime are 
also suggested by heat capacity and IR studies and correlate with
increasing Curie terms in $\chi$ and $S/T$, where $S$ is the electronic entropy.

\subsubsection{Pressure Dependence of $T_c$}

The pressure coefficient of $T_c$ shows a nearly universal correlation in the 
cuprates, when plotted as a function of $T_c$ normalized to its largest value, 
with zero coefficient in optimally doped materials, and negative coefficient in 
overdoped (or electron doped) materials.  This is interpreted as due to a
pressure-induced doping of the CuO$_2$ planes, along with the parabolic doping
dependence of $T_c(n)$, and is hence consistent with the VHS 
model\cite{Univ1,Schil,Chu}.  Fig.~\ref{fig:d14}\cite{pres1,pres2} 
illustrates a representative sample of the results obtained.  The doping
dependence of the pressure coefficient, Fig.~\ref{fig:d14}a, and the isotope
effect coefficient, Fig.~\ref{fig:d14}b, are similar\cite{pres1}, while the sign
of the effect changes for overdoping, Fig.~\ref{fig:d14}c\cite{pres2}.
More direct evidence for the VHS is found in pressure dependences of $T_c$ for 
the Hg-cuprates, Fig.~\ref{fig:e14}\cite{HgChu1}, where the nonlinear slope of 
$T_c(p)$ has been ascribed to the pressure-induced crossing of the Fermi level 
by a VHS\cite{HgChu}.
\par
LSCO provides an exception to the universal scaling behavior: $dT_c/dp$ is large
and roughly doping dependent around the optimum doping\cite{Gugen,Noha2}; in
the overdoped phase, it remains positive in the orthorhombic phase, and 
essentially zero in the tetragonal phase\cite{press,press2}.  This
can be understood readily: for most cuprates, the doping of the CuO$_2$ planes
is delicately controlled by charge transfer from other layers, and the main
effect of pressure is to alter that balance, producing an excess hole doping.
On the other hand, doping in LSCO is strictly chemical, replacing La by Sr, and
the change of T$_c$ with pressure is dominated by other effects -- related to
inhibition/enhancement of the competing structural (tilt mode) instability.

\subsubsection{Hall Effect Sign Reversal(?)}

The detailed analysis of high-T$_c$ cuprates has called attention to this
anomalous phenomenon, which had been observed, but not explained, in 
conventional superconductors: the Hall effect can change sign on passing from 
the normal to the superconducting state\cite{Halrev}.  This is a material 
dependent property, found in some superconductors and not in others.  In the 
cuprates, it seems to be present in a limited doping range near the optimal 
T$_c$.  It is possible that this property also is explained by proximity to a 
VHS.  Thus, in the time-dependent Ginzburg-Landau (TDGL) theory\cite{TDGL}, 
there are two contributions to the Hall coefficient, the normal one, and an 
`anomalous' one due to vortex backflow, which is
proportional to $\partial N(E_F)/\partial E_F$ -- the {\it slope} of the dos.
If this slope has the opposite sign from the sign of the carrier charge (i.e.,
if the slope is negative for holes), the anomalous term has the opposite sign
from the normal term, and can lead to the sign reversal.  Since this slope
changes sign at a VHS, such Hall effect anomalies should be present in some 
doping range.
\par
In light of this, recent experiments by Jones, et al.\cite{JCS} on YBCO are
interesting.  They find a correlation of Hall effect anomaly with TEP.  For
$S\le 0$, there is an anomaly, whose magnitude scales linearly to zero as $S
\rightarrow 0$; for $S>0$, no Hall effect anomaly is found.  This result is
consistent with the VHS picture, since the VHS coincides with $S=0$ in YBCO,
thanks to the chain contribution (see Fig.~\ref{fig:11}c).  
In LSCO\cite{HaLS}, which lacks the chain layer, a Hall effect sign reversal is 
found in underdoped or slightly overdoped samples, but not in strongly overdoped
samples; this was taken as evidence in support of the TDGL theory.

\subsubsection{$T_1$, Penetration Depth}

The nuclear spin-lattice relaxation rate has two anomalous features -- the
lack of an obvious Hebel-Slichter peak and a power-law T-dependence at low
temperatures\cite{T1}.  The latter feature is suggestive of nodes in the gap, 
which will be discussed further in the subsection on s vs d wave symmetry 
(VI.D.11). The lack of a Hebel-Slichter peak can be understood as due to strong 
pair-breaking quasiparticle scattering near $T_c$\cite{HeSl,RM7}.
\par
Similar results hold for the penetration depth: the VHS model can
explain the T dependence near $T_c$\cite{RM7}, while there appears to be an
anomalous power law temperature-dependence at low temperatures, suggestive of 
gap nodes\cite{Hard}.

\subsubsection{$H_{c2}$}

The shape of the superconducting $H-T$ phase diagram is also sensitive to 
structure in the dos.  The role of the VHS has been worked out by Dias and 
Wheatley\cite{DW}. The most striking result is that at low temperatures $H_{c2}$
is greatly enhanced, $H_{c2}(0)\propto T_c^{\sqrt{2}}$, as opposed to the BCS 
value, $H_{c2}(0)\propto T_c^2$.  Such an anomalous enhancement is indeed a 
characteristic feature of those cuprates with low-enough $T_c$'s that 
$H_{c2}(0)$ can be measured\cite{Hc2}. 

\subsubsection{Gap Anisotropy}

Since the electron-phonon coupling $V$ is enhanced by the phonon renormalization
$R$ (Eq.~\ref{eq:0}) in the denominator, $V$ should have a strong anisotropy, 
being largest where $R$ is smallest\cite{anis}.  This can lead to a strongly
anisotropic gap, with the largest values near the VHS's, as found 
experimentally.  It should be noted that, in the calculations of Ref. 
\cite{anis}, only an s-like gap was assumed, so the resulting $\Delta$ is
of `extended-s' symmetry.  It seems likely that, under the appropriate
circumstances (e.g., a repulsive on-site $V$), a d-wave gap could be
stabilized.  Similar anisotropy is found for an extended VHS\cite{Ab2}.

\subsubsection{s vs d Wave}

Recently, a number of experiments have found strong evidence that most of the
high-T$_c$ cuprates have gaps of d-wave symmetry.  The present status of this
issue is briefly summarized here, and it is shown that the VHS model is 
compatible with {\it either} anisotropic-s or d-wave gaps.
\par
Several low-temperature properties of the cuprate superconductors have shown 
anomalous power law dependences on temperature, as if there were nodes of gap
zeroes along parts of the Fermi surface.  These properties include penetration 
depth measurements and nuclear spin-lattice relaxation times, $T_1$, discussed 
above.  In the past two years, several experiments have provided more direct
evidence that the superconducting gap function has d-wave, or mixed s-plus-d 
wave symmetry.  Thus, while early photoemission experiments on
Bi-2212\cite{Pgap1} found an isotropic gap, more recent measurements find large
gap anisotropy\cite{Pgap2,Pgap3,Pgap4,Pgap6,Pgap10,Pgrev}, with the maximum gap 
along the Cu-O-Cu directions (i.e., near the VHS's), and a minimum gap, 
consistent with a zero gap at approximately $\pm 45^o$ to the directions of the 
maxima.  If the gap zeroes fell at exactly $\pm 45^o$, this would be consistent 
with pure d$_{x^2-y^2}$ symmetry.  While some high resolution studies found the 
minima off of the $45^o$ line\cite{PEanis,Pgap6}, this is now understood to be 
due to interference from a superlattice replica band.  The data have now been 
reanalyzed\cite{Pgap8} as a broad minimum gap, consistent with $\Delta =0$,
within $\pm 10^o$ of the 45$^o$ angle. However, another group finds a very sharp
minimum at 45$^o$\cite{Pgap10} -- the difference may be an impurity effect.  
Similar photoemission results have now been reported for YBCO\cite{Scha}.
There is also a report\cite{One} of a highly anomalous temperature dependence of
the anisotropy in Bi-2212, which is found to be large near $T_c$, but much 
reduced at lower temperatures.  This could be consistent with a gap of pure 
d-wave at $T_c$, but with an s-wave admixture at lower temperatures, but it is 
not clear if this is consistent with other experiments.  
\par
Photoemission experiments are insensitive to the
sign of the gap function, and hence cannot distinguish d-wave from
highly anisotropic s-wave superconductivity, wherein the minimum gap can be very
small, or even zero, but the gap function never changes sign.  More recently,
a number of experiments have been reported which are sensitive to the sign of
$\Delta$.  There is particularly strong evidence for the existence of $\pi$
junctions\cite{Pi00,Pi1,Pi2,Pi3}; in a $\pi$ junction, the sign of the gap 
changes across the junction, leading to a phase shift of $\pi$ across the
junction.  For a closed loop containing an odd number of $\pi$ junctions (a $
\pi$ loop), there will be a net phase shift of $\pm\pi$, which can only be
compensated by a spontaneous circulating supercurrent in the loop, in the
absence of an applied magnetic field.  Evidence for such a current, containing
a half quantum of flux, is particularly clear in the experiments of Tsuei,
Kirtley, et al.\cite{Pi2}.
\par
Both theoretical and experimental issues related to possible d-wave 
superconductivity have been recently reviewed by Scalapino\cite{Scal-d}.
At this point, the case for a gap which changes sign on different parts of
the Fermi surface appears to be strong, although the experimental situation is 
by no means fully resolved, and some experiments still favor an s-wave 
gap\cite{Pi0}.  A number of theoretical issues must still be settled, however.
In particular, it was noted that evidence for $\pi$ junctions is generally found
in bilayer cuprates (cuprates with two CuO$_2$ planes per unit cell), and hence 
could be modified by interlayer coupling.  For example, it is possible that the 
gap has s-wave symmetry, but the intra- and inter-layer gaps have opposite 
signs\cite{TwoS1,MaYa,Ubb2,TwoS2,Radt,MGZ}.  
However, the squid-ring experiment has now been repeated on Tl-2201, and
finds identical d-like symmetry in this single-layer cuprate\cite{CCT}.
\par
Much of the analysis described above assumed s-wave symmetry, and should be
repeated for a d-wave gap.  Since most of these features are predominantly 
sensitive to the presence of a peak in the dos, it is not expected that this
reanalysis would significantly modify any conclusions.  Newns and 
co-workers\cite{Newd} have recently generalized their earlier calculations, and
find that the experimental results are equally consistent with d-wave pairing.
On the other hand, the symmetry of the pairing might reveal something about the 
nature of the bosons which are exchanged, as discussed below.

\subsubsection{d Wave Pairing Mechanism}

\par
Given that the gap is d-wave, there are two questions: what does this tell
about the pairing mechanism, and what about the role of the VHS.  The first
question is still being debated, but whatever the answer, the VHS can still play
an important role.  For example, a d-wave gap arises naturally 
in a model in which the superconductivity is due to spin fluctuations: since the
pairing potential is purely repulsive, the gap equation has a solution only if
$\Delta$ has contributions of both signs on the Fermi surface\cite{Scal-d}. 
Indeed, the early calculations of Dzyaloshinskii\cite{Dzy} and Schulz\cite{Sch}
showed that the VHS would greatly enhance spin-fluctuation-induced 
superconductivity in a Hubbard model, leading to a d-wave gap.  More recent 
calculations find a similar result\cite{NewD}.  
\par
However, {\it observation of a d-wave gap does not rule out a phonon
(or any other) pairing mechanism}.  All that is required is sufficient 
anisotropy of the pairing interaction, which follows naturally in a VHS model.
Thus, it is known that an anisotropic pairing interaction greatly enhances
T$_c$, while producing an anisotropic gap\cite{ani1,ani2}.  If the pairing
interaction is expanded in terms of a double set of spherical harmonics, and
the matrix of coefficients diagonalized, then the highest T$_c$ corresponds to
the channel with the largest positive eigenvalue\cite{ani3}, with the gap
having the same symmetry.  In the present problem, the chief ingredients are
the highly anisotropic VHS-coupled pairing and the large on-site repulsion $U$.
By forming a non-s-wave gap, the on-site repulsion is automatically cancelled.
A VHS-based electronic pairing mechanism, based on the charge channel\cite{LiLe}
also can lead to d-wave superconductivity; moreover, the boson frequency could 
be substantially higher for the charge modes (plasmons) than for the spin modes.
\par
In conventional superconductors, the electron-phonon interaction is nearly 
local, due to strong screening effects.  In the cuprates, the {\it ionic}
contribution to the dielectric constant is very large, $\sim 40-80$\cite{diel},
so screening is much less effective, leading to a very anisotropic electron-ion
potential, and hence to predominantly d-wave scattering\cite{Sant}.
Fehrenbacher and Norman\cite{FehrN}
have shown that a gap of d-wave symmetry can appear in a VHS model, as long as
$V(\vec q)$ has a symmetry to couple to the VHS -- i.e., does not vanish at the 
VHS's, $\vec q =(\pi /a,0),(0,\pi /a)$.  Chan and Plakida\cite{MTP} have also 
found (in a model not involving the VHS) that a purely electron-phonon coupling 
model of superconductivity can still lead to a d-wave gap.
\par
Song and Annett\cite{SongAn} have analyzed the various
planar-oxygen-related phonon modes which are most relevant to the cuprates
(Section VIII) -- the breathing and octahedral tilt modes.  They found that the
breathing mode leads to a repulsive contribution in the d-channel, while the
tilt modes lead to an attractive d-wave pairing.  (There is a stronger tilt-mode
contribution of s-wave symmetry, which is overwhelmed by the Hubbard $U$ 
repulsion.)  However, their calculation found a small net d-wave T$_c$ value.
Nazarenko and Dagotto\cite{NaDa} showed that this T$_c$ could be greatly 
enhanced by fixing the Fermi level at a VHS.  Indeed, by using an {\it extended}
VHS, they were able to reproduce approximately the doping dependence of both 
T$_c$ and the isotope effect (maximum $T_c\simeq 30K$).  In their calculation,
the extended VHS was introduced via the remnant (`ghost') antiferromagnetic
superlattice, but since this result depends only on the net dos, it should hold 
for other mechanisms of extended VHS formation as well.  The importance of this
same mode has also been pointed out by Bulut and Scalapino\cite{BuSc} and Sakai,
et al.\cite{SPS}.  Alternatively, mixed magnon-phonon pairing can lead to d-wave
superconductivity with a high T$_c$, in the proximity of a VHS\cite{ShSch}.

\subsection{Magnetic Properties}

Originally, a sharp dichotomy was believed to exist between strong
correlation and Fermi liquid theories of high-T$_c$ superconductivity.  If
strong correlations lead to a spin-charge separation, then the physical
electron will be a convolution of spinon plus holon states, and this convolution
was believed to wash out the Fermi surface.  However, such reasoning does not 
agree with experiment.  It would predict that there are no CDW's in 1D metals, 
where correlation effects are strong and spin-charge separation can be 
rigorously demonstrated.  In contrast, CDW's are almost always found in
lower-dimensional materials.  A closer analysis of the 1D situation 
shows a more complicated picture.  There is still a well-defined Fermi
surface, but it is `softer': the particle distribution function, $n(\vec k)$,
has a power-law discontinuity at the Fermi level, rather than the step 
discontinuity characteristic of a Fermi liquid\cite{OSh}; equivalently, the 
renormalization function $Z$ has a power law zero.  This more general situation,
wherein there is a Fermi surface satisfying Luttinger's theorem (area of
Fermi surface corresponds to number of conduction electrons/holes), but with 
power-law discontinuities, is known as a `Luttinger liquid'\cite{Hald,Lut}, as 
opposed to a Fermi liquid.  While Luttinger liquids exist in 1D, their status in
higher dimensions is not clear.  It has been suggested that a 2D Luttinger 
liquid exists for an artificial, superstrong Coulomb repulsion\cite{MuF}, but 
this remains controversial.  For more physical interactions the 
Fermi liquid state is found to be stable.  
Experimentally, angle-resolved photoemission studies show that the cuprates
do have well-resolved Fermi surfaces, and that even for an insulating cuprate,
there is a reasonably well-defined energy dispersion\cite{Well}.  Most theories
now find that strong correlation effects in 2D lead to Fermi liquids with
renormalized band parameters.
\par
Since the VHS is a topological property, it should still be
present in a Luttinger liquid.  Anderson has now stressed that in fact nesting 
effects are enhanced in Luttinger liquids\cite{And}.
Indeed, the dos divergence leads to a breakdown of both Haldane's Luttinger 
liquid theory\cite{Hald} and the renormalization group (RG) derivation
of Fermi liquid theory\cite{FeRG1}, in the vicinity of a VHS.  The nature of
this breakdown remains an open question.

\subsubsection{Spin-fluctuation Induced Pairing}

\par
Hence, the VHS can be expected to play an important role in any strong
correlation theory of superconductivity, including those in which the
pairing arises from spin fluctuations.  
It was early noted that high temperature superconductivity in the cuprates
borders on an antiferromagnetic Mott-Hubbard insulating state of the half filled
CuO$_2$ band.  This led to a number of suggestions that the superconductive
pairing is mediated by magnetic excitations.  Most of these theories are based 
on a peak in the spin susceptibility at the commensurate wavevector, $Q=(\pi /a,
\pi /a)$.  This peak was initially introduced phenomenologically\cite{MMP}, but 
in any satisfactory theory, it must
ultimately be derived self-consistently -- i.e., beginning with a
susceptibility derived from the electronic band structure.  The only
successful attempts to do this have been those which place the Fermi level
near a VHS.  For instance, Monthoux and Scalapino\cite{MoSc} have 
found a self-consistent, d-wave superconducting transition in a 2D Hubbard 
model, with a peak in the magnetic susceptibility near $(\pi ,\pi )$, of the
form hypothesized for the nearly antiferromagnetic model.  
  From the insert to their Fig. 1, it can be seen that they have chosen a 
Fermi level close to, but not precisely at the VHS.  (Recall
that correlation effects generally move the Fermi level even closer to the VHS.)
\par
These calculations would benefit from a more detailed analysis of the role of
the VHS.  Thus, the reasons for a particular choice of band parameters are not
clearly stated, and the resulting Fermi surfaces are often inconsistent with
other experiments and/or band structure calculations.  Moreover, once an
initial set of parameters is chosen, the theories often have difficulty in
explaining the doping dependence of various effects.  One case in point is
the incommensurate neutron diffraction peaks in LSCO, discussed in the following
subsection.  
\par
In fact, understanding the doping dependences has proven to be a major problem 
for the basic Van Hove scenario.  My own assessment is that the problem
lies not with the VHS, but with the assumption of rigid band filling, which
underlies most of these analyses.  For instance, consider the role of strong
correlation effects.  According to the slave boson calculations (Section VII),
the transition to a charge transfer insulator at half filling is not a nesting
effect, but is driven by a divergent effective mass.  Some theories
turn the transition into a SDW nesting instability, by fixing the VHS at half
filling (e.g., in the Hubbard or tJ model, this happens automatically unless a
second-neighbor hopping parameter $t^{\prime}$ is introduced; it also happens in
the three-band model when $t_{OO}=0$).  However, once the VHS is fixed at half
filling, the doped material has an electron-like Fermi surface and doping moves
it further away from the dos peak; both of these features make it difficult to
understand the physical properties of the cuprates.
\par
Shifting the VHS to a finite hole doping (by choosing nonzero values for $t^
{\prime}$ or $t_{OO}$) produces Fermi surfaces in better agreement with band
structure calculations and photoemission experiments, but now a strongly
non-rigid-band filling is required to simultaneously explain the Mott-Hubbard
transition at half filling.  Such doping-dependent band structure 
renormalizations arise naturally in slave boson calculations of correlation 
effects, Section VII.  Unfortunately, even this may not be adequate.
There is considerable experimental and theoretical evidence that the doping is
not continuous, but involves a (possibly nanoscale) phase separation between the
magnetic insulator and the state of optimum T$_c$ (Sections X,XI).  These 
complications should be kept in mind when analyzing the incommensurate neutron
diffraction peaks, discussed below.

\subsubsection{Incommensurate Diffraction Peaks}
\par
Neutron diffraction studies have found that in La$_2$\-Cu\-O$_4$, the 
antiferromagnetic peaks are commensurate, at $Q=(\pi /a,\pi /a)$.  As the 
material becomes Sr doped, the peaks split and become incommensurate, at $Q\pm
\delta (\pi /a,0)$ or $Q\pm\delta (0,\pi /a)$, with the incommensurability 
$\delta$ increasing proportional to the doping $x$, 
Fig.~\ref{fig:15}a\cite{inco}.  On the other hand, the peak is usually found to
remain commensurate in YBCO as $\delta$ is reduced from 1, although there are
indications of anisotropy, consistent with a non-resolved splitting into a 
four-fold pattern.  Thus, in underdoped YBCO$_{6.6}$ (T$_c$=53K), Tranquada, et 
al.\cite{TGS3} found a pattern with the same symmetry as LSCO, but the
unresolved incommensurability was only about half as large.  In a nearly
stoichiometric crystal, $T_c=92.4K$, Mook, et al.\cite{Mook} have found a
strikingly different result, Fig.~\ref{fig:15}b. Again, a possible non-resolved 
four-fold splitting is found, but rotated by 45$^o$ with respect to LSCO.  
A similar splitting is found in LSNO, Fig~\ref{fig:15}c\cite{LSNO2}.  However,
the feature in stoichiometric YBCO is anomalous: this feature is only seen in 
the superconducting state, and at high energies, $\approx 41meV$, whereas the 
other peaks discussed above are normal-state, low-energy features.  Possible
interpretations of this feature are given in the following subsection. A recent 
reanalysis\cite{STS3} of a YBCO$_{6.6}$ (T$_c$=46K) crystal provides some 
additional evidence: the lineshape is independent of frequency between 2 and
40meV, and is the same both above and below T$_c$.  It appears to be 
incommensurate, but whether the spots were oriented as in LSCO or LSNO (or in
some other pattern) could not be resolved.
\par
A number of attempts have been made to relate these peaks to nesting of the
Fermi surfaces, and to interpret the difference between LSCO and YBCO as due to 
differences in their Fermi surface topologies\cite{TKF,ZLL,LZAM}.  The Fermi 
surfaces are calculated using the three band model, with $t_{OO}\ne 0$.
There are certain difficulties associated with such interpretations, however.
In the first place, in a weak coupling calculation the intensity is 
much larger for inter-VHS scattering (which in these models is
commensurate at $Q$) than the nesting-associated scattering when the Fermi
level is shifted off of the VHS (see Fig. 8 of \cite{BCT}).  In a marginal Fermi
liquid model, this VHS intensity enhancement can be greatly reduced by 
quasiparticle lifetime effects\cite{LZAM}.  
In the doped material, the model predicts an incommensurate peak in the
susceptibility, in accord with experiment; however, the doping dependence is
difficult to reproduce\cite{BCT}.  This is because of VHS nesting: the peak is
commensurate, at $\vec q=\vec Q_0$, when the VHS is at the Fermi level. Thus,
using the LDA band structure, in which the VHS falls near the doping of
highest T$_c$, the nesting model would predict a commensurate peak at optimum
doping, and an incommensurate peak at half filling\cite{BeatL} -- the opposite
of experiment.  Hence, to explain the experimental results, it has been
suggested that the LDA calculation is wrong, and that the VHS's are 
much closer to half filling\cite{LZAM}.  The new slave boson calculations in
the three-band $tJ$ model (Section VII.B) will improve the situation, in that
a magnetic VHS falls at exactly half filling, leading to commensurate
diffraction peaks; however, there will be a second VHS at optimum doping, which
will still pose problems.
\par
Finally, the Fermi surface assumed for YBCO\cite{TKF,ZLL} locates the
VHS far from the Fermi level, in poor agreement with photoemission and LDA
results.  This is because these calculations either neglect the interlayer
splitting parameter, $t_z$ (Eq.~\ref{eq:9}), or match\cite{ZLL} the Fermi 
surface areas to the early calculations of Yu, et al.\cite{YMFK}, which greatly
underestimate the hole doping.  When the interlayer coupling is correctly 
included, the major topological difference, which was supposed to explain the 
experiments, is no longer present: both LSCO and YBCO have VHS's present very 
close to the Fermi level, in optimally doped materials (in agreement with
experiment).  The difference is that in YBCO, this is associated with an 
extended VHS.
\par
On cooling LSCO into the superconducting phase, the four incommensurate peaks 
remain at the same value of $\delta$, but with intensity reduced by $\approx
60\%$\cite{Mas}.  This is a potential problem for the above model: the peaks
should be washed out for an isotropic, s-wave gap.  For a d-wave gap, there can
still be magnetic scattering, but only between the gap nodes, where 
quasiparticles persist at low temperatures.  This internode scattering would 
also lead to a four-fold incommensurate pattern of peaks, but rotated by 45$^o$
with respect to the high-T peaks\cite{rotn}.  This rotation is not observed.
The experimental observations can be explained\cite{Quin}, if there is a
strong (resonant) impurity scattering\cite{HiGo}, which smears out the
quasiparticle distribution in $k$-space.  In a more recent experiment, Yamada,
et al.\cite{Yama} found strikingly different results: in very uniform, high-T$_
c$ crystals, they found that the intensity of the incommensurate peaks 
completely vanished at low T for $\omega\le 3.5meV$.  Since there was no 
scattering intensity at the positions of the gap nodes in the normal state, it
was not surprising that none appeared in the superconducting state.
\par
A correlation between incommensurate peaks and electronic domain structure has 
now been clearly established, both for LBCO at $x\simeq 0.125$, and for the 
closely related compound, La$_{2-x}$Sr$_x$NiO$_4$.  Considerably more evidence 
for such phase separation is discussed below, in Section XI.  
\par
Barzykin, et al.\cite{BPT}, have used a model of {\it discommensurations} to
resolve a problem with the $^{17}O$ spin-lattice relaxation rate.  This rate is 
insensitive to antiferromagnetic effects, which can be understood in terms of a 
matrix element cancellation, but only if the spin-fluctuation peaks are 
commensurate.  They postulate that the system breaks up into domains, within 
which the spin fluctuations are commensurate, separated
by domain walls.  Then the O's inside a domain see only the commensurate 
fluctuations, while neutron scattering sees the average periodicity over several
domains, and hence is incommensurate.  Walstedt, et al.\cite{Wals} objected that
Cu spins in the domain walls would still spoil the good agreement of the
domain model with experiment.  However, this assumes that the walls can be 
treated as incommensurate magnetic states, whereas (in the VHS model), the
walls are closer to a second, conducting electronic phase, where the local
Cu moments are strongly washed out.  In this model, the persistence of the peaks
in the superconducting state would follow if the magnetism and 
superconductivity are associated with different parts of the sample.  It should
be noted, however, that more recently Zha, et al.\cite{ZBP}, have shown that
the desired matrix element cancellation can also be achieved by introducing a 
new term into the transferred hyperfine coupling Hamiltonian, without any 
necessity for domain formation.
\par
There is still a potential problem for the VHS model here in that no
superlattice modulation is expected at optimum doping in the simplest VHS model
for LSCO, whereas experimentally the incommensurability remains 
large\cite{Yama}.  Spin-orbit coupling, by splitting the VHS, can produce
an incommensurate diffraction pattern of the observed form, and can explain
the doping dependence (since the spin-orbit splitting is suppressed by
correlation effects near half filling)\cite{RMXA}.  However, the observed
splitting is larger than expected for the estimated magnitude of the spin-orbit 
coupling.  Another possible explanation for this problem is suggested in 
Section XI.I.
 
\subsubsection{Antiferromagnetic Excitations}
\par
In studying the magnetic properties of the cuprates, the susceptibility, Eq.
\ref{eq:02}, has RPA-like corrections,
\begin{equation}
\chi_{RPA}(\vec q,\omega )={\chi (\vec q,\omega )\over 1+J(\vec q)\chi (\vec q,
\omega )},
\label{eq:12c}
\end{equation}
where $J(\vec q)=J(c_x+c_y)$ is the magnetic exchange\cite{ZLL}.  This
susceptibility has been analyzed by a number of groups, for either the tJ,
tt$^{\prime}$J, or three band models.  While most of the results of these
calculations have been discussed above or in Section IX.A.3, there is one
interesting result which may involve a {\it different kind of VHS}.  Near
optimum doping, the inelastic neutron scattering measurements in YBCO find that
the susceptibility has a peak near 41meV, which sharpens greatly in the 
superconducting state, Fig.~\ref{fig:15}b\cite{m411,m412,Kei2,Mook}.  Lavagna 
and coworkers\cite{Lav} have interpreted this
peak as arising from a d-wave superconducting gap.  The gap renormalizes the
band structure into four hole pockets, centered on the gap zeroes, and the
peak in $\chi$ is associated with the VHS of this renormalized band structure.
\par
This result has been confirmed by a number of groups\cite{MaYa,BSK}.  Mazin
and Yakovenko\cite{MaYa} show that the peak position is given by $E^*\simeq 
\Delta_0+\sqrt{\Delta_0^2+\zeta_{VH}^2}$, where $\Delta_0$ is the
superconducting gap, and $\zeta_{VH}$is the energy between the VHS and the
Fermi level, taken to be $\approx 10 meV$.  As the temperature is raised and the
superconducting gap vanishes, this peak evolves into a weak, broad peak centered
at the VHS.  Blumberg, et al.\cite{BSK} showed that if the VHS is bifurcated, 
the normal state peak is equal to the energy gap between the Fermi level 
and the electronic dispersion at the $X$-point (the saddle between the two humps
of the camelback of Fig.~\ref{fig:b5}c).  This energy can be much larger than
the minimum energy associated with the bifurcated VHS, thus explaining why the
normal state peak is located at $\approx$30meV, even though the bifurcated VHS's
are only $\le$10meV from the Fermi level. In order to explain the 
experimental results, Blumberg, et al. found that they could not use a rigid 
band model, but that they had to adjust the parameters to pin the VHS near the 
Fermi level over an extended doping range.  [In contrast to the above papers, 
Liu, et al.\cite{LZL} were able to model the peak as a structure at $\sim 2
\Delta_0$, even though their $\zeta_{VH}\approx 100meV$.]
\par
In interpreting these features, caution is advised, since there is some
evidence that the gap is related to a pseudogap rather than to the
superconducting gap\cite{Bour}.
\par
None of the above calculations suggest why the peak is incommensurate.  One
possible explanation comes from the band structure predictions of {\it 
bifurcated} (or better, {\it extended}) VHS's, Figs. \ref{fig:a5} and 
\ref{fig:b5}, Sections IV.B.2,V.D.4.  If the VHS's in the superconducting state 
were similarly bifurcated, then they would be split into peaks at $(\pi\pm
\delta_x,0)$ and $(0,\pi\pm\delta_y)$, so inter-VHS scattering would have four 
incommensurate peaks at $(\pi\pm\delta_x,\pi\pm\delta_y)$ and $(\pi\pm\delta_x,
\pi\mp\delta_y)$, which is consistent with the observed symmetry of the 
diffraction peak, Fig. \ref{fig:15}b.  (Note that $\delta_x$ and $\delta_y$ need
not be equal.)  The fact that four well-resolved peaks are not found is 
consistent with the photoemission data,
which find extended VHS's, which spread continuously from, e.g., $(\pi +\delta_
x,0)$ to $(\pi -\delta_x,0)$, so the resulting inter-VHS scattering should
have the form of a smeared-out square, as observed.  The size of the square is
also in better agreement with the extended VHS: the two bifurcated VHS's in 
Fig. \ref{fig:b5}c are separated by 0.83 of the $\Gamma-Y$ distance (length of
arrow), whereas the width of the flat region of the extended VHS in Fig. 
\ref{fig:b5}d is 0.62, and the edge of the diffraction square in Fig. 
\ref{fig:15}b is 0.46 (all in the same units).  The agreement between the latter
two is quite reasonable, particularly since the data on the extended VHS are 
from Y-124, while the diffraction peak is in Y-123.  

\section{Correlation Effects and Slave Bosons}

The above calculations have all assumed a nearly-free-electron model.  How
are they modified in the presence of strong on-site Coulomb repulsion?
To date, most VHS calculations have included
correlations via the slave boson technique.  Hence, this section is devoted to
a detailed summary of that technique, while alternative calculations, which are
just beginning to address the VHS problem, are discussed in Subsection VII.E.
\par
In order to simplify the calculation of correlation effects, the three-band
model is often replaced by a single band, Hubbard or tJ model.  In this
case, the atoms are taken as copper, and the role of the oxygens must be 
carefully included, via effective hopping and exchange parameters.  Zhang and
Rice\cite{ZR} showed that much of the effect of the O's could be included by
hybridizing the Cu d-orbital with a symmetric combination of p-orbitals from
the nearest neighbor oxygens.  However, in the slave boson approach, the 
three-band model is not substantially harder to handle.  Moreover, there are
important electron-phonon couplings involving the O's, and care must be taken to
properly incorporate their effects into a one-band model.  For these reasons, 
one band models of correlation effects will not be discussed, except 
peripherally, in this review.  On the other hand, the oxygen breathing-mode
phonon, in which the four O's adjacent to a given Cu alternately move closer to
the Cu or further away, can be thought of as a dynamic Zhang-Rice effect.  Since
this phonon appears to play a significant role in the cuprates, this analogy 
will be discussed below.

\subsection{Slave Bosons in the 3-Band Model}
\par
On-site Coulomb repulsion $U$ acts to greatly reduce the probability of
double occupancy of an atomic orbital.  If $U$ is strong enough, this repulsion
can lead to a metal-insulator (Mott) transition in a half filled band, where
each atom has exactly one electron, and no electron can move without 
creating a situation of double occupany.  In the cuprates, the Fermi level lies
in a predominantly copper-like band, and is near half filling.  In this case,
the situation is more complicated, since even when $U\rightarrow\infty$ the
Cu can hybridize with the O's, and still have a metallic state.  However, if 
there is a large enough energy cost in hopping -- that is, if $\Delta /t_{CuO}$
exceeds some critical value -- then there can again be a phase transition at
half filling, this time to a {\it charge-transfer insulator}.
\par
This situation is found experimentally in undoped La$_2$CuO$_4$ and in YBCO,
$\delta =1$, where the insulating phase is also found to display 
antiferromagnetic order.  It {\it cannot} be reproduced by LDA calculations,
even if the magnetic order is included\cite{Pickt}.  Hence, a variety of 
techniques have been applied to the problem, to try to incorporate the strong
on-site correlations more accurately.  One popular technique, for both cuprates
and heavy-fermion materials, has been the slave boson technique\cite{slab}. This
method involves self-consistent renormalization of the hopping parameter, and 
leads to a charge-transfer insulator ($t_{CuO}\rightarrow 0$) at half filling, 
above a critical value of $\Delta /t_{CuO}$.  
\par
The simplest version of the theory assumes $U\rightarrow\infty$, where $U$ is 
the Cu Coulomb energy, while the O Coulomb energy $U_p$ can be neglected.
(Near half filling, very few holes are on the O's, so the probability of double
occupancy is already small.)  The earliest slave-boson calculations on the
cuprates\cite{KLR} neglected direct O-O hopping, but this has been included in
more recent calculations\cite{sad,RM3,Gri,New}.
The theory is formally a large-N theory, where N 
is the degeneracy of the electron states on a Cu site, and the calculation is 
usually carried out only to lowest order in $1/N$.  At this order, no magnetic
effects are included.  Moreover, $N\simeq 2$ for the cuprates, so it is 
doubtful whether the lowest-order predictions of the theory are quantitatively 
correct.  Nevertheless, detailed calculations are important -- for example,
to see how close the cuprates are to the metal-insulator transition, for a 
`realistic' choice of parameters.  
\par
Taking $U\rightarrow\infty$ is equivalent to forbidding double occupancy of the 
Cu's. In a slave boson calculation, this constraint is enforced by introducing a
boson to occupy the second orbital of each singly occupied site.  By satisfying
the constraint in mean field, it is possible to reduce the calculation to a
self-consistent renormalization of the bands.  The Cu-O hopping parameter is 
reduced by the renormalization, while $\Delta$ is renormalized to a lesser 
extent, so the antibonding band becomes more Cu-like.
\par
The equations of self-consistency can be written
\begin{equation}
r_0^2={1\over 2}[1-\sum_ku_k^2f_h(E_k)],
\label{eq:13a}
\end{equation}
\begin{equation}
\Delta_0-\Delta ={1\over 2r_0^2}\sum_ku_k^2f_h(E_k)(E_k-\Delta ),
\label{eq:13b}
\end{equation}
where the renormalized parameters are $t_R=r_0t_{CuO}$ and $\Delta$, 
$f_h(E_k)$ is the Fermi function, and $u_k$ is the d-wave amplitude of the 
wave function.  $E_k$ and $u_k$ are calculated using the renormalized 
parameters, and it is assumed that $t_{OO}$ is not renormalized.  Only the case
$T=0$ will be considered in the present paper.  As written, 
Eqs.~\ref{eq:13a} and \ref{eq:13b} are valid 
for a hole picture, so $f_h(E_k)=1$ for $E_k>E_F$, and $=0$ otherwise.  An 
equivalent electron picture can be constructed, but some care must be taken,
since the above sums are then over all three occupied bands\cite{RMXA}.
The sums are normalized per Cu atom, so for example
\begin{equation}
\sum_k(1)=6,
\label{eq:14a}
\end{equation}
that is, that there are three bands (spin 1/2), and
\begin{equation}
\sum_kf_h(E_k)=1+x.
\label{eq:14b}
\end{equation}
\par
Equations~\ref{eq:13a} and \ref{eq:13b} can be generalized to include spin-orbit
and electron-phonon coupling\cite{RMXA}.  These equations are solved 
self-consistently by guessing initial values for the
renormalized parameters and numerically performing the double integrals to allow
calculation of the corresponding bare values, then readjusting the initial 
guesses until the known values of the bare parameters are recovered.  
If $t_{R}$ is small enough, however, the equations simplify
to a scaling form, for which only a single parameter, $\Delta$, need be varied. 
Details are given in Ref. \cite{RMXA}.
\par
One subtle point of the slave boson calculations must be briefly discussed.  In 
order to have a well-defined $N\rightarrow\infty$ limit, some of the parameters 
must be renormalized by a factor $N$ or $\sqrt{N}$.  (Renormalized parameters 
will be denoted with a prime: $V^{\prime}=NV$.)  However, I 
believe that $t_{CuO}$ should be renormalized as
\begin{equation}
t_{CuO}^{\prime}=\sqrt{N\over 2}t_{CuO}.
\label{eq:15}
\end{equation}
The reasoning is that the Cu degeneracy is 2 for spin and $N/2$ for orbital 
states, and Eq.~\ref{eq:15} is consistent with the assumption that the Cu-O 
hopping does not involve spin flip.  This factor of two will be assumed 
throughout the rest of this paper.  A search of the literature shows that this 
factor underlies much of the controversy about whether the slave boson 
calculation produces a metal-insulator transition (MIT) at half filling, for 
reasonable parameter values.  Those papers which include the factor of two find 
a MIT; those which neglect this factor do not.  In the following sections, the 
good agreement with experiment, and with other theoretical approaches, only 
holds when Eq.~\ref{eq:15} is assumed.  A fuller discussion is found in Ref. 
\cite{RMXA}.

\subsection{Exchange Corrections}

\subsubsection{Inclusion of Exchange}

A failing of the conventional slave boson theory is that it does not predict the
antiferromagnetic properties of the insulating phase at half filling.  This is 
because the theory 
is valid in the limit $U\rightarrow\infty$, and in this limit, the exchange 
coupling, $J\propto 1/U$, vanishes.  In a systematic $1/N$ expansion, exchange 
terms only appear in order $1/N^2$.  However, both experiment\cite{Well} and
theory\cite{DagN,Naz1} (Section IV.B.6) have revealed the importance of these 
terms.  Thus, whereas the slave boson calculations predict that the bandwidth 
collapses (effective mass diverges) at the Mott transition when $J=0$, for 
finite $J$ the dispersion remains finite, even in the presence of a charge 
transfer gap, Fig.~\ref{fig:a8}.
\par
A finite $U$ can be incorporated into a slave boson calculation by going to a
four-slave-boson approach\cite{SB4}, assigning separate slave bosons for each
empty or doubly occupied or singly occupied site, with two bosons in the latter 
case to account for the electronic spin.  Alternatively, exchange can be 
approximately included by introducing a Heisenberg $J$ 
term\cite{DiDo,CaGr,CDG}, 
\begin{equation}
H^{\prime}=J\sum_{j,\hat\delta ,\sigma ,\sigma^{\prime}}
d^{\dagger}_{j,\sigma}d_{j,\sigma^{\prime}}
d^{\dagger}_{j+\hat\delta ,\sigma^{\prime}}d_{j+\hat\delta ,\sigma}
\label{eq:1aa}
\end{equation}
with a mean-field decoupling of the $d^4$-term.  Caprara and Grilli\cite{CaGr}
have called this a three-band tJ model.  They find that the dispersion remains 
finite at half filling, even though the optical conductivity has a gap.  

\subsubsection{Magnetic Phases}

\par
A variety of magnetic phases are possible, including uniform, flux\cite{Affl},
and N\'eel phases.  In the {\it uniform} and {\it flux} phases, there is a 
non-zero value of 
\begin{equation}
\Delta_{ij}\equiv \sum_{\sigma}<d_{i\sigma}d_{j\sigma}^{\dagger}>
=\Delta_1e^{i\theta_{ij}}.
\label{eq:13d}
\end{equation}
If the phases $\theta_{ij}$ around a plaquette add up to zero (modulo $2\pi$),
Eq.~\ref{eq:13d} describes a uniform phase; in the flux phase, the phase around
a plaquette is $\pi$ (again, modulo $2\pi$).  For the uniform phase, the 
self-consistent equations are as follows: Eq.~\ref{eq:13a} remains
unchanged; in Eq.~\ref{eq:13b}, the term $\Delta$ on the right-hand side of the
equal sign should be replaced by $\tilde\Delta =\Delta -2\Delta_1(\bar c_x+\bar
c_y)$; and there is a third equation, 
\begin{equation}
\Delta_1={-J\over 2}\sum_ku_k^2f_h(E_k)(\bar c_x+\bar c_y).
\label{eq:13c}
\end{equation}
The resulting energy dispersion is given by Eq.~\ref{eq:2a}, with the
substitution $\Delta\rightarrow\tilde\Delta$.
At half filling, for $\Delta >\Delta_c$, there is an insulating phase with
$r_0=0$, but $\Delta_1\ne 0$.  In the uniform phase, there is no phase
transition: $\Delta_1$ is non-zero for all temperatures, but monotonically
decreases in magnitude as $T$ is raised -- and also as the system is doped away
from half filling\cite{RM12} (in agreement with four-slave-boson 
calculations\cite{TIF}).  In the insulating phase, the energy dispersion is 
\begin{equation}
E=\Delta+2\Delta_1(\bar c_x+\bar c_y),
\label{eq:13e}
\end{equation}
Fig.~\ref{fig:a15}a.

Introduction of a phase change of $\pi$ around an elementary plaquette opens
a gap over most of the Fermi level, with the gap vanishing at one point
(actually, at a set of symmetry-equivalent points).  In these {\it flux phases},
the partitioning of the net phase change among the four links of the plaquette
is a gauge degree of freedom; surprisingly, the particular choice modifies the
electronic energy dispersion.  For instance, if the full phase change of $\pi$
is allocated to a single bond along $x$, this changes the sign of $J$ for that
bond ($e^{i\pi}=-1$), leading to a dispersion
\begin{equation}
E=\Delta+2\Delta_1\sqrt{\bar s_x^2+\bar c_y^2},
\label{eq:13g}
\end{equation}
Fig.~\ref{fig:a15}.  On the other hand, if the phase change is distributed
equally, with $\pi /4$ on each bond, the dispersion becomes
\begin{equation}
E=\Delta+2\Delta_1\sqrt{\bar c_x^2+\bar c_y^2},
\label{eq:13h}
\end{equation}
Fig.~\ref{fig:a15}c.  The contrast of these two cases is instructive: the former
gaps the flat part of the energy dispersion, but leaves the VHS ungapped; the
latter gaps the VHS, while having a zero gap in the middle of the flat
bands, at $(\pi /2a,\pi /2a)$.  Yet both gaps produce the same dos, and hence
the same energy lowering.  Thus, in the flux phases, {\it VHS nesting is
equivalent to conventional nesting}.  Despite this formal equivalence, {\it only
the VHS nesting is observed experimentally,} Fig.~\ref{fig:a8}a.

How can this symmetry breaking be understood?  One possibility is a term like
$t_{OO}$, which breaks electron-hole symmetry, and weakens the flat band 
nesting.  However, at half filling, $t_{OO}$ should not affect the dispersion, 
so a second-neighbor exchange, $J^{\prime}$, would be needed.  Alternatively, I
suspect a spin-Peierls interaction: it is the tilting of the CuO$_6$ octahedra
which introduces spin-orbit coupling and Dzyaloshinskii-Moriya exchange,
thereby stabilizing a long-range magnetic order.  A detailed understanding of 
this process does not yet seem at hand, but a discussion of the issues involved
is in References \cite{Bones}.  It seems likely that the resulting spin-phonon 
coupling will favor a particular flux phase.

There can also be a N\'eel phase, quite similar to the SDW phase of Kampf and 
Schrieffer\cite{KaSch}, but with the energies scaled by $J$ instead of $U$, with
gap function
\begin{equation}
\Delta_2\equiv (-1)^{i_x+i_y}\sum_{\sigma}<\sigma d_{i\sigma}^{\dagger}
d_{i\sigma}>.
\label{eq:13f}
\end{equation}
Note that $\Delta_1$ and $\Delta_2$ may be simultaneously non-zero at mean
field level, both in the uniform and flux phases, Fig.~\ref{fig:a15}d-f.  When 
a small N\'eel gap is added to the uniform phase, the
resulting dispersion matches that found for a single hole in the tJ 
model\cite{LMan,DagN}, Fig. \ref{fig:a8}b, but is in poor agreement with the
dispersion found experimentally in the insulator SCOC, Fig. \ref{fig:a8}a.  On 
the other hand, the dispersion of the flux phase matches that found in SCOC, 
Fig.~\ref{fig:a8}a\cite{Lau1,WeL,RM11}.  This is somewhat surprising, since
the flux phase is generally found to be stable away from half filling, whereas 
the experiments are in the Mott insulating phase, at half filling.  Since 2D
fluctuations are known to eliminate the spin wave gap in the AFM phase, it has 
been hypothesized that the fluctuation-corrected AFM phase will resemble the 
gapless flux phase\cite{Wang0}, but so far fluctuations have not been included
in the theory.  More recently, a good fit to the dispersion has been obtained
in the model of a single hole in an antiferromagnetic background\cite{VoGo}.
However, this model involves a mixture of hopping and exchange parameters, 
whereas the 3-band tJ model would suggest that only the exchange parameters
affect the dispersion near half filling.
\par
The magnitude of the predicted bandwidth can also be estimated.
Using the mean field decoupling, the equilibrium value of $\Delta_1$ in the
uniform phase is\cite{CaGr} $2J/\pi^2=0.203 J$, which is smaller than that found
in Ref. \cite{DagN}, $\Delta_1\simeq 0.55J$.  However, if Eq.~\ref{eq:1aa} is 
instead decoupled via a Jordan-Wigner transformation\cite{Wang0}, $\Delta_1/J=
(1/2+2/\pi^2)=0.703$, in better agreement.  (In attempting to reproduce the 
calculation of Ref. \cite{CaGr}, I found\cite{RM12} $\Delta_1=4J/\pi^2=0.406J$.)
\par
The consequences of these novel magnetic phases for a VHS model 
have yet to be worked out; to understand the competition between magnetic and 
structural instabilities, it will be necessary to include a spin-Peierls 
coupling, Section VIII.E.

\subsection{Numerical Results}

\par
A charge transfer insulating phase can occur only at exactly half filling, just
as in the one-dimensional Hubbard model\cite{LW}.  The transition occurs only if
$\Delta_0$ exceeds a critical value $\Delta_c$.  For $t_{OO}=0$, $\Delta_c$ can
be found analytically\cite{Cast}
\begin{equation}
{\Delta_c\over t_{CuO}}=4\sqrt{{1\over 2}+{2\over\pi^2}}=3.353,
\label{eq:16a}
\end{equation}
or $\Delta_c=4.36eV$ when $t_{CuO}=1.3eV$. 
\par
For $t_{OO}\ne 0$, the critical value must be evaluated numerically\cite{RMXA}
\begin{equation}
\Delta_c=3.353t_{CuO}+2.94t_{OO}.
\label{eq:16b}
\end{equation}
For $\Delta_0>\Delta_c$, the Fermi level jumps discontinuously as $x$ passes 
through zero, with the discontinuity being the charge-transfer gap.
Away from half filling, renormalization pins the Fermi level close to the 
VHS\cite{RM3,New,QSi}.  
\par
These results are illustrated in the numerical calculations of Fig.
~\ref{fig:16}, assuming $t_{OO}=0.25eV$, as $\Delta_0$ increases from $4eV$ 
(solid lines) to $5eV$ (dashed lines) to $6eV$ (dot-dashed lines). Figure 
~\ref{fig:16}a shows that $t_R$ renormalizes to zero exactly at half filling 
($\Delta_0=5eV$ is just below the transition, so $t_R$ has a small but finite 
value at half filling).  Near half filling, the Fermi energy changes rapidly 
(Fig.~\ref{fig:16}b), from a value near $\Delta_0$ (Cu-like 
carriers) when $x<0$ to a much smaller value (more O-like) for $x>0$.  This
change is discontinuous above the Mott transition; this is the charge-transfer 
insulator gap\cite{KLR}, and should be compared to the one-dimensional Hubbard 
model (see Fig. 3 of Ref. \cite{Recon}).  
\par
This result is consistent with the results of other, non-slave boson 
calculations of the three-band model. Thus, the 
studies that produced the parameters of Table III\cite{Hyb,GraM,Esk,MAM} also 
explored the low energy sector of the three-band model, and all find an 
insulating state at half filling with a large charge transfer gap, comparable to
experiment.  See also Refs. \cite{BC}, which find similar results.  
\par
As $x$ increases, $t_R$ grows $\sim\sqrt{x}$ while $\Delta$ decreases, 
ultimately becoming negative.  At the VHS, $\Delta$ is small and positive.
This is consistent with LDA calculations, suggesting that the LDA calculations
should be reasonably accurate away from half filling.  Note the importance of
self-consistency.  By extracting parameters from the LDA calculations which are
not renormalized by correlation effects, a large $\Delta$ value is found.  Using
these parameters in the slave boson calculation, it is possible to explain both
the MIT at half filling and the small renormalized value of $\Delta$ near the 
VHS.  If, on the other hand, one had merely chosen tight-binding parameters
which reproduced the band structure near the VHS, a slave boson calculation
utilizing those parameters would have found a metallic state at half filling.
\par
When $\Delta$ becomes negative, the Fermi level falls directly in the O-like 
band.  In this regime, the three-band model should not be trusted, because of 
the proximity of the $d_{z^2}$ band to the Fermi level.  Indeed, in this doping 
regime an enhanced $d_{z^2}$ character is often observed 
experimentally\cite{DFG}.
\par
The quasi-pinning of the Fermi level to the VHS is illustrated in Figs. 
~\ref{fig:16}c and d.  In a rigid-band filling picture, the bandwidth and the 
position of the VHS would both be independent of doping, so the separation 
between the VHS and Fermi level would smoothly track the variation of the Fermi 
level with doping.  This is not what happens in the presence of strong 
correlation effects.  Figure~\ref{fig:16}c shows that the position of the VHS 
(defined by the hole doping $x_{VHS}$ at which the VHS would coincide with the 
Fermi level) actually changes with doping, while Fig.~\ref{fig:16}d shows the 
energy separation between the Fermi level and the VHS, $\Delta E=E_F-E_{VHS}$.  
Note for example the case $\Delta_0=6eV$: over the doping range $x=-0.1$ to $x=
+0.3$, the Fermi energy varies by $4eV$, while $|\Delta E|\le 10meV$! In 
general, the pinning range extends from half filling to a finite 
doping where the VHS again intersects the Fermi level, Fig. \ref{fig:16}d.  This
doping, which in the VHS model corresponds to optimum doping, is fixed by some 
electron-hole asymmetry parameter, $t_{OO}$ or $t^{\prime}$.  In the limit $t_
{OO},t^{\prime}\rightarrow 0$, this second point collapses to half filling.  The
pinning is further illustrated in Fig.~\ref{fig:a16}\cite{New}, which shows the
dos vs energy for several different dopings.  Clearly there is a large shift in
the absolute Fermi level, $E_F$, while $E_F-E_{VHS}$ has a much smaller change.
Thus, correlations strongly pin the VHS near the Fermi level, especially when a 
Mott transition occurs at half filling.  Note further that the shift
$\Delta E$ is not a monotonic function of doping, but actually vanishes a
second time at a finite $x$ (Fig.\ref{fig:16}d). 
\par
In these calculations, part of the pinning is due to a genuine renormalization 
of the shape of the Fermi surface with doping, but part simply arises from the
collapse of the bandwidth to zero at half filling.  In this sense, the results
of including exchange in a three-band tJ model are even more interesting,
Fig.\ref{fig:19}\cite{RM12}.  A strong pinning of the Fermi level to the VHS is
still found.  But here, the bandwidth remains finite at half filling, so the
pinning is mainly due to changes in the shape of the Fermi surface.  Indeed, 
from Eq~\ref{eq:13e} it can be seen that the VHS falls exactly at the Fermi
level in the uniform phase at half filling.  The double crossing of Fermi level
by the VHS (Fig.\ref{fig:16}d) is also found in the present case, now associated
with a crossover from magnetically-dominated behavior (bandwidth $B\sim J$) at 
half filling to a more metallic behavior ($B\sim t$) in the doped material.
\par
This strong pinning of the VHS near the Fermi level has been found in a
number of slave boson calculations\cite{RM3,New,QSi}, as well as in 
infinite-dimensional calculations\cite{inD,MaKr}, correlation 
calculations\cite{Pines,Dich} which do not involve slave bosons, and 
renormalization group calculations\cite{rgpin1,rgpin2}. Recent 4-slave boson 
calculations of the one and three band Hubbard models find even stronger 
pinning\cite{pinn}.  This pinning can be described as an effective mass 
renormalization\cite{MaKr}, $\Delta E =\Delta E_0/m^*$, where $\Delta E_0$ is 
the bare splitting and $m^*$ an effective mass.  If there is a charge reservoir,
this pinning can even alter the hole density in the CuO$_2$ planes\cite{rgpin1}.
This may explain some anomalous results, such as the absence of BiO pockets in 
the Fermi surface of Bi-2212, which disagrees with band structure calculations.

\subsection{How Good are Slave Boson Calculations?}

\subsubsection{Comparison with Monte Carlo Calculations}

A number of objections have been raised to the slave boson
approach for calculating correlation effects.  The main problem is that the
constraint of no double occupancy is satisfied only on average.  Zhang, et 
al.\cite{ZJE} have suggested that, if the constraint is satisfied exactly, then
the metal-insulator transition would arise for all $\Delta >0$, rather than
only above a critical threshold, $\Delta >\Delta_c$.  It is therefore important 
to test the predictions of the slave boson theory, particularly near half 
filling, where correlation effects are strongest.
\par
One test is to compare the predictions of slave boson theory with predictions of
correlation effects by other means.  A particularly attractive comparison is 
with Monte Carlo calculations on finite sized clusters, since these calculations
are essentially exact, except for being limited to finite size and not-too-low
temperatures.  These calculations assume $t_{OO}=0$, and a large but finite
value for the on-site Coulomb repulsion, $U_d=6t_{CuO}$.  Figure~\ref{fig:17} 
shows a comparison of the Fermi energy vs doping, for several different values 
of $\Delta$, for both cluster calculations (symbols)\cite{Dopf,Scal} and slave 
boson calculations (lines).  Since this is theory vs theory, the
comparison has essentially no adjustable parameters.  The agreement is fairly
satisfactory.  Both theories find evidence for an insulating gap at half 
filling, above a critical value of $\Delta$. The cluster calculations find a
smaller $\Delta_c\simeq 2t_{CuO}$.  The critical value $\Delta_c$ in the slave
boson calculation would be reduced slightly by including exchange
corrections\cite{CaGr}.  Both calculations display the same overall 
doping dependence.  The present comparison would therefore tend to validate both
calculations (the cluster calculation, in that finite size and finite
temperature effects do not seem to limit their validity).

\subsubsection{Comparison to Experiments}

\par
A separate test of slave boson theory is to compare its predictions with 
experiment.  An important test is to see how well theory fits the 
photoemission-derived energy
band dispersions.  Estimating the bare parameters from LDA calculations, 
Fig.~\ref{fig:18}a compares the experimentally-determined dispersion for 
Bi-2212 with both the bare band dispersion, and the slave boson corrected 
dispersion, assuming the bare parameters are the same as those found for LSCO.  
The overall reduction in bandwidth is well reproduced by the slave boson 
calculation, but {\it it cannot account for the extreme flatness of the bands in
the immediate vicinity of the Fermi level} (the extended VHS).
\par
Since the extended VHS's have much stronger singularities, it is important to
try to understand just how they might be produced.  One possibility is that
this is due to interlayer coupling.  For instance, in Bi-2212 there is a 
Bi-derived Fermi surface pocket immediately above the Fermi level at the VHS.  
In principle, level repulsion from this pocket could make the VHS anomalously
flat.  However,
tight-binding models are unable to reproduce such a feature: the dos always
has a logarithmic divergence, independent of the position of the second band,
and the magnitude varies smoothly and monotonically as the degree of overlap
between the two bands is varied\cite{RM6}.  Also, since the extended VHS seems 
to be a common feature of many high-T$_c$ cuprates, it would be preferable to 
have an explanation which depends on the CuO$_2$ planes only.  
Fig.~\ref{fig:18}b suggests one such explanation: a polaronic renormalization of
the electronic bandwidth within a Debye frequency of the Fermi level\cite{RMXB}.
Somewhat surprisingly, the polaronic enhancement is found to be pinned to the
VHS, and not to the Fermi level.  Hence, it can also explain the case of
YBa$_2$Cu$_4$O$_8$, where the extended VHS is found 19meV
below the Fermi level.

\subsubsection{Doping Dependence of Photoemission in YBCO}

\par
The doping dependence of angle resolved photoemission spectra was studied
early on in YBCO\cite{Liu2}, and recently in Bi-2212\cite{BAPS,Biund,Gp1,Gp2}.  
The results in these two experiments are strikingly different.  In Bi-2212, the
spectra are dominated by the opening of a pseudogap in the normal state above
the superconducting transition temperature, while such effects are not seen
in the experiments on YBCO.  With recent improvements in sample quality
and photoemission resolution, I would strongly urge that the experiments on
YBCO be reproduced (note that in these experiments, a feature associated with 
the VHS was not identified).  However, it is striking how well the data on
Bi-2212 agree with the VHS theory of pseudogap formation, while the YBCO data
are for the most part consistent with slave boson calculations in the absence 
of a pseudogap.  The former data will be discussed in Section IX.A.2, and the
latter in this section.
\par
Photoemission in YBCO provides strong evidence for the pinning of the Fermi 
level to the VHS.  Indeed, {\it there is virtually no shift of the Fermi 
surfaces with doping}!  Experimentally, it is found\cite{Liu2} that a strong 
feature near the $X$ and $Y$ points $\approx$1eV below the Fermi level
shifts by 200meV as the oxygen is reduced from O$_{6.9}$ to O$_{6.3}$, 
Fig.~\ref{fig:d19}.  At the same time, the saddle point feature near the
$X(Y)$-point is within resolution limits locked to the Fermi level, shifting by
less than 10meV over the same doping range.  Figure~\ref{fig:a19} shows the
actual photoemission spectra along $\Gamma - Y(X)$, both for Y-124 (a),
YBCO$_{6.9}$\cite{Camp} (b), YBCO$_{6.5}$ (c), and YBCO$_{6.3}$ (d)\cite{Liu2}.
It can be seen that the VHS is nearer to the Fermi level in YBCO than in Y-124,
but that there is no evidence for a shift in this feature with doping in YBCO.
While the intensity is largely lost by O$_{6.3}$, there is still a clear 
shoulder which has the same dispersion as in the more highly doped samples.
\par
This pinning of the Fermi level near the VHS is predicted by the slave boson
calculations, Fig.~\ref{fig:16}d.  Figure~\ref{fig:19} shows the expected band 
dispersion for LSCO as $x$ is varied from the VHS to half filling.  The total
shift of the Fermi level from the VHS is only 16meV -- quite comparable to 
experiment.  However, experimentally it is also found that the bandwidth is 
nearly independent of doping, which is not reproduced by the slave boson 
calculations of Fig.~\ref{fig:16}a (the bandwidth is proportional to $t_R^2$).
This can be very nicely explained by the inclusion of exchange corrections.
Figure~\ref{fig:19} shows the band dispersion calculated self-consistently at a
number of doping levels assuming the uniform magnetic phase (Eq.~\ref{eq:13d}).
All curves are referenced to the Fermi level, taken as $E_F=0$.  For 
electron-doped samples (dashed lines) the calculated bandwidth changes rapidly
with doping, decreasing as half filling is approached, signalling the transition
to an insulating phase.  However, at half filling, a finite bandwidth remains,
due to the magnetic excitations, with bandwidth $\sim\Delta_1\sim J$.  On the
hole doping side, the band dispersion of the {\it filled states} is nearly
doping independent out to a doping beyond the VHS, while the dispersion of the
unoccupied states (not visible in photoemission) increases rapidly with doping.
This behavior is in quite good agreement with photoemission in the metallic
regime in YBCO.  Experimentally, removing oxygen leads to negligible change in 
the band dispersion, until $\delta\simeq 0.65$, at which point the conduction 
bands abruptly vanish\cite{Liu,Liu2}.  While I am comparing the theory for LSCO 
with experiments on YBCO, this is not expected to make any qualitative 
difference in the band dispersions.  
\par
In the slave boson calculation, the absolute position of the Fermi level,
Fig.~\ref{fig:16}b, is calculated with respect to an assumed zero in the
predominantly oxygen-like nonbonding band -- the calculation of the absolute
position of the Fermi level, with respect to all the other bands in the crystal
is beyond the scope of slave boson theory.  However, the general feature of
relatively weak variation in the hole doped phase, with a strong/discontinuous
jump at the insulating state at half filling, would be expected to persist.
The sudden vanishing of the dos for YBCO$_{6.3}$ would be consistent with this
opening of the Mott gap -- recall from Section VI.B that the hole doping of the
planes vanishes near O$_{6.3}$.  However, a photoemission peak at the gap energy
was not observed in these experiments, and another motivation for repeating the
experiments on YBCO would be to identify a feature associated with the charge
transfer gap, and study how it varies with doping.
\par
Similar experiments have been performed near the metal-insulator crossover,
in a number of d=1 Mott-Hubbard systems\cite{Mottran1}, including Ca$_{1-x}$Sr$
_x$VO$_3$\cite{Mottran}, Fig.~\ref{fig:b19}.  Instead of a gap at precisely 
half filling suddenly disappearing and being replaced by intensity at the
Fermi level, what is observed in the spectrum is a 
gradual reduction of the insulating (gapped) part of the spectrum, while the 
metallic fraction at the Fermi level gradually increases.  The intensities of 
the two components change with doping, but the respective lineshapes are 
approximately invariant.  This is not consistent with the slave boson 
calculations, and suggests an alternative interpretation in terms of
localization or phase separation, discussed below in Sections X and XI.  
Thus, near half filling, the spectrum would consist of two components, Fig.~
\ref{fig:d19}b, a metallic component similar to that in optimally doped 
materials, plus a pair of narrow bands, separated by the charge transfer gap, 
characteristic of the antiferromagnetic insulating phase.  The metallic feature
would gradually increase in intensity as doping is increased, having the
appearence of a midgap feature.  This is consistent with Fig.~\ref{fig:b19}, and
suggests 
that phase separation is a common feature near the Mott transition, although it 
is probably not necessary that the second phase be tied to a VHS.  There are,
however, some materials which do show an effective mass divergence at the
metal-insulator transition, as discussed in Subsection VII.D.5 below.
\par
\par
The shift $\Delta E$ of the `1eV' feature with doping (Fig. \ref{fig:d19}c) is 
also interesting, and may provide a clue to the origin of the peak.  This will 
be further discussed in Section XI.A.3 below, when the picture
of phase separation in YBCO has been discussed. 

\subsubsection{Chemical Potential}

\par
Several groups have measured the doping dependence of the chemical potential
$\mu$. However, comparison of the experimental results with slave boson 
calculations is non-trivial since the calculations ignore Coulomb effects.  In 
an elemental material such as Si, the approximation can be justified since the 
dopant ions are assumed to form a uniform background, keeping the material 
neutral on average.  In a layered material, however, this is a more severe 
approximation, since the doping counterions are located on different layers, so 
changing the hole doping shifts the overall potential of the CuO$_2$ planes with
respect to the other planes, in ways which are ignored in slave boson 
calculations.  Moreover, additional screening can be produced by polarizing 
atoms on other layers.  F-or instance, the apical O's move 
closer to the planes with increasing hole doping, and in overdoped materials 
part of the holes go to the apical O's rather than to the planes.
\par
Experimental evidence for this is seen in Bi-2212, where there is clear evidence
for non-rigid-band filling: the valence band, the Bi($5d$) core level, and the 
Pb($5d$) core level all shift by different amounts with 
doping\cite{One2,pot2,pot3,pot4}. Thus, it is difficult to evaluate the 
significance of the finding that the Fermi level is nearly unchanged in absolute
value between hole-doped LSCO and electron-doped NCCO\cite{pot1}.  While one 
would initially expect that the Fermi level would shift by the charge transfer 
gap, the complete elimination of the apical O's should partially compensate for 
the shift, in a way which lies beyond the realm of slave boson calculations.  An
attempt to analyze these level shifts in terms of bond-valence sums is presented
in Ref.~\cite{pot2}, but the dominant effect in Bi-2212 seems to be a chemical 
potential shift\cite{pot3,pot4}.
\par
There is one interesting observation which might find an explanation in the 
slave boson calculations.  In LSCO\cite{pot5,pot6} and NCCO\cite{pot1}, there is
very little shift of the chemical potential with doping, whereas in Bi-2212,
there is a small shift in the metallic regime, but a substantial shift as the
insulating phase is approached.  This behavior should be compared with 
Fig.~\ref{fig:16}b: when there is a Mott transition at half filling, the Fermi
level is very weakly dependent on doping in either the hole-doped or the
electron-doped regimes, but with a discontinuous change at exactly half filling.
However, this discontinuity requires a critical value of the Cu-O energy
separation, $\Delta\ge\Delta_c$.  For $\Delta$ slightly less than $\Delta_c$,
there is no discontinuity in the chemical potential, so {\it it varies rapidly
in the immediate vicinity of half filling}.  This is consistent with the 
suggestion that the AFI phase is harder to produce in Bi-2212 because $\Delta
<\Delta_c$.
\par
An additional complication in interpreting the chemical shifts is the 
possibility of phase separation, which could lead to flat (doping independent) 
segments of $\mu$ for isolated CuO$_2$ planes.  However, nanoscale phase 
separation is actually more correctly viewed as a charge-bunching phase (as in 
a striped phase, Section X.B.3); as such, $\mu$ can still depend on doping.  
Indeed, even these measurements of $\mu$ find indications of phase separation.  
Thus, van Veenendaal, et al.\cite{pot3}, in studying the Cu $2p^{3/2}$ core 
level of Bi-2212, found a broadening which could indicate such an inhomogeneity,
but {\it only in the CuO$_2$ planes}!  This is precisely what would be expected 
from the models of hole-driven phase separation discussed in Section X below.  
Furthermore, there is evidence for two Sr environments in LSCO\cite{pot6}.

\subsubsection{Effective Mass Divergence}
\par
Another experimental observation also points to the possibility of phase
separation.  The slave boson calculation predicts that the Mott-Hubbard 
transition involves the collapse of the electronic bandwidth as Cu-O 
hybridization is eliminated.  Experimentally, this should show up as a 
divergence of the effective mass, and hence of the dos, as half filling is
approached.  Such a divergence is indeed observed in the linear-T coefficient of
the specific heat, $\gamma$, for La$_{1-x}$Sr$_x$TiO$_3$ and Y$_{1-x}$Ca$_x$TiO$
_3$\cite{Kuma}, but is not found in the cuprates.

\subsection{Other Theories of Correlations}

\par
As discussed above, slave boson calculations predicted that correlation effects
produce a strong doping dependence of the electronic bandwidth, in particular, 
with hopping $t\rightarrow 0$ as half filling is approached.  Despite this,
the holes have a well-defined dispersion, and there should be an enhanced VHS,
pinned near the Fermi level.  Recent Monte Carlo calculations of the tJ and 
Hubbard model have substantially confirmed these predictions, and verified the 
important role played by auxiliary parameters such as $t_{OO}$ or $t^{\prime}$
which break the electron-hole symmetry.  These calculations have further
shown the importance of magnetic effects near half filling, which provide a 
finite residual bandwidth $\approx 2J$ in the insulating limit.  Such magnetic
effects can be incorporated in the slave boson scheme via the 3-band tJ model.
\par
Dagotto, et al.\cite{DagN} have shown that when there is an antiferromagnetic 
gap, each of the subbands has a well-defined VHS; they call their theory, which 
combines spin fluctuations with the VHS, ``the best of both worlds''.  They
find that the resulting bands are particularly flat, and can give rise to an
extended VHS -- which in turn can generate a high-T$_c$ d-wave superconductor
from a {\it phonon}-mediated pairing mechanism\cite{NaDa}!   However, Ioffe
and Millis\cite{IoMi} have pointed out a problem:  to generate the requisite
superlattice, the antiferromagnetic fluctuations must be quasistatic, with
energies less than $k_BT$.  This in turn would lead to long-range AFM order in 
the limit $T\rightarrow 0$.  In contrast, near optimum doping, NMR finds spin
fluctuations only at energies larger than $k_BT_c$.  On the other hand, there
is ample evidence (Section IX) for the short-range {\it structural} disorder,
which can also produce extended VHS's and pseudogaps.

\section{Electron-phonon Interaction}

The role of electron-phonon interaction in the high-T$_c$ cuprates has been
vigorously debated. There is considerable experimental evidence for structural 
instabilities and other phonon related anomalies; of particular interest are
anomalies near T$_c$.  However, anomalies at T$_c$ are always expected, and do 
not necessarily indicate that the electron-phonon coupling is strong enough to 
cause the elevated transition temperatures.  Moreover, structural instabilities 
are common to a wide variety of mostly non-superconducting perovskites, and
can be due predominantly to phonon anharmonicity, with little or no electronic
contribution\cite{ZV}.  The purpose of this chapter is to show theoretically how
electron-phonon coupling could play a strong role in the cuprates; the following
chapter will present experimental evidence that this is indeed the case.
\par
A great number of theoretical models of electron-phonon coupling have been 
applied to the cuprates -- some involve particular phonons, most often 
breathing mode or apical oxygen related, while others are more generic models 
of polaronic or bipolaronic coupling.  Several conferences have been devoted to 
this topic\cite{eph}.  In this section, I will concentrate on the 
electron-phonon couplings associated with the VHS -- in particular, VHS nesting 
or Jahn-Teller effects.

As discussed in Section VI.A above, the VHS leads to $ln$ or $ln^2$ 
singularities in the charge and spin susceptibilities.  In this section, the 
singularity in the charge susceptibility will be explored.  In principle, this 
can lead to both structural phase transitions and an enhanced electron-phonon 
coupling parameter $\lambda$.  
\par
This is a long and important chapter because the phonon coupling has proven
to be more subtle than expected.  In conventional superconductors, it will be
recalled, there is a strong competition in the charge channel between
superconductivity and structural instability in the form of a combined CDW
Peierls instability.  In LSCO, the evidence for a CDW was ambiguous.  The
LTO phase was initially identified as a CDW phase\cite{LB,LTOCDW}, but this 
identification was abandoned when it was realized that the VHS degeneracy is not
split in the LTO phase\cite{Poug}.  No more promising candidates appeared in the
other cuprates.
\par
A first hint came from the discovery of the LTT phase in La$_{2-x}$Ba$_x$CuO$_4
$, which was found to strongly suppress T$_c$\cite{LTT1}, and the subsequent 
realization that this phase does split the VHS degeneracy\cite{LTT,LTTB,LTTP}.  
Based on this idea, the model of a Van Hove -- Jahn-Teller (JT) effect was 
introduced\cite{RM8a,RM8b,RM8c}, and it was postulated that
both the LTO and HTT phases could be {\it dynamic} JT phases, with important
short-range order.  It was gradually realized that the main transition, 
corresponding to a CDW in other systems, was the onset of short-range order in
the HTT phase, which could be identified with the experimentally observed
pseudogap phase.  A similar phase is found in YBCO.  Experimental evidence for
the pseudogap is presented in Section IX.A.
\par
Analysis of the LTT transition is complicated because the soft mode, the tilting
of the CuO$_6$ octahedra, couples quadratically to the electrons (non-Migdal 
behavior)\cite{BB,RM8b,SongAn}.  A detailed analysis\cite{RMXB} of the ionic 
model finds that there is strong linear coupling to a variety of in-plane modes,
but as these soften, they couple to the lower frequency tilt modes, and drive 
them soft.  These planar modes, discussed in Subsection D below, can display 
strong {\it polaronic} coupling effects.  Indeed, the extended VHS's may be 
generated by polaronic band narrowing.
\par
This section will attempt to discuss these features in detail.  Subsection
A discusses the special features of VHS nesting which are responsible for the
pseudogap and for nanoscale phase separation.  The various phonons which can 
couple to a VHS are discussed in Subsection B.  Subsection C describes the
quadratic coupling to tilt-mode phonons, and the interpretation of the LTT and 
LTO phases of LSCO as static and dynamic VHS-JT phases, respectively.  In 
Subsection D,
it is shown that, in the ionic model, there is a linear coupling to planar 
stretch modes with ensuing polaronic effects.  Complications due to correlation
effects are pointed out in Subsection E, while an alternative 
structural-instability model, involving an apical oxygen double well, is briefly
discussed in Subsection F.  Following this, Section XI provides a summary of 
experimental data, particularly in regard to pseudogaps, short-range order, and 
anomalies near T$_c$.  

\subsection{VHS Nesting}

\subsubsection{Pseudogap Formation}

Early calculations of VHS-induced structural distortions assumed the presence of
a zone-boundary ($\vec Q$-point) phonon which has a strong linear coupling to 
the VHS.  While no specific phonon was assumed, the gap $\vec q$-dependence is
consistent with that of an O-O stretch or breathing mode (Appendix C).  These 
calculations predicted a pseudogap formation (Subsection IX.A) and a 
phonon-induced phase separation (Subsection X.A) when doping away from the VHS.
Pseudogap formation follows from inspection of the (nesting-modified) Migdal
expression for the electron-phonon-induced self energy\cite{SuMo,RM5},
\begin{eqnarray}
\Sigma (\vec k,\omega)={1\over N}\sum_{\vec q}|g({\vec q})|^2{\omega_{\vec q}
\over\Omega^{R}_{\vec q}}\bigl({f(E_{\vec k-\vec q})+n(\Omega_{\vec q}^{R})
\over\omega +\mu -E_{\vec k-\vec q}+\Omega_{\vec q}^{R}+i\delta}+ \nonumber \\
{1-f(E_{\vec k-\vec q})+n(\Omega_{\vec q}^{R})
\over\omega +\mu -E_{\vec k-\vec q}-\Omega_{\vec q}^{R}+i\delta}\bigr),
\label{eq:rev1}
\end{eqnarray}
where $g(\vec q)$ is the electron-phonon interaction constant, $\omega_{\vec q}$
is the bare phonon frequency, $\Omega_{\vec q}^{R}$ is the phonon frequency 
renormalized by electron-phonon interaction, $n(\Omega )$ is the Bose
distribution function, and $\delta$ is a positive infinitesimal.  The nesting
is accounted for by the renormalization of the phonon frequency, with an
instability arising when $\Omega_{\vec q_0}^{R}$ first vanishes at some nesting
vector $q_0$ (for VHS nesting, $q_0\simeq Q\equiv (\pi /a,\pi /a))$.  To lowest 
order (RPA),
\begin{equation}
\Omega^{R2}_{\vec q}=\omega_{\vec q}^2-4\omega_{\vec q}|g(\vec q)|^2\chi (\vec
q),
\label{eq:rev2}
\end{equation}
where $\chi$ is the bare susceptibility, Eq.~\ref{eq:02}.  The Mermin-Wagner 
theorem can be satisfied by including fluctuations via the mode-mode coupling 
approximation, while a true three-dimensional phase transition can be restored 
by including interlayer coupling effects.  This calculation has previously been 
used to describe the Peierls transition in one dimension\cite{LRA,SuMo}, and was
applied to the  cuprates in Ref. \cite{RM5}.  The resulting pseudogap is 
illustrated in Fig.~\ref{fig:c22}d below.
\par
Thus, VHS nesting requires corrections to Migdal's theorem (the usual Migdal
expression for the self-energy acquires a divergent denominator).  This can be
understood very simply.  The compressibility $\kappa$ is proportional to the
dos, and hence diverges at a VHS -- this is the fundamental basis for nanoscale
phase separation away from the VHS (see discussion in Appendix A).  But in Fermi
liquid theory, a divergence in the compressibiity is related to a divergence in
the static vertex function\cite{AGD}, and hence a breakdown of Migdal's 
theorem\cite{CapaPie}.
\par
In the adiabatic approximation (neglecting $\Omega_{\vec q}^R$ in the energy
denominators), and high-T limit, the self energy becomes
\begin{equation}
\Sigma (\vec k,\omega)\simeq{2k_BT\over N}\sum_{\vec q}|g(\vec q)|^2{\omega_{
\vec q}\over\Omega^{R2}_{\vec q}}\bigl({1\over\omega +\mu -E_{\vec k-\vec q}
+i\delta}\bigr).
\label{eq:rev3}
\end{equation}
The structural instability is driven by the $\Omega^2$ factor in the 
denominator, which can usefully be rewritten
\begin{equation}
\Omega^{R2}_{\vec q}\simeq\omega_{\vec q_0}^2\bigl[R+\xi^2(T)|\vec q-\vec q_0|^
2\bigr],
\label{eq:rev4}
\end{equation}
where $\xi (T)$ is a coherence length\cite{McM1} and $R\equiv (\Omega_{\vec q_0}
^R/\omega_{\vec q_0})^2$ (Eq.~\ref{eq:0}) is the renormalization factor at the 
nesting vector which vanishes at the structural instability temperature $T^*$.
Within the VHS model there is a pseudogap instability at a finite temperature 
at both half filling and at the VHS due to VHS nesting, but as soon as the
material is overdoped, nesting becomes considerably worse, so the structural 
instability (pseudogap) $T^*(x)$ will rapidly vanish at some critical doping 
$x^*$.  This is a quantum critical point (QCP) since, when $T^*$ is
low, quantum corrections become important.  From the observed pseudogaps, Fig.
\ref{fig:23} below, the doping of optimum superconducting T$_c$ falls close to
this critical point in YBCO, and further away in LSCO.  Recently, Perali, et 
al.\cite{PCCG} have explored the consequences of introducing such a critical 
point in the overdoped regime, but without assuming that the structural 
instability is related to VHS nesting.

\subsubsection{Phase Separation}

\par
Phonon-induced phase separation was adduced from a somewhat different formalism,
commonly used in studying alloy phases and structures\cite{Haf}.  Here, the
electronic ground state energy is expanded in a power series in the single
ion pseudopotential $w(\vec q)$:
\begin{equation}
E_e=E_0+E_1+E_2+... ,
\label{eq:rev5}
\end{equation}
with $E_n\propto w^n$ (Eq. 2.12 of Ref. \cite{Haf}).  The term $E_1$ depends on 
the ionic density, but not the arrangement of the ions, so the leading 
structure-dependent term is $E_2$, which can be written as
\begin{equation}
E_2={1\over 2N}\sum_{\vec q\ne 0}{\chi (\vec q)\over \epsilon_e(\vec q)}|w(\vec
q)|^2|S(\vec q)|^2,
\label{eq:rev6}
\end{equation}
where $\epsilon_e$ is the dielectric constant, defined below Eq.~\ref{eq:02},
and $S(\vec q)$ is the structure factor of the ions in a unit cell.  In the
theory of Hume-Rothery phases (discussed in Section 7.2 of Hafner\cite{Haf}),
the series of structural phase transitions as a function of doping is explained
by writing the electronic energy up to order $E_2$, and then noting that each 
phase has a Kohn anomaly (coming from $\chi$), and hence a minimum in the free 
energy when the Fermi surface intersects the Brillouin zone boundary.  When
the free energies are calculated for a series of different structures as a 
function of doping, it is found that the Kohn anomaly minima all occur at 
different dopings depending on the particular Brillouin zones. This leads to
a phase separation, with regions of two phase coexistence determined 
by the common tangent construction (as in Fig. \ref{fig:j23} below).
\par
When the above formalism is applied to the cuprates\cite{RM3}, the logarithmic
divergence in $\chi (\vec Q)$ ensures that $E_2$ will have a minimum at the VHS,
making this a very stable phase,
so that either under- or overdoping will lead to phase separation.  Note that
the formalisms for $\Sigma$ and $E_e$ above are not exactly equivalent. However,
by identifying $g(\vec q)\simeq w(\vec q)$, then expanding Eq.~\ref{eq:rev3} to
lowest order in $|w(\vec q)|^2$ yields an expression approximately equivalent to
Eq.~\ref{eq:rev6}.  In Appendix C, some tight-binding realizations of this
formalism are discussed, showing that similar results hold for both
conventional and VHS nesting.

\subsection{`Umklapp' vs. `Jahn-Teller' limits}

Given that the splitting of the dos peak associated with a VHS can stabilize a
CDW, the question arises as to which phonons are involved.  There are a number 
of phonons which can couple to the VHS, but sorting them out requires some
effort.  In principle, there are two different ways to split the VHS degeneracy,
associated with the `Umklapp' and the `Jahn-Teller' limits of the theory.  The
starting point is the high temperature tetragonal (HTT) phase in which the 
$X$ and $Y$ point VHS's are assumed to be degenerate. 
The `Umklapp' limit involves a phonon of wave vector $Q=(\pi /a,\pi /a)$ which
condenses to produce an orthorhombic cell.  Here, the $X$ and $Y$ points
remain degenerate, but Umklapp coupling introduces equivalent gaps at both the 
$X$ and $Y$ points -- this is the direct 2D analog of the usual 1D Peierls 
distortion.  Umklapp scattering leads to period doubling, with associated zone 
folding, leading to `ghost' Fermi surfaces and `shadow bands', 
Fig.~\ref{fig:c22}.  The stronger the umklapp splitting, the more intense the 
ghosts.  The ghost Fermi surfaces are similar to those observed in Bi-2212 
photoemission experiments, Fig.~\ref{fig:5}b\cite{Aeb}, although these ghosts 
may be associated with the ordinary orthorhombic distortion. Curiously, the 
distortion which produces the low-temperature orthorhombic (LTO) phase in LSCO 
does {\it not} split the VHS -- 
there is an additional, glide symmetry which prohibits umklapp splitting of 
the VHS degeneracy.  Spin-orbit scattering can produce a splitting, but this is 
believed to be a relatively small effect\cite{KYam,RMXA}.  Magnetic long-range 
order can also lead to an umklapp gap.
\par
In the `Jahn-Teller' (JT) limit, the structural distortion splits the degeneracy
of the two VHS's, driving one below and the other above the Fermi level.  The
two cases are illustrated in Fig.~\ref{fig:c19}, and compared to the
experimentally-derived dispersion in underdoped Bi-2212\cite{Gp0},
Figs.~\ref{fig:c19}a and b.  In the `Umklapp' limit (Fig.~\ref{fig:c19}d), there
is an energy gap, whereas the `Jahn-Teller' limit (Fig.~\ref{fig:c19}c)
only breaks the degeneracy of the dos at the $X$ (solid line) and $Y$ (dot-dash
line) points in the Brillouin zone without producing a true gap.  This 
Jahn-Teller type distortion is easy to recognize from its symmetry-breaking 
property: if the equivalence under a symmetry operation of the two in-plane 
oxygens is broken, the corresponding distortion splits the two VHS's in energy. 
Thus, the distortion of the low-temperature tetragonal (LTT) phase, 
Fig.~\ref{fig:21}a, is a JT distortion. However, the LTT tilt mode only couples 
quadratically to the holes, whereas the accompanying in-plane shear mode has a 
linear coupling.  The comparison with experiment will be discussed further in
Section IX.A.2.
\par
From Fig.~\ref{fig:21}, it can be seen that the LTO and LTT distortions both
involve octahedral tilts, and differ only by tilting in different directions:
along the Cu-O-Cu bond in the LTT phase, and at 45$^o$ to this direction in the
LTO.  In some related compounds, tilting at intermediate angles is also
possible, leading to a Pccn phase.  Since the Pccn tilt can be considered as a 
superposition of the LTO and LTT tilts, it also leads to a (reduced) splitting
of the VHS degeneracy.
Other phonons can be similarly classified.  Thus, both Cu-Cu and O-O stretch
modes can lead to an orthorhombic distortion, Fig. \ref{fig:20}, and both 
distortions produce a gap at the Fermi level.  However, the Cu-Cu stretch mode 
does not couple to the VHS, and the energy gap shrinks to zero at that point, 
whereas the O-O mode does couple to the VHS, and has a finite gap throughout the
Brillouin zone (an Umklapp gap).  Particular forms of the electron-phonon 
coupling for these modes are discussed in Appendix C.

\subsection{Tilt-mode Instabilities and Dynamic JT Effects}

The tilt mode which drives the LTT instability also splits the VHS
degeneracy.  Hence, this may be the structural distortion that competes with
superconductivity.  That interpretation is discussed in this section, and it
is shown that the LTO and HTT phases can be interpreted as dynamic JT phases 
with the pseudogap corresponding to the dynamic JT onset.  However, there are
still complications.  Whereas conventional strong coupling models involve only 
linear coupling due to Migdal's theorem, the LTT and LTO distortions in LSCO 
involve a tilting mode phonon which only couples quadratically to electrons.  A
number of recent models have explored the role of this quadratic
coupling\cite{BB,BuH,RM8b}.  
\par
This subsection also briefly describes an analogy to the dynamic JT phases in 
ferroelectrics like BaTiO$_3$, presents an interpretation of the dynamic JT 
phase in terms of solitonic moving domain walls, and points out that these 
anomalous low frequency modes could play a role in superconducting pairing.  In 
the following subsection, it will be shown that there is a linear coupling 
(probably to O-O stretch modes) which underlies the instability.

\subsubsection{Rigid Unit Modes}

Most perovskite structures are sensitive to structural distortions -- 
principally either ferroelectric or associated with octahedral tilt-mode
instabilities.  The low-temperature distortions of LSCO and LBCO (LTO, Pccn, 
LTT)
are associated with octahedral tilts, and there is some evidence of a pyramidal
tilting instability in YBCO.  Similar instabilities are found in BPBO, BKBO,
Section XVI.  In undoped SrTiO$_3$, there is an antiferrodistortive instability 
associated with the tilting of the TiO$_6$ octahedra, while doped SrTiO$_3$ is 
superconducting -- one of the first known members of the family of high-T$_c$ 
perovskites.  
Ferroelectric phenomena play a minor role in the cuprates, and are briefly
discussed in Subsection C.5. The tilting instabilities are part of a much larger
class of phenomena.  Heine and coworkers\cite{Hein1,Hein3,Hein2} have developed 
the concept of `rigid unit modes' (RUM's), originally in the framework 
aluminosilicate compounds, to describe their structural instabilities.  In these
structures, there is an array of weakly coupled rigid units (e.g., the CuO$_6$ 
octahedra in LSCO); there are a certain number of vibrational modes of these 
units which do not distort the units, but flex the coupling between them.  Some 
of these modes (the RUM's) have a very low frequency -- they differ from 
zero-frequency modes only by quadratic coupling terms.  These modes are sparse 
in the phonon spectra of these materials, but their locus in $\vec q$-space can 
be worked out from simple symmetry considerations\cite{Hein3}; these 
modes are `natural candidates for the soft modes that typically drive displacive
phase transitions.'  The octahedral tilt modes of the cuprates 
fall into the category of RUM's.
\par
Dove, et al.\cite{Hein2} list four forces important in driving the structural
phase transition.  The most important is the polarizability of the oxygen ions,
which also plays a large role in the cuprates.  Another is that Si-O-Si bond
angles prefer to be bent, $\approx 145-150^o$, rather than straight (180$^o$); 
the analogous result may also be important in the cuprates.  Most interestingly,
they note that in special cases, electronic instabilities can play an important 
role.

\subsubsection{Tilt-modes in the Cuprates}

\par
The strong anharmonicity of the tilting modes in LSCO is known from LDA
frozen phonon calculations.  Pickett, et al.\cite{LTTP} have carried out 
extensive numerical calculations of the phonon energy surfaces in LSCO and LBCO,
particularly near the soft modes of the LTO and LTT phases.  They analyzed 
undoped La$_2$CuO$_4$, but with lattice constants appropriate to 
La$_{1.9}$Ba$_{0.1}$CuO$_4$, to 
approximate the effects of lattice strain.  They found that the HTT structure is
unstable and the HTO structure metastable, with energy minima associated with 
the LTT phase, Fig.~\ref{fig:a20}a.  The HTT is actually at a maximum of the 
energy surface, $\approx 46meV$ above the LTT minima, while the LTO phase lies 
in a shallow local minimum, $\approx 14meV$ above the LTT minima.  This result 
is consistent with the double well potential derived from the experimental
temperature dependence of the LTO soft mode phonon in LSCO\cite{BuH},
Fig.~\ref{fig:a20}b.  
\par
Given this large anharmonicity, the LTO and LTT transitions can be understood
in a purely ionic model without any electron-phonon coupling\cite{ZV,Plak}: 
there are large interlayer strains due to the lattice mismatch between the 
CuO$_2$ and the LaO planes\cite{Good2}, which can be relieved by tilting of the 
CuO$_6$ octahedra.  However, the question arises as to whether electron-phonon
coupling can play a role in these transitions.  Thus, frozen phonon calculations
automatically include electron-phonon coupling effects, and a Landau theory of
the transition\cite{IBM} cannot distinguish between ionic and electron-phonon 
models since the Landau theory parameters are chosen phenomenologically.  
In a model which simultaneously incorporates both anharmonicity and 
electron-phonon coupling\cite{RM8b,RMXA}, it is possible to explore the 
competition.  While a certain amount of anharmonicity appears to be necessary, 
it is possible to fit both the phonon softening and the doping dependence of the
transitions either by purely ionic models, or by models with a strong
electron-phonon coupling, Fig.~\ref{fig:bb21}.

\subsubsection{Tilt-stretch Mode Coupling}

\par
However, a purely ionic model cannot provide a {\it self-consistent} explanation
of the doping dependence of the LTO transition in the cuprates\cite{RMXB}.  
Detailed analysis of the model shows that the dominant effect on doping is the 
change of the Cu-O bond length which, in an ionic model, is due to a large 
decrease of ionic size with valence, as $O^{2-}\rightarrow 
O^-$\cite{Brad3,GrenO,RMXB}.  Such a large hole-induced variation of an ionic 
size is known to lead to strong electron-phonon coupling terms, so that a model 
which neglects these terms cannot be correct. Hence this linear electron-phonon 
coupling should be included in the model, via the length dependence of the
hopping parameters, in which case the VHS will play a dominant role in the 
physics of the transition.  A model combining strong VHS-related electron-phonon
coupling plus interlayer-strain-induced anharmonicity can explain both the 
temperature dependence\cite{RM8b} and the doping dependence\cite{RMXB} of the 
LTO transition.  Nd-substitution experiments\cite{Buch2} rule out a pure 
electron-phonon model since the Nd changes the atomic size without providing any
charge transfer; hence, a pure electron-phonon model would predict no change in 
the transition temperature.  On the other hand, photodoping (Section XI.A.5)
changes the hole content without modifying the ions, and {\it this too can 
drive the LTO-HTT transition}\cite{pdop1a}, thereby ruling out a purely ionic 
model.  An important result of this analysis is
that there must be coupling not only to the tilt modes, but to in-plane modes
which change the Cu-O bond length.  
\par
The mode coupling is most easily seen at the ionic level: an oxygen with an
excess hole, $O^-$, is much smaller than $O^{2-}$.  Hence, an isolated $O^-$
will cause the two adjacent Cu's to approach as in a Cu-Cu stetch mode. A
coherent pair of O's sharing the hole will lead to an O-O-type
stretch mode.  If the hole further delocalizes along one direction, say along
$x$, then the $x$-axis will shrink (more $O^-$-like) while the $O^{2-}$-like
$y$-axis will expand, coupling to the shear mode.  Finally, since the CuO$_2$
layer is under lateral compression due to interlayer misfit, the larger
$O^{2-}$'s may buckle out of the planes, thereby giving rise to the LTT-type
tilting distortions.  This would explain the curious feature that the lattice
constant along the tilted axis is actually larger than that along the
transverse axis.
\par
While the combined influence of two phonon modes -- high-frequency
bond stretching and low-frequency tilt modes -- complicates the analysis, it
does provide one simplification.  Whereas the tilt modes couple quadratically
to the holes, the stretch modes have a strong linear coupling.  This means
that the coupling can be analyzed in terms of conventional {\it polaronic}
effects\cite{RMXB}.  In fact, the Holstein and Holstein-Hubbard models can also
play a role\cite{Scal1,Schu1,Schu2,Schu3,Mars1,BFal,RM8a}, since the 
associated Umklapp scattering can also split the VHS
degeneracy.  Note that, while these modes are not condensed in the macroscopic
LTO or LTT phases, they could play a role in a dynamic JT or local polaronic
picture.  Polaronic effects are discussed in Section VIII.D.3.

\subsubsection{VHS JT Effect and the LTT Phase}

The Jahn-Teller (JT) effect was initially developed for molecules and applied
to isolated impurities in solids.  In molecules, the JT theorem states 
that whenever a molecule has an electronic degeneracy, there is a molecular 
distortion which splits the degeneracy.  Since the electrons can repopulate into
the lower of the split energy levels, the distortion produces an electronic
energy lowering, which overcomes the potential energy of the distortion.  The
net result is that a molecule with an electronic degeneracy can lower its
energy by distorting spontaneously.  A static distortion is referred to as a
static JT effect.  However, if the energy barrier is small, the molecule can
tunnel between several equivalent distortions, giving rise to a dynamic JT
effect\cite{DJT}.  
\par
A similar effect can arise in a crystalline solid -- the 1D 
CDW with accompanying Peierls distortion being the most familiar example.
In this case, the electronic degeneracy is associated with the two Fermi
surface points.  In the cuprates, there are a number of candidates for JT 
effects.  The distortion of the CuO$_6$ octahedra in LSCO resembles a
classical molecular JT effect: in the undistorted molecule, the Cu $d_{x^2-y^2}$
and $d_{3z^2-r^2}$ orbitals are degenerate -- the elongation of the Cu- apical O
distance is the JT distortion which splits the $d$ hole degeneracy. [However, a
similar distortion is found in La$_2$NiO$_4$, even though Ni is not a JT
molecule.  Hence, the distortion may instead be a purely static ionic effect.] 
A number of theories have been proposed wherein this distortion, in dynamic or 
static JT form, contributes to the high T$_c$ values\cite{AoK}.  Alternative 
dynamic JT models  for the cuprates have been proposed by a number of 
groups\cite{Clou,Web,Engl} -- indeed, the double well models for the apical
O's can be a variant of this idea -- while dynamic JT models have also been 
proposed for other novel superconductors, particularly the fullerenes (Section 
XVII.C), and for a number of perovskites (Section XI.H). 
\par
The VHS-JT effect is distinct from these models in that the electronic 
degeneracy is associated with the two VHS's, at $X$ and $Y$.  Any structural 
distortion which splits that degeneracy will drive a VHS-JT effect.  This was 
initially pointed out in 1977 by Friedel\cite{FriedVHS}, and more recently 
applied to the cuprates\cite{RM8a,RM8b,RM8c}.  Now in LSCO and LBCO, the lowest 
energy phonons which split the VHS degeneracy are the LTT-type tilting modes 
(out of the plane) of the CuO$_6$ octahedra, which are 
anomalously low frequency rigid-unit modes, further softened because the tilting
relieves some of the interlayer strain.  While these phonons differ from 
conventional CDW modes, they do behave like Jahn-Teller phonons.  In the
low-temperature tetragonal (LTT) phase, the apical oxygen tilts toward the x 
or y axis, splitting the VHS degeneracy\cite{LTT,LTTB,LTTP}, Fig.~\ref{fig:21}a.
Hence, the LTT phase can be considered as a static JT phase.  
\par
Pickett has found that the splitting of the VHS degeneracy can lead to a 
significant enhancement of the LTT phase stability\cite{LTTP}. It has been 
pointed out\cite{Axe2,NMNF} that his assumed LTT tilt angle is too large, by 
about 
a factor of 2, and that the VHS splitting effects would be much smaller if a 
more realistic angle is assumed.  In fact, there is now clear evidence that 
there is a critical tilt angle below which the electronic anomalies 
are greatly reduced\cite{Buch}.  There are several considerations which could
enhance the VHS splitting over the LDA value.  Thus, LDA calculations generally 
overestimate electronic bandwidths by $\sim$ a factor of two; a smaller 
bandwidth, due to correlation effects, would enhance the importance of the VHS 
splitting.  In fact, LDA probably overestimates the c-axis dispersion by an even
larger factor\cite{YF}, and it is this factor which smears out the logarithmic
singularity of the VHS.  Polaronic effects (Section VIII.D) considerably enhance
the sharpness of the VHS peak, possible explaining the observed extended VHS's.
Moreover, if the tilting is partly dynamic, the macroscopic average tilt angle 
will underestimate the maximum local tilt angles.  Indeed, recent EXAFS 
measurements on LSCO, at $x=0.15$\cite{Bian3} report evidence for incredibly 
large tilt angles, $\Phi\approx 14-18^o$.  Finally, the distortion which drives 
the LTT instability is not a pure tilt, but includes an in-plane O-O stretch 
mode which leads to a {\it linear} splitting of the VHS degeneracy.
\par
Due to the strong effect it has on T$_c$, there have been many experimental
investigations of the LTT phase, both in LBCO and in LSCO.  Substitution of La
by other trivalent ions (especially Nd) greatly extends the range of hole doping
over which the LTT phase is stable.  This phase has turned out to be 
surprisingly complex.  No strong electronic anomalies are found when the
octahedral tilt is less than a critical value, $\Phi_c\simeq 3.6^o$, but when 
the tilt exceeds this value, there is a metal-insulator transition at the
LTT transition. At lower temperatures T$_c$ is suppressed, being replaced by
a magnetic transition\cite{Buch}.  The exact relation between these three 
transitions has been studied in detail, but remains controversial.  Since 
nanoscale phase separation also seems to play an important role in this phase, a
detailed discussion of the LTT phase will be deferred until Section XIII, after 
phase separation has been discussed.

\subsubsection{Dynamic JT and the LTO and HTT Phases}

In a uniform low-temperature orthorhombic (LTO) phase the apical oxygen tilts 
along the [110] direction, Fig.~\ref{fig:21}b, leaving the two planar O's 
equivalent\cite{Poug}.  Hence, the uniform phase is not JT-related in the 
absence of spin-orbit coupling, and in the cuprates spin-orbit coupling is 
a small effect\cite{RM8c,RMXA}.
\par
Magnetic effects offer an interesting possibility.  In the mean-field theory,
long-range antiferromagnetic order opens an energy gap which also splits the VHS
degeneracy.  However, fluctuations eliminate this gap in a strictly 2D material,
and the residual gaps observed experimentally are again too small (indeed, they
are partly induced by spin-orbit coupling).  Moreover, the long-range ordered 
(3D) antiferromagnetic phase is restricted to lower temperatures and smaller 
doping than the LTO phase.  It is possible that short-range magnetic order could
help stabilize an LTO phase, particularly in the presence of nanoscale phase 
separation.  At present, there are no calculations of this effect.  
\par
Another possibility is that the LTO phase is a 
{\it dynamic} JT phase in which the LTT tilt direction hops back and forth 
between x and y directions, giving an average macroscopic orthorhombic symmetry.
Thus, the local symmetry differs from the global one, stabilized by a dynamic 
splitting of the VHS degeneracy\cite{RM8c}. In such a phase, the macroscopic 
symmetry is orthorhombic, but locally the octahedra have a predominantly 
LTT-like distortion, with tunneling between adjacent minima.  This is consistent
with frozen phonon calculations, which find that the LTO distortion is unstable 
with respect to an LTT configuration\cite{LTTP}, Fig.~\ref{fig:a20}.  
\par
Moreover, the pronounced double well structure in Fig.~\ref{fig:a20} is strong 
evidence that {\it even the HTT phase cannot be a simple uniform, untilted 
phase, but must involve local octahedral tilts}; the existence of the LTO phase 
must also be due to strong `entropic effects' -- i.e., due to the extreme 
softness of vibrations in the direction of the LTT tilts.  Thus, as a function 
of temperature, the sequence of transitions $LTT\rightarrow LTO\rightarrow HTT$ 
can be understood as a series of JT phases, Fig.~\ref{fig:b21}, in which
the system progresses from being trapped in a single potential minimum, to 
hopping between two minima, to hopping between all four.  At still higher 
temperatures, all tilting is lost.  This onset of tilting seems to correlate 
with the experimentally observed pseudogap transition, Section IX.A.
There is now considerable experimental support for such local LTT
order which will be discussed below, Section IX.B.
\par
If the HTT and LTO phases are both dynamic JT phases, then the transition 
between the two would be due to long-range strain forces, but would have 
relatively little effect on the Fermi surfaces.  This would explain why the 
HTT-LTO transition has so little effect on the in-plane resistivity\cite{Ong}; 
however, the transition drives a {\it metal-insulator transition} in the c-axis
resistivity\cite{Kamb}, with $\rho_c$ increasing with decreasing T in the LTO 
phase. 
\par
Experimentally, there are hints that the LTO phase has a different character in 
undoped and doped LSCO (Section XI.I).  This would be consistent with the
picture of a nanoscale phase separation (Sections X and XI) in which the phase
near half filling is dominated by the Mott-Hubbard instability, and that in
doped material by proximity to the VHS.  Thus, while a static LTO phase near
half filling could be stabilized by magnetic effects, that phase in the doped
material could be a dynamic JT phase.

\subsubsection{Analogy with Ferroelectrics}

\par
In BaTiO$_3$, there is a series of structural phase transitions\cite{KAM} as a 
function of temperature which are analogous to the proposed sequence HTT$
\rightarrow$LTO$\rightarrow$LTT in LBCO, Fig.~\ref{fig:a21}.  In the titanate, 
at very low temperatures the structure is rhombohedral, with the Ti distorted 
off of the cube center, pointing 
toward one of the eight surrounding O's.  This appears to be a simple, static
distortion, with net ferroelectric moment.  As the temperature is raised, the
symmetry switches to orthorhombic, as if the Ti were now pointing towards the
center of a cube edge, halfway between two O's.  But microscopically, this is
not what is happening.  Instead, the Ti still points toward one of the O's,
instantaneously, but {\it dynamically hops from one distorted position to
an adjacent one}, in such a way that the average distortion is orthorhombic.
At higher temperatures, there is a transition to a tetragonal phase, in which
the Ti hops among four adjacent minima.  Finally, at room temperature the
material is in a `cubic' phase, but the Ti is still distorted off-center, but
now hopping among all eight possible positions.  This last phase has no net
ferroelectric moment.
\par
The dynamic nature of these phases was first detected from streaking of X-ray
diffraction spots\cite{Com}, which showed that the Ti was tunneling between
equivalent energy minima in all but the lowest-T phase.  EPR 
experiments\cite{MB} on BaTi$_{1-x}$Mn$_x$O$_3$ found that the Mn$^{++}$ were
on the distorted Ti positions in the rhombohedral phase, but no EPR signals
were found in the higher-T phases due to the rapid reorientation of the ions.
It should be cautioned that this `eight-site' model is not universally accepted,
and that recent Rietveld refinements of the neutron diffraction pattern do not 
find 
evidence for the proposed Ti displacements\cite{KLBC}. On the other hand, recent
femtosecond Raman measurements\cite{fRam} have provided strong support for the
model, in that the relevant phonons are found to obey a strongly damped, 
soft-mode dynamics (consistent with an order-disorder transition with small
energy barriers).  Additional low frequency modes following relaxational 
dynamics, which would have been inconsistent with the eight-site model, are not
observed. 
\par
The microscopic origin of the effect is not yet clear, although a dynamic JT 
phase is not ruled out.  Although BaTiO$_3$ is an insulator,
a JT effect is still possible if the Fermi level lies in a hybridization gap --
a gap between two bands which are strongly hybridized\cite{Coh,RM8b}.  This 
appears to be the case in BaTiO$_3$, where the Ti and O hybridize.  Indeed, the 
sequence of phase transitions has been interpreted as a pseudo-JT 
effect\cite{Berz}.  Regardless whether the 
microscopic mechanisms are identical or not, BaTiO$_3$ provides an example of
a material with dynamic local order -- and a perovskite at that.  Many of the
properties of domain wall dynamics, of long range ordering, etc. would be
expected to be qualitatively similar in the two systems.  Unfortunately, much
remains to be learned about the perovskite phases, although there is still
considerable interest in them.
\par
The analogy with LBCO can be best understood by comparing Fig.~\ref{fig:a21} 
with Fig.~\ref{fig:b21}, which illustrates the tilting mode instabilities of
LBCO, interpreted as a dynamic JT effect.
\par
Recent studies are stressing the analogy with the cuprates.  It is found
experimentally that the Debye-Waller temperatures (which give a measure of local
distortions) are very similar for the cations in BaTiO$_3$ and the 
cuprates\cite{KLBC}.  Recent band structure calculations note the importance
of Ti-O hybridization in BaTiO$_3$\cite{hyb1} and similar covalency in KNbO$_3$
and KTaO$_3$\cite{hyb2}.  Photoemission studies find indications of covalency in
the reduced valence of the Ti\cite{hyb3}.  The same problem of electron-phonon
interaction in the presence of strong correlation effects is being studied (1D 
model) for the perovskite ferroelectrics as for the cuprates\cite{Egep}.  It is 
found that, near the point of crossover between a Mott and a charge transfer
insulator, there is a strong enhancement of electron-phonon coupling.

\subsubsection{Solitons}
\par
The dynamic JT phase could be realized as a pattern of LTT-type domains, 
separated by dynamic domain walls which can be described as solitons\cite{RM8d},
Fig.~\ref{fig:c21}a.  Static versions of these solitons are found in 
transmission electron microscopy (TEM) studies of the LTO-LTT transition.  It
was initially thought that, since the LTT phase is tetragonal, this transition
should be clearly marked by the disappearence of twinning in the LTT phase.
Instead, there appeared to be no change in the pattern of twins at the 
transition\cite{LTT02,Buch5}.  In fact, the LTT phase has orthorhombic symmetry
within a single CuO$_2$ layer, but the global symmetry becomes orthorhombic due
to the interlayer stacking.  Thus, the single-layer twin boundaries are
equivalent to antiphase boundaries (APB) of the 3D structure, 
Fig.~\ref{fig:c21}b.  A careful TEM study of the LTT-LTO transition\cite{APB} 
showed that the LTT APB's readily transform into APB's and twin 
boundaries of the LTO phase by a simple O tilting, without changing their 
locations.  Thus, the microtwinning, which is typically observed in doped 
LSCO\cite{LTT02}, may be a direct manifestation of the dynamic JT effect.
In another TEM study of the LTT phase, Zhu, et al.\cite{Zhu} find a domain 
pattern which they interpret in a similar fashion, Fig.~\ref{fig:c21}c.  
\par
Bianconi and Missori\cite{Bian2} have applied a variant of the above model to
explain their EXAFS observations\cite{Bian1} in Bi-2212.  They find two
in-plane Cu-O bond lengths, and assume that the longer one is associated with
LTT domains.  From the relative fraction of long bonds, they estimate the LTT 
fraction, and by assuming that these domains give rise to the incommensurate 
modulation in Bi-2212, they can reconstruct the striped phase `polaronic CDW', 
Fig.~\ref{fig:c21}d.  Similar EXAFS data have now been reported\cite{Bian3} for 
LSCO ($x=0.15$) and again interpreted in terms of alternating LTO and LTT 
stripes.  Bianconi further assumes that the LTO stripes contain all the doped 
holes.  I would have expected the opposite: that the LTO stripes are the
remnant of the undoped AFM phase at half filling, while holes are added to the
LTT stripes.  In LSCO, the LTT domains correspond to $\approx 
32\%$ of the stripes, suggesting that the hole doping in the LTT stripes is 
$0.15/0.32\approx 50\%$.  This seems rather high.  If, however, the domains are 
dynamic, the LTT percentage could be underestimated.  Thus, if the tilting
octahedra have long bonds only about half of the time, the LTT domains would be
$\sim 64\%$ of the population, with $\sim 25\%$ hole doping.

\subsubsection{Low-Frequency Modes}

\par
The possibility should, of course, not be overlooked that these dynamical
Jahn-Teller oscillations, being very low frequency, could themselves be
playing an important role in the superconductive pairing.  Indeed,
as early as 1946, Teller suggested that the dynamic 
JT effect could be responsible for ordinary superconductivity\cite{Tell}.
Hardy and Flocken\cite{HaFl} showed that, if octahedral tilting leads to a
double well, electron-phonon coupling with such a double well could
greatly enhance $\lambda$ due to anharmonic effects.  Variants of this
double well model have often been applied to the cuprates; usually, however, one
assumes that the double well is associated with the apical 
oxygen\cite{MdL,Plakx}.

Similar dynamic Jahn-Teller effects have been predicted in the Buckyball
superconductors\cite{Auer}, and the anomalous properties of `Berryonic matter' 
are just beginning to be explored\cite{MTD,Bu}.  These will be discussed further
in Section XVII.C.

\subsection{Linear coupling to Stretch or Breathing Modes and Polarons}

\subsubsection{Which Modes are Coupled?}

\par
As discussed in Section VIII.C.2, the strong doping dependence of the LTO 
transition temperature is due to the large variation of the oxygen ionic radius 
with valence.  This variation is a manifestation of a strong linear 
electron-phonon coupling to those phonons which modulate the Cu-O bond length. A
number of in-plane phonon modes can be involved, most importantly the O-O and 
Cu-Cu stretching modes, Figs.~\ref{fig:20}a and b. The O-O stretch modes are 
particularly strongly coupled, because they split the VHS degeneracy, and can 
open a full gap at the Fermi level, Fig.~\ref{fig:20}d. However, at half 
filling, correlation effects weaken coupling to the O's, so spin-Peierls 
coupling mainly involves the Cu-Cu stretch mode.  This latter mode does not 
split the VHS degeneracy (due to matrix element effects), but does open up a 
conventional nesting gap over the rest of the Fermi surface, Fig.~\ref{fig:20}c.
\par
Understanding which modes dominate the linear coupling is still an unresolved
question.  One problem is that
the electron-phonon interaction depends not only on the phonon mode, but also on
the nature of the coupling (Appendix C).  For instance, the above analysis
assumes that the coupling arises through the phonon modulation of the Cu-O 
separation, which leads to a modulation of the hopping parameter $t$. In this 
case, when the calculation assumes coupled O-O stretch modes along both $X$ and 
$Y$, the coupling produces two modes:
a breathing mode, where one Cu has 4 short O bonds, and the other 4 long bonds,
and a quadrupole mode, where each Cu has two long and two short bonds, but for 
one sublattice the long bonds are along $X$, and for the other along $Y$.  For
$t$-modulation coupling, it is found that the gap of Fig.~\ref{fig:20}d is
doubled for the breathing mode, and vanishes for the quadrupole mode -- the 
coupling is to a charge modulation of the `effective Cu atom', the Zhang-Rice
hybridization of Cu with the symmetric combination of nearest-neighbor O's.
As discussed below, this appears to be the opposite of what is seen in neutron 
diffraction: the quadrupole mode has an anomalous softening, the breathing mode
does not.  This strong coupling to the quadrupole mode suggests that a different
coupling mechanism is more important, but its particular form has not yet been
determined.  Perhaps a true molecular JT effect involving coupling to the Cu 
$d_{z^2}$ orbitals would work.  A similar coupling has been proposed in the
related La$_{1-x}$Sr$_x$MnO$_3$ compounds\cite{AJM}, Section XI.H.5; indeed,
in the manganates, the charge-ordered phase is accompanied by a combined
tilt-mode JT distortion\cite{LMO1}.
\par
Why the breathing mode does not show strong softening remains a puzzle.
Perhaps correlation effects are strong enough to suppress this $t$-modulation 
coupling.  Nevertheless, a number of 
groups\cite{Yone,Fesh,breat1,breat2,Fehr3,CGS,dHD} have recently found 
theoretical evidence for strong electron-phonon coupling to the breathing modes,
leading to possible CDW or striped phase instabilities.  
Inhomogeneous Hartree-Fock calculations\cite{Yone} on small clusters find that
Cu-O stretch modes can play an important role in modifying the one-hole state
near the AFI at half filling.  If the electron-phonon coupling to this mode is 
strong enough ($\lambda\sim 1$), the one hole state crosses over from a
ferromagnetic polaron to a dielectric (spin-quenched) polaron.  The latter state
is closer to what is found experimentally in lightly-doped LSCO, Section XI.B.

\subsubsection{Experimental Evidence}

Neutron diffraction studies\cite{PiRe} find tantalizing evidence that the 
high-frequency O-O stretch modes soften anomalously in the doped cuprates, LSCO 
as well as YBCO, Figures~\ref{fig:a22} and \ref{fig:b22}.  These figures
illustrate the dispersion of three O-O stretch modes of different symmetry. The 
breathing mode (Fig.~\ref{fig:a22}a, top) is the highest $\Sigma_1$ modes in 
Figs.~\ref{fig:a22}b, c, and d, while the linear stretch mode 
(Fig.~\ref{fig:a22}a, bottom) is the highest $\Delta_1$ mode.  The dispersion 
of the planar quadrupolar mode is illustrated in Fig.~\ref{fig:b22}.  No 
softening is found for the breathing mode in either LSCO (Fig.~\ref{fig:a22}b) 
or YBCO (Fig.~\ref{fig:a22}c) or in the related but non-superconducting 
compound La$_2$NiO$_4$. [While Fig.~\ref{fig:a22}d\cite{PiRe2} shows softening 
of the $\Sigma_1$ mode, the sample had poorly defined modes, and more recent 
work\cite{PiRe} at lower temperatures find that the $\Sigma_1$ mode has a 
dispersion very similar to that of the cuprates.] Infrared studies by Ohbayashi,
et al.\cite{Ohba} do find an anomalous decrease in the intensity of the 
breathing mode (around $670cm^{-1}$) when LSCO and LBCO are doped into the 
superconducting regime.
\par
The planar quadrupolar mode should split the VHS degeneracy, although it appears
that coupling to the Cu JT effect is necessary to open a gap (Appendix C).
This mode was also predicted to 
be important in several other strong electron-phonon coupling theories which
do not involve the VHS\cite{Web,ZBG}.  Figure~\ref{fig:b22} shows that this mode
has an anomalous dispersion suggestive of a very strong softening for a wide 
variety of perovskites\cite{PiRe}.  However, this mode does not have a strong 
doping dependence, and a similar softening is found in many non-superconducting 
perovskites.  The lack of doping dependence may be more apparent than real due 
to mode coupling.  As the mode frequency softens, it pushes down other 
phonon modes with which it couples until the ultimate instability is associated
with the very low frequency tilt modes.  Due to anticrossing phenomena, the
planar quadrupolar branch may be pinned above the next lower branch, and the
doping dependence appear mainly in the lower branches.  Recently, Arai, et 
al.\cite{Ara4} have noted that at T$_c$ there are anomalous
changes in the neutron scattering of a number of zone center and $(\pi /a,\pi 
/a)$ phonon modes, including the planar quadrupole and breathing modes, and they
have suggested that there is a correlation between these high-energy phonons and
the low-frequency buckling modes of the LTO distortion\cite{Ara3,Ara}.
\par
The really strong, doping-dependent softening is found for the planar O-O 
stretch mode, Fig.~\ref{fig:a22}a, lower frame. Note that the softening is 
greatly enhanced in the doped material.  In the same regions of $\vec q$-space, 
there is considerable smearing of the phonon modes. In a recent study, 
the signal of the Cu-O stretching mode was lost near the $X$ (or $Y$) point, 
with its intensity shifted to lower frequencies\cite{zig3}.  The origin of this
softening is unclear.  Splitting of the VHS degeneracy requires a nesting along 
the diagonal coupling the two VHS's, while a linear O-O stretch mode leads to 
nesting along $(\pi /a,0)$\cite{RMXB}.  It has been suggested that 
these anomalies are associated with inter-VHS nesting of the bifurcated 
VHS's\cite{OKA}.  However, bifurcated VHS's are not found experimentally in 
YBCO and are not expected to arise in LSCO and LNO, which show equally strong
mode softening. 
It is tempting to relate this mode softening to the domain structure associated 
with the striped phases.  From the incommensurate neutron diffraction peaks, 
these stripes are aligned along $(\pi /a,0)$) in LSCO.  However, the stripes
are tilted by 45$^o$ in LNO, and possibly also in YBCO, Fig.~\ref{fig:15}.
\par
The importance of electron-phonon coupling in the Cu-O bond stretching
modes at $\vec q=0$ (which are related to both Cu-Cu and O-O stretching) is
revealed by optical studies of a series of related, perovskite-like structures,
which show an anomalously strong dependence of this mode frequency on Cu-O bond 
length\cite{Taji}.  Neutron pair distribution function (PDF) analyses have been 
an important probe of local order in the cuprates.  So far, however, most of the
anomalous behavior has been found to be associated with c-axis displacements,
and not in-plane motions.  However, this may be because the planar motion is at
a higher frequency -- out of the energy window of previous PDF analyses; new
experiments are planned\cite{Eg7}.  Curiously, a low-frequency Cu-O stretch
mode has been observed only in LSCO near $x\simeq$1/8, where 
the material is close to the LTT instability which suppresses T$_c$ (the stretch
mode was not observed at the higher doping $x=0.15$)\cite{Eg8}.

\subsubsection{Polarons}

\par
A number of theories have suggested 
the importance of polaronic effects in the cuprates\cite{pol}. Moreover,
these calculations naturally give rise to a polaronic renormalization of the 
electronic bandwidth which is pinned to the VHS (even when this falls away from
the Fermi level).  This could explain the experimental observation of
extended VHS's\cite{PE2}, Fig.~\ref{fig:18}b.  To fit this extended VHS requires
an electron-phonon coupling constant $\lambda\simeq 3$; this is in good 
agreement with the values found from analyzing the mid-infrared absorption band 
in LSCO as a polaronic band, $\lambda = 4.8$ (for $x$=0.1), and $\lambda =2.4$ 
($x$=0.15)\cite{BEk}, while $\lambda\simeq 2$ is estimated from point-contact 
spectroscopy\cite{Deut}.  A similar renormalization of the VHS is found in a
generalized Migdal-type diagrammatic approach to the Holstein-Hubbard 
model\cite{Schu3}.  The flattening of the bands leads to a large enhancement of 
the electron-phonon coupling parameter $\lambda$.  However, the narrow peak is 
rapidly washed out as $T$ increases, and it may be hard to observe any 
enhancement of $\lambda$ above T$_c$.  A related result was found by Salkola, et
al.\cite{SEK}; in the presence of a charged striped phase of arbitrary origin,
there will be a tendency to flatten the dispersion near a VHS since the
energy gap structure has its largest effects where the dispersion is weakest --
i.e. at a VHS.
It will be interesting to analyze polaronic effects further within the Van Hove 
scenario to see, for instance, whether the polarons are large or small, or if 
there are bound bipolarons.
\par

\subsection{Phonons and Correlations}

Correlation effects at half filling modify the nature of the electron-phonon
coupling in the cuprates.  Slave boson calculations find that for modes which 
involve modulation of the hopping parameter $t$, the coupling is renormalized to
zero\cite{KiT,RMXA} since Cu-O hopping is itself suppressed.  However, this 
conclusion is not consistent with the results of Anderson\cite{And}
who found that correlations actually enhance nesting effects, and 
Eliashberg\cite{Eli} who finds that electron-phonon
coupling can remain strong in the presence of strong correlation effects.
Just as in a 1D system, coupling via a modulation of $t$ can be replaced by a 
coupling by a modulation of the exchange energy $J$, leading to a possible
spin-Peierls transition.  In the one-band 
Hubbard model, the two-dimensional spin-Peierls interaction was analyzed by 
Zhang and Prelovsek\cite{ZP} and by Tang and Hirsch\cite{TH}.
They found that (1) in the very strong coupling limit, the spin-Peierls phase
is suppressed by N\'eel ordering up to a critical electron-phonon coupling
$\lambda_c\simeq 1.175$\cite{ZP}.  Within the Van Hove scenario, the 
electron-phonon coupling parameter is expected to be $\lambda\ge\lambda_c$; 
therefore, even in this extreme limit a spin-Peierls transition should exist. 
(2) However, this strong coupling limit sets in only for $U\ge U_c\simeq 15 
t_{CuO}$\cite{TH}, which
is considerably larger than the on-site Coulomb repulsion $U$ in the cuprates.
For $U\le U_c$, the structural transition temperature is virtually unchanged 
from the $U=0$ Peierls transition.  This behavior is very different from the
one dimensional case, where correlations actually enhance the Peierls 
transition up to some $U_c$.  Note moreover that the magnetic instability
itself (flux or N\'eel phase) can split the VHS degeneracy, Section VII.B.2.
\par
Thus, in the presence of strong correlations, there should be a striking
crossover in the nature of the electron-phonon coupling\cite{RM11}.  
In the doped material, the electron-phonon coupling is Peierls type due
to modulation of $t_{CuO}$ with strong coupling to O-O bond stretching modes.
Near half filling, the VHS can be split by the magnetic instability, and there
can be an additional spin-Peierls coupling due to modulation of the 
exchange $J$, with strong coupling mainly to Cu-Cu stretching modes.  Such
a crossover arises naturally in the context of the three-band tJ models
discussed above, which involve a spin (J dominated) to charge (t dominated)
crossover as a function of doping.  This crossover may have been seen
experimentally in underdoped Bi-2212, Section IX.A.2.
\par
There is one additional factor which can enhance electron-phonon coupling near a
VHS.  Grilli and Castellani\cite{GC1} have shown that correlation effects
reduce electron-phonon coupling only when $\vec q\cdot\vec v_F$ (or more
generally $(E_{\vec k}-E_{\vec k-\vec q})/\hbar$) is $>\omega_{\vec q}$.  Near a
VHS, the opposite limit holds since $v_F\rightarrow 0$, so that the bare 
coupling should be used.  Note, however, that this applies to dynamical phonons.
For a static structural distortion, the phonon frequency $\omega_{\vec q}$ is
renormalized to zero, and correlation corrections remain important.

\subsection{Apical Oxygens}

In addition to the tilting modes, there have been a number of reports in the
cuprates of other structural anomalies, also associated with possible (dynamic)
JT instabilities.  Perhaps the most persistent candidate has been the apical O
for which it has been suggested that the potential has a double well for motion
perpendicular to the planes.  This was originally suggested from EXAFS 
anomalies, including changes in Cu EXAFS at $T_c$\cite{Api}.

However, frozen phonon calculations find no evidence for an apical O double well
potential\cite{Maz,AmDl}, neutron diffraction studies fail to observe the 
splitting of the apical oxygen position\cite{Schw,zig3}, and the Cu-O c-axis 
vibrations are much stiffer and display much less anharmonicity\cite{PiRe} than 
would be expected from the proposed double well potential.  In fact, neutron
diffraction thermal factors for the apical O are much larger in-plane 
(associated with octahedral tilting) than along the c-axis (related to the
proposed double well)\cite{Kata,Brad33}.  Alternative 
explanations for the EXAFS observations have been proposed in terms of either 
an experimental artifact\cite{Roh} or a phase separation\cite{Eme1}. In the 
latter picture, the apical oxygen acts as a source of holes for doping onto the 
CuO$_2$ planes, and the Cu -- apical O bond length decreases with increasing 
hole doping of the planes (see the discussion in M\"uller\cite{MAp}).  In the
presence of a nanoscale phase separation (Section X), there will be local
domains of alternating high and low hole density.  Hence, an EXAFS experiment 
will observe two different Cu -- apical O lengths associated with the two
types of domain, the shorter length correlated with the higher hole density.
Recently, this picture has been confirmed; the 500$cm^{-1}$ phonon mode,
associated with c-axis vibrations of the apical O, was observed to split in
an O-doped LCO crystal with a clear macroscopic phase separation (staging) into
hole rich and hole poor regions\cite{Qui}.  Neutron pair distribution function
studies also find evidence for local (dynamic) phase separation in several of
the cuprates with one domain characterized both by the planar O's having an
enhanced out-of-plane displacement, and by a reduced Cu-apical O 
distance\cite{Eg7}.
\par
Brandow\cite{Bran} has suggested that the Cu-apical O distance $d_A$ modulates 
the Cu-planar O energy splitting $\Delta$; the smaller $d_A$, the smaller is 
$\Delta$.  If so, this could be an additional role played by the apical O.
Since $\Delta$ is close to the critical value needed for the Mott-Hubbard
instability, small changes in $d_A$ could locally shuffle the planes between an
insulating (long $d_A$) and a metallic (short $d_A$) state; this correlation
is consistent with the PDF analyses\cite{Eg7}.

\section{Electron-Phonon Interaction: Experiments}

The VHS-JT model makes a number of predictions concerning the structural phases,
which have nothing to do directly with the superconductivity.  Hence,
experimental confirmation would be a strong support for the VHS model as a
whole.  Recent photoemission experiments directly relate the pseudogap to the
magnitude of the VHS splitting, while other experiments provide highly 
suggestive evidence for local order in the lanthanum cuprates similar to that 
proposed for the dynamic JT state.  The potential importance of these findings, 
coupled with the difficulty of probing short-range order, means that additional 
experiments would be most welcome.  Confirmation of this aspect of the model 
would constitute a major predictive triumph of the VHS theory, and would confirm
that the VHS is playing a crucial role in the low energy physics of the 
cuprates.

\subsection{Pseudogap}

\subsubsection{Experimental Observations}

The Van Hove -- Jahn-Teller model predicts a competition between structural
instability and superconductivity.  However, the fundamental structural 
instability involves the onset of a tilting of the CuO$_6$ octahedra (in LSCO),
which, as a dynamic JT effect, may involve only short-range order.  An analogous
situation is common in 1D conductors, where {\it pseudogaps} are often found 
(see p. 38 of Ref. \cite{ODC}).  Due to strong fluctuation effects, long-range
order cannot exist in a 1D system.  Hence there is typically a wide temperature
range between the mean-field transition temperature $T_{MF}$ and the temperature
of true 3D order $T_{3D}$, driven by weak interchain coupling.  However, below 
$T_{MF}$, there are large fluctuations into the localized state, which are 
reflected in the dos, producing a pseudogap, Fig.~\ref{fig:c22}c\cite{LRA} -- 
the dos is strongly reduced, but does not equal zero until a true gap appears 
below $T_{3D}$.  A similar situation arises in a 2D system, which again cannot 
display long-range order.  Figure~\ref{fig:c22}d\cite{RM5} illustrates a 
CDW-like pseudogap in the cuprates, while Fig.~\ref{fig:c22}e shows a magnetic
pseudogap\cite{KaSch} -- the strong similarity follows because this latter is
calculated in the (weak coupling) SDW limit, which is virtually 
indistinguishible from the CDW theory.
\par
In this subsection, I will discuss the experimental evidence for the tilting 
onset, with concomitant appearance of a pseudogap.  Recently, a `pseudogap' 
phase was found within the HTT phase of LSCO. At a temperature $T^*$, Fig.~
\ref{fig:23}a, the susceptibility peaks and the anomalous temperature dependence
of the Hall coefficient $R_H$ begins\cite{Hwa}, in good agreement with 
prediction\cite{Hallb}.  While there is no long-range order associated with 
$T^*$, there is a correlation with the LTO phase, in that, as the doping is 
varied, $T^*\simeq 2T_{LTO}$.  
\par
A similar `pseudogap' phase is found in YBCO, Figs. \ref{fig:23}b and c: the 
magnetic susceptibility $\chi_Q$ peaks at a finite temperature\cite{MP}, and 
then decreases at lower temperatures, as if a gap (the `pseudogap') were
opening up in the dos.  This pseudogap is also observed in the dynamic
structure factor\cite{R-M} ,the resistivity\cite{tr123}, and the heat 
capacity\cite{Lor}, showing that this is not simply a magnetic effect, 
but is consistent with a dos change.  Moreover, the magnetic susceptibility
shows a pseudogap not only at the `antiferromagnetic' wave vector $(\pi /a,\pi
/a)$, but also at $\vec q=0$\cite{MehrX}.  The pseudogap transition temperature
is also correlated with the onset of the anomalous T-dependence of the Hall 
coefficient\cite{Chen}, just as in LSCO.  It has also been suggested that there
is an enhancement in the TEP below the pseudogap onset\cite{Coop}, but this
claim has been disputed\cite{ZGy}.  The similarity of the pseudogap in YBCO to
that in LSCO has been emphasized by Batlogg, et al.\cite{Bat1}, Fig.
\ref{fig:23}d.  While there is no long-range structural phase transformation 
associated with the pseudogap, recent neutron diffraction data is suggestive of 
a local tilting instability\cite{Schw}, consistent with a dynamic JT model. 
\par
YBa$_2$Cu$_4$O$_8$ (Y-124) is underdoped, and similar in many of its properties 
to YBCO$_{6.67}$, but additional doping, provided by substituting Ca for Y can
enhance T$_c$ to $\approx 92K$. It also shows evidence for a pseudogap in Knight
shift measurements\cite{Kn124}, spin-lattice relaxation times\cite{Kn124}, and
resistivity and Hall coefficient\cite{tr124}.  Neutron pair distribution 
functions\cite{Send} find local distortions of both planar, apical, and chain 
related O's, which have the same temperature dependence as the pseudogap.
When the hole doping in Y-124 is varied by cation substitution, the doping 
dependence of the pseudogap is found to be similar to that found in 
YBCO\cite{Buck1}.  Whereas no elastic or heat capacity anomalies are found in 
pure YBa$_2$Cu$_4$O$_8$ near the pseudogap, in Ca substituted samples there are 
clear anomalies in both properties at 150K, strongly suggestive of a 
second-order phase transition\cite{TF3,TF4}.  A curious feature of the 
resistivity in Y-124 is its anisotropy\cite{tr124}: a clear pseudogap dip is 
observed below 160K, in the a-axis resistivity $\rho_a$ only, but not in $\rho
_b$.  Suprisingly, both $T_c$ and the temperature onset of the dip increase 
under pressure\cite{ZGp}.  While such
behavior is difficult to reconcile with a magnetic model for the pseudogap, the
authors suggest that it is associated with enhanced c-axis coupling, which could
shift the whole phase diagram to higher temperatures.
\par
In YBCO and Y-124, the optical conductivity shows direct evidence for a 
pseudogap, both in the a,b plane\cite{opt1,opt2} and along the 
c-axis\cite{Tim7,Tim2}.  For example, in optimally doped YBCO (T$_c$=93K), the 
a-axis conductivity starts to decrease near T$_c$, as if a gap is opening up.
At low T, $\sigma_a$ is close to zero for $\omega\le 400cm^{-1}$, and is 
depressed up to $\approx 1000cm^{-1}$ (the b-axis behavior is similar, but
complicated by the presence of an extra chain contribution).  This was initially
taken as evidence for the superconducting gap, but it was found that as the 
oxygen content is reduced, the conductivity continues to decrease over the 
same frequency range, {\it but the falloff begins at a higher temperature}.
Indeed, $\sigma_a$ has the same temperature and doping dependence as found from
the NMR relaxation, $1/T_1T$ -- showing that the pseudogap couples to both spin
and charge degrees of freedom.  Furthermore, in both YBCO$_{6.7}$ and Y-124
at low frequencies, the c-axis conductivity $\sigma_c$ is also depressed with 
decreasing T, again closely following the T-dependence of the Knight shift. This
pseudogap has a flat bottom out to
200$cm^{-1}$ in YBCO$_{6.7}$ (180$cm^{-1}$ in Y-124), independent of T, and the 
zone of depressed $\sigma$ persists out to about $400(350)cm^{-1}$.  
\par
The normal state Raman spectrum of most cuprates shows evidence for a broad
electronic continuum which can be understood in a marginal Fermi liquid 
picture.  Cooling optimally doped materials below T$_c$ leads to the opening of
what appears to be a gap at low frequencies, but with a substantial spectral
weight persisting
in the gap at low T, suggestive of d-wave pairing.  Just as with the optical 
gap, this was initially interpreted as a superconducting gap, but was found
to persist at higher temperatures in underdoped material.  Thus, in YBCO, the 
gap terminates in a broad electronic peak which in $B_{1g}$ symmetry is 
centered near $500cm^{-1}$ ($\approx$60meV).  Even in optimally doped YBCO$_
{6.952}$, the peak starts to appear about 40K above T$_c$\cite{opt3}, while 
in underdoped materials, it begins to grow at even higher 
temperatures\cite{opt4,Rama}, suggesting that the peak corresponds not to the
superconducting gap, but to the opening of a pseudogap above $T_c$.
\par
Furthermore, the frequency shift of several infrared
and Raman phonons, which begins at T$_c$ in optimally doped YBCO, actually
starts at temperatures higher than T$_c$ in YBa$_2$Cu$_4$O$_8$ and 
oxygen-deficient YBCO, close to the (pseudogap) temperature at which the 
susceptibility peaks\cite{Lit,Alten}.  
\par
Most recently, neutron scattering revealed yet another feature which appears to
give a sharp indication of the superconducting gap onset in YBCO.  A broad, weak
magnetic feature at $Q_0$ sharpens into a well-resolved peak near 41meV just
below T$_c$\cite{m411,m412,Kei2,Mook}.  Perhaps not surprisingly, in Zn doped 
samples, this sharpening is found to occur at a {\it higher temperature}, even 
though the Zn suppresses T$_c$\cite{Bour}.
\par
The $T^*$ phase boundary can be estimated within the VHS-JT model as a function 
of doping\cite{RM8a}; in this calculation, it was assumed that the VHS remained
pinned to the Fermi level at all dopings.  This was intended to approximate the
effects of nanoscale phase separation.  However, when strong correlation effects
near half filling are included via the three-band $tJ$ model (Section VII.B), it
is found that this condition is automatically satisfied, without assuming
phase separation.  There is a crossover from spin dominated to charge dominated
response, with the VHS intersecting the Fermi level at two different dopings,
and remaining very close at intermediate doping.  Accompanying this crossover,
the electron-phonon coupling crosses over from spin-Peierls-like at half filling
to ordinary Peierls-like (this is a mean field approximation to the dynamic JT 
phase) in the doped material.  Since the nesting is better at half filling, the 
structural transition temperature is highest there, and superconductivity is 
suppressed.  As doping increases, the nesting worsens, and the structural 
transition temperature decreases.  At the same time, the superconducting 
transition is less sensitive to nesting, so that the $T_c$ increases until the 
two transitions meet at the VHS phase.  The resulting calculations give a good 
description of Loram's heat capacity data,
Fig.~\ref{fig:23}b\cite{RMPRL}, and also explain the fact that the 
superconducting T$_c$ increases as the LTO transition temperature decreases.
\par
One further prediction relates the $T^*$ transition with the pseudogap onset in 
YBCO.  Based on the Uemura plot, Figs.~\ref{fig:c14} and \ref{fig:b14}a, the 
doping at optimum $T_c$ is higher in YBCO ($\approx 0.25$ hole per planar Cu) 
than in LSCO ($x\simeq 0.16$).  Within the VHS model, this suggests that the 
bare VHS is closer to half filling in LSCO, which in turn implies better nesting
and hence a higher tilting instability onset temperature.  This is what is found
experimentally, and may explain why additional long-range ordering 
transitions are found in LSCO, but not in YBCO.  Since the tilting instability
competes with superconductivity, this may in turn explain why $T_c$ is so much
lower in LSCO than in YBCO.  (Note that this interpretation is not consistent
with the alternative finding, Fig.~\ref{fig:b14}b, that the optimal T$_c$ occurs
at the same hole doping in both YBCO and LSCO.)
\par
In Y-124, Zn doping is found to strongly suppress T$_c$, and to inhomogeneously
fill in the gap states of the pseudogap, but without affecting the gap 
energy\cite{Zngap}.  This seems to occur by a {\it local} suppression of the 
pseudogap, affecting primarily Cu's which are nearest or next-nearest neighbors 
to a given Zn.
\par
Clear evidence for a pseudogap is also found in Tl-2212\cite{Tlps} from a peak
in the susceptibility.  The variation of $T^*$ with O-doping is remarkably 
similar to that found in YBCO, even though O-doping does not give rise to 
long-range antiferromagnetic order.  Similar effects are found in 
Bi-2212\cite{Bips}.

\subsubsection{Pseudogaps in Photoemission}

While Fig.~\ref{fig:4} shows that the VHS's are close to the Fermi level in all
the cuprates, the separation $E_F-E_{VHS}$ is rather larger than expected for
the Van Hove scenario.  In this subsection, I explore the possibility that
the feature observed in photoemission as the VHS is actually {\it a 
pseudogap-split VHS}, as in Fig.~\ref{fig:c22}.  For instance, in NCCO, if $E_F
-E_{VHS}$ is identified as $\Delta^*$ for the pseudogap, while the pseudogap 
transition temperature is estimated from Fig.~\ref{fig:23}a as that of LSCO for 
$x\rightarrow 0$, then $\Delta^*\simeq 220meV$, $T^*\simeq 1050K$, and $2\Delta
^*/k_BT^*\simeq 4.9$.  A similar calculation for Y-124 yields $\Delta^*\simeq 19
meV$, $T^*\simeq 120K$, and $2\Delta^*/k_BT^*\simeq 3.7$.  Both numbers are 
comparable to (but smaller than) the corresponding ratio for the superconducting
transition.  
\par
It is not clear why a similar pseudogap is not seen in underdoped YBCO (Section
VII.D.3).  In underdoped Bi-2212, it is found that the separation $E_F-E_{VHS}$ 
systematically increases with underdoping\cite{Biund,Gp0} (Figs.~\ref{fig:c19}a
and b) as predicted by the VHS model
of the pseudogap.  Recent studies\cite{Gp1,Gp2} have found two
remarkable features: the splitting (or shifting) of the VHS increases 
monotonically with underdoping, while a {\it true gap} $\Delta^g$ (reduced
photoemission intensity near the Fermi level) of {\it d-wave symmetry} is 
found in all samples below a pseudogap transition temperature $T^*$.   The two 
features are clearly distinct; the maximum gap is $\approx 25meV
$\cite{Gp1,Gp2}, while the peak shifts by $>200meV$ from the Fermi 
level\cite{Gp0}.
Remarkably, $T^*$ scales with the VHS splitting, whereas the pseudogap features
(gap energy and anisotropy) are virtually indistinguishible from sample to 
sample, and indeed coincide with the feature identified as the superconducting 
gap in optimally doped Bi-2212. Evidence that the pseudogap has d-wave
symmetry has also been found in heat capacity\cite{Gp3} and NMR\cite{Gp4} 
studies.  There are some differences in these studies; for several cuprates, 
NMR finds two d-wave gaps -- a pseudogap which varies strongly with 
doping, and a superconducting gap with a zero temperature limit $\Delta^g(0)
\simeq 8T_{c,max}$ which is independent of doping (where $T_{c,max}$ is the 
maximum $T_c$ in a given compound, as a function of doping). This latter is 
similar to the second photoemission gap.  On the other hand, heat capacity sees 
a single d-wave gap with its peak coincident with the VHS.
\par
As illustrated in Fig.~\ref{fig:c19}, the VHS splitting can be understood as the
opening of a (possibly dynamic) gap, with a crossover from a structural gap
near optimum doping (Figs.~\ref{fig:c19}c and d) to a magnetic gap near half 
filling (Fig.~\ref{fig:c19}e).  The key feature is that, in the absence of a gap
opening, the VHS would be pinned to the Fermi level over the full doping range
from half filling to optimal doping, consistent with slave boson predictions
(Fig.~{\ref{fig:16}d).  Note that if the photoemission peak at $X$ is 
associated with the onset of a gap, then there
must be a corresponding pileup of dos {\it above} $E_F$.  This constitutes a
strong prediction of the model, and I would greatly encourage any inverse
photoemission or equivalent studies to search for such a feature.  While
angular resolution is not necessary for detecting the dos pileup, the theory
makes very definite predictions about where in $\vec k$-space this dos should 
reside.  Much additional information can be extracted from direct photoemission
studies about the nature of the pseudogap phase by a careful comparison of
the $X$ vs $Y$ points of the Brillouin zone.
\par
In principle, two different types of gap-opening are possible, which can be
associated with the `Umklapp' (Fig.~\ref{fig:c19}d) and the `Jahn-Teller' 
(Fig.~\ref{fig:c19}c) limits of the theory.  The theory has been analyzed for 
the tetragonal phase, assuming a degeneracy of the $X$ and $Y$ point VHS's in 
the absence of a structural distortion.  In the `Umklapp' limit, the degeneracy 
is maintained, but Umklapp coupling introduces equivalent gaps at both the $X$ 
and $Y$ points; this is the direct 2D analog of the usual 1D Peierls 
distortion.  In the `Jahn-Teller' limit the structural distortion splits
the degeneracy of the two VHS's, driving one below and the other above the
Fermi level.  Such a splitting is expected for the orthorhombic distortion in 
YBCO, where the VHS is clearly seen along $Y$, but not along $X$. However, the
magnitude of the splitting is not clear due to complications associated with
the chain Fermi surface, which is expected to interact strongly with the plane
Fermi surface near $X$.  Such a splitting would be a clear signature of a 
JT-like effect.  However, if the experiment averages over many domains (as in
a dynamic JT effect), the photoemission would appear to be a superposition of 
the two Fermi surfaces, with x and y axes interchanged, thereby complicating the
interpretation. 

In the `Umklapp' limit, there is a true gap\cite{RM8a} with the maximum gap (at
$X$ or $Y$) equal to the VHS splitting, in agreement with the heat capacity 
data\cite{Gp3}.  Note that the `Umklapp' form of the dispersion due to O-O
stretch modes (Fig.~\ref{fig:c19}d) gives a good description of the
experimental data (Figs.~\ref{fig:c19}a and b) including the camelback 
curvature.
As discussed in VIII.E, correlation effects are expected to lead to a striking
crossover from charge-dominated behavior at optimum doping to spin-dominated
behavior at half filling.  This crossover was taken into account in the model
calculation of the pseudogap, Fig.~\ref{fig:23}b\cite{RM8a}.  Hence, it is
gratifying to note that with increased underdoping the electronic dispersion,
including VHS splitting, evolves smoothly\cite{RBL} into the form found in the 
AFM insulator, SCOC, Fig.~\ref{fig:c19}e.

The interpretation of the other gap-like feature, $\Delta^g$, is much more
problematic.  The maximum gap falls at the same point of the Brillouin zone as 
the VHS splitting, but is much smaller, $\approx 25meV$, and doping independent.
This gap does not appear to be associated with a conventional structural 
instability; the gap seems to open over the entire Fermi surface, which would
require an exactly filled band below the gap -- i.e., the gap must fall at
exactly half filling of the band, which is not the case here.
\par
I suggest two possible explanations.  First, there could be a `dynamic JT' 
gap; the static JT tilt does not produce a gap, but leads to a 4-fold degeneracy
of the symmetry-broken state.  Quantum tunneling between these states can lead 
to a dynamic restoration of the ground state symmetry -- this is the dynamic JT 
effect.  While in the molecular limit there should be a true gap\cite{RM8c}, the
nature of the broadening of this gap in the solid state is less clear.  It is
plausible to suppose that such a gap should be of d-wave symmetry; positive 
along $Y$ and negative along $X$, to preserve the JT distortion.  This would 
explain why the gap is so much smaller than the VHS splitting.
\par
Alternatively, if it turns out that there are really two d-wave gaps 
simultaneously present,
one due to a pseudogap and the other to superconductivity, an interpretation in
terms of phase separation might be possible.  (For a uniform system with two
competing s-wave gaps, once the pseudogap wins out, the superconducting gap is
reduced to zero\cite{RM8a}.  For d-wave gaps, both could coexist, if the one had
maxima where the other had minima; however, in the present case, both gaps
seem to maximize near the location of the VHS's.)  The fact that the zero-T
superconducting gap remains constant\cite{Gp4}, while $T_c$ and the amplitude of
$\Delta C$ both decrease with increased underdoping\cite{Gp3} would tend to
support this idea.  Indeed, in the strongly underdoped regime ($p\le 0.11$, with
optimum $T_c$ corresponding to $p=0.16$) a `filling in of the normal state
pseudogap' has been observed, and taken as an indication for phase 
separation\cite{Gp5}.  However, the onset temperature for $\Delta^g$ is equal to
the pseudogap temperature rather than the superconducting T$_c$, which seems
inconsistent with this second model.
\par
Continued photoemission studies of the underdoped cuprates are highly desirable.
Among the many questions to be addressed, some have to do with electron-doped
materials, e.g., is there a gap near the Fermi level in NCCO, and do the 
residual Fermi surfaces have electron-like symmetry?
\par
In all of this, there is one important caveat; if one adds a d-wave symmetry gap
to a structureless, parabolic band, then the point of the maximum gap
automatically becomes a saddle-point VHS!  Hence, it is important to distinguish
whether the VHS exists in the underlying band structure or not.  Evidence for
this was discussed in Section VI.C.1.  Note that, to explain the pseudogap, 
the VHS must be pinned to the Fermi level over an extended doping range, as
predicted in the presence of strong correlation effects.

\subsubsection{Magnetic Models of the Pseudogap}

\paragraph{\bf General Considerations}
\par
These pseudogaps have been alternatively interpreted in terms of a spin gap,
arising within a purely magnetic model of the cuprates.  A number of 
inequivalent magnetic calculations have been applied to this pseudogap, which 
are summarized in Ref. \cite{IoMi}.  Thus, Kampf and Schrieffer\cite{KaSch} used
a weak coupling SDW approach to model the effect of quasi-long-range AFM 
fluctuations, Fig.~\ref{fig:c22}e.  In RVB-like, or gauge field calculations, a 
spin-charge separation is assumed, and it is found that there should be two 
independent transitions in mean-field, one at $T_{BE}$ associated with Bose 
condensation of the holons and the other ($T_D$) with the appearance of a finite
RVB order parameter on the spinons\cite{SHF,NagL}, Fig.~\ref{fig:a23}.
Only when both spinons and holons are condensed will there be a 
long-range superconducting order.  The onset of the spinon pairing is taken as 
the pseudogap onset, $T^*$. This onset is higher than the superconducting 
transition in the underdoped regime, and coincides with it in the overdoped
regime.  While this reproduces experimental observations in the underdoped 
regime,  Tallon, et al\cite{Coop} find that the pseudogap temperature is
clearly below the superconducting temperature in the overdoped regime,
Fig.~\ref{fig:23}c, although other groups find a variety of behaviors; see
the data of Hwang, et al.\cite{Hwa}, Fig.~\ref{fig:23}a and of Chen\cite{Chen}, 
et al., Fig.~\ref{fig:23}c (the filled triangles are from overdoped samples).  
The frequency shift of the 340$cm^{-1}$ phonon also begins considerably below 
T$_c$ in overdoped YBCO\cite{Alten}.
\par
Moreover, within this RVB model, fluctuations were found to eliminate the 
pseudogap phase\cite{Ubb}, and it was suggested that the pseudogap might be 
caused by interlayer coupling\cite{Ubb2}.  This was originally proposed in the 
nearly-antiferromagnetic Fermi liquid model of the cuprates\cite{MiMo}.  The 
idea is that there are major differences in the magnetic
properties of bilayer cuprates, such as YBCO, which have two CuO$_2$ planes per
unit cell, and single layer cuprates, such as LSCO.  Some properties, such as
the pseudogaps, would be associated with interlayer exchange coupling, and hence
would be absent for single layer cuprates.  However, this result is inconsistent
with the experimental observation of the pseudogap in LSCO.  Barzykin and 
Pines\cite{BaP} have recently reanalyzed the experimental situation, and find 
that the pseudogap onset depends only on hole doping, and ``bilayer coupling 
plays little or no role in determining spin pseudogap and scaling behavior''.  
They find that the pseudogap onset can be found from a scaling argument in the 
nearly antiferromagnetic Fermi liquid model.  Ioffe and Millis\cite{IoMi} have
shown that the fluctuation problem can be overcome {\it if the Fermi level is
close to a VHS}.  In a weak-coupling calculation, the pseudogap is also found to
be VHS-related\cite{HotFuj}. On the other hand, Vilk\cite{Vilk} finds that the 
shadow bands cannot be understood as precursors to the AFM band. 

\paragraph{\bf Interpreting the Underdoped Cuprates}
\par
It is instructive to see how the recent photoemission and heat capacity
results on underdoped cuprates can be understood in the magnetic model.  I
claim that the data constrain the theory in two ways: (1)  the pseudogap
is driven by a splitting of the VHS degeneracy and (2) the experimental 
observations require VHS pinning, and therefore a nonzero $t^{\prime}$ in the 
one-band model.  (In all of this discussion, I am concerned with the observation
of the VHS splitting, which is the large energy in the problem; I do not 
understand the origin of the smaller $\Delta^g$.)  
\par
In the VHS model\cite{RM8a}, correlation effects renormalize $t\rightarrow 0$ at
half filling so the residual dispersion is magnetic, with perfect nesting
(square Fermi surface).  In this case, the pseudogap arises from a nesting 
instability, as in the flux phase or a spin-Peierls transition\cite{RM8a,RM11}, 
Fig.~\ref{fig:a15}.  In contrast, the RVB model interprets the pseudogap as a 
Bose condensation of holons, which should have nothing to do with a Fermi
surface.  However, in their recent work, Wen and Lee\cite{WeL} show that their 
gauge-field model is closely related to the flux phase, and hence produces the 
same VHS splitting,  Fig.~\ref{fig:c19}e.
\par
Given this, the magnetic model becomes virtually indistinguishable from the
present VHS model (based, e.g., on the 3-band tJ model) at half filling.
However, the models diverge away from half filling.  In particular, since the
RVB calculations are based on a simple $tJ$ or Hubbard model with electron-hole
symmetry ($t^{\prime}=0$), there is no pinning of the VHS to the Fermi level.
Thus, with doping the Fermi level moves away from perfect nesting, and the gap 
will quickly close.  Moreover, when the gap
just closes, the VHS will be far away from the Fermi level.  Thus, the model
cannot explain the experimental results that the VHS is close to the Fermi
level near optimum doping.  To be definite, consider the results of
Fig. \ref{fig:b10} for LSCO: when the pseudogap just closes, for $x\simeq 0.26$,
the VHS is at the Fermi level.  We can compare this with the expected result for
the $tJ$ model.  At this doping, the dispersion should be dominated by $t$.
Assuming $t=0.125eV$, then for $x=0.26$ the VHS would be 70meV away from the
Fermi level.  This is clearly incompatible with observations.  Hence point (2)
above: to pin the VHS near the Fermi level requires a nonzero value of 
$t^{\prime}$.
\par
At this point, the magnetic model looks more and more like the VHS model.  
Indeed, the only possible difference lies in the role of phonons.  The data that
I have assembled in this section strongly suggest that phonon effects play a
large role, particularly away from half filling.  The remaining question is how
large a role do magnetic effects play near optimal doping.  Since the magnetic
Fermi surface nests at half filling, magnetic effects would be expected to
decrease rapidly with doping, and there is theoretical (Section VII.B.2, Refs. 
\cite{TIF,RM12}) and experimental (Section XI.D, Fig.~\ref{fig:27}) evidence 
that this is the case.  Moreover, Tranquada and coworkers\cite{Tran,Tran2} have
shown very clear evidence for magnetic and charge striped phases in Nd
substituted LSCO, and provided strong circumstantial evidence that the stripes
exist as fluctuations in the Nd-free materials.  The evidence is that the 
striped phases compete with superconductivity, with the magnetic and 
(presumably) charge
ordering temperatures increasing with decreasing doping -- just like the
pseudogap temperature!  Such a result does not follow naturally from the 
gauge-field models, but was indeed predicted in the VHS model, while such a 
magnetic-to-charge crossover has {\it already} been incorporated in the theory 
for the pseudogap\cite{RM8a}.

While a magnetic model\cite{MP} can explain the anomalous 
susceptibility, it seems difficult in such a model to explain evidence for a 
simultaneous gap in the charge excitations\cite{Hwa,Gp3} and the direct 
modifications in the phonons, unless there is (at the minimum) a strong 
spin-Peierls coupling\cite{NKF}.  Hence, a model combining strong correlations 
and strong electron-phonon coupling is required -- as in the Van Hove -- 
Jahn-Teller model.
\par
Finally, it should be noted that a similar pseudogap is found in the C$_{60}$
superconductors\cite{gp60}; a phononic, nesting-related gap can be easily
generalized to describe this case, whereas a magnetic gap seems improbable.

\subsection{Short-range Order and the Dynamical JT Phase}

\par
Beyond the presence of a pseudogap, a number of recent experiments have provided
highly suggestive evidence for local order in the lanthanum cuprates very
similar to that proposed for the dynamic JT state.  Thus, by studying La NMR, 
Hammel, et al.\cite{Hamm} have found that in La$_2$CuO$_{4+\delta}$ there is 
not a single, well-defined octahedral tilt angle in the LTO phase, but rather an
extremely broad distribution, from essentially zero tilt up to some maximum 
value, Fig.~\ref{fig:c23}.  The experiments could not distinguish static from 
dynamic disorder, but are consistent with dynamically tilting octahedra.  (The 
dynamic JT effect is likely to be sensitive to disorder, so that the local tilt 
angles may be pinned by impurities.)  Comparable results are found in LSCO; the 
Cu NQR line width increases by a factor of $\approx$20 from its value in 
undoped La$_2$CuO$_4$\cite{Imai}.
\par
Similar spectral broadening effects are found in studies of the c-axis
polarized phonons in underdoped YBCO\cite{Tim1} and in Y-124\cite{Tim2}.
In YBCO$_{6.6}$, the optical conductivity displays three sharp features at room 
temperature (Fig. \ref{fig:e23}) which are associated with an O(2)-O(3) plane 
bending vibration (315$cm^{-1}$) and the apical O(4) vibration (570 and 610$cm
^{-1}$ -- the mode is split in underdoped samples due to chain O vacancies).  
When cooled below 150K, these three peaks weaken, particularly the 315$cm^{-1}$
mode, and are replaced by an extremely broad feature at $\approx 400cm^{-1}$.  
The inset shows the T-dependence of the intensity of the 400$cm^{-1}$ feature.  
The curve has a point of inflection close to T$_c$.  In Y-124, this feature
seems to appear at a temperature ($\approx$100K) significantly below the 
pseudogap onset near 180K.
It must be kept in mind that these c-axis phonon anomalies are not universal
in the cuprates -- they are not found in optimally doped YBCO, nor in
LSCO\cite{Tim3} or Bi-2212\cite{Tim4}, but similar anomalies are found in
Pb$_2$Sr$_2$RCu$_3$O$_8$ (R = rare earth)\cite{Tim5}.
\par
Billinge, Egami, and coworkers\cite{BEg} have developed the technique of
studying pair density functions (PDF's) derived from neutron powder diffraction
data as a means of probing details of local order in the cuprates.  These PDF's
have supplied evidence for (possibly dynamic) local tilting in 
Tl-2212\cite{Eg2}, LSCO\cite{Eg3}, NCCO\cite{BEg}, and the infinity-phase 
compound, Ca$_{0.85}$Sr$_{0.15}$CuO$_4$\cite{Eg4}.  Billinge, et
al.\cite{Bill}, have used this technique to determine the local tilt of a single
CuO$_6$ octahedron in LBCO.  They found that {\it there was no change in the 
local tilt direction in crossing the LTT-LTO phase boundary} (Fig.~
\ref{fig:d23}) -- the tilt in both phases was found to be consistent with that 
expected for the LTT phase! 
Above the transition, the LTO order builds up over an extended 
cell of dimensions $\approx 10\AA$.  These findings have since been confirmed 
by XAFS analysis in LSCO\cite{xafs}, and are in good agreement with the present 
dynamic JT model.  Also, in measurements of the elastic moduli near the HTT-LTO
transition, the critical fluctuations could only be understood if there were
deviations from pure $Bmab$ symmetry in the LTO phase, suggested to be 
associated with local deviations of the tilt axes from $[1,1,0]$ or $[1,\bar 1,0
]$ directions\cite{TF2}.  PDF's also find that the local symmetry in the HTT 
phase has finite octahedral tilts with random directions\cite{Eg3,HTO1,Eg7}.
\par
Electron diffraction experiments find similar results.  In LBCO, $x=0.125$, LTT
superlattice spots are found up to the LTO-LTT transition at 70K, but diffuse
spots, associated with local LTT order, persist in the LTO phase almost up to 
the HTT phase boundary at 180K\cite{ED1,ED2}.  In La$_{1.875-y}$Sm$_y$Sr$_{0.125
}$CuO$_4$, long-range Pccn (or LTT) order is found for $0.2\le x\le 0.4$ (the 
highest doping studied), but short range Pccn/LTT order was seen, as either 
diffuse spots or streaks, over the entire range $0\le x\le 0.4$ and up to room 
temperature\cite{ED3}. In this material, there is again a strong anticorrelation
between Pccn/LTT order and superconductivity.  Neutron diffraction studies of 
the phonon dispersion in LSCO\cite{PiRe} had initially been interpreted in terms
of `extra' branches.  It is now believed that these branches are due to the
orthorhombic splitting, but {\it are found to persist well into the HTT phase}.
Raman measurements\cite{Sug} also find evidence of local orthorhombic order in
LSCO for $x>0.2$.
\par
Measurements of elastic properties show consistent results.  There is a
substantial Curie-Weiss-like softening of $c_{66}$ as the HTT-LTO transition is 
approached from above, and the quality factor $Q$ is also substantially reduced.
This was initially interpreted in terms of critical fluctuations\cite{Migl2}.  
However, the softening persists to very high temperatures, 80K above the 
HTT-LTO transition, much higher than a theoretical estimate for fluctuations.
After considering a number of models, Migliori, et al.\cite{Migl3} concluded 
that the most probable explanation involves a {\it linear coupling} between the
order parameter and strain.  This is not allowed in tetragonal symmetry, so it 
is concluded that the {\it local symmetry} is non-tetragonal.  As further
evidence, it is noted that a number of mode-anticrossing phenomena are present
in the resonant ultrasound experiment, which would not be allowed in
tetragonal symmetry.  An alternative possibility is that nonlinear coupling is
anomalously strong\cite{BuH}.
\par
Moreover, throughout the LTO phase the elastic properties remain anomalous;
$c_{66}$ remains soft, and $Q$ is reduced by a factor of $\sim 100
$\cite{Migl2,LLN}.  This is assumed to be caused by martensitic-like
domain wall motion.  While it was related to the presence of twinning in the LTO
phase, I wonder whether it might not be rather associated with dynamic LTT
fluctuations; it would be interesting to repeat the experiment with an 
untwinned sample.  Most interestingly, there is a partial restiffening of $c_{
66}$ in LBCO ($x =0.15$) below 63K\cite{Foss}, associated with the onset of the
LTT phase.  In LSCO, the observation of this stiffening was taken as evidence
for an LTT onset below T$_c$\cite{LTTfl}.  
A similar effect is found in La$_2$NiO$_{4+\delta}$\cite{Bril}.

\subsection{Anomalies at T$_c$}

Given the above picture of a dynamic JT effect competing with superconductivity
for the VHS dos, it is hardly surprising that, when the superconducting 
transition does occur, the loss of dos would lead to striking modifications
in the local structure -- in particular, modifications of the local tilting
structure.  Thus, in the Bilbro-McMillan theory for A15 compounds\cite{BM}, 
the onset of superconductivity freezes in, or even reduces the order parameter 
of the competing structural instability.  A similar effect is predicted for the
cuprates (Fig. 4 of Ref. \cite{RM8a}).  There is
considerable experimental evidence for such structural anomalies in the 
immediate vicinity of T$_c$, which will be briefly reviewed here.   
\par
These experiments do not necessarily imply that
electron-phonon coupling plays a dominant role in superconductivity.  
Anomalies at T$_c$ are expected due to very general thermodynamic relations
because the free energy in a superconductor depends on strain.  Small
anomalies are found in conventional superconductors, and these should be larger
in the cuprates since they scale with $N(0)T_c^2$\cite{PBA}.  Thus, when the
superconducting gap opens, the linewidths of most phonons will be affected.
If the phonon frequency $\hbar\omega_{ph}$ lies below the gap energy $\Delta$, 
the normal-state electron-phonon scattering will be cut off, and the phonon 
width will decrease. But, if $\hbar\omega_{ph}\ge\Delta$, there will be 
additional scattering associated with pair-breaking\cite{ZZ}.  This change in
scattering has been observed, and used to estimate the superconducting 
gap\cite{phgap}.  However, it does not require a very large electron-phonon
coupling to explain these results, $\lambda\approx 0.6$ is adequate. (This 
result is based on analysis of c-axis polarized phonons, which may only have a 
quadratic electron-phonon coupling.)
\par
On the other hand, some phonon anomalies observed at T$_c$ in the 
cuprates\cite{Moo1} are so large that they cannot be explained by conventional 
Eliashberg theory\cite{Zeyh} unless nesting effects are included\cite{Mars}. 
Alternatively, some of the following anomalies may be associated with the 
pseudogap, which generally falls at a temperature close to T$_c$ in optimally 
doped cuprates.  Whether the superconducting or the pseudogap dominates can be 
determined by studying underdoped samples, where the two gaps are well 
separated.
\par
A summary of these studies is as follows: (1) in most of the cuprates,
there is a local, probably dynamic order, which is different from the average,
long-range order. (2) This local order consists predominantly of deviations of
the oxygens (those in the CuO$_2$ planes and those adjacent to these planes)
from their equilibrium positions. (3) These planar oxygen motions are 
predominantly tilts and shears, which change local O-O distances while leaving
Cu-O distances fairly constant. (4) At T$_c$, there is a sudden decrease of
this local disorder.  (Within the VHS model, superconductivity `eats' part of
the dos peak which was driving the dynamic JT effect.)
\par
Evidence for these effects has been summarized by Egami and 
coworkers\cite{Eg8,EgGins}, and is briefly recapitulated here.  PDF analysis 
finds evidence for dynamical tilting,
which decreases at $T_c$ in Tl-2212\cite{Eg2} and NCCO\cite{Eg6,BEg}.  Electron 
diffraction studies in Tl-2223 find diffuse scattering that increases 
dramatically below T$_c$\cite{Tled}; the symmetry of the scattering is 
consistent with LTT-type tilts.  There are also anomalies in thermal
expansion coefficients, which may be associated with local structural
distortions\cite{Eg5,Brad}.  In LSCO, the anomalies in the (anisotropic) thermal
expansion coefficients are consistent with a freezing in of the octahedral tilt
below $T_c$, whereas above $T_c$ the tilt angle is changing significantly with
temperature\cite{Brad}.  In O-doped La$_2$CuO$_{4+\delta}$, a shift of the La 
NQR frequency near T$_c$ has been interpreted as a sudden change of tilt 
angle\cite{Hamm2}.  A similar freezing in of local tilt disorder also
appears to be found in YBCO; thermal expansion measurements find an anomaly in
the orthorhombic strain, associated with freezing in of the orthorhombic
deformation\cite{Mein1}, while a similar freezing in is inferred from the 
temperature dependence of the Debye-Waller factors in a neutron diffraction 
study\cite{Schw}, and from changes in the Cu-Ba distances observed via 
EXAFS\cite{Roh}. Indeed, in an untwinned single crystal, there is striking
evidence that the orthorhombicity is greatly reduced immediately below $T_c
$\cite{Mein2}.  Neutron diffraction measurements of the phonon density of states
in LSCO\cite{Ara2} find an anomalous sharpening of the O-related peaks below T$_
c$, as if a dynamic fluctuation were quenched; Zn doping is found to have the 
same effect as increasing T.  Related effects are found in YBa$_2$Cu$_4$O$_8
$\cite{Kal11} and YBa$_2$Cu$_{3.5}$O$_{7.5}$\cite{Kal12}, where the b-axis 
length becomes T-independent below T$_c$.  Finally, Arai, et al.\cite{Ara}, in 
an inelastic neutron scattering study of YBCO, found evidence for a local $<110
>$-type buckling distortion, with a dynamic correlation length which {\it 
diverges} at $T_c$; similar anomalies are found in LSCO\cite{Ara3}.
\par
Some ion channeling studies\cite{Eg1} find a sudden increase in ordering below 
the superconducting transition in YBCO and Bi-2212, associated predominantly 
with a,b-plane vibrations of Cu's and apical O's; in an O-deficient sample, the 
anomalies were absent\cite{Eg13}. However, while a second group also reported 
channeling anomalies\cite{Eg12}, the two studies are not mutually consistent.  
A more recent study on films found no evidence for channeling 
anomalies\cite{Heck}.  Clearly, more work is required to resolve this issue.
\par
An intriguing possibility arises from a study of the c-axis optical conductivity
of LSCO\cite{TNU}; below T$_c$, a sharp plasma edge appears, which is associated
with {\it superconducting carriers}.  That is, since the c-axis plasma frequency
is smaller than the superconducting gap, the reflectivity is controlled by the
former.  For present purposes, the relevant point is that the observation of the
plasma frequency signals a coherent c-axis charge transport which is absent
above T$_c$.  Such normal-state incoherent c-axis transport was predicted in 
connection with spin-charge separation models\cite{WHA}.  Within the present 
model, this could be associated either with nanoscale phase separation or with 
the dynamic JT effect. If this is purely 2D, then the tilts in different layers 
will be uncorrelated, preventing coherent c-axis hopping.  At T$_c$, the tilting
is frozen out, and coherent c-axis transport appears.  Alternatively, if the 
holes are confined to domain walls, Coulomb repulsion will cause walls in 
different layers to avoid one another, leading to a very incoherent c-axis 
transport.
\par
M\"ossbauer studies\cite{Bool} of YBa$_2$Cu$_3$O$_8$ find motional narrowing of
the nuclear resonance above T$_c$ due to {\it dynamically fluctuating EFG's},
associated with vibrations of the planar and apical O's.  Below T$_c$ there is
significant line broadening associated with greatly reduced fluctuations.  
This is in striking agreement with the present model of dynamic JT fluctuations,
which are partially quenched below T$_c$.  It may also provide an explanation
for a puzzling feature of EFG measurements in the cuprates.  The asymmetry
parameter $\eta$ should be proportional to the orthorhombic splitting in the
LTO phase; instead NQR measurements find an anomalously small $\eta\approx 
0$\cite{SaHo,NQR0}.  This result would be consistent with 
motional narrowing.  Below $T_c$, an excitation energy of 9.3meV is found; this 
is comparable to the inter-(LTT)-well excitation energy $\approx 15$meV 
calculated by Pickett, et al.\cite{LTTP,Pickx} for octahedral tilts in LBCO.
\par
In Bi-2212, Mook, et al.\cite{Moo1} used neutron resonance absorption 
spectroscopy (NRAS) to measure the average kinetic energy of the Cu related to 
the phonon density of states.  They found a significant softening just above 
$T_c$, which is arrested by the superconducting transition. This softening is 
related to a large decrease in multiphonon scattering (indicative of strong 
anharmonicity) in the same temperature range for both Cu and O 
modes\cite{Moo2}.  Below T$_c$, the oxygen breathing modes are broadened over a
large range of $\vec q$ space near $(\pi /a,\pi /a)$\cite{Moo2}, suggestive of 
strong electron-phonon coupling, perhaps enhanced by nesting effects\cite{VCr}.
No anomalous NRAS softening was found for Cu in YBCO\cite{Moo3} (although, due 
to the presence of chain sites, the expected change is close to the limits of
resolution of the measurement); below $\approx$100K, the Cu kinetic energy 
stopped
changing with T, similar to Bi-2212.  A similar softening -- here of the elastic
constant $(C_{11}-C_{12})/2$ -- is found in LSCO ($x\simeq 0.14$), which is
replaced by a mode hardening below T$_c$\cite{Noha}.  The symmetry of the
softening is consistent with an incipient LTT or Pccn phase. The superconducting
T$_c$ was reduced by an external magnetic field, and it was found that the mode
softening persisted to $T_c(B)$, Fig.~\ref{fig:b23}.  The solid line is fit to 
the theoretical expression\cite{Luthi}
\begin{equation}
{\Delta (C_{11}-C_{12})\over 2} =-2d^2\int_0^{\infty}N(E)(-{\partial f(E)\over
\partial E})dE,
\label{eq:99h}
\end{equation}
where $d$ is the deformation potential, $\Delta E_{\vec k}=2d(\epsilon_{xx}-
\epsilon_{yy})$, and it is assumed that the dos is given by a 2D VHS.  This
anomaly is much larger, and begins at a higher temperature ($\approx$50K) in 
optimally doped LSCO than in either underdoped ($x$=0.09) or overdoped ($x
$=0.19) material\cite{Noha2}.  The magnitude of the effect is much larger than 
in conventional superconductors, but that is expected since $\Delta C_{ij}/
C_{ij}\propto (\Delta (0)/E_F)^2$.  What is unconventional is the sign of the
effect; for conventional superconductors, the lattice usually softens below
T$_c$\cite{Noha2}.  As we have seen, the stiffening can be a signature of a
competing structural instability.
A similar hardening is found in Y$_{0.9}$Ca$_{0.1}$Ba$_2$Cu$_4$O$_8$; however,
it is displaced about 10$^o$ below T$_c$ and resembles a first-order
transition\cite{TF5}.
\par
Measurements of the elastic properties of the cuprates find a large number of
anomalous changes in sound velocities and attenuation peaks, but there is
tremendous variability in the results of different groups.  In polycrystalline
samples, the results may depend on grain size or on void fraction.  Recent
reviews\cite{TF2,TF1,Cann,Ultra} have concluded that there is considerable 
evidence for a structural instability a few degrees above T$_c$ in both the La- 
and the Y-cuprates. For some elastic constant studies\cite{Eg5,Migl} there 
appears to be an anomaly at a temperature corresponding to the optimum T$_c$ for
LSCO, extending over a wide composition range, including insulating samples.  
The form of the anomaly -- a minimum in the sound velocity -- is suggestive of a
low-frequency relaxation process, such as domain wall motion\cite{Migl}.  A
resistive anomaly is found in the same regime ($\approx 37K$) in overdoped 
LSCO\cite{press2}.  Similar anomalies are reported {\it above} T$_c$, at 95K 
in LSCO ($x=0.2$) and at 120K in YBCO\cite{Bhat}.  In high-resolution dilatometry
studies of twinned and untwinned LSCO crystals, instabilities in the range 
40-60K (not reproducible in detail) were found in the {\it twinned} samples 
only, reinforcing the suggestion that these features are `related to the 
relaxation of twin-induced internal stresses'\cite{Gugen}.
The Cu NQR frequency also displays an anomaly at T$_c$, suggestive of a 
structural modification\cite{Ries,NQR,Hamm2}; however, such anomalies are found 
in conventional superconductors\cite{Ries}, although their origin is not
understood.
\par
Besides the evidence for a structural anomaly just above T$_c$, additional 
information can be gleaned from a study of the discontinuities in elastic
parameters {\it at} T$_c$.  Millis and Rabe\cite{MiRa} provided a detailed
analysis of the experimental data as of 1988 -- unfortunately mostly on 
polycrystals.  They found that, whereas the first strain derivatives of T$_c$
were comparable to those of conventional superconductors, some second
derivatives are anomalously large.  They suggest that T$_c$ is extremely 
sensitive to some form of shear distortion, which by symmetry cannot couple to 
T$_c$ in first order.  An analogous case occurs in V$_3$Si, where 
superconductivity couples to a CDW which reduces the lattice symmetry from cubic
to tetragonal. In the cuprates, this mechanism could also work, but only if 
``the transition is ... not the observed orthorhombic-tetragonal transition but 
instead a potential transition to a still lower symmetry phase, which is 
inhibited by the presence of the superconductivity''\cite{MiRa}.

In light of the evidence that, at optimum T$_c$, the pseudogap and 
superconducting transitions virtually coincide, it is important to analyze the
data on local structural anomalies away from optimum doping to see which of
the two transitions they are associated with.

\subsection{Anomalies at T$_N$}

In the VHS model for the pseudogap, it is suggested that near half filling
the pseudogap and LTO transitions are of a spin-Peierls nature, involving
a modulation of the exchange coupling $J$\cite{RM11}.  In this case, it might
be expected that there should be some structural anomalies at the N\'eel
transition.  There appears to be some experimental evidence for such anomalies.
Thus in LSCO:Gd, Rettori, et al.\cite{Rett} used Gd EPR to detect the local
magnetic field near the N\'eel transition.  For a quantitative fit to the field
dependence, they were required to assume that, below T$_N$ there is either a
lattice distortion or a Heisenberg-type interaction between the Cu and Gd spins.
In La$_2$CuO$_{4+\delta}$, there are enormous transport anomalies near T$_N$ at 
small $\delta$, which show prominent hysteresis and are correlated with thermal
expansion anomalies\cite{CHTF}; however, these anomalies are probably
associated with phase separation, and not with the N\'eel transition.  Clear 
phonon softening at the N\'eel transition is also found in the related compound 
CuO\cite{CuO}.

\subsection{Transitions off of the CuO$_2$ Planes}

\subsubsection{VHS Related}

As described in Section VIII, the VHS-JT model is considerably more flexible 
than the conventional CDW model in that a large number of phonon modes -- any 
mode which splits the VHS degeneracy -- could be involved.  In this subsection, 
I briefly note that the cuprates often display structural instabilities 
involving atomic displacements completely {\it off of the CuO$_2$ planes}, often
involving quite complex (incommensurate) distortions, and yet these distortions 
also split the VHS degeneracy.  Thus for instance, in YBCO, the chain ordering 
splits the degeneracy of the $X$ and $Y$ point VHS's.  In Tl-1201 and Tl-1212, 
the structural distortions have been analyzed in detail\cite{GGL}. The O-atoms 
show a very large displacement from their nominal positions, and this distortion
breaks the symmetry of the two in-(CuO$_2$-)plane O's, thereby lifting the VHS 
degeneracy.  However, these off-CuO$_2$ plane distortions do not show any
sudden changes at T$_c$, as opposed to the in-plane (dynamic) 
distortions\cite{Eg7}.  Since Bi-2212 has a different symmetry from LSCO, it
may be that the orthorhombic distortion splits the VHS degeneracy -- certainly
the `shadow Fermi surfaces' seen in photoemission\cite{Aeb} are exactly those 
which would be produced by VHS nesting.  
\par
These distortions can have two different roles in VHS theory: (1) 
it is possible that the transitions are driven by the VHS instability; (2)
they could have independent origins, but by splitting the VHS
degeneracy, they could weaken CDW-like instabilities of the CuO$_2$ planes
which compete with superconductivity.

In a related vein, Phillips\cite{JC1} has suggested that defects off of the 
CuO$_2$ planes can act collectively to enhance the strength of a dos peak 
associated with the VHS.

\subsubsection{CDW's on the Chains}

In assessing the importance of electron-phonon interaction in the cuprates, it
is important to not overlook the clues offered by the CuO chains in YBCO.  There
are certainly many differences: the chains are far from half filled, so 
correlation effects are weaker; the chains are nearly 1D, so the theory should 
be easier; the chains are very sensitive to oxygen disorder.  Moreover, it is
only recently that the use of untwinned single crystals has allowed the
unambiguous extraction of the chain contribution to transport.  Hence, the signs
that there are CDW's on the chains are most intriguing, and should be followed 
up with further study.  The most striking results are Edwards, et al.'s STM 
studies\cite{deLo} of in-situ cleaved surfaces of YBCO.  They found a modulation
along the chains with wave vector close to the expected value of $2k_F$
for the chains.  They also found evidence of a $\approx 20meV$ energy gap; this
seems to be more consistent with an induced superconducting gap (albeit rather
large). Near an oxygen vacancy, the modulation amplitude increases, as if the
CDW is pinned, but the gap is reduced.  This would be expected for 
superconductivity since oxygen vacancies or disorder quench superconductivity.
However, the question then arises as to why no CDW gap is seen.  A neutron
diffraction study\cite{nchai} has now found evidence for incommensurate
fluctuations with wave number consistent with the chain $2k_F$.
\par
In analyzing transport properties of untwinned YBCO, Fehrenbacher\cite{Fehr}
assumed that the difference between the conductivities along the $b$ and $a$ 
axes could be identified with the chain conductivity, $\sigma_{ch}$.  [One note
of caution: the VHS dos peaks are associated with quasi-1D motion along the
respective Cu-O-Cu links in the plane.  Even if the chains were insulating, the
conductivities along $a$ and $b$ could differ simply because the $X$ and $Y$ 
VHS's can be shifted by different amounts from $E_F$.]  He found that he could 
describe $\sigma_{ch}$ in terms of disorder scattering, but only by postulating
an unreasonably large disorder.  A much more satisfactory fit was found in terms
of a strong polaronic coupling to the {\it chain} Cu-O stretch modes.  Near
half filling, he found evidence that the chains form incommensurate CDW's with
the charge motion associated with moving domain walls.  The similarity of this
picture with the model I have described for the planes is striking.
\par
There has also been evidence that the chain O's have a double-well potential
with two minima displaced off of the chain axis in the $a$ direction\cite{zig1}.
This is consistent with frozen phonon calculations of Cohen, et al.\cite{CPK},
which find such a potential.  It has been proposed that some low temperature 
phase transitions identified in elastic constant measurements could be related 
to ferroelectric or antiferroelectric ordering of these O's\cite{Cann,TF4}.
However, inelastic neutron scattering finds no anomalous or anharmonic behavior
associated with the chain zigzag mode\cite{zig2,zig3}.  Recent neutron 
pair-distribution function studies in Y-124\cite{Send} have found evidence for
ferroelectric domains associated with chain-O displacements on a scale of
$\approx 10-20\AA$.  It was postulated that these domains were associated with
polaronic effects of holes in the CuO$_2$ planes.

\subsection{Other Evidence for Strong Electron-Phonon Coupling}
\par
In addition to the above experiments, which have a direct bearing on VHS theory,
there are a number of other results which testify to the importance of 
electron-phonon coupling in the cuprates, but either do not single out a 
particular phonon mode, or are associated with modes which do not split the VHS
degeneracy.
\par
Tunneling studies are a traditional means of measuring electron-phonon coupling
strength and, in fact, of determining the full spectrum of coupling in terms of
Eliashberg's $\alpha^2F(\omega )$.  While many studies have been carried out,
the results of different groups -- or even of the same group on different
samples -- are not always consistent.  Rather than attempting to provide a
detailed survey, I will here merely list some suggestive results.  $\alpha
^2F$ has been measured for the low T$_c$ material, NCCO\cite{Zasa}, and shown to
predict the correct T$_c$ values.  In BKBO, the results are mixed: Dynes, et 
al.\cite{Dyn} finds coupling to a number of low frequency phonons only ($\hbar
\omega\le 8meV$), which leads to $\lambda\simeq 1$, T$_c\simeq$4-5K, much less
than the experimental value (11-13K).  Zasadzinski, et al. \cite{Zasa} find an 
additional coupling to a high-energy optical phonon, and are able to produce the
correct T$_c$.  Similarly, in YBCO, Dynes, et al.\cite{Dyn} find coupling only
up to $\approx 30meV$ (below most of the oxygen vibrations), which cannot fully
account for the experimental T$_c$'s.  Nevertheless, this measured $\alpha^2F$
yields $\lambda\simeq 2$, T$_c\simeq$60K, which means that electron-phonon
coupling is by no means negligible!  Finally, recent measurements on Bi-2212
have found a phonon coupling spectrum which leads to T$_c$=87K, $\Delta =
22$meV\cite{A2F}.  Strong coupling is not restricted to any one mode, but is
spread out over a large number of phonons, including a number of high frequency 
phonons -- in particular, phonons at 57meV (axial apical O mode) and 72meV 
(Cu-O stretch mode).  On the other hand, the direct contribution to T$_c$ of the
high frequency phonons is small.  For example, eliminating the 57 (72) meV peak 
from the analysis reduces T$_c$ by only 6.6 (3.3) K.
\par
Point-contact spectroscopy of YBCO also finds modulations associated
with particular phonon frequencies\cite{pcs}.  The modulations are very 
selective; only certain phonons show strong coupling, generally those
which show large frequency shifts at T$_c$ (i.e., at the pseudogap).  
\par
The $a,b$-plane optical conductivity of several of the cuprates shows a series 
of notches at frequencies corresponding to {\it c-axis} polarized longitudinal
optical phonons\cite{cnotch}.  This suggests that the phonons, associated with O
stretching and bending, couple strongly to the electrons via some 
symmetry-breaking mechanism, perhaps associated with incipient CDW/SDW 
formation.
\par
While the isotope effect on T$_c$ is very small in op\-tim\-ally-doped YBCO, 
there appears to be a large O isotope effect on the penetration depth, $\lambda
$, due to an isotope effect on the effective mass, with 
$-d\ ln(m^*)/d\ ln(M_O)\simeq-0.6$\cite{ZhMo}.  This was interpreted as evidence
for strong phonon anharmonicity and/or breakdown of the Migdal approximation 
(associated with polaron formation).

\section{Phase Separation}

Phase separation is one of the most complicated and controversial
issues in the study of the high-T$_c$ cuprates.  There is thought to be
a phase separation {\it of the holes}, which can be limited to very short range 
by Coulomb forces.  Many experiments have now found evidence for phase 
separation in the cuprates, often on a very fine (nanoscopic) scale. Such a 
phase separation arises in a number of different theories involving strong 
correlations, including the VHS theory.  Indeed, it has been hypothesized that 
phase separation and superconductivity are closely related 
phenomena\cite{GRC,Rai,DaRC}.  This can be
seen quite easily: the smallest scale possible for phase separation occurs
when the hole-rich domain contains exactly two holes.  But this can be thought
of as a state of real-space pairing.  If larger-scale phase separation is
inhibited, and the pairs become mobile, then one can envision a Bose
condensation of these pairs into a superconducting state.  (While conceptually
appealing, this is not what appears to happen in the cuprates, but nevertheless
it motivates the proposed connection between phase separation and 
superconductivity.)  For calculations on finite clusters, it is easier to
detect evidence of phase separation, so the recommended procedure is to
delineate the domain of phase separation, then look near this region for
signatures of superconducting pair correlations\cite{DaRC}.
\par
Emery and Kivelson (EK)\cite{EK1,EK2} have shown that low-carrier-density
superconductors, such as the cuprates, are generically sensitive to phase
fluctuations, both classical and quantum.  Classical phase fluctuations are
governed by a phase fluctuation temperature $T^{max}_{\theta}\propto\hbar^2n_s
\xi /4m^*$, where $n_s$ is the superfluid density and $\xi$ a coherence length.
Fluctuations of the phase of a superconductor become important when $T_c\simeq
T^{max}_{\theta}$, and are negligible when $T_c<<T^{max}_{\theta}$.  EK find
that the latter condition holds for conventional, A15, and heavy Fermion
superconductors, but that $T_c\simeq T^{max}_{\theta}$ for the cuprates and some
organics, with BKBO and the C$_{60}$ superconductors in an intermediate
position.  Quantum fluctuations\cite{EK2}, on the other hand, become important
when the conductivity $\sigma(T_c)\le\sigma_Q\equiv 4e^2/hb$, where $b$ is a 
microscopic length.  Again, quantum fluctuations are negligible for conventional
superconductors, $\sigma(T_c)/\sigma_Q>>1$, but this ratio falls to $\sim 10$ in
the optimally doped cuprates, and is smaller in the underdoped materials.  If 
the ratio becomes less than one, superconductivity can be destroyed.  
\par
Despite the experimental evidence for phase separation, despite the many
theoretical investigations, despite two international conferences\cite{phassep} 
(with a third imminent), there is still considerable reluctance to seriously
consider the role of phase separation in the cuprates.  This is presumably 
because in these complex, multi-component systems, great care is needed to
ensure that the materials are in thermodynamic equilibrium (phase separation
could arise from incomplete mixing/oxygenation).  In optimally-prepared
samples, the `phase separation' is usually restricted to such nanoscopic scales 
that it ceases to resemble the more familiar phase separation in other 
materials: thus, the superconducting transition temperature can change smoothly 
with doping, since the domain size is comparable to the coherence length.  
Further complicating the experimental detection, this nanoscale phase separation
can be {\it dynamic}, so a particular probe can detect a homogeneous average if 
it averages over many fluctuation lifetimes.  Moreover, particularly in the low
doping regime, effects of phase separation can be confused with localization
effects, particularly as the nanoscale domains can easily localize on
potential fluctuations.  In this and the following sections, 
I will provide a detailed summary of the evidence, both theoretical and 
experimental, for such a phase separation.  In particular, there 
are circumstances in which more conventional, large-scale phase separation can
occur.  Table V summarizes the experiments I have found that provide evidence 
for such phase separation.  At this point, one should begin to go
beyond gathering such data, and attempt the harder task 
of showing that {\it all} experiments are consistent with this picture. As a
first step in this direction, Section XI.J briefly summarizes experiments which
are suggested to provide evidence against phase separation.

\subsection{Phase Separation in the VHS Model}

Hole phase separation arises in a number of theories.  As the hole density is
varied, the free energy is found to have a region of concave curvature 
(negative compressibility), for which the state of uniform hole density is
unstable.  When doped into this region, the holes will tend 
to spontaneously phase separate into domains of low and high hole density, 
outside of the unstable range (the exact densities are found by a Maxwell 
construction). The various theories differ in the forces which act to stabilize 
the two end-phases, but once phase separation has occured, all predict similar 
physics for the resulting domains, dominated by strong charging effects.  
\par
In the VHS model, it was found\cite{RM3} that one end-phase always lies exactly 
at the VHS, stabilized by strong electron-phonon coupling (see Section VIII.A). 
Essentially, this is due to the diverging compressibility near a VHS (Appendix
A) manifested as a strong electron-phonon coupling. The Van Hove -- 
Jahn-Teller effect, by splitting the dos peak, lowers the free energy of the 
combined system of electrons and phonons, but {\it only when the Fermi level 
coincides with the VHS!}  The free energy lowering has a $ln$ or $ln^2$
divergence, Eqs.~\ref{eq:O17} and ~\ref{eq:CC4}.
The associated free energy minimum leads to a remarkable state of 
affairs: while the hole doping precisely at the VHS is thermodynamically stable,
both the underdoped and overdoped regime are unstable, and lead to a phase
separation between the VHS phase and a second, stable phase with very different
hole doping\cite{RM3}.  On the underdoped side, the second phase is the 
antiferromagnetic insulator (AFI) with zero hole doping (half filled band); in 
the overdoped case, this phase appears to be a normal metal, but its exact 
nature is still unclear.  (Bok\cite{Bok} has suggested that the superconducting 
condensation energy can stabilize the VHS phase. This may be possible, but would
require very strong coupling, since phase separation starts near room 
temperature, far above T$_c$.)
\par
A similar cusp-like minimum of the free energy can
arise from either VHS nesting (Eq.\ref{eq:O17}) or conventional nesting
(Eq.\ref{eq:CC4}).  Grilli and Castellani\cite{GC1}, and Becca, et al\cite{CGS} 
also find that strong electron-phonon coupling can enhance the instability 
towards phase separation.
\par
It should be noted that, within the Van Hove scenario, phase separation is an 
added complication: in the {\it basic} scenario, which most VHS theorists have 
followed, the idea of phase separation is not addressed.  Even in the {\it 
generalized} Van Hove scenario, the occurence or non occurence of phase 
separation depends sensitively on material parameters (e.g., the low frequency 
dielectric constant).  It does not seem to play an essential role in enhancing 
$T_c$, but rather competes with it.  In short, if it exists, it can be 
understood (and was predicted) within the VHS model, but it is not a necessary 
consequence of the model.  

\subsection{Nanoscale Phase Separation}

\par
Mott\cite{Mott} has pointed out that for a general metal-insulator transition,
the number of free carriers should jump discontinuously at the transition.  A
consequence of this is that there should be a region of phase separation near
the transition.  Two special cases of this are the Mott-Hubbard transition near
an AFI, and the CDW transition in BKBO/BPBO.
Phase separation associated with doping away from an AFI phase at
half filling was recognized prior to the discovery of high-temperature 
superconductivity\cite{AFI,Nag}.  Nagaev's book\cite{Nag} provides a detailed
description of the nanoscopic nature of such a phase separation for a 3D AFI.  
More recently, Nagaev has reviewed phase separation in the cuprates on the basis
of this model\cite{Nag2}.
Since the {\it holes} are trying to phase separate by bunching up, the resulting
domains will not be electrically neutral unless the doping ions are sufficiently
mobile so as to be able to follow the hole motion. In the extreme limit that 
these ions are unable to move (the Sr in LSCO seems to be a good approximation),
the phase separation is purely associated with hole bunching, and Coulomb 
effects restrict the domain size to a scale of a few nanometers (the exact size 
depends sensitively on the value of the low-frequency dielectric constant and 
the level of doping\cite{RM3}). While Nagaev speaks of islands of the minority 
phase in a background of the majority phase, I have found that a lower energy 
state is associated with a phase in which the minority phase is segregated into 
grain boundaries between large domains of the majority phase\cite{RM3}.  Such 
grain boundary phases have also been found in Hartree-Fock calculations of the 
doped Hubbard model\cite{gbph}.  Fig~\ref{fig:j23} shows schematically how
complicated the resulting phase diagram can become: Fig~\ref{fig:j23}a shows the
free energy vs doping for the bulk phases, suggesting a macroscopic phase
separation.  However, Coulomb repulsion arrests this macroscopic phase 
separation, stabilizing a variety of finite domain structures, 
Fig~\ref{fig:j23}b.  Fig~\ref{fig:j23}c sketches a possible resulting phase
diagram drawn with the cuprates in mind.  Commensurability effects between the
domain size and the underlying atomic lattice can lead to a large variety of
phases (for examples, see Ref.~\cite{Low}, and the classical analogs cited in 
Section XII.A), including striped phases and staging effects, which will be
discussed in detail below.

\subsubsection{What's in a Name?}

In a sense, the phrase `phase separation on a
nanoscopic scale' is a contradiction in terms.  These
are single-phase materials that have the intrinsic property that the charge (or
spin) is inhomogeneously distributed on a nanoscopic scale.  Perhaps a more
neutral phrase, such as `charge bunching phase', would be preferable.  However,
at this point, our knowledge of the detailed structure of these phases is too
vague, and moreover, there seems to be a continuum of states from the
charge bunching limit to true macroscopic phase separation.  Hence, I will 
continue to use the technically incorrect but striking phrase, nanoscale phase
separation.
\par
Recently, it has become popular to designate these as CDW phases.  I think this
is also potentially misleading.  While any charge bunching of the holes will of
necessity be accompanied by local lattice distortions, I believe that underlying
driving force (two preferred hole densities) is distinct from the
driving force for conventional, Peierls-distortion-cum-CDW formation (Fermi
surface nesting), and designating both phenomena by the same name will only lead
to confusion -- especially as both phenomena may be present in the cuprates.

\subsubsection{Phase Separation vs. Localization}

\par
This nanoscale phase separation can easily be confused with localization 
effects.  For instance, in LSCO, even if there were no tendency for either holes
or Sr to phase segregate, there would still be a random distribution of Sr ions,
with either single ions or small clusters.  This random clustering will induce
a random potential on the CuO$_2$ planes, which could lead to hole localization.
Now if the localization potential is random, some holes will be strongly 
localized, others more weakly.  Thus, while localization leads to an 
inhomogeneous distribution of holes on the CuO$_2$ planes, it also leads to a
broad distribution of hole densities.  The characteristic feature of nanoscale
phase separation is that there are {\it only two preferred hole densities}, one
high and one low (zero near the AFI), and all of the holes are distributed
between these two types of sites.  Of course, this is an idealized situation.
If the size of the dense hole domains is truly nanoscopic, i.e., containing
only a few holes, then finite size (including charging) and proximity effects 
will cause the equilibrium hole density to depend somewhat on domain size.
Furthermore, such small domains will feel the background potential due to, e.g.,
Sr inhomogeneity, and can in turn become localized.  Fortunately, it is possible
to dope LCO by either Sr or Ba substitution or by excess oxygen.  Because the
interstitial O's have a high mobility, the phase separation expands to a
macroscopic scale.  Despite this, many of the resulting phenomena, including
superconductivity, depend only on the local hole density, independent of the 
doping method, providing strong evidence that they are dominated by the physics 
of hole bunching.  

\subsubsection{Phase Separation Realized as a Form of CDW}

\par
Recent cluster calculations, involving a variety of tJ and Hubbard models, but
including a long-range Coulomb interaction\cite{HDN}, have found that the phase
separation is replaced by a regime of CDW's.  Now, these CDW's are not a phase
separation in the conventional sense; the energy remains concave as a function
of hole density, and the hole domains have a finite equilibrium width rather 
than diverging to infinity.  However, these CDW's are precisely a form of 
nanoscale phase separation, as illustrated in Fig.~\ref{fig:f23}c.  Instead of 
the weak charge fluctuation typically associated with CDW's
(Fig.~\ref{fig:f23}a), these domains show maximal modulation, with all holes 
confined to one domain of the CDW, and the other domain remaining an AFI.  This 
`phase-separating CDW' bears some resemblance to the discommensurations 
introduced by McMillan\cite{McMd}, Fig.~\ref{fig:f23}b; however, the usual 
discommensurations are controlled by the Peierls distortion which accompanies 
the CDW, while in the present case, it is the charge modulations which play the 
dominant role.  For example, a common realization of discommensurations arises
with atoms or molecules adsorbed onto or intercalated into graphite.  In this
case, the graphite planes provide preferred sites for the adsorbed/intercalated
atoms to occupy, so there is a set of filling factors for which these atoms are
exactly commensurate with the substrate.  For other filling fractions, a
uniform, periodic array of atoms would be incommensurate.  However, the atoms
can lower their energy by forming commensurate patches, with incommensurate
domain walls in which the atomic density is either higher or lower, to 
accomodate the atomic excess or deficit.  If the adsorbed or intercalated layer
is charged, the electronic density will also be modulated, but the modulation
is controlled by atomic commensurability effects.  In the nickelates, a similar,
but electronic mechanism seems to arise, Fig.~\ref{fig:31}, below: there are 
charge stripes in which every other or every third atom has an excess hole.  In 
the cuprates, however, the charge bunching is believed to 
be a hole doping effect, due to the existence of one (or two) preferred hole 
densities, and not due to commensurability effects (most of the calculations do 
not consider the ions explicitly).
\par  
Variations in doping lead to changes in the periodicity and width of the
hole-rich domains, with relatively small changes in the hole density in those
domains.  The distinction between these CDW's and the conventional notion of 
CDW's as weak modulations can be illustrated by a particular example discussed 
in Section XVII.B.  In UPt$_3$, a CDW phase is found to be associated with two
separate superconducting transitions.  Based on the present view of CDW's, we
might have assumed that the two T$_c$'s were associated with the high and low
density components of the CDW; instead, the CDW has been assumed to involve only
a negligibly weak electronic density modulation, with its main role being to
lower the lattice symmetry.
\par
Despite this distinction between the electronic striped phases and conventional
discommensuration-related CDW phenomena, the former will nonetheless involve
a significant contribution from lattice distortions.  First, within the VHS 
model, the phase separation is driven by the strong electron-phonon coupling, 
via Hume-Rothery effects (Section VIII.A).  Secondly, on any model, Coulomb
effects oppose the phase separation, and nanoscale phase separation is possible
only if the Coulomb effects are sufficiently screened by the background
dielectric constant.  In the cuprates, the low frequency dielectric constant is
quite large, due to strong phonon polarizibility.  Hence, charged stripe
formation will be accompanied by significant lattice relaxation.
\par
It should be noted that in this nanoscale, CDW phase, the breakup into hole-rich
and hole-poor domains is automatically accounted for, so $E(n)$ remains a 
concave function, and the compressibility is positive.  Equivalently, the
{\it chemical potential $\mu$ is not flat} (independent of hole density $n$).
This is important, because the chemical potential has been measured, and the
fact that it usually has no flat regions as a function of $n$ has been taken as
evidence against phase separation.  These results can readily be understood.
The free energy is a local property.  If the phase separated domains exist on a
sufficiently nanoscopic scale, then a free energy which is an average over a 
local cell can be defined.  If, on the other hand, the domains can grow to
macroscopic size, then no local free energy can be defined, and a phase 
separation occurs to tie together the free energies of the two limiting phases.
The calculations in the following subsection provide examples of such stripe or
CDW phases; note in particular Castro Neto's local phase separation regime
`above the spinodal'.

\subsubsection{Calculations of Nanoscale Phase Sep\-aration}

\par
L\"ow, et al.\cite{Low} developed a spin 1 Ising model to study the effect of
Coulomb effects on retarding phase separation.  They found clear evidence for 
nanoscale phase separation, into a complicated array of stripe and checkerboard
phases.  Again, the resulting domain structures have positive compressibility.  
The size of the individual domains depends on the strength of the Coulomb 
repulsion, i.e., on the background dielectric constant\cite{RM3,Eme1}.
Chayes, et al.\cite{Chay} have recently explored the competition between CDW
order and superconductivity in such a model.
Mehring and coworkers\cite{Mehr} have also applied Ising, as well as 
Ginsburg-Landau models to study the phase separation.  The effect of Coulomb
repulsion in arresting a phase separation at nanoscopic scales arises in many
situations that have little to do with superconductivity.  Some of these are
discussed in Section XII.  The numerical simulations\cite{rlax4,rlax3} 
typically lead to stripe ($\sim$domain wall) or island phases.  Castro Neto and
Hone\cite{CNH} modelled the stripe phase via a real space renormalization
approach as an anisotropic nearest neighbor Heisenberg exchange model, and
were able to reproduce the magnetic phase diagram of LSCO.  Zaanen, et 
al.\cite{ZHS} studied the dynamics of a domain-wall fluid, suggesting that this
could describe the low-frequency magnetic dynamics of the cuprates, and
showing that there could be a crossover from a classical to a quantum fluid.
\par
Castro Neto\cite{CaN} showed that if the holes form a Fermi liquid, then 
the phase separation can be described in terms of Landau Fermi liquid theory.  
He found that even in the absence of long-range Coulomb effects, there is a
regime of local-scale phase separation (charged stripes).  The condition for 
long-range phase separation is the vanishing of the compressibility (the 
spinodal line), when the Landau parameter $F_0<-1$, and he 
derived formulas for the rate of domain growth.  Coulomb effects limit the 
domains to a finite size.  The compressibility becomes wavenumber dependent,
\begin{equation}
\kappa(q)={\kappa_0\over 1+F_0+{2\pi e^2N(0)\over\epsilon q}},
\label{eq:99}
\end{equation}
where $\kappa_0$ is the unrenormalized compressibility and $\epsilon$ is the 
background dielectric constant.  Thus, phase separation is limited to sizes 
smaller than
\begin{equation}
R\simeq{\epsilon |1+F_0|\over e^2m^*}.
\label{eq:99a}
\end{equation}
Note that near the Mott-Hubbard transition at half filling, $m^*\rightarrow
\infty$ and $R$ takes its minimum size.

\subsubsection{Phase Separation Realized via Staging}

There is yet another way to introduce a modulated structure which allows the
hole gas to separate into hole-rich and hole-poor domains: that is for the
doping ions to form a 1D modulated structure in the direction perpendicular to 
the CuO$_2$ planes.  A similar phenomenon has been thoroughly
studied in graphite intercalation compounds (GIC's)\cite{GIC}, where it is known
as {\bf staging}.  Staging was initially reported for a related compound, La$_2
$NiO$_{4+\delta}$ (LNO), when doped by excess oxygen\cite{stag}, but has more
recently been found in the cuprates\cite{Custag,Cust2}.  {\bf Intercalation} is 
a phenomenon in which the dopant atoms or molecules form {\it intercalated} 
layers, with a fairly uniform composition, between the layers of the host 
molecule (graphite, LNO).  Increasing the level of doping corresponds to adding 
extra layers of intercalant.  These layers repel one another, and tend to form 
with a regular spacing between layers.  {\bf Staging} is the name given to this 
phenomenon: the {\it stage number} is the number of host layers between 
successive intercalant layers -- i.e., a stage 3 K-intercalated GIC has three 
layers of graphite separating every layer of potassium.  Since the composition 
of the intercalant layer is (fairly) constant, a stage $n$ compound corresponds 
to a particular doping level. For doping levels intermediate between two stages,
the extra intercalant can be incorporated in a number of different ways: by 
adjusting the intercalant density in a given layer, by inserting extra layers, 
either at random or in a regular pattern, or by phase separating into two 
regular staged regions\cite{Saf}.
\par
For stage index $n\ge 3$, the hole doping is nonuniform, with the majority of 
holes concentrated on the layers adjacent to the intercalant layer\cite{PiSt}.  
Thus, if the holes have a natural tendency to phase separate, then staging is 
another method for accomodating this phase separation on a nanoscopic level.  In
light of this, it is interesting to note that in both LNO and LCO, the lowest 
observed stage is 2: for $n\ge 2$, the hole density in the boundary layers is 
nearly independent of stage, whereas for stage 1, the hole density would nearly 
double.  
\par
In LCO, there is a well-defined transition to a phase-separated state, near
room temperature.  The nature of the high-T phase is not well understood, and it
may be pertinent to recall the Daumas-Herold\cite{DH} model of staging in GIC's.
They suggest that there is actually the same number of intercalant molecules in
every layer, but below the phase separation temperature, these molecules phase
separate into high and low density regions in each layer.  Interlayer coupling
then causes the formation of well-staged domains.  Thus, the transition from a
uniform to a well-staged compound (or an interstage transition, as extra 
intercalant is added) can take place without any molecules having to hop between
layers.  Note also that, within an individual layer, the phase separation is
identical to the traditional view of phase separation in LCO: the special
features associated with staging involve interlayer coupling only.
\par
Since the intercalant tends to nestle into geometric depressions in the adjacent
host layer, the host's stacking sequence can change across an intercalant layer.
Thus, for K in GIC, the K fits into the hole in the benzene-like rings of the 
graphite, so the K-layer changes the graphite stacking sequence from AB, where 
rings on consecutive layers do not line up, to AA, where they do; the 
intercalant forms {\it ordered antiphase domain boundaries}.  The antiphase 
boundaries lead to additional spots in the diffraction pattern.  Such antiphase
boundaries are seen in LNO and LCO, and cause what is essentially a 
one-dimensional phenomenon to appear as a complex 3D ordering.  Hence, much of 
the early work, which did not recognize the staging phenomenon, needs to be
reinterpreted.  Clearly, staging presents additional complications when the
hole doping is achieved by simultaneously varying both $x$ and $\delta$; the
resulting phases are independent functions of both $P$ and $\delta$.
\par
It should be noted that in a layered material subject to nanoscale phase
separation, Daumas-Herold domains should form {\it even if the doping molecules
are completely immobile.}  All that is required is that holes which bunch on one
layer avoid bunched holes on an adjacent layer.  Such interlayer screening will
lower the cost in Coulomb energy for growth of the charged domains.  It will
also tend to confine the charges along the c-axis, since charges on one layer
will tend to see an insulating phase on adjacent CuO$_2$ layers.

\subsection{Macroscopic Phase Separation}

\par
Phase separation also exists in an opposite limiting case, in which the doping 
counterions are mobile.  The canonical example is La$_2$CuO$_{4+\delta}$ (LCO),
in which holes are introduced by adding excess oxygen.  This oxygen occupies an 
interstitial site, and remains highly mobile at temperatures below room
temperature.  Hence, when the holes phase separate, {\it the interstitial oxygen
can follow}.  Since the two equilibrium phases are now electrically neutral,
the resulting domains can now be macroscopic\cite{JDJ,LCO}.  Experiments on 
oxygen-doped LCO will be further discussed in Section XI.A.1.
\par
The experimentally observed 
contrast between Sr-doped and O-doped La$_2$CuO$_4$ is striking.  In the former
case, the direct evidence for phase separation is weak and controversial; the
domains are so small that $T_c$ varies continuously with doping, due to
proximity effects, while the Meissner fraction tracks $T_c$ (at extremely low
magnetic fields, proximity effects can lead to intergranular Josephson coupling,
and the Meissner fraction is found to increase).  In O-doped samples, however, 
there is clear two-phase coexistence, with one phase the AFI and the second 
phase a high-$T_c$ superconductor; varying the O-doping, $\delta$, only involves
changing the relative fraction of the two phases.  Yet despite these striking
differences, the optimal superconducting phase is very similar, in both
hole doping and $T_c$, for both O-doped and Sr-doped material.  This would
amount to an incredible coincidence, if it had not been predicted by the
Van Hove theory\cite{predict}.
\par
Double doping experiments (both Sr and O) provide further evidence of the 
essential universality of the superconducting phase\cite{Cho}.  For each $x$ 
between 0.01-0.15, there is an oxygen doping $\delta$ which produces a compound 
with T$_c\simeq$40K.  Iodometric titration studies find $P=0.16$ at optimum 
T$_c$, independent of $x$ or $\delta$ individually.  
\par
Most of the other cuprates can also be hole doped by insertion of highly mobile
oxygen atoms in sites off of the CuO$_2$ planes.  These phenomena are generally
considered independently in each material, but it is informative to consider
them together; in particular, there is evidence for phase separation, or at
least density-wave formation of the excess O's in almost all the cuprates, and
for most there is a connection with optimum T$_c$.  This evidence will be 
discussed in Section XI.A.1-4.

\subsection{Other Models of Phase Separation}

\par
It should be noted that phase separation is also predicted in several different 
theories, totally unrelated to the Van Hove scenario; indeed there have been
a number of international conferences to discuss phase separation in the 
cuprates\cite{phassep}.  All versions sound superficially similar, since the
role of long-range Coulomb repulsion is essential in limiting the phase
separation to a nanoscopic scale.  The theories vary, however, in the driving
force for the phase separation.  These alternative theories mainly fall into two
groups, depending on whether the driving force is magnetic, associated with 
doping away from the AFI\cite{HiSi,Eme,Eme1}, or electronic, associated with 
Cu-O valence fluctuations\cite{CGK}.  In addition, Phillips\cite{JC1} has a 
model in which the physics is dominated by defects off of the CuO$_2$ planes.
This model also involves nanoscale phase separation, with superconducting
islands and insulating boundaries, and his papers provide numerous additional
examples of experiments which can be interpreted in terms of such a local phase
separation.

\subsubsection{Magnetic Models}

\par
One point must be kept clear from the start.  Magnetic phase separation -- that
is, a phase separation which sets in when the material is doped away from a
Mott-Hubbard insulating state at half filling -- is present in the VHS model, 
and appears to be a very general feature of the Mott transition.  Indeed, it was
well-known in the 3D problem, long before cuprate superconductivity was
discovered.  However, in the VHS model, the underdoped material contains
domains of the Mott-Hubbard and VHS phases, and the latter are crucial for
superconductivity.  By contrast, there are strictly magnetic models, where the
nature of the conducting phase is not discussed, and superconductivity arises
in the phase separated regime, from the presence of the nanoscale size,
fluctuating domains.  It is to this class of theory that I now turn.
\par
The magnetic models are basically similar to the ferron phase discussed by 
Nagaev\cite{Nag}.  Here, a hole can delocalize over a finite domain by flipping
the spins of the neighboring Cu's, forming a small ferromagnetic island.  A
second hole can lower its energy by localizing on the same island.  In 
principle, this looks like a real space pairing mechanism, but at higher doping
levels, additional holes are found to be attracted to the same island, and the
phenomenon is seen to be one of nanoscale phase separation.  In the 3D materials
that Nagaev studied, this phase is not clearly seen, because it competes with
trapping on point impurities.  In this case, the cuprates offer a better
possibility for experimental realization, since the dopant ions are on a 
separate layer, so the Coulomb interactions are considerably weakened.
\par
Since this magnetic phase separation is expected to be present in the pure
tJ model, there has been considerable theoretical study of the conditions for
its occurence, recently reviewed by Dagotto\cite{ElDa}.  It is clearly expected
if the exchange is large, $J/t>1$\cite{WFB}, but the physical regime of smaller 
$J$ remains controversial\cite{AnSo,Eme1}.  The majority of studies find that 
the phase separation is arrested at the smallest possible scale -- individual 
pairs.  Such single pair states could be interesting, but do not appear to be 
consistent with experimental findings.  Phase separation close to half filling 
has been found in the large-N\cite{NtJ} and large-S\cite{ALa} limits of the tJ 
model, and in Kondo lattice models\cite{Kond}.
\par
Since the ferron islands constitute local regions which violate time reversal
symmetry, they should have been found by the various probes for anyons, so it
can be taken as experimentally determined that ferrons are not present in the
cuprates.  Interestingly, however, antiferromagnetic to ferromagnetic 
transitions are found in some other perovskites, notably La$_{1-x}$Sr$_x$MnO$_3
$ (Section XI.H.5), and Nagaev has suggested that the transition is a
manifestation of the ferron phase\cite{Nag2}.
Emery has stated that the hole-doped domains in his model\cite{Eme1}
need not be magnetic, but no details have been published.  In the VHS
model, the state near half filling differs in that the hole-doped
regions are non-magnetic metals at the VHS concentration.
\par
Note that even if the purely magnetic model for {\it phase separation} should 
turn out to be correct, the fact of phase separation probably rules out many
magnetic models for {\it superconductivity}, which are based on the properties
of holes in a magnetic background, and assume a smooth evolution with doping
within a homogeneous phase.  It remains a possibility that the domain 
fluctuations themselves could play a role in Cooper pairing, as discussed below.

\subsubsection{Valence Fluctuation Models}

The origin of phase separation in the valence fluctuation model follows from
the last term of Eq.~\ref{eq:8}.  This term depends self-consistently on the 
renormalization, and hence can lead to an instability against phase separation
when $V$ exceeds a critical value $V_c\approx 0.88t_{CuO}\approx 1.1eV
$\cite{Rai}.  The phase separation should persist in the presence of long-range 
Coulomb effects, as long as there is some screening\cite{MHF}.
This value of $V_c$ is somewhat larger than the value estimated from most 
LDA-type analyses, Table III.  Moreover, Feiner, et al.\cite{FJR} find that the
use of more realistic models ($U_d<\infty$, $U_p>0$) further suppresses the
tendency towards phase separation.  Phase separation has also been found in an
extended Hubbard model, in the weak coupling but $d\rightarrow\infty$ limit
(with $d$ the dimensionality)\cite{Dong}.  The condition for the onset of the 
valence fluctuation instability can be shown\cite{HRS} to coincide with the 
onset of the charge transfer instability\cite{VSA}, and hence phase separation 
precludes this latter possibility.

While these models can produce regions of nanoscale phase separation, I do not
see how either theory can explain the phase diagram, with a very narrow region
of single phase, and domains of phase separation for either under- or 
overdoping.  In particular, in the valence fluctuation model, there is no reason
for the underdoped phase to coincide with the AFI\cite{Eme1}.  
Caprara, DiCastro and Grilli\cite{DG,CDG} have recently addressed this issue,
and find that they can approximately reproduce the phase diagram by including
{\it both} mechanisms -- magnetic in the underdoped material, Cu-O charge 
transfer for overdoped, Fig. \ref{fig:i23}.  Their model thus is becoming quite 
similar to the VHS model, although it still differs, in that it is a {\it 
purely electronic} model for both phase separation and high T$_c$.  While this 
is quite an interesting possibility, the model does not explain why the optimum 
superconductivity exists in such a narrow region between two phase separated 
regimes, and it is not clear how this model would explain such properties
as the strong electron-phonon coupling or the presence of pseudogaps.

\subsubsection{Fluctuation Driven Superconductivity}

\par
Within both models, the presence of two degenerate states has led to a proposal
of a novel superconducting mechanism, based on the anomalous fixed point of
the multi-channel Kondo model\cite{Eme3,Varm}.  While these models are 
interesting, they do not seem to be consistent with the experimentally observed
phase separation in the cuprates.  Thus, the fixed point is the point where the
two phases are degenerate in energy, which should approximately correspond to
the percolation crossover, where the two phases are in most intimate contact.
On the other hand, superconductivity seems to optimize in a single-phase
regime, where there are relatively few domains of the second phase.  The
existence of macroscopic phase separation (e.g., La$_2$CuO$_{4+\delta}$) is 
particularly hard to explain on these quantum fluctuation models.

\section{Phase Separation: Experiments}

There is considerable evidence for phase separation in the cuprates, which has 
been summarized in several reviews\cite{RMrev,Eme1,Mehr2,Szy}, and has been 
extensively discussed in the procedings of two international 
conferences\cite{phassep}. Recently, Goodenough and Zhou\cite{BGood} have 
reanalyzed 
the experimental situation, and find strong support for a double transition -- 
two regimes of phase separation separated by a narrow pure phase which 
corresponds to the optimum $T_c$.  They find this pure phase to be associated 
with `correlation polarons', where both strong electron-phonon coupling and 
correlation effects play a significant role.  This would appear to be a fair
description of the dynamic JT phase.
\par
In this section, I will attempt to give an overview of the variety of 
experimental evidence for phase separation in the cuprates.  Table V summarizes
the experimental observations, while the following subsections give brief
descriptions of the individual cases.  Figures \ref{fig:g23}-\ref{fig:24},
\ref{fig:25}, \ref{fig:30}-\ref{fig:31}, and \ref{fig:34}
show the experimentally determined phase diagrams for a series of cuprates
(LCO, LSCO, YBCO) and related compounds (LNO, LSNO, BKBO).

\subsection{Underdoped Regime: Macroscopic Phase Separation}

\subsubsection{O-doping of La$_2$CuO$_{4+\delta}$}

In La$_2$CuO$_{4+\delta}$, the interstitial oxygen phase separates below room
temperature into O-rich domains of macroscopic size\cite{JDJ,LCO}. Jorgensen, et
al.\cite{JDJ2} estimated the domain size in a polycrystalline sample from the 
neutron diffraction linewidths; in all cases, the magnetic phase was found to be
correlated over distances $>5000\AA$, while in the metallic phase the domain 
size was $\approx 3000\AA$ ($>5000\AA$) for a sample which was 30\% (60\%) 
metallic.   For a single crystal, about 35\% metallic, both phases had 
correlation lengths of $\approx 500\AA$\cite{Vak}.  Ryder, et
al.\cite{Ryd} carrier out TEM studies of these materials, and found a striking
herringbone pattern of domains; in their sample, the domain width varied from
$300-1500\AA$.  A magnetic field dependence of the diamagnetic fraction is taken
as evidence for the electronic nature of the phase separation\cite{Krem}.
\par
The phase separation begins at about 300K, and is arrested around 200K, when the
oxygen diffusion becomes too slow to stay in 
equilibrium\cite{LCO,Chail,Krem,Hamm2}.  Figure~\ref{fig:g23} is a composite 
phase diagram based on the work of several groups\cite{Rad3,GrenX}.  As 
discussed in Section VI.B, there is some uncertainty about relating the O 
content $\delta$ to the hole doping, $P$, with different groups finding 
relations from $\delta =2P$\cite{GrenX} to $\delta =P$\cite{Rad2}.  Here, we 
assume the 
intermediate result, Eq.~\ref{eq:30b}.  In addition, there seems to be some
sample or preparation dependence of the phase separation: thus, while the 
present phase diagram suggests that phase separation is absent for a range of
$\delta$'s near $\delta =0$, phase separation was initially discovered through
the observation of an infinitesimal superconducting fraction in nearly
stoichiometric material, $\delta\simeq 0.017$ (see Refs. \cite{Hamm1,Krem}).
[Note: in early work, it was believed that the low-O phase corresponded to
$\delta =0$; by conservation of mass, this moved the O-rich phase boundary to 
$\delta =0.08$, with $P=2\delta =0.16$\cite{JDJ2}, the hole doping of optimal 
T$_c$.  Newer work, both on the phase diagram and on the relation between $P$ 
and $\delta$, has led to the revised view discussed in the text.]
\par
Recent neutron diffraction studies of electrochemically oxygenated LCO crystals
find that the O-rich phase is organized into well-staged regions\cite{Custag}, 
similar to graphite intercalation compounds and LNO (discussed below).  Stages
from 2 (interstitial O's between every other CuO$_2$ plane) to 6.6 (presumably
an average over several layer spacings, 6, 7, etc.) were found.  Each pure
stage corresponds to a particular doping, $\delta$, and for intermediate doping
regions, phase coexistence between several different stages (2, 3, 4) or between
a stage $n$ and the undoped AFI phase were found.  Wells, et al.\cite{Custag}
have surveyed the earlier literature, and suggest that most of the findings
are consistent with staging, in which the weak superlattice spots were not 
observed.  Similar staging has also been found in the two-phase coexistence
regime $\delta\approx 0.015$\cite{Cust2}, with an onset near 250K, 
which displays some ($\approx 10K$) hysteresis.  Most interestingly, at $\approx
220K$ the staged region itself undergoes a structural transition to a phase 
with superlattice order $\approx 70\AA$ along the $a$-axis -- perhaps a stripe 
phase of the Daumas-Herold domains.
\par
By electrochemically oxygenating LCO, it is also possible to
produce samples with O doping higher than the miscibility gap ($\delta\simeq 
0.08-0.12$).  These samples are found to be approximately single phase, but with
large-period superlattices\cite{Rad2,Gren2,Gren}.  For a given $\delta$, there 
can be several different superlattices\cite{Lagu}.  While the superlattice 
structures have not yet been worked out, they appear to involve modulated 
octahedral tilts\cite{Rad2} or regular shear planes\cite{Lagu,Galy}.
Much clearer superlattice effects are found in the analogous La$_2$NiO$
_{4+\delta}$ compounds, discussed further below.
\par
In electrochemically doped La$_2$CuO$_{4+\delta}$, there is a range of $\delta$ 
over which two superconducting phases co\-ex\-ist, with T$_c$'s of 32K and 
$\approx 40-45K$\cite{Wat,GrenX,Chou,Gren,Rad}.  Grenier, et al.\cite{Gren2} 
suggested that these two phases correspond to $\delta$ = 1/16 and 1/8, 
respectively, while Johnston\cite{DCJ} finds the two-T$_c$ coexistence between
$0.06\le\delta\le 0.11$, with a single T$_c\simeq$43-45K for $0.11\le\delta\le 
0.12$.  There are strong quenching/annealing effects in this second two-phase
coexistence regime\cite{Krem,DCJ}, but not for $\delta =0.11$\cite{DCJ}.
The two superconducting phases have a characteristic difference in pressure
dependence: for the 32K phase, $\partial T_c/\partial p$ is large and positive,
while for the 43K phase, it is nearly zero\cite{DJpres}.
\par
The transition temperature of the higher-T$_c$ phase depends somewhat on the
nature of the sample: 40K in a single crystal, 43-45K in polycrystalline 
powders, and 47K in thin films\cite{Gren3}.  Moreover, in the films, there is
evidence for {\it yet another, higher T$_c$ phase}, this time with T$_c\simeq$
55-60K\cite{Gren3}.
\par
Recent experiments\cite{Hor1,Hor2} have added yet another wrinkle, suggesting
that electrochemical oxidation produces metastable phases, and that, for
samples annealed at 110$^o$C, the phase diagram looks very different, Fig.~
\ref{fig:g23}b. The phase boundary of the lower two phase region extends to 
much lower $\delta$, $0.006\le\delta\le 0.047$ (note that $P=2\delta$ is 
assumed). At room temperature, there is a second two-phase regime, for $0.075
\le\delta\le 0.17$, between two orthorhombic phases, with the $\delta =0.17$ 
phase having a large orthorhombicity, 0.019.  [The phase at $\delta\simeq 0.17$
is inferred by extrapolation; it has not proven possible to synthesize bulk
LCO for $\delta >0.12$.]  At low temperatures, Meissner 
effect measurements suggest that this splits into two adjacent
two-phase regions, $0.06\le\delta\le 0.085$ and $0.085\le\delta\le 0.12^+$.  All
three end-phases are superconducting, with the $\delta =0.085$ having a reduced
$T_c$, reminiscent of the LTT phase in LBCO, LSCO (in fact, $P$ is reported to 
be 0.125 for this phase, although how this was estimated is not stated).  Since
the x-ray analysis found no evidence for this extra phase, it is assumed to be
due to an electronic phase separation.
\par
There are two ways to experimentally study the crossover from macro- to
nanoscopic phase separation: by quenching La$_2$\-Cu\-O$_{4+\delta}$ or by 
double doping, La$_{2-x}$\-Sr$_x$\-Cu\-O$_{4+\delta}$.
By quenching La$_2$CuO$_{4+\delta}$ through the region of high O 
mobility\cite{Krem,Hamm2}, it is possible to reduce the domain size, approaching
the domain of nanoscale phase separation.  Thus, the superconducting fraction
and T$_c$ both decrease with degree of quench, although T$_c$ seems to have a 
series of plateaus, near 40-45K, 32K, and 28K.  These various $T_c$'s seem to
be associated with different structural instabilities near the quench 
temperatures\cite{KOIY}.  In quenched polycrystalline
samples, a Cu EPR signal is observed (although none is seen for a single 
crystal)\cite{EPR1}.  This signal is anticorrelated with the superconducting 
signal, and was interpreted in terms of the formation of spin-polarized
clusters.  Similar results are reported for underdoped YBCO\cite{EPR2}.
In microwave absorption of La$_2$CuO$_{4+\delta}$, a dip is observed at zero 
applied field (called the low-field signal, LFS)\cite{Mehr2}.  This signal is 
similar to that seen in granular superconductors, due to weak links between
superconducting domains; consistent with this, it is found to vanish at T$_c$.
The amplitude of the LFS is sensitive to cooling rates, being largest for a
slow cool.  Similar LFS are found in LSCO, but only in the metallic island
regime ($x\ge 0.05$); the signal height is independent of cooling rate, 
suggesting that the islands form at temperatures higher than room temperature.
\par
In double doping experiments, Sr doping is found to inhibit the O-related phase
separation, with evidence for phase separation disappearing at $x=0.03$ 
(interestingly, in the phase separated regime, T$_c$ is a strongly {\it
decreasing} function of $x$)\cite{Cho}.  For finite $x$, the low-$\delta$ phase 
is forced away from the AFI phase at $P=0$, and this is believed to destabilize
the phase separated state.
\par
In studying these phase separations, it will be important to develop new probes
which are sensitive to the fine spatial scales and slow fluctuations.  Thus, 
Mehring, et al.\cite{Mehr} studied the La nuclear spin-lattice relaxation rate
in La$_2$CuO$_{4+\delta}$; they found that the ordinary longitudinal ($T_1^{-1}
$) and transverse ($T_2^{-1}$) relaxation rates were insensitive to the phase
separation onset, whereas a stimulated echo relaxation rate ($T_1^{*-1
}$), specially chosen for sensitivity to slow fluctuations, had a strong peak
at 230K, the onset of phase separation.  (On the other hand, Hammel, et 
al.\cite{Hamm3} found a large peak in ($(T_1T)^{-1}$).)

\subsubsection{O-doping in YBa$_2$Cu$_3$O$_{7-\delta}$}

\par
It is instructive to compare the effects of oxygen doping in La$_2$CuO$_{4+
\delta}$ and in YBCO.  In the former case, T$_c$ of the superconducting fraction
is independent of doping over the full range from $\delta =0^+$ to $\delta =
0.08$.  Similarly, there are two plateaus in $T_c(\delta )$ for YBCO, near $T_c
=90K$ and $70K$, but in other ranges, $T_c$ varies continuously with $\delta$.
The 70K plateau is associated with phase separation, due to two types of 
O-ordering in the chain layers.  Fully oxygenated YBCO$_7$, with all chains
complete, forms the Ortho I phase, while at YBCO$_{6.5}$, the stable phase, 
Ortho II, has every other chain occupied\cite{DdF}.  The doping dependence of
T$_c$ can be understood by the competition between these two types of 
domain\cite{Poul}, by assuming that T$_c$ is 58K in the ideal Ortho II phase, 
and 93K in ideal Ortho I, and that charge transfer takes place only from domains
which exceed a minimal size.  The same model\cite{Poul1} can also explain the
room-temperature annealing studies\cite{Vea} as due to a gradual growth of the
ordered domains.  At exactly $\delta =0.5$, the Ortho II phase transition
occurs at 398K, Fig.~\ref{fig:24}; however, domain growth is logarithmically 
slow, due to random field effects\cite{OrtII}.
\par
A number of experiments now find that, as the O-doping is varied, YBCO is 
composed of microscopic islands of three compositions, corresponding to the
undoped ($\delta =1$) AFI phase, the Ortho II phase, at $\delta =0.5$, and the 
Ortho I, at $\delta =0$.  This is seen in the splitting of 
crystalline-electric-field (CEF) transitions, when a few percent of a rare earth
is substituted for the Y\cite{Pekk,CEF}, or from Tm NMR in TmBCO\cite{Bakh}, 
and in the resonance Raman spectrum of a YBCO$_{6.4}$ crystal\cite{Ili}. 
Two anharmonic relaxation peaks (labelled PH1, PH2) have been hypothesized to
be associated with Ortho II or Ortho I domains\cite{Cann}, and a plot of their
intensity vs $\delta$ yields a similar distribution plot.
\par
However, electron microscopic studies have also reported a number of other
intermediate phases.  One of these, the Ortho III phase, has now been 
confirmed by neutron diffraction to exist at $\delta\simeq 0.23$, for $T\le
375K$\cite{OrtIII}.  Figure~\ref{fig:24} shows a theoretical\cite{Ced} phase 
diagram of YBCO, along with phase boundaries determined by electron 
diffraction\cite{Yang}.
\par
Inelastic neutron scattering studies\cite{R-M2} find, in addition to the 
pseudogap onset at high temperatures, a true gap in the spin response spectrum: 
the imaginary part of the susceptibility, $\chi (\vec q,\omega )$, vanishes for 
$\omega\le E_g$.  This gap seems to be related to superconductivity, in that 
$E_g$ is found to be proportional to $T_c$ (see Ref.~\cite{TaW}).  However,
the constant of proportionality changes with doping, being 3.5 in the mixed
Ortho I -- Ortho II regime, and only 1 in the Tetrag -- Ortho II regime ($T_c
\le 60K$).  This suggests that the antiferromagnetic domain 
walls in the latter case act as pairbreakers.
\par
While this O phase separation is usually analyzed completely in terms of ionic
(O-O repulsion) effects, hole bunching could also provide the driving force, and
it may be that the two effects are complementary.  Thus, at YBCO$_{6.5}$, with 
every other chain fully occupied, adding more O  (here, reducing $\delta$)
leads to exactly the same phase separation as in La$_2$CuO$_{4+\delta}$,
except that the excess O's are restricted to sites on the non-fully occupied
chains.  The annealing studies suggest that the O loses mobility in about the
same temperature range in YBCO\cite{Vea} as in La$_2$CuO$_{4+\delta}$.  Pekker,
et al\cite{Pekk} concluded that in the Ortho II phase there are {\it hole 
chains} in the CuO$_2$ planes running parallel to the occupied chains, as in a
CDW, and this is consistent with the fact that the Meissner fraction is
substantially lower in the Ortho II phase than in the Ortho I\cite{Graf,AlSc}.
\par
Recent experiments in Ca or La substituted Y$_{1-x}$A$_x$Ba$_2$Cu$_3$O$_{7-
\delta}$ ($A$ = Ca, La) have
found a very striking confirmation of this idea, Fig. \ref{fig:T10}\cite{Tal10}.
It is found that the 70K plateau is present for each value of $x$ (as $\delta$
is varied), but {\it the plateau falls at a fixed hole doping $P\simeq 0.12$, 
and not at a fixed value of $\delta$}!  It should be noted that this doping is 
approximately the same as the $\approx 1/8$ doping where $T_c$ is suppressed in
LSCO and LBCO (Section XIII); indeed, in the plot of $T_c$ vs $x$ for LSCO,
the dip at 1/8 makes the curve (Fig.\ref{fig:25}) look very similar to those in 
Fig. \ref{fig:T10}.
\par
Further evidence that hole phase separation is important is the finding that 
YBCO$_7$ is overdoped\cite{over}, and shows evidence for phase separation
(Section XI.F).  Such behavior would be unthinkable if the phase separation is 
due to O-ordering, but is expected in a
VHS model, since there is no reason for the VHS to fall exactly at $\delta =0$.
\par
Hence, it is plausible to assume that the Ortho II phase is also driven by hole 
phase separation, perhaps a stripe CDW phase similar to those found in La$_{2
-x}$Sr$_x$NiO$_4$ (discussed below).  The chain ordering would then be the
associated Peierls distortion stabilized at $\delta\simeq 0.5$ by 
commensurability effects.  Franck\cite{Franck} has suggested an alternative 
possibility, that the Ortho II phase is associated with a second VHS.  He finds
that the Cu isotope effect becomes vanishingly small {\it twice}: both at 
optimal doping $\delta\simeq 0$ and on the plateau $\delta\simeq 0.3$.  As 
discussed above (Section VI.D.4), a VHS is known to produce a 
minimum in the isotope effect.

\subsubsection{A Test Case: 1-eV Feature in YBCO}

The present model of nanoscale phase separation should have predictive or at
least explanatory power.  Here, it will be used to propose an explanation for
the sharp peak found at $\approx$1eV below the Fermi level by photoemission
experiments in YBCO, Fig. \ref{fig:d19}.  This feature is found to shift
significantly with doping, Fig. \ref{fig:d19}c, but to be nearly independent of 
doping in the range 6.7-6.5.  Comparison with the phase diagram of YBCO
strongly suggests that this feature is associated with the Ortho II phase: it
is independent of doping when the Ortho II is the majority phase, and shifts,
due to proximity effects, when the Ortho II is a minority phase.  
\par
Moreover, the present model suggests the origin of the feature.  The Ortho II
phase is considered to involve hole segregation on the planes in parallel with
chain ordering: alternating stripes of hole-doped and undoped unit cells.  The
doped stripes are superconducting, but with reduced T$_c$ and Meissner fraction.
Similarly, the undoped stripes are insulating, {\it but with a reduced
charge-transfer gap}.  The photoemission spectrum has the expected form, Fig.
\ref{fig:d19}b.  Thus, the 1eV feature may be the signature of the reduced
charge-transfer gap.  In underdoped material, it smoothly evolves into the
larger gap of YBCO$_6$\cite{Liu2}.  
\par
One complication in interpreting this feature is that it may also be related to
a surface feature.  Thus, if the surface cleaves preferentially on the chain 
layer, it is possible that half of the chains go with each surface, to maintain
approximate charge neutrality.  In this case, each surface would be left with
approximately an Ortho II surface layer, which would display the 1eV feature. 

\subsubsection{O-doping in Bi- and Tl-Cuprates}

\par
Extra, interstitial oxygen is also found in the Bi-cuprates.  This does not
lead to phase separation, but to an incommensurate CDW, which Bianconi and
Missori\cite{Bian2} interpret in terms of a nanoscale (polaronic) phase 
separation.  Doping on the Ca site seems to lead to a phase separation into
two phases with different (commensurate) superlattice periods\cite{CaBi}.
\par
Finally, while the hole-doping process in the Tl-cuprates is not completely 
understood, it is known that Tl$_2$Ba$_2$CuO$_{6+\delta}$ (Tl-2201)is overdoped 
when fully oxygenated, and must be reduced in argon to become superconducting 
(T$_c\ge$85K).  The fully oxygenated samples are again found to contain 
interstitial oxygen\cite{Shim,ASAM}, which may lead to a phase 
segregation\cite{ASAM}.  In orthorhombic Tl-2201, incommensurate
modulations are observed in electron\cite{Tlel} and x-ray\cite{Shim} 
diffraction, accompanied by a splitting of the Cu-NQR peaks\cite{Shim1}.

\subsubsection{Photodoping}
\par
A macroscopic phase separation, similar to O-doped La$_2$CuO$_4$, also arises 
from photodoping of YBCO and LSCO.  This is discussed in detail by Emery and 
Kivelson\cite{Eme1} (their Section 4.1) and by Yu and Heeger\cite{YH}.
Photoexcitation produces a population of holes, which migrate to the CuO$_2$
planes, and electrons, which localize on a different layer, assumed to be at O
vacancies on the chain layer in YBCO.  These
electrons are much more mobile than any chemical species, so if the holes tend 
to bunch up, the electrons follow, leaving charge-neutral domains.  In La$_2
$CuO$_4$, there is a striking change in the spectrum of Raman active phonons: 
the Raman peaks associated with the undoped material are bleached, while those 
peaks associated with the optimally doped material appear\cite{pdop1a}.  This is
just what would be expected in a phase separation scenario: the photodoping 
produces islands of optimally doped material, thereby reducing the volume of 
undoped material.  Also, the electronic part of the photoabsorption spectrum 
matches that of optimally doped 
LSCO, independent of the laser intensity\cite{pdop1} -- that is, the holes 
preferentially bunch up to the doping of optimum T$_c$.  Just as in La$_2$CuO$_
{4+\delta}$, this is very strong evidence for the VHS model, as opposed to a
model in which the upper limit of phase separation is random, or one in which
the superconductivity arises in the two phase regime. In a sense, this 
experiment provides the stronger evidence, since the electrons can 
follow the hole motion to lower temperatures than the interstitial O's, which 
freeze out near 200K.
\par
Similar effects are found in YBCO\cite{pdop2,pdop2b} and Tl 
compounds\cite{pdop2b,pdop2c}.  Strikingly, there is a dip
in the resistivity suggestive of a superconducting transition on the metallic
islands.  While the transition is not evident at the lowest photon intensities
(perhaps due to proximity or charging effects in the smallest drops), once the
transition appears, the onset temperature is $\ge 100K$, independent of
light intensity\cite{pdop3}.  This is actually {\it higher} than in optimally 
doped YBCO -- particularly when it is realized that the chains remain badly 
broken, due to the large oxygen deficiency ($\delta =0.7$).  Could the 
photoexcited electrons be superconducting?  A similar T$_c$ enhancement (to 
$\simeq 40K$) is found in photodoped La$_2$CuO$_4$\cite{pdop1}.
\par
The appearance of the Raman phonons associated with the optimally doped material
provides additional information on the role of electron-phonon coupling.  
Chemical doping changes the phonon spectra in two ways: first, due to 
metallic screening, and secondly, due to tetragonal-orthorhombic transitions in 
both LSCO and YBCO.  Photodoping also produces {\it both} of these
changes\cite{pdop1a,pdop4}!  This is strong evidence against the purely ionic 
model of the HTT-LTO transition in LSCO.
\par
The above experiments have all dealt with transient photoconductivity after a
short laser pulse is absorbed by the sample.  The sample's response to a
long-time continuous irradiation reveals an additional feature: a persistent
photocurrent, which can last indefinitely if the sample is kept well below
room temperature, but which anneals away near room 
temperature\cite{Persist}.  Some echo of
this phenomenon is found in the pulsed experiments, in that at high pulse
intensities the photocurrent decay becomes very slow\cite{pdop3}.  This is the
opposite of what is usually seen in semiconductors, where photoirradiation
enhances disorder.  It is believed to be associated with better oxygen ordering
in the chain layer -- larger domains of the Ortho I and II phases.  This can 
also be driven by the electronic phase separation: if the oxygen follows the
holes, then the photoinduced phase separation leads to the growth of larger
Ortho I domains.  When the light is turned off, part of the hole excess decays,
but the oxygen does not have enough energy to diffuse and disorder the domains. 
The better domain order leads to larger hole transfer to the planes, and hence 
improved superconductivity\cite{Poul}.  This residual hole transfer is small 
compared to that induced by the direct photoexcitation; T$_c$ is much less 
than the $\approx 100K$ inferred from the pulsed experiments.  On warming to 
room temperature, entropic effects become important, the oxygen disorders, and 
the excess photocurrent decays.  

\subsubsection{H-doping}
\par
A similar macroscopic phase separation may be found in H-doped YBCO, although 
the experimental situation is confused at present.  In bulk powder samples, H 
NMR shows evidence for a number of inequivalent H environments\cite{H1,H4}, one
associated with antiferromagnetism with a high transition temperature, $T_N
\simeq 420K$, another with superconductivity\cite{H1}.  The superconducting 
onset temperature remains constant at 92K throughout the
doping regime, until superconductivity is completely destroyed\cite{H2}.
However, in epitaxial films, a sharp transition is found at all H dopings, and 
$T_c$ falls smoothly to zero as the H concentration increases\cite{H3}.  
Moreover, it is believed that the phase separation has more to do with the H
than with the holes\cite{H1}.  On the other hand, in Tl-2201, there is a
correlation between the H-induced phase separation and the degree of 
hole-doping\cite{Nied2}: there is clear phase separation in underdoped 
materials, while there appears to be a continuous H-doping in the (hole) 
overdoped regime.

\subsection{Underdoped Regime: Single Hole}

\par
Nanoscale phase separation is much more difficult to prove, since the scale is
comparable to that of fluctuation effects.  Thus, in LSCO, T$_c$ varies
continuously with doping in the underdoped regime, so any superconducting 
domains must be comparable in size to the superconducting coherence length, 
$\simeq 16\AA$.  Indeed, at this point, it is more correct to think of this as a
unique phase characterized by charge bunching, rather than in terms of a phase
separation.  Nevertheless, there is considerable evidence for this unique
phase.
\par
Indeed, when even a single hole is doped into the half filled AFI state, there 
is evidence for localization effects.  Since a Mott insulator requires exactly
one electron per atom, adding even a single electron or hole to a half filled
band should result in an insulator-to-metal transition.  This is confirmed
by detailed calculation in both 1D and 2D systems, but is not observed
experimentally in the cuprates.  Instead, it is found that the insulating phase 
exists over an extended doping regime, indicating that the first doped holes are
localized\cite{YuLe}.  Localization can arise from a number of causes,
including disorder or (charge or spin) polaronic effects.  Such effects arise 
naturally in a model of nanoscale phase separation: the added hole would tend to
form a compact region of well-defined hole density, rather than spreading out to
change the average density by a small amount over a large spatial area.  This 
compact charged region would then be sensitive to localizing on a charge 
fluctuation.  This would be particularly true in LSCO, where the dopant Sr forms
an immobile charged impurity.
\par
In a study of lightly doped La$_2$CuO$_{4+\delta}$ ($\delta =0.014$), Falck and
coworkers\cite{Falc} found two peaks in the reflectivity, at 0.13eV and at 
0.5eV.  These features correspond to the two components generally found in the
mid-infrared absorption of the cuprates, at 0.13-0.16eV and 0.5-0.75eV.  Now
since these mid-infrared features are found at essentially all dopings in all 
the cuprates, they must be intrinsic features of the CuO$_2$ planes, insensitive
to the structure and doping of the other layers.  In the lightly doped
sample, Falck, et al.\cite{Falc} suggest that the 0.5eV feature may be an
exciton associated with Cu $d_{x^2-y^2}\rightarrow d_{3z^2-r^2}$ transitions.  
Why these are not seen in the undoped material is unclear; however, if their
interpretation is correct, this transition is not of immediate interest. 
The other transition, at 0.13eV, is potentially very interesting.
Falck, et al. interpret this as due to the ionization of a {\it localized 
polaron}, in agreement with Bi and Eklund\cite{BiE}.  
\par
The strong T-dependence of this peak is due to
thermal ionization, with activation energy 0.035eV; the large difference between
this thermal ionization energy and the photoionization energy, 0.13eV, is 
ascribed to a Frank-Condon effect: it is the energy of the local lattice
distortion associated with strong coupling to $\approx$43meV optical phonons. 
For a simple {\it molecular} JT effect, the net energy is a combination of 
elastic and electron-phonon terms, of the form
\begin{equation}
E={K\over 2}x^2-Gx,
\label{eq:99d}
\end{equation}
where $x$ is the JT distortion.  Minimizing $E$ with respect to $x$ gives the
thermal excitation energy, $E_{JT}=G^2/2K$ at distortion $x_0=G/K$.  If this is 
$\simeq 35meV$, then the optical excitation energy is the energy for a vertical
transition between the two JT-split states, $E_{opt}=4E_{JT}\simeq 140meV$, in
good agreement with observation.  Thus,
the initially doped holes are readily localized because of strong polaronic
effects.  This is precisely what is predicted by the VHS phase separation
scenario.  Note that, in the magnetic phase separation model, the first holes
would form {\it magnetic} polarons -- small ferromagnetic patches -- and not
the {\it dielectric} polarons observed experimentally.
\par
There are also experiments which have been interpreted as evidence for
magnetic polarons.  For instance,
in a magnetic neutron scattering study of YBCO, Burlet, et al.\cite{Burl}
found a coexistence of magnetic Bragg peaks with rods of magnetic scattering,
for $y\equiv 1-\delta =0.37$, a doping too low for any superconductivity to
exist.  Further, for $y$ = 0.3, 0.33, the magnetic Bragg peaks have a 
reentrant feature: below a temperature $T^{\prime}\approx 50-100K<<T_N$, the
intensity of the Bragg peak starts to decrease with decreasing temperature.
They attribute both of these features to AF-polarons\cite{AFp}.
\par
However, these magnetic polarons should be local domains of ferromagnetic
order -- i.e., regions in which time-reversal symmetry is violated.  Now there
was an extensive search for just such domains, in the context of anyon theory,
and the demise of that theory is directly correlated with the inability of
experiments to detect the relevant domains.  Hence, I think that it is unlikely
that ferromagnetic domains exist in equilibrium in the cuprates.  On the other
hand, the quenching experiments discussed in Section XI.A.1 showed that a 
metastable phase with non-vanishing EPR signal, suggestive of 
spin-polarized clusters, can be quenched in (see also Section XI.F.3).

\subsection{Underdoped Regime: AFM Islands}

\par
Nanoscale phase separation presents a rich phase diagram as a function of 
doping, Fig. \ref{fig:j23}.  For $x<x_{VHS}$, there is a two-phase regime 
involving a magnetic insulator and an `anomalous metal' (VHS) phase, while for 
$x>x_{VHS}$, the two phases are the VHS phase and a normal metal. Near the 
midpoint of each of these regimes, there will be a percolation crossover.  Thus,
the underdoped material should evolve from an antiferromagnetic metal to a phase
with antiferromagnetic islands separated by grain boundaries of VHS-like phase,
to a metallic phase with `intrinsic' weak links (VHS phase islands with
magnetic grain boundaries) to a pure VHS phase.  The AFM-island phase should
display superparamagnetic behavior, and can presumably be identified with the
`spin glass' phase in lightly doped LSCO; indeed, recent experimental evidence
has been presented for just such a phase\cite{Cho}.  Of course, the physical
properties of the grain boundary material will be strongly modified by proximity
effects.  This is presumably why the threshold for superconductivity is at $x
\simeq 0.05$, close to the upper limit of the spin glass phase.

\subsubsection{Finite-size-corrected N\'eel Transition}
\par
The magnetic properties of single-layer cuprates have recently been reviewed in
detail by Johnston\cite{HaMM}, including evidence for phase separation.  From 
the peak in the (neutron scattering) dynamic spin structure factor,
$S(\vec q,\omega )$ at $\vec q=(\pi /a,\pi /a)$,  Keimer, et al.\cite{Keim}
deduced the magnetic correlation length $\xi (x,T)$, and showed that it could
be fit to the form
\begin{equation}
\xi^{-1}(x,T)=\xi^{-1}_0(x)+\xi^{-1}(0,T),
\label{eq:99b}
\end{equation}
where $\xi^{-1}_0(x)=\xi^{-1}(x,0)$.  Emery and Kivelson\cite{Eme1} pointed out
that this is just the form expected for phase separation, if the magnetic 
domains have size $\xi_0(x)$.  This allows an estimate of the magnetic
domain size, $\xi_0(0.02)=200\AA$, $\xi_0(0.04)=40\AA$.
\par
Cho, et al.\cite{Cho} showed that the doping dependence of the N\'eel 
temperature in LSCO, Fig.~\ref{fig:25}, could be understood as due to finite 
size effects in the AFM-island phase:
\begin{equation}
T_N(x)=T_N(0)[1-({x\over x_c})^n],
\label{eq:99c}
\end{equation}
with $T_N(0)=301K$, $x_c=0.0212$, $n=1.9$.  From mean-field theory, this 
suggests that the AFM domain size $L$ scales like $L(x)\propto x^{-n/2}\propto 
1/x$,
which implies 1D domains (stripes).  Consistent with this model, they showed 
that the N\'eel peak in the susceptibility could be scaled onto a single curve
for all samples with $x\le 0.02$, and that $L(x)$ could be quantitatively 
equated to Keimer, et al.'s $\xi_0(x)$.

\subsubsection{Roughening Transition}
\par
The above analysis holds for $T\ge 50K$, when the holes are mobile.  Below this 
temperature, the holes become localized\cite{DJ1}, presumably on the potential
disorder associated with the random Sr distribution.  At $T_f\simeq(815K)x$ 
(about 8-16K), there is a large peak in the ${}^{139}$La nuclear spin relaxation
rate\cite{DJ1}, similar to one seen in LBCO\cite{DJ3} and La$_2$CuO$_{4+\delta}
$\cite{DJ2}, and $L(x)\propto x^{-1/2}$ for $T<50K$.  
This has been interpreted as follows\cite{DJ4}: the random
potential is assumed to completely break up the domain walls into isolated 
holes, and $T_f$ represents a freezing of the transverse components of
spins of Cu's adjacent to these O holes.  The resulting transition appears to
be totally unconnected with the spin glass phase found at slightly higher 
dopings (see the next paragraph), and I would like to suggest an alternative
picture.  As the holes localize, the domain walls start to meander and collide, 
the stripes break up into more equiaxed domains, leading to $L(x)\propto x^{-1/
2}$.  Thus, $T_f$ would represent a disorder-induced roughening 
transition\cite{THwa}, with the resulting domains identical to those found in 
the spin glass phase, Fig~\ref{fig:26}.

\subsubsection{Spin-Glass Phase}
\par
For higher doping, $0.02\le x\le 0.08$, there is no long-range N\'eel order, and
the spin freezing transition is replaced by a spin-glass-like transition, $T_{gl
}$, Fig.~\ref{fig:25}.  However, the presence of a specific heat peak at the 
transition suggests that it is not a conventional spin glass\cite{ABLT}.  This 
transition is also interpreted in terms of the AFM-island phase, with domain 
size $L(x)$ $\propto x^{-1/2}$\cite{DJ5}, again in agreement with the AF 
correlation length, $\xi_0$.  The molecular field transition predicts $T_{gl}
\simeq J^{\prime}(\xi_0/a)^2\propto L^2\propto 1/x$, in good agreement with 
experiment, Fig. ~\ref{fig:25}.  

\subsection{Underdoped Regime: Metallic Islands}

\par
From M\"ossbauer measurements on ${}^{57}Fe^{3+}$ in LSCO (0.5\% Fe) and ${}^
{170}Yb^{3+}$ in YBCO, Imbert, et al.\cite{Imb} found that the spectra at 
arbitrary hole doping could be decomposed into a superposition of a magnetic 
part, similar to the undoped state, and a metallic part, similar to optimum 
doping.  Figure~\ref{fig:27} shows the magnetic fraction found in the two 
materials.  The magnetic domains are quasistatic, but the local probe cannot 
determine their size. In LSCO, the low $x$ ($0.06\le x\le0.16$) appears to be a
spin glass phase, while higher-$x$ samples ($0.16-0.25$) display no spontaneous 
spin freezing.  For $0.14\le x\le 0.16$, the two components overlap.  Similar 
results are found by Kobayashi, et al.\cite{Koba}, from La $T_1$ measurements, 
in which a rapidly decaying component is associated with magnetic domains, and 
from  $\mu$SR measurements by Weidinger, et al.\cite{Weid}, which also measure 
the magnetic fraction and local magnetic field.  Both results from local probes 
are seen to be in good semiquantitative agreement. More recent measurements find
that the magnetic fraction vanishes at the optimum T$_c$ in LSCO\cite{Kief}.   
Niedermayer, et al.\cite{Nied} reported similar $\mu$SR results, finding a
common scaling for both LSCO and YBCO, when the doping is
expressed as $T_c/T^{\dagger}$, where $T^{\dagger}$ is the optimum $T_c$ for
LSCO, and is approximately equal to $T_c$ for the Ortho II phase of YBCO
($T^{\dagger}=60K$, $T_{cII}=58K$).  However, their more recent work has failed
to confirm the presence of a separate magnetized fraction\cite{NiedBud}.
They suggest that this result is not incompatible with the presence of a striped
phase, since fluctuations could average out the magnetization on the slow
time scale of the $\mu$SR measurement.
\par
In a more recent study, Imbert, et al.\cite{Imb2} have gone to ultralow Fe
concentrations, by doping LSCO with radioactive ${}^{57}Co$, which decays to 
form ${}^{57}Fe$.  They found characteristic differences from their earlier
study, which suggest that the Fe is not simply a passive probe.  I would
interpret the changes (in a slight modification of their interpretation) as
showing that the Fe locally pins the insulating regions.  Thus, for the
ultra-low-Fe-doped samples, the M\"ossbauer found only Fe associated with the 
insulating domains, up to $x=x_t=0.19$ (in these samples, the measured $x$ may 
be too large by $\delta x\approx 0.02-0.04$\cite{Imb2}), and there was no doping
range in which both spin glass and metallic domains coexisted. I would suggest 
that the latter behavior
is more representative: in the lightly doped samples, the isolated Fe always
pinned the insulating domain wall locally, whereas in the higher-doped samples,
most walls were pinned, so some Fe remained in a metallic domain. The
concentration $x_t$ would then represent the highest doping at which any
insulating domain walls were present.  These results further suggest that, at
least near an Fe impurity, there is not much difference in the insulating
phase, whether it is in a large AFM island, or the domain wall on a metallic
island.

\subsection{Crossover at Optimum T$_c$}

\subsubsection{Meissner Effect}
\par
Some properties show a striking crossover as the doping is varied from the
underdoped to the overdoped regimes.  Thus, early evidence for nanoscale phase 
separation came from measurements comparing the doping dependence of T$_c$ and 
the Meissner fraction of LSCO.  Both were found to depend strongly on $x$, 
maximizing at the optimum doping and falling off for either underdoped or 
overdoped materials\cite{Meiss1,Meiss2,Rad}, Fig.~\ref{fig:28}a.  Allgeier and 
Schilling\cite{AlSc} find that the proportionality between the Meissner fraction
and $T_c/T_{c,max}$ holds in LSCO, YBCO, and several Tl and Bi cuprates as well,
Fig.~\ref{fig:28}b.
More recent measurements in LSCO find that the Meissner fraction is strongly 
field dependent, and can reach large values even in underdoped samples, if the 
field is low enough ($\approx 0.1G$)\cite{Kamb,Meiss3}.
In fact, this is what would be expected for nanoscale phase separation: the
superconducting regions are close enough to be Josephson or proximity effect
coupled, but this coupling can be destroyed by a very weak field.  Yoshimura,
et al.\cite{Meiss2} also found that, in all compositions studied, a small
fraction of the sample went superconducting near the optimum T$_c$, and 
concluded that only the sample with $x=0.16$, $\delta =0$ is single phase.  The 
field dependence shows a characteristic difference between the underdoped and
overdoped regimes\cite{Meiss3}: the underdoped sample ($x=0.11$) consists of
superconducting grains separated by weak links, while the overdoped sample is
mostly nonsuperconducting, but the superconducting regions can still couple at
low fields (suggestive of proximity-effect coupling over long distances via
normal metal regions).
\par
The discussion by Kitazawa, et al.\cite{KTNK} shows how careful one must be in
estimating the Meissner fraction, in the presence of pinning and weak links.
However, they are trying to prove that their samples are homogeneous, and I
believe that their measurement cannot rule out the {\it nanoscale} phase 
separation we are discussing.  In terms of the present model, I would interpret 
their results as follows. (1) It is extremely difficult to produce a uniform 
nanoscale phase separation; most preparation techniques lead to some larger
scale phase separation, with evidence for domains with distinct T$_c$'s. (2) In 
sufficiently low magnetic fields a 100\% Meissner effect is found, but this is 
broken down by quite low fields, due to the domain walls acting as weak links.  
For example, a magnetic field of only 0.061Oe was needed to reduce the Meissner 
fraction of a single crystal by nearly 50\%!  [In stoichiometric YBCO, a similar
effect is found, but for a much higher field, $\approx$30Oe\cite{K-E}.]  This 
suggests that, in the uniform nanoscale domain regime, the domain walls are less
than $\xi_0\simeq 16\AA$ thick.  Other
experiments suggest that the domain walls are only about one unit cell in 
thickness. (3) The range of doping over which a 100\% Meissner fraction can be
found should be an estimate of the percolation crossover between the VHS phase 
and the second phase.  These limits  are $x=0.085$ on the underdoped side and
$x=0.25$ for overdoped material, both reasonable values. (4) One should keep 
in mind that for YBCO, a large Meissner fraction is found over a broad doping
range even though it is clear that there is nanoscale phase separation between
the Ortho I and Ortho II phases\cite{Cava}.

\subsubsection{Susceptibility}
\par
A similar crossover behavior is found for the susceptibility\cite{over,AlSc}. 
In underdoped cuprates, the susceptibility above $T_N$ can be scaled into a
{\it universal form}\cite{susc1}.  For overdoped cuprates, this same form holds,
if a Curie-Weiss term is first subtracted off\cite{susc2}.  In LSCO and LBCO,
the Curie constant grows linearly in $x-x_{VHS}$, where $x_{VHS}$ is the doping
of optimal T$_c$, up to $x\simeq 0.26$.  The origin of these paramagnetic
centers is not clear, but for all dopings, they correspond to less than 1\% of
the Cu spins.  Tallon and Flower\cite{over} point out that these spins can
act as pairbreakers, thereby reducing T$_c$ in the overdoped cuprates, but 
caution that they could be extrinsic, due to impurity phases.
\par
Neutron scattering experiments find a large falloff in magnetic scattering on
going from optimum doping, O$_{6.92}$ to a small overdoping, O$_{6.97}
$\cite{spinover}.

\subsubsection{Spin-Lattice Relaxation}
\par
Whereas the nuclear spin-lattice relaxation rate, $T_1^{-1}$ shows a $T^3$
falloff below $T_c$ in YBCO, suggestive of a d-wave energy gap, $T_1^{-1}$
shows a `saturation' effect at low temperatures in LSCO, which changes its form
between the underdoped and overdoped regimes\cite{NQR1}.  In underdoped samples
($0.1\le x\le 0.15$) $T_1^{-1}$ saturates to a T-independent value, which is
interpreted as evidence for the presence of localized magnetic islands.  The
saturation can be suppressed by an applied magnetic field.  Unfortunately, the
interpretation is complicated, since the saturation is not observed in a sample
with $x=0.085$, and the observed effects may be associated with local regions
of LTT phase, centered near $x=0.115$, similar to the LTT phase of LBCO.  For
overdoped samples ($0.2\le x\le 0.24$), a very different behavior is observed:
after a dip at $T_c$, the Korringa law, $(T_1T)^{-1}=const.$ is obeyed, 
suggesting the presence of normal carriers in the superconducting state.  While
this was interpreted as due to pairbreaking effects in a d-wave 
superconductor\cite{NQR1}, it could also be a result of nanoscale phase 
separation in the overdoped regime, where the second phase is an ordinary metal.

\subsubsection{Transport Effects}
\par
The insulating grain boundaries between metallic grains should behave as 
`intrinsic weak links', greatly reducing the ability of these materials to carry
critical currents.  Evidence for such weak links had been found in many early
experimental studies, and some more recent studies have placed this picture on a
firmer basis.  Most of these studies, however, involve either YBCO in the mixed
Ortho I -- Ortho II regime or overdoped LSCO; in other words, the two domains
consist of `VHS' and metallic islands.  Thus, Jones, et al.\cite{ECJ1} studied
the critical current on the 90K plateau of YBCO.  They found that $J_c
\rightarrow 0$ as $\delta\rightarrow\delta_c$, with $\delta_c\simeq 0.2$
marking the edge of the 90K plateau.  This was taken as evidence that the
current flows by a percolation path of Ortho I material.
\par
In measuring magnetization hysteresis, a characteristic `fishtail' or
`peak effect' is often observed.  While a number of models have been proposed,
one possibility is that it is associated with domains of a second, metallic
phase.  In LSCO, this effect is found only in overdoped samples, $x\ge 
0.14$\cite{fish}.  However, in overdoped YBCO, it can be eliminated by proper O
annealing\cite{fish2}.
\par
Finally, it should be expected that these domains would play a role in the weak
link problem associated with grain boundaries\cite{wklnk}, particularly since
the grain boundaries could act as nucleation sites for domains.  Hence, it is 
most interesting that evidence has been found for regions of hole depletion
near the weak link boundaries\cite{wklnk2}.  The width of the depletion region,
$\sim 80-600\AA$, while highly sample dependent, is much larger than the scale 
of the structurally disordered region $\approx 10\AA$ or of metallic cation 
disorder, $\approx 50\AA$, but is comparable to the typical sizes of Ortho I, II
domains.

\subsection{Overdoped Regime}

\subsubsection{Miscibility Gap}
Overdoped LSCO was one of the original three cuprate materials in which 
Jorgensen, et al.\cite{JDJ} reported a miscibility gap.  The second (higher hole
density) phase is found to be $x\simeq 0.5$\cite{JDJ} (for samples annealed
below $\approx$1100K, a different phase, La$_2$SrCu$_2$O$_6$, is 
formed\cite{JDJ1}).  Radaelli, et al.\cite{Rad} find single-phase 
behavior only in samples which are annealed at $T\ge 1170^oC$ and then quenched 
to room temperature. The samples need to be fired at 1170$^o$C for only 24 hours
or less to produce homogeneous samples.  However, if these samples are 
subsequently annealed in 20\% oxygen at temperatures between 950-1050$^o$C for
24 hours and then quenched, evidence for two-phase behavior appears.  This is 
strong evidence that at T=1170$^o$C, entropy wins out and produces a homogeneous
phase, but that {\it phase separation spontaneously develops at lower 
temperatures}, proving that the phase separation is an equilibrium phenomenon.  
Note that this is a macroscopic phase separation; the samples quenched from 
T=1170$^0$C would be ideal candidates for studying a purely electronic 
(nanoscale) phase separation at low temperatures.  
\par
Studies on LCO doped with both Sr and Nd\cite{Buch9} find similar results. Even
in samples carefully preannealed and then quenched from 1150$^o$C, there is
evidence for Sr inhomogeneity.  Strikingly, high-resolution ($200\AA\times 200
\AA$) EDX analysis found that the Nd is uniformly distributed, whereas there
are significant variations in the Sr content (and hence in hole doping).  This
suggests a stringent test of the phase separation models.  The Sr distribution
should be highly non-random.  Ideally, there should be a bimodal distribution,
but the possible existence of striped phases would allow the presence of
intermediate Sr dopings.  Additional EDX studies have been promised, and it is
hoped that these distributions will be measured. 
\par
A similar study by Takagi, et al.\cite{Tak,Taka} reported quite different 
results, finding evidence for `compositional fluctuations' in their as-prepared 
overdoped samples.  While they ascribed their results to random fluctuations, 
the Meissner signal clearly showed two well-defined plateaus, with one 
corresponding to the optimum T$_c$.  However, long term annealing (one month at 
1000$^o$C) reduced the amount of this optimally doped fraction, and added 
curvature to the Meissner signal, as if there were now a broad distribution of 
T$_c$'s.  If the equilibrium state of the overdoped system involved macroscopic
phase separation, then very different behavior would be expected, with the two 
plateaus sharpening with increased annealing.  What can one say in the light of 
such incompatible results?  Clearly, more experiments are needed to sort out
the situation.  A number of groups have disputed a main conclusion of Takagi, et
al., that the HTT phase is nonsuperconducting, noting that this is inconsistent 
with pressure\cite{press,press2} and Bi substitution\cite{BiLCO} studies, in 
which the LTO phase disappears long before superconductivity does.  Contrawise,
by substituting Pr for La, the HTT-LTO transition can be shifted to much
{\it higher} Sr doping, but the superconductivity still disappears at $x\approx 
0.24$, well within the LTO phase\cite{Buch4}.  
\par
Moreover, to explain the residual T$_c$'s found in the HTT phase, Takagi, et al.
suggested that there are residual regions of LTO phase present, which again 
would be consistent with a nanoscale phase separation.  I propose therefore that
the best interpretation of their data is in terms of nanoscale
phase separation, involving a VHS phase of LTO symmetry and a metallic phase of 
HTT symmetry.  Since the phase separation is nanoscopic, a macroscopic probe
will resolve only a single lattice structure, which will cross over from LTO to
HTT very close to the point at which the holes have a percolation crossover,
from LTO domains with HTT walls to HTT domains with LTO walls.  Hence, in 
Takagi's macroscopic HTT
phase, the superconductivity would arise from the LTO domain walls, with T$_c$
reduced by proximity effects (the second plateau would be associated with larger
grains of LTO phase).  Annealing would cause the HTT islands to grow, stretching
out the LTO domain walls between adjacent islands.  The larger domains of 
metallic phase would lead to a stronger proximity-effect reduction of T$_c$, in 
accord with the data.  Note that the fact that the pre-annealed samples already 
show a proximity effect means that the average scale of the LTO domains is only 
$\sim \xi\simeq 16\AA$, the superconducting correlation length.  The fact that a
month-long anneal does not homogenize the material on this length scale strongly
suggests that the phase separation is an equilibrium phenomenon.  Moreover, 
neutron derived radial distribution functions find that the local symmetry in
the HTT phase for $x>0.2$ is orthorhombic\cite{HTO1} (similar results were found
earlier for lower $x$\cite{Eg3}.
\par
A further hint of phase separation in overdoped LSCO comes from 
positron-annihilation spectroscopy\cite{PASt}.  In overdoped samples, a smearing
of the data is found, which obscures the shape of the Fermi surface, leading to
poor agreement with band structure calculations.  After ruling out a number of
possible origins for the smearing, one remaining candidate would be `a 
distribution of segregated phases'.
\par
A significant difference should be noted between these results and the O-doping
discussed earlier: the temperature range.  Whereas in O-doped material, 
macroscopic phase separation begins near room temperature, for Sr the separation
is found near 1000$^o$C.  This difference may be due to the zero-point motion of
the oxygen.  The characteristic energy scale of hole phase separation should be
$\sim J\simeq 1300K$, the exchange constant.

\subsubsection{Cu NQR}
\par
Cu NQR, a very local probe, sees evidence for two distinct Cu sites in {\it the 
metallic phase} of LSCO\cite{NQR,NQR0,NQR1}!  (The signal from the AFI phase is
shifted to much higher frequencies by the large hyperfine field of the ordered 
Cu moments.) The relative intensities are T-independent, but vary in an inverse 
sense with doping, one site ($B$) increasing, the other ($A$) decreasing with 
increasing $x$.  The two sites coexist in the range $0.12\le x\le 0.4$; above
$x=0.4$ the $A$ peak disappears and a new ($C$) peak appears\cite{NQR2}.  This
suggests that the metallic phase is approximately $x=0.4$, close to the value
$x\simeq 0.5$ found by Jorgensen, et al.\cite{JDJ}.  
The splitting cannot be caused by proximity to a Sr impurity,
since the same two NQR frequencies are found in La$_2$CuO$_{4+\delta}$ as in 
LSCO\cite{Hamm3}.  However, there are some features which make it difficult to
interpret these two sites as regions of superconducting and normal metal phase.
Thus, while the two sites coexist primarily in the overdoped regime, site $B$
persists down to $x=0.12$; while it seems to be absent at $x=0.10$\cite{NQR1}, 
the $A$ peak also disappears at slightly lower doping, due to a very short 
spin-spin relaxation time.  Moreover, the relaxation rate of both peaks is 
modified almost identically by the onset of superconductivity, with the {\it 
same T$_c$ at each site}.  This is not what I would have expected for a 
nanoscale phase separation, where the normal islands may have a lower or zero 
$T_c$.  Hence, the origin of these two peaks remains puzzling.  It should also
be noted that LDA calculations have been unable to reproduce either of the EFG
values for the planar Cu's, whereas the same calculations work well for the
other constituents, including the chain Cu's in YBCO\cite{SSB}. Martin\cite{RLM}
analyzed this problem via quantum chemical calculations on clusters, and 
concluded that the second peak could be due to Cu neighbors of a {\it localized}
hole, and that $\approx$25\% of all the holes are localized.  If this is the
case, there may be a relation between these holes and the anomalous
paramagnetic sites discussed in the following subsection.  However, their
connection with the nanoscale phase separation remains unclear.  In a recent
experiment, when Goto, et al.\cite{Got} made a special effort to homogenize 
their samples (to see the suppression of T$_c$ at $x\simeq 0.115$), they found
only a {\it single} Cu NQR site in LSCO at $x\approx 0.15$, although two sites
were still found in similarly prepared LBCO.
\par
${}^{89}Y$-NMR finds evidence for broadened peaks in overdoped YBCO\cite{BrA};
at lower temperatures ($\approx 100K$) two well resolved features are seen.  The
second peak is, however, ascribed to O deficiency.

\subsubsection{EPR}

For completeness, I should briefly comment on reports of EPR signals in the
cuprates, which have often been taken as evidence for phase separation in these
materials.  First, it is important to note that EPR is usually {\it not} seen in
either the optimally doped or the antiferromagnetic insulating cuprates, the
signals which are occasionally reported being due to impurity phases\cite{EPR0}.
As discussed by Mehran and Anderson\cite{MeAn}, this is a very surprising 
result, 
and has led to a number of careful searches, up to 1150K in La$_2$CuO$_{4+\delta
}$\cite{EPR01}, and for low concentrations of Cu in La$_2$NiO$_4$\cite{EPR02},
which failed to find any signals.  It is tempting to think that this might be a
signal of dynamic tilt modes, similar to what is found in BaTiO$_3$,
discussed above; however, while this would work for the optimally doped
cuprates, it is not clear whether there is dynamic tilting in the AFI.
\par
In the intermediate doping regime, there have been persistent reports of EPR 
signals, which might be associated with phase separation.  EPR in quenched 
samples was discussed above.  EPR has also been seen in underdoped YBCO, where
it has been assumed to be associated with paramagnetic centers on CuO chain
fragments\cite{MeAn,Tepl}.  In a detailed analysis of the specific heat of YBCO,
Phillips and coworkers\cite{Phil} have attributed the linear term in the
specific heat to a sample dependent density $n_2$ of Cu$^{2+}$ moments, which
also produce a Schottky anomaly in $C_v$ in a magnetic field.  The 
superconducting fraction $f_s$ is found to be suppressed by $n_2$, 
as\cite{Phil2}
\begin{equation}
f_s\simeq 1-n_2/0.012.
\label{eq:99g}
\end{equation}
If $n_2$ is associated with chain fragments, the correlation with $f_s$ is
consistent with the Poulsen\cite{Poul} model of $T_c(\delta )$, which
assumes that the plane is locally nonsuperconducting if the adjacent chain is
too short.  This would also explain why there is no simple correlation between
$n_2$ and $\delta$\cite{EPR03}.  Teplov, et al.\cite{Tepl} have also determined
a paramagnetic center density from its effect on Tm spin-lattice relaxation.
They find that this density is correlated with the fraction of highly-disordered
regions in their sample, and is related to $f_s$ by Eq. \ref{eq:99g}.

\subsubsection{Superconductivity}

While the overdoped phase has been characterized above as an essentially normal
metal, there is some evidence that it can also be a superconductor, usually with
a reduced $T_c$ value.  Thus, as mentioned above, in electrochemically 
oxygenated La$_2$CuO$_{4+\delta}$, there is a series of supercondicting phases, 
with increasing T$_c$'s.  The 40-47K phase seems to correspond to $P=0.16$, as 
in optimally doped LSCO.  At even higher doping levels (so far, only found in
thin films) an even higher, 55-60K superconductor is found\cite{Gren3}.  In
overdoped LSCO, Zhou, et al.\cite{press2} find several doping ranges in which
there are two coexisting superconducting phases, even though x-ray analysis
finds only one phase present.
\par
Finally, in YBCO, the optimum doping (maximum T$_c$) occurs at $\delta\simeq 
0.1$.  In the overdoped regime, $\delta\le 0.09$, there is evidence for the 
simultaneous presence of two superconducting phases\cite{Kald,Clau}.  This new 
phase is associated with a structural phase transition (probably first order) in
which the dimpling of the CuO$_2$ planes suddenly decreases\cite{Clau,Kald2}.  
In the same doping range, two peaks are seen in the specific 
heat\cite{Juno3,Choy,Lor1,Juno2}.  From combined specific heat and 
susceptibility measurements, Janod, et al.\cite{Juno2} show that the two T$_c$'s
are split by $\approx 4K$ at YBCO$_7$, and smoothly merge near the optimum 
T$_c$, $\sim$YBCO$_{6.92}$, and then {\it another} split peak arises in 
underdoped samples! (The conversion between annealing temperature, used in 
\cite{Juno2}, and $\delta$ is discussed in \cite{Juno4}.)
\par
Janod, et al.\cite{Juno2,Juno4} interpret the split transition in terms of 
limited oxygen diffusion: a thick shell near the grain surface is fully
oxygenated, impeding diffusion into the grain interior.  Conder, et 
al.\cite{Cond} have shown that this interpretation is unlikely, and that 
nanoscale phase separation is a more probable interpretation.  
However, recent experiments on high quality YBCO crystals find
only a single peak in the heat capacity in the overdoped regime\cite{Breit}.

While considerably more experimental work is necessary to sort out the details
of this new superconducting phase, a few speculations about possible origins can
be made.  One possibility is that the other phases are associated with
subsidiary VHS peaks, such as might be generated by interlayer coupling or
spin-orbit coupling effects, discussed above.  Since split $T_c$'s are not 
observed in overdoped Bi-2212 or Tl-2201, it might be thought that the
second peak is somehow associated with CuO chains.  However, Ca substitution
shows that the appearence of the second peak correlates with hole overdoping in
the planes, and not with $\delta$\cite{overT}.

A different possibility is that
these phases are due to proximity effects between superconducting and normal
domains.  This is similar to the Allender-Bray-Bardeen (ABB) model of excitonic
superconductivity\cite{ABB}, arising from proximity effects associated with 
intimate contact between a metal, with large carrier concentration, and a 
semiconductor/semimetal with strong electron-phonon coupling but low carrier 
density.  In the present case, superconductivity could arise in the overdoped 
normal metal due to intimate proximity with the VHS phase.  Interestingly,
the dominant thermodynamic weight is associated with the lower $T_c$ phase (this
phase correlates with the transport $T_c$).  This would suggest that the lower
$T_c$ phase is the VHS phase, and the metallic phase has a higher $T_c$.  If
true, this could confirm ABB's prediction that such excitonic coupling could
enhance $T_c$.

However, it must be cautioned that several recent studies of the heat capacity
of YBCO have failed to observe a split peak in the overdoped 
regime\cite{Breit,Gp3}.

\subsection{Why are there Several Discrete T$_c$'s?}

In La$_2$CuO$_{4+\delta}$, there seem to be discrete superconducting phases with
T$_c$ $\simeq$ 32K, 43K, 55K, and possibly 28K, while the Ortho II phase in
YBCO has T$_c$=59K.  Such a series of T$_c$'s arises naturally in a model of
nanoscale phase separation, either as a result of staging\cite{Custag} or
due to CDW formation.  In the case of staging, the different stages resemble
the layering seen in, e.g., Bi-22(n-1)n, and the different stages would be
expected to have different characteristic temperatures.  In analogy with 
intercalated graphite, most of the holes would be attracted to the layers
bounding the intercalated O's.  Hence, a single stage could display two $T_c$'s,
one associated with the bounding layers, one with the interior layers.
\par
Alternatively, a similar effect could arise from commensurability effects in a 
CDW phase: as the doping is varied, the CDW periodicity changes, and certain
periodicities will lock in to the underlying crystal lattice periodicity.  
In the absence of disorder, they will tend to form a devil's 
staircase\cite{Low,Mehr}.  Since each CDW is composed of 
alternating layers of a VHS phase and either a magnetic or metallic phase, a
range of T$_c$ values will arise due to proximity
effects between the two domain types\cite{prox}.  In fact, if the domains are 
wide enough, it is possible that one CDW will have two superconducting 
transitions, one for the VHS domain, another for the metallic domain.

\subsection{La$_{2-x}$Sr$_x$NiO$_4$: An Analogy with Well-defined CDW's}

\par
La$_2$NiO$_4$ is isostructural to La$_2$CuO$_4$, and can similarly be doped
either by Sr substitution or by excess O, which occupies the same interstitial
site.  The phase diagram is also very similar, with an AFI at half filling,
crossing over to a spin glass phase, and finally a conducting phase.  However,
the conducting phase arises only for much higher doping ($x\simeq 1$), and no 
superconducting phases are found.  The intermediate doping regime is most 
relevant to the present discussion, since nanoscale phase separation seems to be
manifest in a complicated series of CDW's, associated with rational filling 
fractions.  Most interestingly, these phases show evidence for strong polaronic 
effects.
\par
However, much work still needs to be done to answer some very basic questions 
concerning the nickelates.  For instance, when holes are added, which states are
occupied?  Isolated Ni$^{2+}$ is a spin 1 system, with two holes, one $d_{x^2-y^
2}$, one $d_{3z^2-r^2}$, with spins parallel to satisfy Hund's rule.  In the 
nickelate, the $d_{x^2-y^2}$ hole will tend to hybridize with O $p_x$ and $p_y$ 
orbitals from the planar O's, while the $d_{3z^2-r^2}$ hole will hybridize with 
the apical O's $p_z$ orbital, forming two bands which cross the Fermi level.
At half filling, this hybridization is quenched by the large Hubbard $U$,
leading to a charge transfer gap of $\Delta\simeq 4eV$\cite{Saw}, about twice
as large as in the cuprates.  The first holes doped in are about 30-40\% Ni, 
60-70\% O\cite{Tan}, comparable to NiO\cite{NiO} (in the cuprates, the doped 
holes are mostly O character).  The O holes are approximately an equal mix of 
$p_x$, $p_y$, and $p_z$\cite{Saw}.
\par
The composition of the $d$-hole has not been determined, and different groups
have proposed different symmetries.  Thus, some suggest that it is $d_{3z^2-r^2}
$-like, at least up to $x=0.5$\cite{Cav} in La$_{2-x}$Sr$_x$NiO$_4$ (LSNO).   
For undoped LNO, $x$=0, there is a 
large transferred hyperfine field on some of the La nuclei\cite{Wada}, which 
remains nearly constant in magnitude up to $x=1$, whereas the fraction of La 
nuclei experiencing this field falls linearly to zero at $x=1$.  The large 
magnitude of the field can be explained by a transferred hyperfine interaction 
between the La and the Ni $d_{3z^2-r^2}$, coupled via the apical 
O\cite{tranhyp}.  In this picture, doping $0\le x\le 1$ predominantly eliminates
the $d_{3z^2-r^2}-p_z$-band, so that doping $x> 1$ would be similar to the 
cuprates.  This could explain why the holes for $x\le 1$ are so much more 
strongly localized -- the material remains insulating at all dopings, and 
antiferromagnetic interactions persist out to much larger dopings.  Also, the 
material with $x>1$ has a metallic conductivity down to low temperatures, $T\ge 
100K$, while for $1.2\le x\le 1.5$, it remains metallic to the lowest 
temperatures\cite{TakeGood,LSNO5}.  
\par
However, there are a number of objections to this
picture.  Thus, the Cu-apical O distance decreases with $x$, so that the
NiO$_6$ octahedron is almost undistorted at $x=1.4$, suggesting that the
$d_{x^2-y^2}$ and $d_{3z^2-r^2}$ orbitals are nearly degenerate\cite{TakeGood}.
Moreover, band structure calculations find the first holes to be of $b_1$
($x^2-y^2$) symmetry\cite{McMa}.  Also, the `mid-infrared band' of the optical
conductivity has an identical $x$ dependence in both LSNO and LSCO\cite{BiE},
suggesting that the doped holes are very similar (i.e., of $x^2-y^2$ symmetry).
The fact that the $p$ component of the holes is an equal mix of $x$, $y$, and
$z$ states suggests that a similar mix holds for the $d$ states.
Indeed, much more complicated behavior is possible; certainly, there appears to 
be an anomaly near $x=0.5$\cite{TakeGood,Cav}, (the resistivity has a 
maximum), which could signal a charge transfer to the $d_{x^2-y^2}-p_{x,y}$
band\cite{Cav}.  Also, it has been suggested that, since the NiO$_2$ planes
become more compressed as $T$ decreases, there could be a high-spin ($S=1$) to
low-spin ($S=1/2$) transition at low $T$, even near $x=0$\cite{TakeGood}.
\par

\subsubsection{La$_2$NiO$_{4+\delta}$ (LNO)}
In La$_2$NiO$_{4+\delta}$ (LNO), the oxygen doping leads to a series of phases,
closely related to the LTO, LTT, HTT, and Pccn phases of LSCO; however, in LNO,
these phases are separated by at least four regimes of phase separation in the 
doping range $0\le\delta\le 0.18$\cite{LNO1,LNO4}, and there are a number of 
additional phase transitions below room temperature.  Several of the phases are 
antiferromagnetic, often with a weak ferromagnetic component due to spin 
canting\cite{LNO3,LNfer}.  
\par
An added complication, which has only recently been recognized\cite{stag}, is
that the oxygen interstitials form well-defined staged compounds, analogous to
graphite intercalation compounds (GIC's)\cite{GIC}.  Staging as a means of
achieving nanoscale phase separation is discussed in Section X.B.5.  The phase 
diagram of LNO shows evidence for several distinct stages, 
Fig.~\ref{fig:30}\cite{stag}.
Moreover, when the doping densities are intermediate between the dopings 
associated with pure phases, there is a well-defined regime of phase separation 
(e.g., between stage 2 and stage 3).  This regime of phase separation was 
predicted for GIC's\cite{Saf} and has been taken as evidence that the 
interstitial O's have a preferred density, which may be controlled by a hole 
phase separation\cite{stag}.  A similar staging has recently been reported for
La$_2$CuO$_{4+\delta}$\cite{Custag}.
\par
Whereas neutron and X-ray diffraction studies typically find evidence for phase
separation, electron microscope investigations generally find a 
complicated series of CDW's, which appear to be associated with commensurate 
ordering of the O interstitials, and are associated with rational doping 
fractions, $\delta =1/2 n$, $n=1,2,3,...$\cite{LNO2}.  Incommensurate CDW's 
are also observed, which seem to gradually evolve between two neighboring 
commensurate CDW's.  More recent neutron and X-ray studies do find evidence for 
superlattices, but only below room temperature\cite{LNO4}, while the electron 
diffraction studies find these superlattices at room temperature.  A 
number of possibilities exist for explaining this difference\cite{stag} -- the 
thinner samples may stabilize superlattices, or the locality of the probe may 
better reveal local structure, or the vacuum or electron-beam heating may effect
subtle adjustments to the oxygen concentration or mobility.  In any case, the 
electron diffraction studies reveal a tendency which is now clearly revealed in 
low-T diffraction studies.
\par
While the phase diagram of these materials is still being worked out, 
Fig.~\ref{fig:30} provides a clear impression of its complexity.  
Note that the staging is found for $0.03\le\delta\le 
0.11$; for higher doping, the commensurate magnetic order disappears, and is
replaced by an {\it incommensurate} magnetic order\cite{LNO6,LNO7}, similar to 
LCO for $\delta >0.06$.  Whereas for lower doping, the N\'eel temperature is 
decreasing with $\delta$, for $\delta\ge 0.11$, $T_N$ increases with increasing 
$\delta$.  There is long-range incommensurate order below the N\'eel temperature
and short-range incommensurate order above, extending beyond $2T_N$.
\par
This incommensurate phase is most clearly defined at $\delta =0.125$, and has
much in common with a phase found in LSNO at a similar doping, $x\simeq 0.2$
(discussed below).  Both have peaks in the susceptibility when the magnetic
field is parallel to the a,b planes (for lower doping, the susceptibility peaks 
in a perpendicular field).  In both, the incommensurability has the same 
symmetry, and is interpreted in terms of hole-rich domain walls\cite{LNO6}.  The
incommensurate {\it magnetic} ordering is accompanied by incommensurate {\it
charge} ordering, at half the periodicity\cite{LNO7}.  This provides strong
evidence for the charge domain wall model: the magnetic domains have a period
twice as long, because the holes invert the phase of the antiferromagnetic
spin order, thereby acting as antiphase domain boundaries.  This is
expected if the holes are predominantly on oxygens, as originally pointed out by
Aharony, et al.\cite{Aha}: the hole, located
on a planar O, causes both adjacent Cu spins to line up antiparallel with the
hole spin and hence parallel with each other.  Moreover, the incommensurability
is found to be strongly temperature dependent, with a tendency to lock in at
rational filling fractions.  This is interpreted as due to a competition
between hole and O ordering, and may lead to a devil's staircase of ordered
phases\cite{LNO7}.  A similar combined spin and charge ordering is found in
antiferromagnetic Cr\cite{CrAF}, and more recently, in LSNO, with $x=
0.2$\cite{LSNO3}.  It should be noted that in LNO, unlike the cuprates, the hole
doping is assumed to follow $P=2\delta$; the close similarity found between
LNO, with $\delta\simeq 0.125$ and LSNO with $x=0.2$ would tend to confirm
this assignment.
\par
At higher doping, a compound LNO$_{4.25}$, prepared by electrochemical 
oxidation, is found to have long-range order of the interstitial O's\cite{LNO5},
accompanied by CDW's in the NiO$_2$ planes.  

\par

\subsubsection{La$_{2-x}$Sr$_x$NiO$_4$ (LSNO)}
If the situation is confused for LNO, it is even more confused for La$_{2-x}
$Sr$_x$\-NiO$_4$ (LSNO).  This is in part because LSNO readily takes up extra 
oxygen, particularly for $x\le 0.2$, and many studies do not measure the value 
of $\delta$.  Not only does this mean that the hole content is not known, since
$P=x+2\delta$, but it may strongly influence the properties of superlattices.
One electron microscopic examination found that, when $x$ and $\delta$ were both
non-zero, superlattices were observed only when $\delta\ge 0.06$, independently 
of $x$\cite{Saya}.
\par
Surprisingly, the magnetic properties\cite{LSNO1,LSNO2} and the HTT-LTO
transition temperature\cite{LSNO3} of LSNO are very similar to those of LSCO:
long-range antiferromagnetism is eliminated at a quite low doping, $x\simeq 
0.04$, being replaced by a low-temperature spin-glass transition, Fig.~ 
\ref{fig:31}\cite{LSNO1}.  More recent $\mu$SR experiments\cite{Chow} have found
onsets of magnetic ordered states at considerably higher temperatures, 
throughout the doping range $0.4\le x\le 1.2$, with a local maximum at $x=0.5$ 
(open squares in Fig.~\ref{fig:31}).
The spin glass phase has been interpreted in terms of
a hole-rich domain wall model\cite{LSNO2,LSNO6}, very similar to that 
subsequently applied to LSCO.  In LSNO (specifically studied at $x$=0.2), this 
phase is associated with {\it incommensurate modulations of the 
antiferromagnetic scattering peak} -- thereby making a connection between domain
walls and incommensurability which has only been theoretically assumed in LSCO. 
The modulations differ, in that the incommensurate peaks in LSNO are at 
$(\pi\pm\delta\pi ,\pi\pm\delta\pi )$ and $(\pi\pm\delta\pi ,\pi\mp\delta\pi )$,
while in LSCO they are at $(\pi\pm\delta\pi ,\pi )$ and $(\pi ,\pi\pm\delta\pi 
)$, but the values of $\delta\pi$ are quite comparable\cite{LSNO2}, Fig.~
\ref{fig:15}c.  The domain size is also similar, $\approx 9\times 17\AA
$\cite{LSNO2}.  The incommensurability also appears to be dynamic: at 2K, the 
fluctuations are slow compared to the characteristic neutron-diffraction time 
scale, $10^{-11}s$, but by 80K, the fluctuation time has fallen to $\approx 10^
{-12}s$.  One curious feature is that the susceptibility has a peak at the 
transition for the B-field oriented within the a-b plane\cite{LSNO4}.
\par
The incommensurate phase near $x=0.2$ bears a strong resemblance to the phase
in LNO at a comparable doping, $\delta =0.125$.  As might be expected (since
the Sr is much less mobile than the O), the incommensurate phase in LNO
displays true long-range order below 110K, whereas in LSNO there is only a
short-range spin freezing in the same temperature range\cite{LSNO3}.  The fact 
that this phase is so similar for both Sr and O doping shows
that it is driven predominantly by hole ordering, as opposed to ordering of
the interstitial O's.  As further evidence, the correlation length actually
increases with increased Sr doping\cite{LSNO3}, and hence is unrelated to the
spacing between Sr impurities.  Finally, since there is nothing comparable to
staging for Sr doped samples, this incommensurate phase persists to lower 
doping, $x\simeq 0.135$, with the degree of incommensurability scaling with 
$x$\cite{LSNO3}.
\par
Similar charge-ordering phases have been observed in recent electron diffraction
studies\cite{LSNO}, but again there are characteristic differences with neutron
data, perhaps exacerbated by a lack of knowledge of the oxygen doping $\delta$
(although relatively small differences were found between samples prepared in
air and those oxygenated at 200 bars of oxygen at 850$^oC$).  Thus, for a sample
with $x=0.2$, the ordering temperature was considerably higher, $\approx 220K$, 
and with a different incommensurability, although with the same symmetry. These 
studies were carried out over a wide range of hole doping, and observed clear
lock-in effects to $\approx 1/3$ near $x\simeq 0.33$ and to commensurate at $x
\simeq 0.5$.  This latter value is close to the point at which there is a
significant crossover in many properties of the system (maximum c-axis and
minimum a-axis expansions, maximum resistivity, with a sharp drop in $\rho$ for
larger $x$), and Chen, et al.\cite{LSNO} interpret this as a polaronic lattice,
with one hole in every other cell, and the 1/3 anomaly as due to a lower-density
commensurate polaron lattice (stripe phase) which is presumed to be associated
with an incommensurate magnetic structure.  In subsequent work, anomalies in 
resistivity and susceptibility have been found in the vicinity of these dopings,
at $x$=0.33, 0.5\cite{LSNO8}.  There is a clear transition at T=235K (1/3) or
340K (1/2), nearly independent of whether the dopant is Sr, Ba, or Ca.  For
$0.2\le x\le 0.4$, the 1/3 transition occurs, with only the amplitude of the
anomaly changing, as if the volume fraction of the commensurate phase were 
changing.  Near $x=0.5$, the optical absorption feature associated with an
O-stretching phonon at 670$cm^{-1}$ suddenly grows in intensity, just before
disappearing at higher doping\cite{LSNO7}.
The Seebeck coefficient is found to change sign {\it three times} as $x$ changes
from zero to 0.5\cite{LSNO1}, suggestive of a series of CDW-like 
transitions, which continually remodify the Fermi surface.
\par
Finally, synchrotron x-ray studies of LSNO, $x=0.2$ and LSCO, $x=0.075$ have
found large, incommensurate diffuse scattering\cite{LSNO9}.  By scanning near
the Sr K edge, the authors were able to show that the scattering was not 
associated with Sr atoms, and hence must be hole related.  More recent 
studies\cite{LSNO9A} have clarified the interpretation of these results.  The
scattering is indicative of local structural disorder, possibly originating from
electron-lattice interaction -- polaron-lattice formation.  The apparent
incommensurability is due to the fact that the scattering is dominated by 
transverse, or shear, displacements, for which the scattered intensity is weak
in the commensurate orientation.

\subsubsection{Polarons}
The electron-phonon interaction seems to be about three times stronger in LSNO 
than in LSCO.  Thus, Bi and Eklund\cite{BiE} have identified polaronic peaks in
the optical conductivity, at $\approx 18meV$ in LSCO and 50$meV$ in LNCO.  
[This interpretation for LSCO has been challenged\cite{Cran}, but is supported 
by studies in much more lightly doped Cu materials\cite{Falc}.]  Similarly, in 
neutron scattering (Fig.~\ref{fig:a22}) the softening of the O-O stretch 
mode in the undoped materials is considerably larger for the nickelates than for
the cuprates (although some caution must be exercised, since the LNO sample 
seems to be slightly O-doped\cite{PiRe2}).  A strong coupling of the $x=0.5$
phase with an O-stretch mode\cite{LSNO7} was noted above.  A related compound,
Nd$_2$NiO$_4$ has been studied, because it has a similar sequence of phase
transitions to LBCO, HTT$\rightarrow$LTO$\rightarrow$LTT; its 
orthorhombicity is about 5 times larger than the cuprate, consistent with
stronger electron-phonon coupling. This may explain the characteristic 
differences found in its microscopic properties\cite{APB}.
\par
Correspondingly, polaronic effects in LSNO should be stronger, and the mixed 
charge and spin domain states discussed above have been interpreted in terms of
polaronic phases\cite{LSNO,LSNO8}.  These phases are very similar to those
postulated above for the cuprates.  Thus, whereas phase separation in the 
cuprates is mostly short-range and dynamic, LSNO has a well-developed series of 
striped phases\cite{LSNO}, with varying spacing of the (second-phase) domain 
walls. These phases can display {\it simultaneous} polaronic and magnetic order,
as has been found in the LTT phase in LBCO.  Moreover, the polaronic effects 
keep LSNO in the small polaron limit: the materials remain insulating to much 
higher doping levels, and there have been no unambiguous demonstrations of 
superconductivity in these materials.
\par
The polaronic behavior has been theoretically analyzed in terms of a three-band 
Hubbard-Peierls model\cite{ZaaP}, assuming coupling to O-O breathing modes,
similar to the above discussion for the cuprates, Section VIII.D.  However,
these modes do not show strong signs of softening\cite{PiRe}; for both cuprates
and nickelates, it is important to try to understand the role of the quadrupolar
and 1D stretch modes, which do display strong softening.

\subsubsection{A VHS Theory for LSNO?}
Qualitatively, the compounds based on LNO share many features in common with
the cuprates, and it will be important to try to understand their similarities 
and differences.  Given the uncertainty in the symmetry of the Ni component of 
the doped holes, it would be difficult to develop a very quantitative theory at
this point.  Nevertheless, one can always speculate on how a VHS model might 
look.
\par
Can a VHS theory be applied to these materials?  Since there are two Fermi
surfaces, there will be two sets of VHS's, with a potential for very interesting
interplay -- although it should be noted that the $d_{3z^2-r^2}-p_z$ band may be
considerably more three dimensional.  In principle, a five-band Hubbard model
should describe the bands, with correlations accounted for by the slave boson
model.  A preliminary set of band parameters is available\cite{McMa}.  From
the doping dependence of various properties, I would guess that the Ni 
contribution has approximately equal admixtures of $d_{x^2-y^2}$ and $d_{3z^2-
r^2}$ symmetry, and that doping from $x=0$ to $x=1$ involves mainly a transition
from high-spin to low-spin.  
\par
The staging phenomenon in O-doped LNO provides a clue to a preferred hole
doping, close to $x=0.2$, since all stages $n\ge 2$ have approximately this same
hole density.  Moreover, the material at $x=0.2$ is very similar, for both O and
Sr doping, and there seems to be a clear crossover in behavior from underdoping 
to overdoping about this point.  (Also, samples with $x<0.2$ tend to easily take
up extra O, while for $x>0.2$, a slight oxygen deficiency is often found.)  For
all these reasons, I would suggest that $x=0.2$ corresponds to a VHS in this 
material; the value is slightly larger than found in LSCO, but comparable to
YBCO.  The fact that LSNO is non-superconducting could then be explained as due
to too strong polaronic coupling, possibly complicated by the smaller 
bandwidth associated with a $d_{3z^2-r^2}$ component.  In the underdoped 
material, the staging would be a manifestation of nanoscale phase separation.
The persistence of magnetism would follow from the fact that, for $n\ge 3$, some
layers are weakly doped -- the reduction in $T_N$ following from reduced
interlayer coupling.
\par
The enhanced electron-phonon coupling in the nickelates favors the VHS model
for phase separation, as opposed to a magnetic model.  Unfortunately, whereas 
the VHS model was cited in a number of early experimental studies of phase
separation\cite{JDJ,Ryd,Juno3,Bian2}, it fell into disfavor, so that recent 
nickelate studies cite only the magnetic models for phase separation.
\par
At higher doping, there is clear evidence for charged stripe phases at rational
fillings, $x\simeq$1/2, 1/3.  These would follow from the same phase separation
model, Fig.~\ref{fig:j23}a, provided that the free energy has a minimum
associated with $x=1$.  This is very plausible chemically; indeed many 
materials cannot be continuously doped, but instead form a macroscopic phase 
separation, with a miscibility gap between the two end members.  Note, however,
that the observation of rational fractions $x\simeq$ 1/2, 1/3, shows that the 
two lowest free energy minima are at $x=0,1$.  In this case, doping should just 
mix these two phases, and other phases with shallower minima (e.g., the VHS) 
should not be stable -- unless the phase separation is sufficiently nanoscale 
that properties evolve fairly smoothly with doping.

\subsubsection{Another Anomalous Perovskite: La$_{1-x}$Sr$_x$MnO$_3$}

The colossal magnetoresistance compounds La$_{1-x}$\-Sr$_x$\-MnO$_3$ came to my
attention too late to be adequately described in this survey, but their
properties are so similar to the nickelates that just a brief mention should be
made.  While these nearly cubic compounds are the best studied, there are also 
layered structures with similar properties\cite{MAT1,MAT2}, including in 
particular La$_{1-x}$Sr$_{1+x}$MnO$_4$.  Just as in the nickelates, there are 
charge ordered phases at a commensurate doping, $x\simeq
$1/2\cite{CMR8,SRBC,MAT1} 
(see also \cite{CMR7}).  The most interesting physics is the colossal 
magnetoresistance, associated with a crossover from an AFM insulating phase to a
conducting ferromagnetic phase, near $x\simeq0.2-0.45$, with the transition 
temperature tunable by a magnetic field. This conducting phase bears a close 
resemblance to the ferron phase discussed above\cite{Nag2}.
\par
The physics has traditionally been analyzed in terms of the double exchange
model: three electrons are strongly coupled in $t_{2g}$ states to form a
localized $S=3/2$ state on each Mn ion.  The remaining $1-x$ electron is in a
potentially mobile $e_g$ state, also with a strong Hund's rule coupling to $S$.
This coupling means that the hopping rate depends strongly on $\vec S\cdot\vec
S^{\prime}$, where $\vec S$ and $\vec S^{\prime}$ are the spins of the two atoms
involved in the hop.  At half filling, LaMnO$_3$ is an AFI: hopping is greatly
reduced, since $\vec S$ and $\vec S^{\prime}$ point in opposite directions, so
the hopping electron loses its Hund's rule stabilization.  By creating a local
FM domain the electron can delocalize, thereby lowering its energy.  At a 
critical doping, the system goes ferromagnetic, possibly by a percolation
crossover.
\par
Millis, et al.\cite{AJM} have shown that the double exchange model cannot
be the entire picture -- the localization effects are too weak -- and that a
ferron (or magnetic polaron\cite{mp}) model is unlikely to improve the
situation.  They propose a dynamic JT model\cite{Good0}, involving predominantly
a quadrupole O-O stretch mode (Appendix C) (see also Ref. \cite{RZB}).  
\par
Recently, much evidence has accumulated that there is strong coupling between
the magnetism and lattice distortions\cite{CMR1,CMR2,CMR3,CMR4,CMR5}. 
Thus, within an LDA model, JT distortions
of the planar quadrupole form are essential to produce an AFI 
groundstate\cite{SPV} (see also Ref. \cite{SHT}).
Particularly striking is a giant
negative oxygen isotope effect on the magnetic transition 
temperature\cite{ZCKM}, which has been interpreted in terms of a JT 
polaron\cite{JTpol}.  Photoemission studies\cite{CMR9} find that the insulating
behavior is due to strong small polaron effects, which contribute to a charge
fluctuation energy of $\approx$1.5eV, and which are reduced below the 
ferromagnetic
transition, which restores a finite dos at the Fermi level.  Interestingly, the 
JT polarons can contribute only a small fraction of the charge fluctuation
energy, and the main small polaron effect is due to the large difference in 
ionic size ($\approx 20\%$) between Mn$^{3+}$ and Mn$^{4+}$.  Neutron 
pair-distribution function studies\cite{Bill3} find a local lattice formation
occuring very close to the metal-insulator transition.
Some of the largest polaronic anomalies are associated with a doping
$x\simeq 1/8$\cite{CMR5}, and a charge ordered phase at that doping has now been
found\cite{CMR10}.  There is also evidence for $T,x$ regimes of multiple phase
coexistence\cite{CMR3,CMR6}.  In La$_{0.5}$Ca$_{0.5}$MnO$_3$, electron 
diffraction studies find evidence for charge ordering\cite{ChCh}, with a 
commensurate-to-incommensurate(!) transition coinciding with the AFM to FM
transition.

\subsection{Relation to Dynamic JT Effect}

The previous sections have summarized evidence for two different classes of
anomalies in the cuprates: na\-no\-scale phase separation and local (dynamic)
structural disorder.  Aesthetically, it would be preferable if these two
anomalies were aspects of a single underlying phenomenon.  The prospects of
such a unification are clearly present: the domain walls of the local LTT
phase are exactly the LTO structure which is stable in the undoped regime.
However, this identification will require some modifications of the previously
developed picture.  This section will describe these modifications, and the
experimental support for them.

\par
Such a connection arises naturally in the VHS model of phase separation: the VHS
phase is stabilized by a strong electron-phonon interaction, so it would be 
expected that the VHS phase is associated with a particular lattice structure, 
and for this structure to persist in the hole-rich domains.
Looking back at the picture of a dynamic JT soliton, Fig.~\ref{fig:c21}a, it is 
striking how much it resembles a nanoscale phase separation, with the LTO 
soliton walls as AFI phase, and the LTT domains, with tilt less than the 
critical value, as the superconducting phase.  This impression is further 
enhanced by comparing to the experimentally derived models of the LTT phase, 
Fig.~\ref{fig:c21}c, and the Bi-2212 domains, Fig.~\ref{fig:c21}d.  Thus, it is 
an interesting possibility that the dynamic JT phase is itself a form of 
nanoscopic two-phase mixture.  This is consistent with the picture of 
LSNO presented above.
\par
The full implications of this possibility have not yet been worked out, but it 
would lead to some modifications of the interpretations given earlier.  One 
striking modification has to do with the fact that LSCO at
optimum doping is in the LTO phase.  If this is a dynamic JT phase, then
even at optimum doping there is a two-phase mixture, with
the residual (LTO) domain walls constituting a remnant of the AFI phase.
This would, however, explain the observation of incommensurate peaks
in the neutron diffraction of a sample with $x\simeq 0.14$, i.e., close to 
optimum doping\cite{Yama}, or even $x=0.2$ in Nd substituted 
material\cite{Tran2}, which is otherwise very puzzling.
\par
Further evidence can be seen in the La NMR experiments, 
Fig.~\ref{fig:c23}\cite{Hamm}: 
while the superconducting phase has a broad distribution of tilt angles, the
AFI phase has a well-defined LTO tilt (the splitting of the AFI peaks is due
to the magnetic field).  Hence the AFI appears to be correlated with a {\it 
static LTO phase}, like the solitons in Fig.~\ref{fig:c21}.  Neutron scattering
studies of the HTT-LTO transition\cite{Brad2} confirm this picture.  The
transitions in the doped and undoped materials are strikingly different.  While
the undoped material has critical fluctuations near the transition, in the doped
material ($x=0.13$), fluctuations and anharmonic effects are present over an
extremely broad range of temperatures.  The two samples have {\it different 
values of the critical parameter $\beta$} (orthorhombic diffraction peak
intensity $\propto (T_{T-O}-T)^\beta$), and the doped material shows evidence 
for a central peak, similar to SrTiO$_3$ and suggestive of the proximity of a 
tricritical point.  (However, even in the undoped samples, diffuse scattering
suggests that the local symmetry is lower than the long-range 
symmetry\cite{Brad3}.)  
\par
Another clue is provided by the LTT phase (Section XIII).  Ordinarily, the
dynamic LTO phase crosses over to a static LTT phase as the magnitude of the
tilting (depth of the LTT potential wells) increases.  However, for Ba or Sr
underdoping, the material crosses over to the LTT phase near $x\simeq 0.125$,
but for lower doping a LTO phase is reestablished. This suggests that there are
two distinct LTO phases, a dynamic one near the VHS, and a static one, perhaps
stabilized by magnetic umklapp scattering, near half filling.  Finally, it 
should be noted that the pseudogaps, which I have interpreted in terms of a 
dynamic JT effect, have also been interpreted in terms of dynamic domain 
formation\cite{Mehr2}.  Direct observation of charge and magnetic stripe
ordering in Nd-substituted LSCO\cite{Tran,Tran2} find that the magnetic ordering
temperature is less than the charge ordering temperature, and both are much less
that the pseudogap temperature.  Nevertheless both the magnetic stripe order 
temperature $T_{ms}$ and the pseudogap temperature $T^*$ have a similar doping
dependence, with $T_{ms}$ vanishing near $x=0.2$\cite{Tran2} and $T^*$ near
$x=0.27$\cite{Gp3}.  Also, the stripes are more stable in the Nd substituted
material than in pure LSCO, just like the LTT phase.  This is suggestive that
both pseudogap and stripe phases are related phenomena, with the pseudogap being
the dominant factor.
\par
Finally, it should be noted that the charge stripe phase in Pr$_{1-x}$Ca$_x$MnO$
_3$ is accompanied by a combined tilt mode and JT structural 
distortion\cite{LMO1}.

\subsection{Arguments Against Phase Separation}

The previous experiments have provided considerable evidence in favor of a
(nanoscale) phase separation.  
My own feeling is that the evidence for some kind of phase separation is quite
strong.  The early experimental results were by and large arrived at 
independently, without much guidance from theory or even awareness of others'
findings.  Indeed, one purpose of the present survey is to gather all these
results into one place.  The neutron diffraction evidence for striped
phases seems decisive\cite{Tran,Tran2}.  

Nevertheless, at this point the harder question must be asked:
can such a phase separation account for {\it all} experimental observations, or
are there some experiments which argue against this picture.
Many experiments have simply been interpreted in terms of a uniform system;
they should now be reanalyzed to see whether they could also be interpreted in 
terms of phase separation -- particularly in the nanoscale limit, when it is 
more appropriate to speak of a novel charge bunching phase.  Here, I will 
concentrate instead on those experiments which have been taken as evidence {\it 
against} such phase separation.

Evidence in favor of a uniform phase for the doped cuprates falls into
three categories: (a) smooth variation of properties with doping;(b) too much 
structure as a function of doping; and (c) absence of evidence for phase 
separation from microscopic probes.  In the following paragraphs, I will expand 
on these arguments, and show how they might be understood within a nanoscale
phase separation model (better: in a charge-bunched phase).
\par
Many properties of the cuprates are observed to change smoothly with doping, 
whereas in a {\it macroscopic} phase separation model, they would be expected 
to take on the two discrete values of the end phases.  The most obvious examples
are the superconducting transition temperature $T_c$ and the pseudogap 
transition $T_g$.  While these quantities are somewhat sensitive to quenching 
vs. slow cooling, the overall reproducibility of these quantities within many 
different experiments on many different samples suggests that there is a 
well-defined, thermodynamic state that the system evolves into as phase 
separation is confined to ever finer scales.  The hypothesis of this review is 
that charge bunching remains an essential feature of this state.
\par
Once the phase separation is arrested on a sufficiently nanoscopic scale, then
the physics is dominated by pattern formation: how to arrange the domains to 
minimize surface tension -- in this case, the excess Coulomb energy.  Hence, a
certain regularity is expected in different experiments.  Moreover, properties
should not be expected to change linearly with doping; in particular, many
properties will change suddenly at a percolation crossover, Fig.~\ref{fig:j23}c.
\par
In the absence of a detailed model of such a state, perhaps an analogy would be 
useful (the following section will discuss a variety of analogies.)
In analyzing Type I and Type II superconductors in a magnetic field, one treats 
the system as two phase only initially, to justify the separation into Type I or
Type II behavior in terms of whether the {\it interfacial surface tension} is
positive or negative.  After that, one treats vortices in Type II 
superconductors as a form of elementary excitation, and largely ignores ideas of
phase separation.  For Type I superconductors in the intermediate state, the 
field domains are much larger, so ideas of phase separation remain important.  
It is important to note that a given superconductor can be interconverted 
between Types I and II by appropriate materials processing: introducing disorder
or thinning down a sample in the direction parallel to the applied field pushes 
the material toward Type II behavior.
\par
In the cuprates, there also appears to be a distinction between Type I (more
macroscopic) and Type II (nanoscale) phase separation, with Type I behavior
promoted by slow cooling from high temperatures and by thinning down samples for
electron microscope examination.  I think that the great confusion displayed by
the data in this section stems largely from incomplete control of these 
processing steps, and that real progress in this area will come with systematic 
studies of the effects of processing on the degree of phase separation.  
\par
The nanoscale limit of a `charge-bunching phase' is the most interesting, but
unfortunately the hardest to understand.  A possible model is the charged
soliton domain wall model of Section XI.I or (?equivalently) the `Berryonic
phase' of doped C$_{60}$, Section XVII.C.  At present, its properties can only
be suggested by analogy with Type II superconductors.  In these materials, there
are not discrete T$_c$'s for the superconducting and normal (T$_c$=0) phases,
but a single T$_c$ which evolves smoothly with magnetic field (i.e., with
fraction of the normal phase).  Something similar would be expected in the 
cuprates.  In the same way, long mean free paths are quite compatible with Type
II superconductivity.
\par
The third type of evidence is the most difficult to handle: there are
essentially local probes, which would be expected to show the presence of two
different populations of, say, Cu atoms.  Indeed, some probes do show such
multiple environments, as the NQR and M\"ossbauer data described above.  Yet
what is one to make of it if these domains are not seen?  (For example, the NMR 
Knight shifts in LSCO and YBCO are not split at intermediate dopings.)
\par
Again, the analogy of the Type II superconductor is useful.  Muon spin resonance
is a useful probe of the local magnetic field, but in a flux lattice a single
sharp response is seen. Detailed analysis shows that the lineshape function is
given by the dos of the 2D flux lattice, and the sharp peak is the associated
VHS (!)\cite{EHB}.   Similarly, NMR linewidths are not split, but can be
significantly broadened\cite{BrA}.  Thus, in the presence of a charge-bunching 
phase, one needs a full microscopic theory to understand just what a local probe
should see.  At present, we are far from having such a theory.
\par
In conclusion, these experimental observations constrain the possible form of a
charge-bunching phase, and detailed calculations will be required to see if such
a phase can be made consistent with all experiments.  I strongly urge additional
experiments, to further test the ideas of charge bunching.  Meanwhile,
evidence in favor of such a phase -- often in the form of striped 
phases -- continues to mount.  Clearly, a better theoretical understanding of
these phases is needed -- in particular, the compatibility between these phases 
and the pseudogaps in underdoped cuprates.

\section{Analogs for Arrested Phase Separation}

Arrested phase separation in the presence of Coulomb effects is not a very well
understood phenomenon, and in order to develop model calculations for the
cuprates, it is useful to have some analogous systems which have been better 
studied.  Indeed, there is considerable literature on {\bf modulated phases},
which has recently been reviewed\cite{modp}: a common formalism can be applied
to a number of disparate systems.  One should also recall the comparison with
CDW's, in Section X.B.3.  Indeed, the charge ordered phases seen in the
nickelates and manganates (Section XI.H) are very good analogues for the 
cuprates, and hence eminently suitable for further study.

\subsection{Dipolar Systems}

\par
One class of analogs includes a group of dipolar fluid or polymeric systems
characterized by short-range attractive and long-range repulsive interactions.
These systems include Langmuir monolayers\cite{Lang1}, charged colloidal
suspensions\cite{Lang2}, block copolymers\cite{Lang3}, and 
ferrofluids\cite{Lang4}.  In all cases, the phase separation is arrested on
a finite scale, but the scale is sufficiently macroscopic that the resulting
domain pattern can be visually inspected (the repulsive force typically falls
off faster than the bare Coulomb force).  Both island and grain boundary (or 
stripe) phases are common.  In a 2D dipolar model, Ng and Vanderbilt\cite{NgV}
find stripe phases near half filling and a hexagonal lattice of circular drops
at other dopings.  A theory of the dynamics of the phase separation has
been presented\cite{Lang5}.  Chemical reactions can also inhibit phase 
separation at finite length scales\cite{Lang6}. The resulting domains (Figs. 2
and 3 of Ref. \cite{Lang6}) resemble the mixed phase in a Type I
superconductor\cite{Hue}, discussed below.

\subsection{Superconductors}

The crossover between the mixed and intermediate states of a superconductor in 
a magnetic field provides an interesting example of the unusual physics that can
arise when domains are restricted to a sufficiently nanoscopic scale.  In a 
finite sample of a Type I superconductor, demagnetization effects at the sample 
surface lead to a mixed state, with superconducting domains in the Meissner 
state and normal domains in which the field penetrates. These can form stripe or
domain wall phases, but indeed there is a rich variety of possible structures, 
strongly dependent on the history of how the field was applied, beautifully 
illustrated in the classical book by Huebener\cite{Hue}.  These domains are 
typically of macroscopic size (there are no charging effects), but are closely 
related to the intermediate, or vortex phase of a Type II superconductor.
\par
The vortex phase can again be looked upon as a two phase system, but one phase,
the normal magnetic phase, is reduced in size to the smallest possible value,
a single flux quantum per vortex.  Whether the mixed or the vortex phase is 
found in a given superconductor depends on the sign of the surface tension 
between superconducting
and normal domains: for a positive surface tension, the system tries to
minimize the size of the superconducting-normal interface, leading to a
perfect Meissner state for an infinite sample, or to the mixed phase in a
finite sample.  If the surface tension is negative, the system maximizes the
amount of interface, by making the size of the normal domains as small as
possible.  Generally, the sign of the surface tension is taken as the defining
characteristic as to whether a superconductor is Type I or Type II.  However,
in an intrinsically Type I superconducting film in a perpendicular field, 
there is a crossover from Type I to Type II behavior as the thickness of the
film is decreased\cite{Tink}.  This behavior follows from Maxwell's equation, $
\vec\nabla\cdot\vec B=0$: for a thin film in the Meissner state, the magnetic
field would also be excluded from a significant volume of space outside the
superconductor, above and below the film.  As the superconductor is made
thinner, there is not enough condensation energy to compensate for the
magnetic energy, so the film enters the mixed state.  As the film gets even 
thinner, the magnetic field must become more and more uniform, so the normal
domains shrink until they contain only a single vortex, and the film is 
effectively Type II.  

\subsection{QHE}

\par
For either the integer or fractional quantum Hall effect (QHE), there is a 
series of incompressible phases at increasing magnetic fields, associated with 
fixed (integral or fractional) values of flux to charge density ratio.  The 
intermediate regime, when the field moves between two of these fixed values, has
often been described as some sort of two-phase regime, where either the magnetic
field or charge bunches up to keep the ratio of charge to flux fixed at the 
incompressible values.  Thus, the free energy of a 2D electron in a magnetic 
field has a series of downward-pointing cusps corresponding to integral filling
of Landau levels, reminiscent of the cusp in Fig.~\ref{fig:j23}a; 
similar cusps are found for the fractional QHE, due to
electron-electron interaction\cite{Halp}.  In 3D electron gases, the integral 
cusps have long been known to be responsible for a phase separation of {\it 
magnetic domains} (magnetic interaction)\cite{MOM}.  A variety of 
percolation\cite{Qperc1,Qperc2} or dHvA magnetic interaction\cite{RMQHE,QdHvA} 
models have been proposed.  Again, a crossover is found\cite{RMQHE}, from 
uncharged magnetic domains in thick samples, to charged domains with uniform 
field in thin samples.  In the fractional QHE regime, Laughlin's quasielectrons 
and quasiholes\cite{LauQHE} play the role of charged vortices: 
topological excitations corresponding to the smallest unit of a second phase.
Chalker and Coddington\cite{CC} developed a model of this percolation which
has been utilized by a number of other researchers\cite{Huck}, including for the
fractional QHE\cite{ShAu}.
\par
Recently, there has been some experimental evidence of the instability of the
uniform phase.  Eisenstein, et al.\cite{Eisen} have measured the compressibility
$\kappa$ of the 2D electron gas, and find that it is {\it negative} over broad 
ranges of magnetic field.  At the fractional steps (filling factor of 1/3, 2/3) 
they find a peak in $\kappa^{-1}$, consistent with the broadening of an
incompressible state.  On either side of the peak are prominent negative
minima, interpreted as a broadened negative divergence in $\kappa^{-1}$
associated with a low density of quasiparticles or quasiholes.  
\par
Apparently, the compressibility $\kappa$ is negative due to the large
negative contributions to the free energy from the exchange and correlation 
energy of a 2D electron gas\cite{Eisen,Efros}, and a similar effect is predicted
to occur in zero field, if the density is low enough\cite{RuW}.  These 
contributions to the Coulonb energy are both negative, and become more negative 
as the density increases.  Hence, the Coulomb energy would spontaneously lower 
if the electron gas compressed -- the origin of a negative compressibility.  So 
why doesn't the gas collapse?  The reason is that there is a hidden, positive
contribution which is actually infinite.  That is, the system is overall
charge neutral, and the infinite repulsion of the electron gas is exactly 
compensated by an infinite attraction to a background positive charge.
\par
There are two possible situations.  If the positive background is effectively
fixed (e.g., due to impurity atom doping), then a spontaneous compression of the
electron gas would lead to a net charge imbalance, and an infinite increase in
direct Coulomb energy. However, if the electron gas breaks up into neutral
cells, and the charge redistributes within each cell in such a way that there is
no net dipole moment, then there will be a net energy increase, but diverging
only weakly, at the quadrupolar order.  The balance of the exchange and
quadrupole energy could then lead to a net inhomogeneous electron gas, which
would be a form of Wigner crystal.
\par
On the other hand, if the compensating charge can move, then both electron gas
and compensating charge can collapse in parallel, until they reach the
density corresponding to the minimum of the combined exchange plus correlation
energy.  Such a state can be realized in a photoexcited semiconductor, in the
form of an electron-hole liquid\cite{EHL}.  In this case, the phase separated 
droplets can attain macroscopic dimensions.  Which case applies to the QHE?
In GaAs, the compensating positive charge is usually fixed, due to a doping
layer, and hence the nanoscale phase separation model is more appropriate.
On the other hand, in a Si inversion layer, the compensation is due to highly
mobile charges on the metallic gate. In this case, macroscopic charge islands
might form\cite{Qperc2}.  
\par
In the integer QHE, something similar has been
observed.  When a sandwich structure of 2D electron gas above 2D hole gas is
formed, macroscopic strips of compressible material form and spread away from
the edge states, as the doping is varied\cite{edgch}.  An interesting 
possibility is that this phase separation into compressible and incompressible
domains is reflected in the energy dispersion in the form of {\it flat 
bands}\cite{flab}!
\par
While the above picture seems to clarify Eisenstein's data, puzzles remain.  The
most curious feature is that these are experimental results, even though a 
negative compressibility would be a violation of le Chatlier's principle, and 
hence of the second law of thermodynamics.  Presumably, the domains are somehow
stabilized by Coulomb effects associated with screening charges in the metallic 
gates.  Thus, while the QHE provides a very interesting analogy to the cuprates,
its own physics still needs to be better understood.
\par
In a spin-polarized QHE step, similar textures can arise as islands of reversed
spins, or `skyrmions'\cite{skyrm}, when the 2D electron gas is doped away from 
exact rational filling.  In this case, the Zeeman effect from the applied
magnetic field plays a role analogous to Coulomb forces, making the 
spin-reversed domains more costly in energy, and leading to a finite domain
radius.

\subsection{`Relaxor' Ferroelectrics}

\par
Another possible analog is provided by `relaxor' ferroelectrics\cite{rlax}, 
which display diffuse phase transitions extending over a range of temperatures.
There is evidence for nanoscale domains, 30-100$\AA$ in diameter, some of which 
are charged.  The suggestion that these nanodomains are stabilized by long-range
Coulomb repulsion\cite{rlax2} has led to numerical simulations of this
nanoscale phase separation\cite{rlax3}.

\section{The LTT Phase in LBCO, LSCO}

\par
While the experimental discovery of the LTT phase helped rekindle interest in
the VHS model, there was always something `not quite right' about this phase.
In particular, it falls at the wrong hole doping: the VHS seems to be clearly
associated with the doping of optimum T$_c$ $x\simeq 0.15-0.17$ in LSCO
or LBCO, but the LTT phase arises at a much lower density, $\approx$0.125 in 
LBCO and $\approx$0.115 in LSCO.  The range over which the LTT phase is stable 
can be
greatly extended by cosubstituting a rare earth (R) ion on the La site, as in 
La$_{2-x-y}$R$_y$Sr$_x$CuO$_4$ (LSCO:R).  While many properties of the LTT phase
are independent of cosubstitution, there are some differences.  Thus, the
LTT phase is present over a broad doping range in the cosubstituted materials, 
mainly on the overdoped side of optimum T$_c$.  In the non-cosubstituted LSCO, 
LBCO, the LTT phase exists only in a very narrow doping range in underdoped 
material; it is essentially a line compound, now believed to be a striped 
phase pinned by commensurability effects, with separate magnetic and charge 
superlattice\cite{Tran}.  Hence, it will be convenient to discuss these two 
situations separately.
\par
This section will attempt to summarize what is known about this unusual phase
and show how it fits into the VHS model.  There has been much experimental
study of this phase, because superconductivity is found to be suppressed in its 
vicinity\cite{Axe,Sato}.  In fact, there are four anomalies which are found in
the same general doping range: the LTT transition, superconductivity 
suppression, normal state transport anomalies, and antiferromagnetism.  Detailed
multiple doping studies on the system La$_{2-w-x-y-z}$R$_w$Ba$_x$Sr$_y$Ca$_z
$CuO$_4$ have been carried out, with R = Nd, Sm, Gd, or Tm, to determine
whether all these anomalies are aspects of a common phenomenon or are
independent of one another.  While magnetism and T$_c$ suppression seem to be
most closely related, the relation of the other two phenomena is less clear,
and remains somewhat controversial.  
\par
A recent series of experiments on LSCO:Nd has provided a fundamental insight
into this problem: superconductivity is suppressed (and a magnetic phase 
appears) only when the octahedral tilt is sufficiently large, $\Phi\ge\Phi_c
\simeq 3.6^o$\cite{Buch,Buch3}. For smaller tilt angle $\Phi$, superconductivity
can persist in the LTT phase with essentially no change in T$_c$.  As Nd content
$y$ is reduced, the line $\Phi =\Phi_c$ extrapolates to $y=0$ at $x\simeq 
1/8$, thus explaining the anomalous behavior of the LTT phase in LBCO.
This behavior follows naturally within the VHS model: the 
splitting of the density of states peak, due to the lifting of the VHS 
degeneracy, is proportional to the square of the LTT tilt angle.  Only when the 
angle is large enough would a significant drop in the dos be expected.  (Note
however that this can only work if the VHS is pinned to the Fermi level in the 
LTT phase.  In the following sections, evidence will be presented for the 
nanoscale phase separation which produces this pinning.)  In light of these
experiments, the cosubstituted material with R=Nd is better understood than the
$x$=1/8 phase, and will be discussed first.

\subsection{LTT in LSCO:Nd}

\par
Rare earth (R) cosubstitution in LSCO (La$_{2-x-y}$\-R$_y$\-Sr$_x$\-CuO$_4$) 
greatly extends the doping range over which the LTT phase is present, and allows
some separation of hole doping vs structural effects.  Starting from the 
doping of optimum T$_c$, the LTT phase is favored by enhancing the interlayer 
mismatch, either by decreasing the doping (less Ba,Sr) or by decreasing the 
average ionic radius at the La site (by substituting Nd, Eu, Gd, Dy, 
Tb)\cite{Buch2}.  Many experiments have used R=Nd, since this can be doped up to
$y=0.75$ before converting to the $T^{\prime}$ structure (the structure of Nd$_2
$CuO$_4$, the parent compound for the electron-doped superconducting cuprates, 
with no apical O's). In this compound, while many anomalous properties are found
at arbitrary doping, two particular hole compositions, $x$ = 1/8 and 0.15, 
appear to be special. 
\par
Thus, in LSCO:Nd the doping $x\simeq 1/8$ displays much stronger T$_c$
suppression, onset of localization, and Hall and thermopower 
anomalies\cite{NaUc,Buch6}.  At $x\simeq 0.15$, which has the highest T$_c$ 
when $y=0$, there is a sharp, first-order transition when $y\ge 0.18
$\cite{Buch7}, which is both an LTO-LTT transition and a 
superconductor-insulator transition.  For $0<y<0.18$, the LTT transition, which
would occur at a lower T than the superconducting transition, is arrested at 
T$_c$, and there appears to be a phase separation, with the Meissner fraction 
being approximately 1 - (fraction of LTT phase)\cite{Buch2}, while there is a 
small residual fraction of superconducting material at larger $y$.  
\par
For $x<(>)0.15$, there is a wide range of Nd doping for which a Pccn (LTT) phase
is stable.  The most interesting results have to do with overdoped samples, 
$x>0.15$.  In these samples, there is a wide range of Nd doping for which T$_c$ 
is also suppressed, but only if the octahedral tilt exceeds a critical value, 
$\Phi\ge\Phi_c\simeq 3.6^o$\cite{Buch}.  For smaller tilts, the LTT phase is 
stable, but superconductivity is not reduced, and there are no resistive 
anomalies.  At the same
value of critical tilt that suppresses T$_c$, the resistivity has a sharp jump 
at the LTT transition, and increases with decreasing T in the LTT
phase.  Moreover, at approximately the same Sr doping, magnetic anomalies 
suggestive of local antiferromagnetic order (but with much larger transferred 
hyperfine fields than in antiferromagnetic LCO) appear below $\approx 32K
$\cite{Buch8}.  The magnetic transition may be driven by a critical tilting,
due to spin-orbit coupling\cite{Bones}.  There also seems to be a change in the 
nature of the LTO-LTT transition, at a critical tilt similar to $\Phi_c
$\cite{Buch2}: for 
larger $\Phi$ values, the transition is sharp and first-order: the orthorhombic 
splitting of LTO X-ray peaks remains unchanged, while an unsplit LTT peak 
appears between them, and grows in intensity as the transition procedes.  For 
smaller $\Phi$, the transition is smeared or continuous, with the orthorhombic 
splitting continuously decreasing as the transition is approached.  From the 
available data, it is not possible to tell if this change occurs precisely at 
$\Phi_c$ or not.
\par
The two special dopings, $\approx 1/8$ and 0.15, also seem related to the 
critical
value of $\Phi_c$.  The first doping is at the intersection of the lines $\Phi =
\Phi_c$ and the line $y=0$.  At $x=0.15$, the sharp superconductor-insulator
transition is due to the coincidence that at this doping the LTO-LTT transition
intersects the $\Phi_c$-line.  By using a different RE from Nd, this 
intersection is shifted to a lower doping, and it is possible to have bulk 
superconductivity in the LTT phase, even at $x=0.15$, when $\Phi <\Phi_c$.
\par
Thus, in LSCO:Nd, the superconducting, resistive, and magnetic anomalies all
appear to be correlated, and all occur in the LTT phase when the octahedral tilt
exceeds a critical value.  While it is possible that the large tilt drives the
magnetic transition, which in turn quenches superconductivity, this picture does
not explain the resistive anomaly at the LTT transition, which rather suggests a
large modification of the Fermi surface.  Indeed, the entire sequence of
anomalies is just what would be predicted by the VHS-JT effect, in which the 
structural distortion splits the VHS degeneracy.  However, this explanation can 
work only if the VHS remains pinned at the Fermi level over this extended doping
range.  In fact, overdoped LSCO is believed to phase separate, with one phase 
remaining pinned at the VHS (Section XI.F). Can such a model also apply to 
LSCO:Nd?  The materials preparation technique used in these studies is known to 
produce significant local variability of Sr content (i.e., of hole 
content)\cite{Buch9}, while there is clear evidence of phase separation in the 
regime $\Phi>\Phi_c$\cite{Buch}.  Even in this regime, there is evidence for a 
superconducting transition, with zero resistivity; however, the Meissner 
fraction is very small and the resistivity is restored by a small magnetic 
field.  These superconducting domains appear to be fairly large: the onset of 
the resistive transition falls at nearly the same temperature over an extended 
range of Sr doping (see Fig. 3 of Ref. \cite{Buch}), as expected for macroscopic
phase separation.  
\par
Given the picture of nanoscale phase separation discussed in Seciton XI.F,
the results on LSCO:Nd can be readily understood in terms of the splitting of
the VHS degeneracy in the end phase at $x\simeq 0.15$.  What is learned in these
experiments is that (a) the same VHS splitting (same critical tilt angle) holds
throughout the overdoped regime; and (b) even at the VHS, $x =0.15$, as Nd is
added there is a phase separation.  This is associated with the LTT-LTO 
transition, but is highlighted by the fact that superconductivity is destroyed
only in the LTT phase (even though both phases have the same {\it magnitude} of
$\Phi$).
\par
One note of caution should be sounded: while there is a clear correlation
between the LTT transition and the metal-insulator transition in Nd-doped
material, the correlation is less clear when other ions are substituted.  In 
mixed La$_{1.875}$Ba$_{0.125-x}$Sr$_x$CuO$_4$, a resistivity minimum is found at
higher T than the LTT transition\cite{MaOd}.  In another study, the LTT phase
was probed by studying La$_{1.775}$R$_{0.10}$Sr$_{0.125}$CuO$_4$, with R = La, 
Nd, Sm, Gd, or Tb\cite{LTT6}.  The results fall into two groups, 
with Nd very similar to La, and the other R's showing very different behavior: 
for these, there is a well-defined LTT transition, with the transition 
temperature varying strongly with R (up to $\approx$150K for R=Tb), but the 
anomalies in $\rho$ and $S$ fall at $\approx 80-100K$, independent of R.  Thus 
the electronic and structural anomalies are not completely correlated.  Again, 
there may be a problem of critical tilting for the VHS splitting to affect the 
electronic properties.  This would go in the right direction: the smaller the 
ionic radius of R, the worse is the interlayer mismatch, so the LTT transition 
occurs at a smaller VHS splitting.  

\subsection{LTT Phase as a Striped Phase}

\par
What is special about $x\simeq 1/8$?  In the VHS model, I believe there are two
competing effects.  First, an LTT-type structural instability is favored by the 
VHS-JT effect when the Fermi level is at the VHS, $x\simeq 0.15-0.17$.  However,
at this high doping the interlayer mismatch is small, so the octahedral tilts 
are small, $\Phi <\Phi_c$, with small barriers between adjacent potential 
minima. This leads to a dynamic LTO phase near the VHS, with local LTT symmetry 
and dynamic domain walls; experimental verification of this prediction was 
discussed in Section IX.B.  When the interlayer mismatch increases, by, e.g. 
reducing the hole doping, $\Phi$ gets larger and the potential barriers 
increase, leading to a more static LTT phase.  Extrapolating the LSCO:Nd data, 
$\Phi =\Phi_c$ at $x\simeq 0.125$, and the LTT phase appears near this doping, 
with reduced $T_c$. However, for smaller $x$, the AFI phase (with LTO-type 
tilting stabilized by AFM umklapp scattering) tends to dominate.  Thus, there is
only a narow window where a static LTT phase is stable.  Within this window, 
commensurability effects act to pin the phase at a fixed doping.
\par
The T$_c$ suppression is closely correlated with an antiferromagnetic phase, and
both are pinned at a commensurate doping value, $x=1/8$, in LBCO\cite{Axe,Sato},
but at $x\simeq 0.115$ in LSCO\cite{LTTmag2,Got}. Double doping 
experiments\cite{LTT2} have demonstrated that the T$_c$ suppression occurs at a 
fixed hole content $x$, and does not depend on, e.g., Ba concentration.  Similar
results are found in LBCO, by introducing oxygen defects (i.e., $\delta <0$, 
which occur predominantly in the CuO$_2$ planes)\cite{Ta-MR,Mood2}.  Again, the
largest T$_c$ suppression occurs when $P\approx x+2\delta =0.125$, but as 
$|\delta
|$ increases, so does the value of this minimum T$_c$.  At the same time, the 
LTO-LTT transition temperature is enhanced.  Finally, hydrostatic pressure 
suppresses the LTT transition and enhances T$_c$ and the Meissner fraction, 
but these still remain small compared to other doping values\cite{LTTpress}.
Near the critical filling, the Meissner fraction is reduced\cite{Got}, but there
is not a perfect correlation between Meissner fraction and reduced T$_c$.  It
appears that the Meissner fraction is reduced over about the same doping range 
in which the antiferromagnetic phase exists.  This range is considerably less 
than the range over which the LTT phase appears (in LBCO), but on the other 
hand, the T$_c$ reduction is confined to an even narrower range, particularly 
in LSCO.
\par
The magnetic ordering was detected by $\mu$SR\cite{LTTmag,LTTmag2} and by Cu and
La NMR\cite{Got}. The transition appears to be antiferromagnetic, with Cu 
moments pointing in the a,b-plane, as in La$_2$CuO$_4$, with a N\'eel 
temperature $\approx$30K.  The internal field is smaller than in La$_2$CuO$_4$,
but different groups find different values, in the range of $\sim 0.1-0.26\mu_B$
per Cu\cite{Got,Wad,Tou}.  The small size of the moment is not uncommon for an
antiferromagnet with a low T$_N$.  Magnetic order at such a high doping is quite
surprising for the cuprates, and may be a sign of a commensurate phase 
separation, such as the stripe phase found in LSNO, Section XI.H -- perhaps 
similar to the soliton model discussed in Section VIII.C.6, which involves 
domains of LTT phase, with domain walls of LTO phase, Fig.~\ref{fig:c21}. 
Indeed, recently, Tranquada, et al.\cite{Tran} found neutron diffraction 
evidence for a coexisting magnetic and charge domain striped phase in
Nd-substituted LSCO; this will be discussed in the following subsection.
\par
The LTT phase in LBCO resembles the Ortho II phase of YBCO, in that it 
seems to be associated with a fixed filling fraction $\approx$1/8 (see the 
discussion of Fig.~\ref{fig:T10}).  Near this filling, earlier studies found
that the holes form two types of domains, one of LTO phase with optimal T$_c$ 
and the other of LTT phase with reduced T$_c$\cite{LTT01}, with domain size 
$\approx 300\AA$\cite{LTT02}.  (The orthorhombic domains have greatly reduced 
orthorhombicity, and may be of Pccn rather than LTO symmetry\cite{APB}.) In the 
mixed LTT-LTO regime, T$_c$ of the LTT phase depends on doping, possibly due to 
elastic strain effects between the two types of domain\cite{LTT01}.  There are 
contradictory results on the doping range for coexisting LTT and LTO order, with
some studies finding a large LTT fraction only at the doping of maximum T$_c$ 
reduction\cite{Axe,Kata}, while others find that the LTT transition is nearly 
complete ($<5\%$ LTO phase) over a wider range, at both $x=0.125,0.15$, while 
only at the former value ($x=1/8$) is superconductivity suppressed\cite{BKLTT}. 
This is consistent with the LSCO:Nd results, since $\Phi <\Phi_c$ at $x=0.15$.
\par
The situation in LSCO is even less clear.  The T$_c$ suppression is confined to
an extremely narrow range of $x$.  Acoustic measurements suggest the onset of
LTT-type order in this vicinity\cite{LTTfl}, but long range order has not been 
found by x-ray or neutron diffraction. In a study where Sr or Ca was substituted
into LBCO in such a way as to keep the hole doping fixed at $x=0.125$, the onset
of T$_c$ depression was found to coincide with the onset of the LTT 
phase\cite{LTT5}.  At $x=0.115$, an electron diffraction study\cite{ED4,ED2}
finds evidence for the LTT transition in LSCO, with onset at 110K -- that
is, at a higher T than in LBCO, $x=0.125$, for which the onset is at 70K.
The electron diffraction study also finds superlattice spots; when
the structure is modified by Sm addition, as La$_{2-x-y}$Sm$_y$Sr$_x$CuO$_
4$\cite{ED2}, the superlattice spots seem to occur only at $x\simeq 0.115$. 
Such a combination of magnetic and structural anomaly resembles those seen in 
La$_{2-x}$Sr$_x$NiO$_4$ (Section XI.H) and in LBCO at $x=0.125$\cite{Tran}.  
However, some caution is advisable.  It is possible that 
the superlattice spots are an artifact, due to double diffraction\cite{Zhu}.
In addition to the electron microscopy, there is an anomaly in the specific heat
of LSCO, $x=0.09$, at T=70K, which resembles the LTT transition in 
LBCO\cite{Wri}.  Also, at $x=0.12$, Raman scattering\cite{Sug} finds evidence 
for Raman forbidden phonons, as if very close to this composition, the inversion
symmetry of the crystal is lost, due to local distortions.  There are also 
anomalies of the low-T Raman shift for several Cu-O related phonons\cite{Crawf}.
Thus, more work is required to determine if this interesting superstructure is 
really present in LSCO, $x=0.115$.
\par
The VHS model also provides an explanation of why this anomalous phase occurs at
different dopings in LBCO and LSCO.  In a VHS model, commensurability can refer
to a {\it rational fraction of the VHS density, $x_{VHS}$}, rather than a 
rational hole per atom value.  Thus, in LSCO, the VHS
falls at a doping $x=x_{VHS}\simeq 0.14-0.17$.  If $x_{VHS}=1/6$, then a
commensurate superlattice with three cells doped to $x=x_{VHS}$ and the fourth
undoped would fall exactly at $x=1/8$.  On the other hand, if $x_{VHS}<1/6$, the
supercell would have to have discommensurations to exactly match the VHS doping,
and would have $x<1/8$.

\subsection{Combined Magnetic and Charged Stripes}

\par
In Nd-substituted LSCO, La$_{1.48}$Nd$_{0.4}$Sr$_{0.12}$CuO$_4$, Tranquada, et 
al.\cite{Tran} have found neutron diffraction evidence for a phase 
consisting of alternating magnetic and charge stripes. The 
charge periodicity is four cells, and the magnetic periodicity is twice as 
large.  In the presence of Nd, the distortion is static, but it is hypothesized 
that the same striped phase is present dynamically in the Nd-free material, 
since the degree of incommensurability $\epsilon$ is the same in both materials.
The results have now been extended to higher doping, $x=0.15,0.2$\cite{Tran2}.
The charge ordering cannot be studied in much detail, given the weakness of the
diffraction peaks, but the magnetic phase transition decreases with increasing
doping, falling below $\approx 15K$ at $x=0.2$.  The doping dependence of 
$\epsilon$ matches that of LSCO.  In the VHS model,
it is surprising to find stripes persisting to $x=0.2$, above optimum doping.  
However, it should be recalled that the pseudogap in LSCO only closes around
$x\simeq 0.27$, Fig.~\ref{fig:b10}\cite{Gp3}.  Moreover, the slope of $\epsilon
(x)$ has a sharp break near $x=0.12$, suggesting a change in the nature of the 
stripes.  In general, the stripe ordering temperature has a doping dependence 
parallel to that of the pseudogap, but is much smaller in magnitude.

A simple model of the stripes is as follows: there are alternating charge 
stripes of width $M_0$ (in units of the lattice constant $a$) and magnetic 
stripes of width $M-M_0$, so the total charge stripe repeat distance is $M$.  
The magnetic stripes are assumed to be undoped, while the hole doping in the 
charge stripes is $x_0$.  Thus, at a doping $x$, we must have $xM=x_0M_0$.  The 
degree of incommensurability is then $\epsilon =1/M$.  For $x\le 0.125$, the 
experiments find $\epsilon =1/2x$ implying that $x_0M_0=1/2$.  This is
compatible with the assumption that both the charge stripe doping $x_0$ and the 
width of the charged stripes $M_0$ are independent of doping.  
\par
This model can explain why the magnetic periodicity is twice the charge 
periodicity.  If the exchange coupling across the charge stripes is
antiferromagnetic, then the rows of spins separated by the charge layer are
out of phase.  If the magnetic stripe contains an odd number of (AFM coupled) 
rows, the first and last row will be identical.  Hence, the first rows of 
adjacent cells will have opposite spin order, so the magnetic periodicity is 
doubled.  Note, however, that if the magnetic cell had an even number of rows,
there would be no period doubling.  Hence, to explain the experimental 
observation that the magnetic periodicity evolves smoothly with doping, it must
be assumed that the magnetic stripes can only have an odd number of layers.
Alternatively, the magnetic stripes would all have to be even if the
intercell exchange were ferromagnetic.

On analogy with the nitrates, Tranquada, et al.\cite{Tran} propose that the 
stripes consist of three cells of undoped, antiferromagnetic phase, and one cell
of hole-doped boundary phase, $M_0=1$, with $x_0=0.5$, leading to an average 
doping of $0.5/4=1/8$.  However, the analogy with LSNO is far from exact.
Thus, in LSNO, the stripe phases extend throughout the full doping 
range, with strong commensurability effects found at all commensurate filling 
factors, with the largest effects at the lowest factors, 1/2 and 1/3.  In 
comparison, if $x_0=0.5$ in LSCO, then the strongest anomalies would be expected
at $x\simeq 0.166, 0.25$, instead of the near complete destruction of 
superconductivity at 0.125, with nothing comparable at the higher filling 
fractions.  

Moreover, in the Nd-substituted LSCO, the stripe phase periodicity deviates 
strongly from that expected for constant-density charge stripes above a doping
$x\simeq 0.12$.  Whereas for $x\le 0.12$, $\epsilon\simeq x$, for higher doping,
$\epsilon$ saturates near a value $\epsilon_m\simeq 1/6$.  
I suggest that this saturation can be best understood if $M_0>1$.
Thus, assuming that the stripe width is $M_0=3$, the corresponding 
hole density in the stripe domains becomes $x_0=(4/3)\times (1/8)=(1/6)$, close 
to the doping of optimum T$_c$ for LSCO.  Or, if $M_0=2$, $x_0=1/4$, somewhat 
large for the optimum doping, but close to the doping at which the pseudogap is 
found to close, Fig.~\ref{fig:b10}\cite{Gp3}.  A case could be made for either 
value, and since the interpretation of the data is different in the two cases, I
will briefly discuss both.

First, assume $M_0=2$.  The saturation of $\epsilon$ can be understood as 
follows: if $M_0=2$, then the at $x=0.12$, the magnetic stripe is two cells 
wide.  From our earlier discussion, the magnetic cell width must always try to 
be even, so $x=0.125$ is the highest doping at which this phase is stable.  For
higher doping, the magnetic stripes become only one cell wide, yielding $M=M_0+
1=3$, or $\epsilon =1/6$.  Because of the reduced stability of the magnetic 
cell, the magnetic transition temperature drops rapidly.  
For higher doping, the magnetic cells would be eliminated, leaving a
pure charge phase, which does not contribute to the magnetic diffraction peak.

For $M_0=3$, the saturation of $\epsilon$ is easily understood: the magnetic
stripes are eliminated, leaving a uniform charge-ordered phase.  But then, why
is there an incommensurate diffraction peak at all?  This seems to require two
effects.  First, the charges themselves must be ordered.  Something like this 
happens in La$_{2-x}$Sr$_x$MnO$_3$, where a $x$=1/8 phase has now been 
found\cite{CMR10}, Section XI.H.5.  The phase consists of alternating layers of 
undoped material and a uniform charge-ordered phase of one hole per 4 sites.
For Nd-substituted LSCO, there would have to be an ordered array of one hole per
six lattice sites.  Moreover, this hole phase would have to be magnetic.  This
would be consistent with the findings of B\"uchner, et al.\cite{Buch}, that
a new, insulating magnetic phase appears in this doping regime when the
tilt angle is larger than a critical value.

From either of the above models, the expected doping of the critical phase 
$\epsilon_m=1/6$ is $x=1/2M=1/6$.  However, $\epsilon$ only asymptotically 
approaches this value for $x\ge 0.2$.  This suggests that at high Sr doping, 
some holes dope 
sites which are off of the CuO$_2$ planes.  This could help explain a number of
anomalous features of overdoped LSCO: for instance, susceptibility and heat
capacity studies find that the VHS gap only closes at $x\simeq 0.27$\cite{Gp3};
the phase diagram in Fig.~\ref{fig:23}a also suggests that the gap does not 
quite close at $x=0.17$, but remains fairly constant for an extended doping 
range.  Furthermore, the thermopower\cite{ZGx,Tal10,ZGy} and the pressure 
dependence of T$_c$\cite{Gugen,Noha2} both deviate from the `universal' curves 
in overdoped LSCO.

One residual problem is to reconcile the magnetic phases found by Tranquada, et
al\cite{Tran,Tran2} with those found by other groups.  The transition 
temperatures found by Tranquada, et al.\cite{Tran2} ($T_m\simeq$ 50, 43, and 14K
for $x=$ 0.12, 0.15, and 0.2, respectively) are close to the temperature found 
by Breuer, et al.\cite{Buch8} ($T_m\simeq 32K$ at $x=0.12$) in Nd-substituted 
LSCO, and to the magnetic ordering temperature found in LBCO, Fig.~\ref{fig:25},
$T_m\simeq 30K$ at $x\simeq 0.12$.  However, the magnetic transitions in LBCO 
do not
evolve smoothly to the N\'eel temperature as the doping is reduced, but fall off
below $x=0.12$, while the magnetic phase seen by Breuer, et al. is associated
with a critical tilting angle, $\Phi >\Phi_c$, which again only holds for $x\ge
0.125$\cite{Buch}.  Hence, the magnetism seems more probably associated with the
hole-doped stripes, and not with the `magnetic' stripes.

\section{The Renormalization Group Theory: Relation to 1D Metals}

As seen above, the VHS plays an important role in {\it both} superconductivity
and structural instability.  This same competition is found
in the old high-T$_c$ superconductors, the A15's, and in the g-ology
theory of 1D metals\cite{Soly}.  In 1D perturbation theory, it is found that if 
both superconducting and charge-density instabilities are treated on an equal
footing, they exactly cancel each other out.  The renormalization group (RG) 
proved to be a powerful technique for studying a system with such competing 
instabilities.  It was found that, in exploring
the full parameter space of the electron-electron interaction parameters, $g_i$,
$i$=1,4, the leading divergence can be either CDW or spin density wave (SDW) or
singlet or triplet superconductivity.  
\par
Dzyaloshinskii\cite{Dzy} and Schulz\cite{Sch} pointed out that the VHS
theory also contains competing divergences -- this time $ln^2$ divergences --
and suggested that a similar renormalization group (RG) calculation could be
applied to the VHS.  The idea is that, since the singularities all arise in the
immediate vicinity of the VHS's, the dominant singularities are properly
described if the entire Fermi surface is replaced by a set of points associated
with each VHS.  Now, in the usual Brillouin zone scheme there are four VHS's,
at the $\pm X$-points and the $\pm Y$-points of the zone.  However, the points
$+X$ and $-X$ are separated by a reciprocal lattice vector, so are really the
same point.  Hence, the reduced Fermi surface consists of {\it two points}, at
$X$ and $Y$.  Near the $X$ ($Y$) point, the Cu-O hybridization is essentially
1D, with hopping only in the x (y) direction along Cu-O-Cu-O bonds (in the CuO$_
2$ planes).  In the 1D metal, the Fermi surface also consists of two points,
but at $\pm X/2$, for the case of half filling.
This picture of a two-point Fermi surface bears a close relation to the band 
Jahn-Teller (JT) effect, discussed in Section VIII.C.3,4, where the two VHS's 
provide the electronic degeneracy.  

\subsection{Renormalization Group Calculations}

The early work of Dzyaloshinskii and Schulz considered only the most singular 
case, corresponding to a square Fermi surface at half filling ($t_{OO}=0$ or $t^
{\prime}=0$.)  In this case, there are $ln^2$ singularities both in the Cooper
channel, supporting superconductivity, and in the zero-sound channel, favoring
CDW or SDW instability\cite{GSST}.  Dzyaloshinskii and Yakovenko\cite{Dzy2}
showed that there could be a very rich phase diagram, depending on the values
of the electron-electron coupling constants, $G_i$.  In the Hubbard model, all
$G_i$'s are equal, $G_i=U/4\pi t$, and the density wave instability
will dominate -- SDW or CDW depending on the sign of $U$.  If,
following Schulz\cite{Sch} we assume that the on-site Coulomb repulsion $U>0$
dominates the electron-electron interaction, the singularity at half filling
will be SDW.  A large $U$ suppresses both CDW and sSC (s-wave superconducting)
responses, while enhancing dSC (d-wave superconducting) response, although the
SDW response still dominates.
\par
Schulz\cite{Sch} showed that, if the material is doped away from half filling,
the SDW response is suppressed, while the dSC response is unaffected, so at a
critical doping the material will cross over to a d-wave superconducting state.
I would suggest a slight variant on this calculation.  The VHS actually falls 
away from half filling, at a finite hole doping, due to $t_{OO}\ne 0$, although 
correlation effects can keep the VHS pinned near the Fermi level over an
extended doping range (Section VII.B).  
The resulting competition between CDW and sSC was analyzed in 
mean-field theory\cite{RM8a}, by assuming that the pinning is perfect, so that 
for all dopings, the curvature of the Fermi surface changes in such a way that 
the VHS is always at the Fermi level (Section VII.B). For a square Fermi 
surface, at half filling, the nesting dominates, leading to a CDW instability.  
Once $t_{OO}\ne 0$, for intermediate dopings, nesting becomes weaker, and the 
$ln^2$ divergence of the CDW susceptibility is cut off to $ln$.  In this regime,
the CDW and superconducting instability can coexist, with the superconducting 
T$_c$ increasing with doping, Fig. \ref{fig:23}b.
\par
This VHS pinning model can readily be applied to the RG calculation.  When the
VHS Fermi surface is not square ($t_{OO}\ne 0$ or $t_1\ne 0$), the Cooper
pair diagram still diverges as $ln^2T$, while the zero sound diagram is cut off 
at low temperatures, leaving only a $lnT$ divergence.  Figure~\ref{fig:33} shows
a numerical integration of the RG equations (details of the calculation are in
Appendix D), both for $t_{OO}=0$ and for a finite $t_{OO}$.  In both cases,
all $G_i$'s are assumed to be equal initially.  In the former case 
(corresponding to half filling), $G_{SDW}$ is the first coupling to diverge,
suggesting that SDW order is dominant.  This is consistent with experiment on 
LSCO, where a weak interlayer coupling stabilizes the strong antiferromagnetic
(AFM) fluctuations of the cuprate planes.  In the latter case (corresponding to 
doped LSCO), as soon as the zero sound diagrams are cut off the SDW divergence 
is weakened, and $G_{dSC}$ is the first term to diverge, suggesting a transition
to (d-wave) superconductivity with doping.
\par
A number of remarks are in order. (1) A strong electron-phonon coupling would
expand the range of CDW and sSC fluctuations\cite{VS}, acting in many ways like 
a negative-$U$ contribution. If a negative $U$ is assumed, the same calculation
would lead to a transition from CDW to sSC as a function of doping. (2) This
latter RG calculation is in good agreement with the mean-field calculation of
the same model\cite{RM8a}.  The latter calculation provides some additional 
insight.  In particular, for a 2D model the dominant singularity does not 
necessarily eliminate a second phase transition involving a subdominant 
singularity, since the first transition need not wipe out the entire Fermi 
surface.  Hence, one might expect to find a doping regime in which both density 
wave (at higher T) and superconducting transition (at lower T) both occur.
\par
So which situation corresponds better with experiment?  While LSCO has both an 
SDW (AFM) instability at half filling and a superconducting transition at finite
doping, the two transitions do not overlap, but are separated by a doping regime
which is `spin-glass-like'.  On the other hand, if I am correct in identifying 
the pseudogap as a short-range CDW, then the overall CDW-SC phase diagram is in 
good agreement with the predictions of the mean field model.  Furthermore, the 
fact that the pseudogap onset temperature in LSCO is always
higher than the AFM temperature suggests that CDW effects are stronger.  Once 
the carriers are localized in the CDW state, there could still be a transition 
to an AFM ordered state at lower temperatures near half filling.

\subsection{Bosonization and the RG Theory of the Fermi Surface}

Stimulated by the suggestion that strong correlation effects could lead to a
breakdown of Fermi liquid theory in the cuprates, a number of researchers have
been exploring the microscopic derivation of Landau Fermi liquid theory, or
of some generalized form, such as Luttinger liquids.  The main approach has
been to seek a generalization of the 1D bosonization technique\cite{bos1} into
higher dimensions\cite{bos3,bos4}, based on the early work of 
Luther\cite{Luth} and Haldane\cite{Hald}.  These bosons (electron-hole pairs) 
behave as sound waves which propagate on the Fermi surface\cite{bos4}.  In this 
case, the vanishing of the quasiparticle residue ($Z=0$) is equivalent to the 
vanishing of the surface tension on the Fermi surface, and signals a phase 
transition at the Fermi surface.  This suggests that the dynamic JT state could 
be interpreted as a quasi-Bose condensation, since the Fermi surface becomes a 
slowly varying function of real-space position and time.
\par
A perturbative RG theory of Fermi liquids has been 
developed\cite{FeRG1,FeRG0,FeRG2,FeRG3}.  Unfortunately, most of the 
calculations to date have avoided the highly singular behavior in the immediate
vicinity of the VHS's, but many findings are still relevant.  For
instance, the Fermi liquid picture is very stable, and it has proven difficult
to make a Luttinger liquid in any dimension greater than one.  The 1D Luttinger 
fixed point is a result of the lack of phase space, i.e., {\it the fact that 
the Fermi surface consists of just two points}!  Hence, the 
Dzyaloshinskii-Schulz models of the VHS suggest a new possibility: to the
extent that the VHS Fermi surface {\it almost} consists of just two points,
does this mean that, at that particular doping the holes {\it almost} form a
Luttinger liquid?  A start of a RG theory of the VHS has now been undertaken by
Gon\'zalez and coworkers\cite{rgpin2}.  They find that the self energy contains
nonlocal divergences, creating a situation reminiscent of string theory.  They
are able to show the existence of a nontrivial, unstable fixed point, and that
above a critical coupling, $U/t$, the renormalization scales to strong 
coupling. 

\section{A Puzzle: the Electron-Doped Cuprates}

From the Fermi surface shown in Fig.~\ref{fig:6}, one might think that the 
electron-doped cuprate superconductors were completely understood: the VHS
is far away from the Fermi level, so $T_c$ is low; case closed.  But this is
far from the complete story.  In the first place, 22K is a low $T_c$ only in 
the context of the cuprates.  More importantly, the Fermi surface of Fig.
~\ref{fig:6} is {\it hole-like}, whereas most transport measurements reveal the 
electron-doped cuprates to be electron-like.  The possible presence of surface 
states in the vicinity of the Fermi energy\cite{LiBa} would complicate the 
interpretation of photoemission spectra in terms of bulk bands.
\par
There is considerable evidence in these materials for multiple phases, with
the superconductivity associated with a phase which is stable only in a narrow 
composition range near optimum $T_c$. For example, in Nd$_{2-x}$Ce$_x$CuO$_4$,
Billinge and Egami\cite{BEg} find evidence for two domains of local structural 
distortions (domain size $\approx 6\AA$).  In one domain, the CuO$_2$ planes 
appear
to be flat, as expected from the macroscopic symmetry.  In the other domain, the
{\it planes are buckled}, having a symmetry similar to the LTO phase of LSCO!
This happens even though the CuO$_2$ plane appears to be under tension, and
such buckling would be expected to be energetically unfavorable.  Moreover, the
superconductivity seems to be associated with these tilted domains.
\par
Near the optimal Ce doping of NCCO, the properties of the material are sensitive
to excess oxygen\cite{NC0,NC1,NCCO}.  This oxygen tends to
localize carriers, favoring magnetic over superconducting order.  Even in
deoxygenated samples, superconductivity coexists with a large volume fraction
($\approx 50\%$) of magnetic clusters, $\approx 25-250\AA$ in 
diameter\cite{NC1,NC5}.
With decreasing oxygen, the Hall coefficient changes from electron-like to
hole-like, with optimum T$_c$ falling near the crossover point\cite{NC2,NC3}.
Jiang, et al.\cite{NC3} speculate that the sign change is associated with an
electronic heterogeneity, with the hole-like regions associated with the tilted
domains found by Billinge and Egami. M\"ossbauer studies, on a sample with 
$\approx
20-30ppm$ ${}^{57}Co$ substituted for Cu, found that the superconducting
fraction is correlated with a particular five-fold coordinated site -- i.e.,
having one excess oxygen present at an apical site\cite{NCCO}.  If the holes
superconduct (see also Ref.~\cite{NC6}), then these compounds are more similar
to the conventional cuprates.  However, the LDA band structure cannot be trusted
in detail, since the hole content at optimum doping is not known, and the local 
structure differs from the macroscopic symmetry, due to oxygen tilts and a 
possible apical oxygen.
\par
In the related Gd$_2$CuO$_4$ compound, there is a structural distortion of the 
planar O's, which could be responsible for the non-observation of 
superconductivity in this compound, and the simultaneous appearance of weak
ferromagnetic order\cite{Oser}.  While the distortion was reported to
involve motion of the O's along the c-axis\cite{NC4}, a detailed analysis finds
long range order involving rotations of the O squares about the 
c-axis\cite{Brad33}. The same structure is found in Nd$_{2-x-y}$Tb$_y$Ce$_x$CuO$
_4$, when $x=0$, $y=1.62$ (chosen to have the same average ionic radius as Gd).
Hence, the competition of superconductivity with structural instability could be
probed by fixing $x=0.15$ and varying $y$.  Increasing $y$ from zero clearly
suppresses the superconducting instability, and there is a range of $y$ for
which neither superconductivity nor long range structural order is present.
\par
The recent series of experiments on the pseudogap phase in underdoped
cuprates\cite{Gp0,Gp1,Gp2,Gp3} raise an interesting possibility: that the shift
of the VHS from the Fermi level is actually the manifestation of the opening of
the pseudogap.  If one extends this result to NCCO, then there may be a gap near
the Fermi level, leaving residual ungapped electron-like pockets.  It would be
very interesting to search for such effects.

\section{The 3D Version: BKBO/BPBO}

Within the Van Hove scenario, the bismuthates resemble 3D analogs of the 
cuprates.  In BaBiO$_3$, the Fermi level is predicted to fall very near the 
VHS's of the cubic phase (Figs.~\ref{fig:c8} and \ref{fig:AB}), but a CDW leads 
to a monoclinic distortion, producing a 1.9eV gap at the Fermi level. In a 
tight-binding model of the Fermi surfaces, keeping only nearest neighbor Bi-O 
hopping, the Fermi surface at half filling is perfectly nesting, leading to a 
strong CDW instability.
\par
This CDW was initially discussed in terms of breathing mode distortions ,
accompanied by 2Bi$^{4+}\rightarrow$ Bi$^{3+}$+Bi$^{5+}$ disproportionation --
a 3D version of striped phases.  However, there is also an {\it octahedral 
tilting mode distortion} in BaBiO$_3$, and recent evidence suggests that this 
distortion is even more important for the structural instabilities.  Thus, LDA 
calculations find that it is necessary to include both distortions to reproduce 
the experimental gap\cite{BBO1}.  A pure tilt distortion accounts for $\approx 
76\%
$ of the energy gap.  The charge disproportionation amounts to less than 0.1e 
per Bi; this is consistent with the non-observation of a split Bi peak in 
photoemission (see Ref.~\cite{BBO5} and references therein).  The small degree 
of charge disproportionation suggests that the CDW electrons are localized on 
the short Bi-O bonds (a bond-order CDW) rather than on the Bi (a site-diagonal 
CDW)\cite{BBO16}.  The role of the tilting is clear: the LDA finds 
significant distortions from a perfectly nested Fermi surface at half filling, 
due to more distant neighbor hopping.  The tilting reduces this hopping, 
bringing the Fermi surface closer to nesting.  The breathing mode may still
play an important role in producing the strong electron-phonon coupling
required for superconductivity\cite{BBO1} -- again, in close analogy to the
situation in LSCO (Section VIII.D).
\par
Once Pb is substituted for the Bi, or K for Ba, in Ba$_{1-x}$K$_x$Bi$_{1-y}$Pb$_
y$O$_3$, the breathing distortion is rapidly eliminated, near $x=0.1$ or $y=0.23
$\cite{BBO2}, leaving an orthorhombic structure ($Ibmm$) with a pure tilt mode 
distortion.  In both BKBO and BPBO, superconductivity seems to be associated 
with a metastable phase.  In BPBO, superconductivity is found for $0.70\le y\le
0.80$, in a tetragonal phase ($I4/mcm$) which is metastable with respect to the
orthorhombic phase below $\approx 450K$\cite{BBO2,BBO3}.  {\it Both the 
tetragonal
and orthorhombic phases have nearly the same magnitude of tilt distortion, but
the direction of distortion is different}, and yet only one phase is 
superconducting!  In the two-phase regime at lower temperatures, the phase 
separation cannot be macroscopic, since the lattice parameters of both 
tetragonal and orthorhombic phases vary smoothly with doping.
\par
In BKBO, superconductivity occurs in a cubic phase, but at K concentrations 
above the room temperature solubility limit, so a two step process (high
temperature anneal to incorporate K, lower T anneal to fill O vacancies) is
required to prepare the superconducting phases\cite{BBO12}.  Within the cubic 
phase, T$_c$ increases with decreasing $y$, having its maximum value at the 
cubic-orthorhombic phase boundary, which is also a metal-insulator transition.
In mixed compounds (both $x$ and $y$ nonzero), T$_c$ appears to be a smooth
function of $x+y$ only, with T$_c$ increasing monotonically as the doping is
reduced towards half filling, where the CDW instability is 
largest\cite{BBO2,BBO4}. 
\par
A major puzzle in these materials is to understand why they are insulating over
such a broad doping range; a nesting transition should only occur in the
immediate vicinity of half filling.  One suggestion is a phase separation 
between insulating and metallic domains\cite{BBO6,Slei}, Fig.~\ref{fig:34}. 
However, a {\it macroscopic}
phase separation can be ruled out by, e.g., the continuous variation of the
plasma frequency\cite{BBO7} with doping.  A number of early theories (based,
however, on the breathing mode CDW) postulated a local, or incommensurate CDW
order\cite{BBO8}.  (For a 2eV gap, the coherence length would only be a few
$\AA$'a.) Yu, et al.\cite{BBO21} did numerical calculations on a finite lattice,
and found that doping proceded by the formation of a local polaron, then
bipolaron, with a tendency for cluster formation: i.e., a local phase separation
of the added holes.  Experimentally, electron diffraction has found evidence for
an incommensurate modulation throughout the full doping range of BKBO, but this
is not found in neutron diffraction studies, and seems to be enhanced by
electron beam heating.  A full discussion, with additional references, is found
in Ref.~\cite{BBO9}.  A small fraction of superconducting phase is found in
insulating BPBO materials\cite{BBO10}.  Optical studies find clear evidence for 
a {\it pseudogap}: a gap is found, peaked at 1.9eV, in pure BaBiO$_3$. As the 
material is doped, with either Pb\cite{BBO7} or K\cite{BBO11}, the peak broadens
and shifts to lower energy, but remains present for virtually all dopings.
Just as with the mid-infrared peak in the cuprates, there is some controversy on
this issue, at least in BKBO.  Sato, et al.\cite{BBO13} were able to fit the
optical conductivity in the metallic regime by a single Drude peak.  However,
they found a scattering rate of $\approx 1eV$, considerably larger than the 
expected pseudogap energy at this doping.  In BPBO, a Raman active phonon at 
$570cm^{-1}$, identified with the breathing mode, is strongly resonant with the 
gap electrons\cite{BBO17}.  
\par
A pseudogap is also seen in the photoemission of BKBO\cite{BBO14}, and it is 
proposed that this is due to a {\it dynamical lattice distortion}, as observed 
in neutron diffraction pair-distribution-function analysis\cite{BBO15}.  
Evidence for (possibly dynamic) short-range order also comes from EXAFS 
studies\cite{BBO18}.  The two different Bi-O bond lengths found in BaBiO$_3$
are found to persist over the full doping range in BPBO and through the 
semiconducting range for BKBO, whereas metallic BKBO has a single, but
anomalously broadened Bi-O length.  On the other hand, neutron diffraction
studies find that the oxygen disorder is largest perpendicular to the Bi-O bond
axis, suggestive of tilt disorder\cite{BBO19}.
\par
In heavily doped BKBO, inelastic neutron scattering studies find soft modes 
associated with BiO$_6$ octahedral tilts, similar to those in LSCO\cite{Brad0}. 
Moreover, the one-dimensional $Bi-O$ bond stretching mode softens, but no 
anomalous softening of the three-dimensional breathing modes is found.  
Taken together, these results suggest that in BKBO the structural and 
superconducting anomalies may be associated with a 3D vHs, as in the 
Bilbro-McMillan model\cite{BM}, which has
been applied to the A15's and BaPb$_{1-x}$Bi$_x$O$_3$\cite{BPBO}.
\par
Dzyaloshinskii\cite{Dzy} has shown that for a 3D (BCC) cubic metal, in the
perfect nesting limit, the VHS's still lead to a weak divergence in the Cooper 
and zero charge channel susceptibilities, $\propto ln^3$, and to similar RG 
equations.  A $ln$ divergence has been found for a tight-binding BiO 
lattice\cite{BBO16,BBO20}.  Note that the divergence is in the (density-density)
correlation functions, while the dos is flat near half filling.

\section{Extension to Other Superconductors}

\subsection{A15's and Martensitic Transformations}

We have already discussed, in Section III, that high temperature 
superconductivity has regularly been associated with peaks in the density of 
states and strong electron-phonon coupling, and that the best place to look for 
new high T$_c$ materials has been believed to be just on the edge of a 
structural instability.  Daalderop, et al.\cite{Muel} showed that both 
superconducting and magnetic instabilities were often associated with a narrow
peak in the dos, caused by the proximity of an $M_1$ and an $M_2$ VHS -- which
resembles a 2D saddle point with a weak coupling in the third dimension, 
Fig.~\ref{fig:a2}c (inset)!  Hence, a VHS model may play a role in other, more 
3D superconducting materials as well.  We have seen that, in the A15 compounds, 
Weger found clear evidence for the important role of (3D) VHS's\cite{Weg}, and
the Bilbro and MacMillan\cite{BM} model for A15 superconductivity involves the 
3D VHS.  This model was applied to BaPb$_{1-x}$Bi$_x$O$_3$ (BPBO)\cite{BPBO} 
prior to the discovery of the cuprates.  Sleight showed that in BPBO, there is a
miscibility gap, and the optimum superconducting $T_c$ is associated with the 
composition of greatest instability, Fig.~\ref{fig:34}\cite{Slei}. He points out
that many of the cuprates are also {\it thermodynamically unstable} near room 
temperature, due to the mismatch in interlayer lattice constants.
\par
Many of the A15 compounds have a martensitic JT phase transition just above 
T$_c$, which competes with the superconductivity\cite{mart}. Krumhansl\cite{JAK}
points out that the HTT-LTO transition in the cuprates is also a martensitic 
transition, which is {\it common to virtually all known high-T$_c$ 
superconductors}, as well as to a variety of other perovskite, shape-memory, and
related alloys.  A common feature of these transitions is a precursor tweed 
phase, which is a nanoscale segregation of the low-temperature phase domains
persisting into the high-temperature phase well above the martensitic transition
temperature.  Motion of the domain walls is extremely soft, and typically leads
to anomalous softening of a shear modulus and enhanced attenuation, even in the
absence of a soft mode phonon!  The tweed structure can persist into the low-T
phase, or can appear even in the absence of a long-range structural phase
transition.  In YBCO, the `tetragonal' phase produced by substituting Co or Fe
for chain Cu's is now understood to be a nanoscale orthorhombic phase, 
associated with tweed structure\cite{twee}.

\subsection{Heavy Fermions}

\par
The heavy Fermion compounds also provide interesting food for thought.  These
compounds are associated with extreme flat bands, with an f-level having a
resonance near the Fermi level, which may be related to a correlation enhanced
VHS.  In fact, the Fermi surfaces have been calculated for CeRu$_2$Si$_2$, and
found to be in good agreement with dHvA measurements\cite{HeFe0}.  The heaviest
carriers are associated with a `pillow-shaped' hole Fermi surface, whose cross
section bears a striking resemblance to the cuprate Fermi surface (except for
interchanging electrons and holes).  The heavy Fermion superconductors are 
widely believed to have non-s-wave pairing, which may be due to coupling between
the order parameters of
superconductivity and a competing order, either magnetic or structural.  
The compound UPt$_3$ has evidence for structural inhomogeneity on a scale of
$\approx 100\AA$; on prolonged annealing at high temperatures (1200$^o$C), this
can be converted into a well-developed incommensurate modulation, which appears
to be a CDW plus Peierls distortion\cite{HeFe}.  Analysis of the specific heat
shows two sharp superconducting transitions in the annealed sample, while the
as-grown sample had a single, broadened peak in $C_v$, with reduced T$_c$.
\par
Coming to this problem from the high-T$_c$ cuprates, I am struck by the 
similarity to the split-T$_c$ peak in the overdoped cuprates.  There are
differences, however, in that there is a coexistent (weak) antiferromagnetic
phase.  A simple model, consistent with observation, might be the
following:  From the relative heights of the $C_v$ discontinuities, the two
superconducting phases are in a volume ratio of about 4:1, with the larger
volume fraction associated with the higher $T_c$.  Thus, the higher T$_c$ phase
might be associated with the domains, the lower with domain walls, where it
might coexist with the weak antiferromagnetism. (Fisk and Aeppli\cite{HeFe}
suggested that the magnetic phase might be associated with domain walls.)
\par
Whether such a model can explain the complete phase diagram of UPt$_3$ is
unclear at this point.  However, it is interesting to note how differently
the CDW has been treated by heavy Fermion theorists who are unfamiliar with
the cuprate analogy\cite{HeFe2}. They interpret the modulation as a conventional
CDW, ignore the (presumed weak) modulation of the electronic density, and
assume that the dominant effect is the change in the local lattice symmetry
associated with the Peierls distortion.

\subsection{Buckyballs}

\par
There are also interesting similarities with the buckyball superconductors.
In this brief space, I can only touch upon one or two salient features.  
A VHS model would seem improbable if superconductivity occurs exactly at
stoichiometry (e.g., K$_3$C$_60$), since the VHS will probably fall off of
half filling, as in the cuprates.  It has been suggested\cite{Cmott} 
that, just as in the cuprates, the half-filled band is a Mott insulator, and
superconductivity is associated with metal vacancies.  However, recent 
experiments on cubic families of fullerides, in which the doping can be 
continuously varied, finds the $T_c$ peak very close to half filling\cite{Yild}.
The situation may be similar to that found in the electron-doped cuprates and
Y-124: only a relatively narrow doping range is stable, and this range does not 
include the optimum stoichiometry for superconductivity.  Consistent with this,
evidence has been found for a pseudogap in K$_3$C$_{60}$\cite{gp60} (high
resolution studies find an extra, sharp feature near $E_F$\cite{gp60a}).
\par
In electron-phonon coupling models, a remarkable separation is achieved: the
electron-phonon coupling constant $\lambda =N(0)V_{intra}$, where $N(0)$ is the
dos, determined by {\it inter-ball coupling}, while $V_{intra}$ is an effective
electron-electron attraction mediated by {\it intra-ball} vibrations. The strong
electron-phonon coupling is found to be predominantly of a {\it dynamic JT} 
nature\cite{VZR,LBST,AGL}!  These calculations were classical, but since the JT
energy $E_{JT}\approx 40meV<\hbar\omega_{ph}$, quantum corrections, associated
with tunneling between equivalent JT minima, should be important.  
Auerbach\cite{Auer} solved a semiclassical model of the dynamic JT effect for an
isolated molecule of C$_{60}^{n-}$, finding a unimodal distortion for $n$=1,2,4,
5, but a {\it bimodal} distortion for $n$=3, the case of most interest for 
M$_3$C$_{60}$.  Particularly near the strong coupling limit, $E_{JT}/\hbar
\omega_{ph}>>1$, this problem is best understood in terms of the Berry phase
which develops from the combined electronic-molecular vibration.  Auerbach, et 
al.\cite{Auer2} generalized these calculations for more realistic, weaker 
couplings, by a combination of exact diagonalization and perturbation theory 
calculations.  They found that the JT effect greatly enhanced the pair binding 
energies over those calculated by standard Migdal-Eliashberg theory.
\par
However, interball hopping can interfere with this dynamic JT effect in a manner
which is not clearly understood.  Manini, Tosatti, and Doniach\cite{MTD} have
created a model system, a lattice of Berry phase molecules, in order to explore
this competition.  Here, an odd number of electrons on a molecule introduces
a Berry phase, changing the orbital angular momentum from integer-valued to
half-integer-valued (the dynamic JT effect).  Intermolecular hopping of an
electron thus carries an extra penalty, in that the orbital angular momenta of
two molecules must adjust.  This favors a correlated hopping of two electrons,
leading to a net pairing interaction.  The model seems to strongly favor a
superconducting instability, with no tendency towards phase separation\cite{Bu}.
This is a very exciting development in the study of strong electron-phonon
coupling.
\par
The {\it inter}-ball part of the problem, the dos, is also interesting.  Band
structure calculations find that the Fermi level in the superconducting M$_3$C$
_{60}$ (M=K,Rb,Cs) compounds is very close to a peak in the dos\cite{EP,NGF}.
Indeed, Mele and Erwin\cite{ME} find that it falls at what appears to be a
{\it diabolical point}\cite{RMXA,Dia} of the band structure, which allows the
possibility of a highly anomalous, nodeless anisotropic superconducting gap.
Experimentally, the dos is found to have a `Stoner' enhancement of a factor of
$\approx$2\cite{SPBZ}.  The dos (and hence T$_c$) also depends sensitively on 
the lattice constant $a$; since $a$ is a strong function of temperature, this 
gives an apparent T-dependence to many properties of the material.  Thus, the 
T-dependence of the Hall coefficient can be understood as a universal scaling,
$R_H(a)$\cite{LuCre}.  Interestingly, $R_H$ is found to change sign in the Rb
doped material; such a sign change is expected near a VHS.
\par
In the polymeric compounds, MC$_{60}$ (M=K,Rb), the Fermi level is again found 
to be near a dos peak, and it is suggested that an antiferromagnetic phase can
be stabilized by splitting this dos peak\cite{EKM}.  In M$_4$C$_{60}$, the
suggestion has been made that the insulating phase is associated with a
nesting-driven density wave instability\cite{EB} or a JT effect\cite{LukKi}.  It
should be cautioned that these fullerenes have considerable orientational 
disorder, which can smear the Fermi surface and broaden dos features.  Even in 
the presence of this disorder, a `relatively well-defined ``Fermi surface'' ' 
persists, Fig.\ref{fig:d8}\cite{ME2}, and the role of the dos could be more 
important if there is significant short-range order\cite{Eg60,Dav} -- perhaps 
akin to nanoscale phase separation in the cuprates.  On the other hand, much of 
the disorder can be eliminated under modest hydrostatic pressures\cite{Dav2}, 
which should allow the observation of sharp Fermi surfaces.

\subsection{Other Superconductors}

\par
In the Chevrel phases, MMo$_6$S$_8$, the superconductivity is associated with a 
rhombohedral phase which is metastable to an insulating triclinic phase at low 
temperatures.  For M = Sn,Pb, the transition to triclinic begins at 100K, but
is not complete at low temperatures, and superconductivity is found in the
residual rhombohedral phase.  If M = Ba, the transition begins at 350K, and is 
complete by 100K, leading to a non-superconducting material\cite{Chev}.

\par
In the new borocarbides, band structure calculations find that the Fermi energy
is close to a peak in the dos\cite{boro}, and this has been confirmed by heat 
capacity measurements\cite{boro2}.  Recent studies suggest two phase
behavior\cite{boro3} and pseudogaps\cite{boro4}.
\par
The Uemura plot shows a universal relation between $T_c$ and $n/m$ for the 
cuprates; it is found that the bismuthates, heavy fermions, organic, and 
Chevrel-phase superconductors all fall on the same Uemura plot\cite{Uem4}, Fig.~
\ref{fig:c14}a.  Brandow\cite{Bran} shows that there are a large number of
familial similarities among these various `exotic superconductors', putting
particular emphasis on the anomalous $H_{c2}$ behavior, Section VI.D.9.

\section{Why are there VHS's?}

Mathematically, this question was answered by Van Hove, who showed that a 
certain minimum number of VHS's must be present in any band structure.  And yet 
it is tempting to speculate on why the VHS is so closely associated with 
instabilities.  One point to note is that the VHS only diverges in 
lower-dimensional materials.  This seems to be associated with the general
instability of ordered structures in dimensions less than three.  For instance,
the 2D VHS is formally unstable against arbitrarily small interlayer coupling, 
which cuts off the $ln$ divergence of the dos.  (I hasten to add that the 
observed broadening of the VHS is small, and has little practical effect;
for instance, the superconducting transition temperature depends only on the
dos integrated over an interval $\sim 4T_c$ larger than the expected 
broadening\cite{RM6}.)
\par
Beyond this, it is important to recall that the 2D saddle point VHS's are 
associated with a crossover from electron-like to hole-like conduction.  While 
this seems fairly trivial, physically, it can be extremely important {\it
chemically}.  Thus, the crossover is associated with the saturation of one
valence state, and the opening of another -- which, for example, determines 
whether Fe is magnetic or not.  In the cuprates, it involves half filling of
the Cu $d_{x^2-y^2}$ band, and the doping of an oxygen $p$ band.  It is
precisely at such points that chemistry happens: where chemical valency plays a
large role, and simple band filling ideas run afoul of `correlation effects'.
In this light, it is interesting to note that recent studies of surface
chemistry have concluded that strong surface reactivity is associated with a 
peak in the local surface dos\cite{MHC}.
\par
Indeed, there seems to be a general connection between the VHS and the
crossover between covalency and ionicity\cite{RMXB}.  This has to do with a
breakdown of the Born-Oppenheimer approximation, that the electronic motion
can instantaneously follow the nuclear motion.  Near a VHS, the opposite limit
may be more appropriate, since $v_F\rightarrow 0$.  Thus, a material may
appear covalent away from the VHS, in that holes can hop rapidly enough that
the ions have an average, non-integer valence, whereas near the VHS, hopping can
be so slow that a well-defined valence can develop.  For the O's, the large
difference in ionic radius between O$^{2-}$ and O$^-$ means that in the ionic
limit, there will be significant relaxation of the positions of neighboring
ions, leading to the local structural disorder (and enhanced electron-phonon
coupling\cite{GC1}). 

\section{Summary}

The original draft of this survey was submitted for publication prior to the
Houston 10th Anniversary Meeting and the 1996 March Meeting.  Yet some of the
results presented at these meetings were so exciting and relevant that I felt
compelled to incorporate them into the final paper.  Recapitulating these 
results provides a rather fitting summary of this paper as a whole.  
\par
One group of papers\cite{Gp0,Gp1,Gp2,Gp3} has to do with the opening of a
pseudogap in the underdoped cuprates.  This evidence clearly shows that the
gap couples to both spin and charge excitations; it is not a spin gap.  
Furthermore, the data would seem to {\it rule out the basic VHS scenario}.  One
cannot describe the physics of doping the cuprates in terms of shifting a VHS
into coincidence with the Fermi level, with or without correlation-induced
VHS pinning.  On the other hand, the data are in excellent agreement with (and 
indeed were largely anticipated by) the generalized Van Hove scenario.  The
opening of the pseudogap is manifested by a splitting of the VHS degeneracy,
and the separation of the VHS from the Fermi level is a measure of the 
pseudogap.  As doping is increased, the pseudogap reduces to zero, and, in LSCO,
is replaced by an unsplit VHS peak, Fig. \ref{fig:b10}, still pinned at the
Fermi level.  (The situation in YBCO is somewhat less clear, due presumably to
the presence of additional bands.)  To complete the picture, it would be very
important to directly observe the second VHS peak, above the Fermi level, and
thereby determine how closely the Fermi level is locked to the average VHS 
position (the heat capacity data\cite{Gp3,Gp6} rules out the possibility that 
the peaks are asymmetrically split by a large amount, since that would be seen
as a double peak).
\par
At the same time, the Buckley Prize Symposium at the 1996 March Meeting, 
honoring 
C.P. Slichter, was devoted to the observation of phase separation and striped
phases in the cuprates\cite{Buck}.  Just how the striped phases can be 
reconciled with the pseudogap remains a most difficult problem in the physics of
the cuprates. (A start at this is presented in Sections XI.I, XIII.B,C.)  Here,
the Van Hove scenario has an advantage, in that both phenomena naturally follow
from the proximity of a VHS to the Fermi level.  Nevertheless, it remains very 
difficult to generate a coherent picture of the simultaneous occurence of
both effects.  Perhaps the best model is a solitonic model, in which the
domain walls of local structural disorder are charged, and hence provide the two
nanoscale phases.

\section{Conclusions}

This has been such a long and involved review because the VHS plays an important
role in most aspects of the physics of the cuprate superconductors: both normal
and superconducting states, both magnetic, structural, including phase 
separation, and superconducting properties.  A major conclusion is that the 
simple VHS model seems to work well near optimum doping: it can explain why 
there is an optimum T$_c$, with linear-in-T electron-electron scattering rate
and minimum isotope effect, etc.  However, the doping dependence must be handled
more carefully.  Thus, the VHS model 
should not be well approximated by a rigid-band theory: correlation effects
enhance the VHS (Section VII), electron-phonon coupling enhances the VHS, 
possibly leading to extended VHS's (Section VIII.D), and defects may further 
enhance the VHS\cite{JC1}.  At least in the case of
electron-phonon coupling, this is due to the slowness of carriers near the VHS:
$v_F\simeq 0$.  This leads to a breakdown of the Born-Oppenheimer approximation,
and indeed to possible antiadiabatic behavior as the electrons may actually move
more slowly than the phonons.  (See the discussion in Ref. \cite{RMXB}.)  The 
same conclusions follow from the (nanoscale) phase separation:
the VHS is pinned to the Fermi level over an extended doping range, so any
theory which assumes a simple rigid band filling must fail.  It is an open 
theoretical problem to properly incorporate this nanoscale phase separation in 
analyzing the doping dependence of the cuprates.
\par
This same problem arises in attempting to interpret experiments in terms of a
VHS theory: while T$_c$ is optimized at a particular doping, the superconducting
properties appear to persist over too broad a doping range, {\it if a rigid
band model of the VHS is assumed.}   It is instructive to look at two classes of
phenomenological theories from the point of view of a VHS model.  Of these 
theories, 
P.W. Anderson\cite{PWAPT} has stated ``I have serious problems with the two 
`hyphenated' Fermi liquid theories -- the marginal and the antiferromagnetic --
because I think they both have inconsistencies.  But they have the great
advantage that they are experiment based, even though they are a bit `myopic'
in that each looks at only a fraction of the experimental data.''
\par
In the {\bf marginal Fermi liquid}
theories, the property of a resistivity linear in $T$ and $\omega$ is abstracted
into a guiding principle.  Varma, et al.\cite{MFL} postulated a phenomenological
form of the hole self-energy, $Im\Sigma\simeq -\omega$, Eq.~\ref{eq:10d}, which 
leads to a scattering rate linear in $T$, $\omega$, and describes a number of 
other anomalous normal state properties.  This form of 
$\Sigma$ follows from the VHS model, but only at particular $\vec 
q$-values\cite{RM4}, while the original MFL self-energy is $\vec q$-independent.
Recently, a number of groups have provided an `improved' MFL theory, by putting 
in a strong $\vec q$-dependence, to be able to explain the magnetic anomalies 
near $\vec Q_0=(\pi /a,\pi /a)$\cite{LZAM,AFMFL}!  Hence, the $\vec q$-dependent
MFL model could follow from a VHS theory, but only if the Fermi level coincides 
with the VHS.  Clearly, pinning the VHS near the Fermi level would greatly 
extend the domain over which a VH-MFL model is valid.
\par
The {\bf antiferromagnetic Fermi liquid (AFFL) theory} introduces an empirical 
susceptibility peaked at $\vec Q_0$, and shows that many of the magnetic 
properties follow from this model.  However, the only successful attempts to 
generate this susceptibility peak from first principles require the VHS to fall 
near the Fermi level\cite{MoSc}.  For a rigid band model, this would
lead to too great sensitivity of the peak to doping.  However, it is now known
that correlation effects pin the Fermi level near the VHS, so the required
peak will persist over an extended doping range.  Previously, any structure at 
$Q_0$ was automatically assumed to be due to AFM fluctuations.  But if the Fermi
level is pinned near a VHS, a peak at $Q_0$ could equally well be due to 
inter-VHS
fluctuations; indeed, in the spin channel, the two will be largely equivalent.
On the other hand, the VHS can also couple to {\it charge} degrees of freedom.
Hence, more care must be taken to determine whether a particular structure is
(e.g.) magnetic or phonon related.  Thus, in fully oxidized YBCO, a $Q_0$ 
neutron scattering peak at 40meV, was found to not be due to magnetic 
scattering, as previously assumed, but to phonon effects\cite{Kei2}.  Magnetic
structure is also present, slightly shifted in frequency, which peaks sharply 
below the superconducting T$_c$\cite{Kei2,Mook}.
\par
If the nanoscale phase separation is adopted as a fundamental principle, 
another corollary almost immediately follows: theories of lightly-doped
antiferromagnets cannot provide a correct picture for the superconducting state,
since the superconductivity first appears well into the phase separated regime,
and hence is more likely to be associated with the hole-doped uniform phase 
(if not driven by the two-phase fluctuations).  Thus, theories of a gradual 
doping of holes into an
AFM background are important, to provide the metastable energy vs doping curve
upon which the phase separation is superposed.  However, superconductivity is
best understood by studying the unique composition of optimum T$_c$, and only
then extrapolating to the very complex mixed-phase regime of intermediate
doping.  
\par
One important role envisaged for this review is to establish a data base for
those special anomalies in the cuprates which can be associated either with
phase separation (Section XI) or strong electron-phonon coupling (Section IX).
There is considerable evidence for both classes of effect, but it has 
been scattered throughout the literature.  An advantage of the Van Hove
Scenario is that both features follow rather directly from the model, and the
model makes detailed predictions which seem to be largely verified.  I would
like to apologize in advance to those authors whose works I have overlooked,
particularly on these issues.  I would ask them to let me know of their results,
so that I might incorporate them at some future date.
\par
{\bf Acknowledgments:} I would like to thank Dan Dessau and Kazimierz Gofron for
providing me with copies of their theses; Frans Spaepen for the thesis of 
Randi Haakenaasen; Dennis Newns, Patrick Lee, and Bob Laughlin for stimulating
conversations; J.B. Goodenough for some useful comments; and Jeff Tallon for 
several preprints.  This paper has greatly benefitted from the comments of 
referees.  Publication 666 of the Barnett Institute.

\appendix
\section{Thermodynamics near a Saddle Point}

Near a 2D VHS, anomalous thermodynamic properties follow from the logarithmic
divergence of the dos\cite{Lif}.  These properties can be calculated for a
noninteracting Fermi liquid in the grand canonical ensemble, as integrations 
over the carrier distribution.  Thus, the average number of electrons is
\begin{equation}
N=2\sum_if(E_i),
\label{eq:A1}
\end{equation}
the internal energy is
\begin{equation}
U=2\sum_if(E_i)E_i,
\label{eq:A2}
\end{equation}
and the grand canonical potential is
\begin{equation}
\Omega =2k_BT\sum_iln(1-f(E_i)),
\label{eq:A3}
\end{equation}
with the factor 2 coming from a sum over spins, the Fermi function
\begin{equation}
f(E_i)={1\over 1+e^{-(E_i-\mu )/k_BT}},
\label{eq:A4}
\end{equation}
and $\mu$ the chemical potential.  The sum over states can be replaced by an
energy integral
\begin{equation}
2\sum_i\rightarrow\int_{-\infty}^{\infty} N(E)dE.
\label{eq:A5}
\end{equation}
\par
For the 2D saddle point, we can take
\begin{equation}
N(E)=N_1ln({B\over 2|E|}),
\label{eq:A6}
\end{equation}
where the energy of the VHS is fixed at $E_v=0$, and the limits on the integral 
of Eq.~\ref{eq:A5} are -$B/2$ to $B/2$.  This is not the most general form for 
the dos.  The VHS need not be at the band center, and the dos need not vanish at
the band edges.  However, it is a convenient form for evaluating the integrals, 
and the thermodynamic functions will display a generic form near the VHS.  In 
this case, $N_1=2V/B$, where $V$ is the sample volume.
\par
The low temperature values of the thermodynamic functions can be calculated,
following standard techniques\cite{Feyn}.  Care must be taken, due to the
nonanalytic behavior near the VHS.  The number of electrons is $N=nV$, where $n$
is the electron density
\begin{equation}
n=1+{2\over B}[\mu ln({eB\over 2|\mu |})+k_BTF({\mu\over k_BT})]
\label{eq:A7a}
\end{equation}
where
\begin{equation}
F(y)=\int_0^{\infty}ln|{y-x\over y+x}|{dx\over e^x+1}.
\label{eq:A7b}
\end{equation}
The function $F(y)$ must be evaluated numerically.  Figure~\ref{fig:A1}a shows
both $F$ (solid line) and two approximations to it: for $y\rightarrow 0$, $F(y)
\rightarrow y(lny-1)$ (dotted line), and for $y\rightarrow\infty$, $F(y)
\rightarrow -\pi^2/6y$ (dashed line). The temperature dependence of the chemical
potential follows from keeping $N$ fixed, independent of $T$:
\begin{equation}
\mu ln({eB\over 2|\mu |})+k_BTF({\mu\over k_BT})
=\mu_0ln({eB\over 2|\mu_0|}),
\label{eq:A8}
\end{equation}
where $\mu_0$ is the chemical potential at $T=0$.  Figure~\ref{fig:A1}b shows
the calculated $\mu$ vs $\mu_0$, for $T=10meV$, $B=1eV$, using either the
exact or approximate $F(y)$ functions.
\par
From Eq. \ref{eq:A7a}, a T-dependent dos can be defined,
\begin{equation}
\tilde N(\mu )=V{\partial n\over\partial\mu}
=N_1[ln({B\over 2|\mu |})+F^{\prime}({\mu\over k_BT})],
\label{eq:A8b}
\end{equation}
Fig. \ref{fig:A1}c.  As $T\rightarrow 0$, $\tilde N(\mu )\rightarrow N(\mu_0)$,
Eq. \ref{eq:A6}, while for $\mu\rightarrow 0$, the small-$y$ limit of the 
function $F(y)$ can be used in Eq.~\ref{eq:A8}.  Then
\begin{equation}
\tilde N(\mu )=N(\mu_0){\partial\mu_0\over\partial\mu}\rightarrow
N_1ln{B\over 2k_BT}.
\label{eq:A14}
\end{equation}
This function directly enters the expression for the compressibility $\kappa$
and the $\vec q=0$ susceptibility $\chi$.  Thus,
\begin{equation}
\kappa =-{1\over V}({\partial V\over\partial p})_{T,N}={1\over n^2}{\partial n
\over\partial\mu}
\label{eq:A12}
\end{equation}
where $p=-\Omega /V$ is the pressure, and 
\begin{equation}
\chi=\mu_B^2\tilde N(\mu )
\label{eq:A13}
\end{equation}
with $\mu_B$ the Bohr magneton.  In a conventional Fermi liquid theory, 
$\kappa$, $\chi$, and $C_v$ are all proportional to the dos.  At a VHS, this is
also true, as will be demonstrated for $C_v$ below.
\par
The other thermodynamic functions can be similarly computed.  The internal 
energy is
\begin{equation}
U(T)=U(0)+\Delta U(T),
\label{eq:A9a}
\end{equation}
\begin{equation}
U(0)={V\over 2}[{B\over 4}+{\mu |\mu |\over B}ln({eB^2\over 4\mu^2})],
\label{eq:A9b}
\end{equation}
\begin{eqnarray}
\Delta U(T)=N_1k_BT[k_BT(F_U({\mu\over k_BT})-{\pi^2\over6}lnk_BT)  \nonumber \\
+\mu F({\mu\over k_BT})]\equiv{(\pi k_BT)^2\over 6}\tilde N_U,
\label{eq:A9c}
\end{eqnarray}
\begin{equation}
F_U(y)=\int_0^{\infty}{xdx\over 1+e^x}ln({B^2\over 4|y^2-x^2|}),
\label{eq:A9f}
\end{equation}
\begin{equation}
\tilde N_U\simeq N_1ln({B\over 2e\bar E_U}),
\label{eq:A9d}
\end{equation}
\begin{equation}
\bar E_U=\sqrt{\mu^2-2\alpha_U^{\prime}\mu k_BT+(\alpha_U k_BT)^2},
\label{eq:A9e}
\end{equation}
where a constant shift has been included to make $U=0$ for the empty band, and
$\alpha_U=0.756$, $\alpha_U^{\prime}=0.336$ are numerical constants, evaluated
at $T=10meV$, $B=1eV$.  Equation~\ref{eq:A9d} is an approximation to the 
numerically calculated $\tilde N_U$ (Fig.~\ref{fig:A1}d); note that $\tilde N_U
$ has the form of a thermally broadened dos, Eq.~\ref{eq:A6}.  The scaling of 
$\bar E_U$ should be compared with that of the resistivity, $\rho$, 
Eq.~\ref{eq:rho}.
\par
To calculate the grand canonical potential, it is convenient to 
integrate Eq.~\ref{eq:A3} by parts, yielding
\begin{equation}
\Omega =V(B-2\mu )-\int_{-B/2}^{B/2}{\cal N}(E)f(E)dE,
\label{eq:A10a}
\end{equation}
\begin{equation}
{\cal N}(E)=\int_{-B/2}^EN(E^{\prime})dE^{\prime}.
\label{eq:A10b}
\end{equation}
Now, ${\cal N}(E)=Vn(\mu =E,T=0)$, with $n$ given by Eq.~\ref{eq:A7a}, so
\begin{equation}
\Omega =\Omega_0-U,
\label{eq:A10c}
\end{equation}
where
\begin{equation}
{\Omega_0\over V}=B-2\mu -\int_{-B/2}^{B/2}[1+{2E\over B}]f(E)dE
\label{eq:A10d}
\end{equation}
is a nonsingular term.  Then the heat capacity $C_v$ and the entropy $S$ should 
both display the divergence of the dos:
\begin{equation}
C_v=\bigl({\partial U\over\partial T}\bigr)_{\mu ,V}=\gamma T,
\label{eq:A11a}
\end{equation}
where 
\begin{equation}
\gamma ={\pi^2\over 3}k_B^2\tilde N_{\gamma}
\label{eq:A11b}
\end{equation}
is the Sommerfeld constant, and $\tilde N_{\gamma}$ has the form of 
Eq.~\ref{eq:A9d}, but with $\alpha_{\gamma}=1.23$, $\alpha_{\gamma}^{\prime}=
0.60$.  
\par
Note that $\alpha_{\gamma}/\alpha_U\simeq\alpha_{\gamma}^{\prime}/\alpha_U
^{\prime}\simeq e^{1/2}$.  This can readily be understood.  For $T<<\mu$,
$\Delta U\propto ln(B/\mu )$ is independent of $T$, so $C_v=d(\gamma T^2/2)/dT=
\gamma T$.  But for $\mu =0$, $\Delta U\propto ln(B/2e\alpha_Uk_BT)$.  In this 
case, $C_v$ picks up an extra term from differentiating the logarithm, which can
be included as $\gamma\propto ln(B/2e\alpha_Ue^{1/2}k_BT)$. Similarly, the 
entropy
$S=\int_0^T(C_v(T^{\prime})/T^{\prime})dT^{\prime}$ is equal to $C_v$ when $\mu
>>k_BT$, but differs in the opposite limit.  Table VI provides a
summary of the logarithmic divergences of various quantities, in the form
\begin{equation}
X=X_0ln({B\over 2\gamma_XY}),
\label{eq:A14b}
\end{equation}
with
\begin{equation}
Y=\cases{|\mu |&if $|\mu |>>k_BT$,\cr
          \alpha_Xk_BT&if $|\mu |<<k_BT$.}
\label{eq:A14c}
\end{equation}
In the expressions for $\Delta U(T)$, 
\begin{equation}
z={12\over\pi^2}\int_0^{\infty}{xlnxdx\over 1+e^x}\simeq 0.546.
\label{eq:A14d}
\end{equation}
The above form for the specific heat was found by Hirsch and
Scalapino\cite{HiSc,Goic}, while anomalous behavior in $S$ was pointed out by
Wohlfarth (quoted in Rice\cite{Ric}).  

\par
From the large-$y$ limit of $F(y)$, 
\begin{equation}
\mu\simeq\mu_0[1+{(\pi k_BT)^2\over 6\mu_0^2ln{eB\over 2|\mu_0|}}],
\label{eq:A15}
\end{equation}
showing a tendency towards phase separation in the region near a VHS.  
That is, when $\mu\ne 0$, {\it the chemical potential is always repelled away 
from the VHS} ($|\mu |>|\mu_0|$)!  This can easily be understood.  When $T$ 
increases from zero, the Fermi function $f(E)$ deviates symmetrically from a 
step function, reducing the occupation probability below the Fermi level and 
raising it above.  But the total electron concentration $n$ is an integral of 
$f(E)$ times the dos $N(E)$. For $E_F<E_v$, the dos is an increasing function of
energy, so at finite T more electrons are created than holes, and the chemical 
potential must shift below $E_F$ (i.e., away from the VHS) to compensate, with 
a larger shift the closer $E_F$ is to the VHS.  
Similarly, for $E_F>E_v$, more holes are created, so $\mu >E_F$.  This same
argument, however, shows that {\it if $E_F$ is exactly at the VHS, the chemical
potential does not shift} (at least to this order).  This result is particularly
obvious in the case of a square Fermi surface at half filling, where $E_F$ must 
be at the VHS for all temperatures.
\par

\section{$\tau (\lowercase{x_c})$ in the One Band Model}

In the VHS model, the doping of optimum T$_c$, $x_c$, is the doping $x$ at which
the Fermi level crosses the VHS.  Now $x_c$ can be determined experimentally,
and theoretically, it is controlled by the parameter $\tau$, which determines 
the curvature of the Fermi surface.  Hence, it is convenient to determine $\tau
(x_c)$, the value that $\tau$ would have to have to reproduce the observed
optimal doping.  This is easily done by calculating the carrier density for 
which the Fermi level coincides with the VHS.
The carrier density is proportional to the area of the Fermi surface:
\begin{equation}
1-x={2\over\pi^2}\int_0^{\pi}\phi_xd\phi_y,
\label{eq:B1}
\end{equation}
with $\phi_i=k_ia$, and
\begin{equation}
\bar c_x=-\bigl({2\tau +\bar c_y\over 1+2\tau\bar c_y}\bigr).
\label{eq:B2}
\end{equation}
The resulting $\tau (x_c)$ is shown as the solid line in Fig.~\ref{fig:B1}.
\par
An approximate formula can be derived by assuming that the area enclosed by
dashed lines in the inset of Fig.~\ref{fig:B1} is an ellipse.  Then, if the 
point where
the ellipse intersects the zone diagonal is $\phi_x=\phi_y=\pi /2-\delta$,
\begin{equation}
x=\delta /2,
\label{eq:B3}
\end{equation}
\begin{equation}
sin\delta ={\sqrt{1-4\tau^2}-1\over 2\tau}.
\label{eq:B4}
\end{equation}
For small $\tau$, this becomes
\begin{equation}
\tau=-x(1-{7\over 6}x^2),
\label{eq:B5}
\end{equation}
the dashed line in Fig.~\ref{fig:B1}.  The dot-dashed line is an empirical fit,
\begin{equation}
\tau =-0.52tanh(2.4x).
\end{equation}

\section{Electron-Phonon Coupling}

\subsection{One-band Model, Cu-Cu, O-O Stretch, Octahedral Tilt Modes}

In a tight-binding model, the electron-phonon interaction can be calculated from
the distance-dependence of the tight-binding parameters.  For example, there can
be an on-site coupling, to the hole density on site $i$, $n_i=c_i^{\dagger}c_i$,
by, e.g., modifying the Cu energy level at a particular atomic site, $R_i$: $
\Delta\rightarrow\Delta (R_i)$), or intersite coupling to the hopping term, 
$c_i^{\dagger}c_{i\pm 1}$ by modifying the hopping parameter, $t\rightarrow 
t\pm\delta t$ or the exchange, $J\pm\delta J$ (spin-Peierls coupling).  
\par
In analogy to a 1D metal, a first choice for the relevant phonon is the 
dimerization mode, $a\rightarrow a\pm\delta a$.  However, in 2D metals, the 
coupling depends sensitively on the particular phonon mode.  Following Tang and 
Hirsch\cite{TH}, we will first study the coupling to a Cu-Cu bond stretching 
mode, Fig.~\ref{fig:20}a.  In this case, the analysis can be carried out in a 
one-band model with $t_1=0$. The strength of the coupling can be estimated by
analyzing a static distortion of the same symmetry.  Expanding,
\begin{equation}
t_{ij}=t_0(1\pm\delta ),
\label{eq:CC1}
\end{equation}
with $t_{ij}>(<)t_0$ if the i-j atomic separation is less than (greater 
than) $a$.  The energy dispersion becomes 
\begin{equation}
E_{\pm}=\pm 2t_0\sqrt{(\bar c_x+\bar c_y)^2+\delta^2(\bar s_x+\bar s_y)^2}.
\label{eq:18}
\end{equation}
The total electronic energy, the integral of $E$ over 
the occupied states, is lowered by the distortion.  By a change of variable 
($k_{\pm}=(k_x\pm k_2)/2$), the electronic energy can be written
\begin{equation}
{\cal E}={-16t_0\over\pi^2}\int_0^{\pi /2}(cos^2\nu +\delta^2sin^2\nu )^{1/2}d
\nu\simeq{\cal E}_0-{16t_0\over\pi^2}\delta^2I,
\label{eq:CC4}
\end{equation}
with
\begin{equation}
I\simeq\int_0^{\pi /2-\delta}{d\nu\over cos\nu}\simeq{1\over 2}ln{\phi_m\over
\delta},
\label{eq:17}
\end{equation}
where $\phi_m$ is a cutoff angle.
Now the elastic energy increase like $\delta^2$, while the electronic energy
decreases like ${\cal E}-{\cal E}_0\propto\delta^2ln|\delta |$.  Hence, the 
electronic energy lowering always wins out, just as in the 1D Peierls 
distortion; the resulting transition temperature has the form of Eq.
\ref{eq:00}.  Indeed, the integrals in Eq.~\ref{eq:17} are {\it the same} as in 
the 1D problem.  Note the contrast with the molecular JT effect, 
Eq.\ref{eq:99d}.  Note also that the electronic energy varies like $\delta^2
|ln\delta |$, and not like $\delta^2ln^2\delta$, the form expected for a VHS.
The reason for this can be seen from the dispersion relation, Eq.~\ref{eq:18}.  
For $\delta =0$, the VHS's are at $k_x=0$, $k_y=\pi /a$, and equivalent points.
The coefficient of the $\delta^2$ term is proportional to $\bar s_x$,
which vanishes at all of the VHS's.  Thus, the distortion does not split the
VHS degeneracy, and hence behaves more like a conventional nesting instability.
\par
On the other hand, an O-O stretch mode {\it does} split the VHS degeneracy, and
hence produces a much larger energy lowering.  This mode was analyzed in the
three-band model in Ref. \cite{RMXB}.  In a one-band model, it does not appear
directly, since the O's are not explicitly included.  However, following Zhang
and Rice\cite{ZR}, we can replace the Cu's by effective Cu's, hybridized with
a linear combination of the four surrounding O's.  Then the O-O stretch modes
with orthorhombic symmetry become breathing modes, leading to a difference of
the on-site Cu potential on the two sublattices, $\pm 2t_0\Delta^*$.  To best
correspond with the 3-band result, $\Delta^* =n\delta$ (with $n$=2 for the 1D
stretch mode of Fig.\ref{fig:20}b).  Thus, the dispersion becomes
\begin{equation}
E_{\pm}=-4t_1\bar c_x\bar c_y\pm 2t_0\sqrt{(\bar c_x+\bar c_y)^2+\Delta^{*2}},
\label{eq:O18}
\end{equation}
where, for variety, a second-neighbor term, $t_1$ was included.  The net
energy lowering becomes, for $t_1=0$,
\begin{eqnarray}
{\cal E}={-16t_0\over\pi^2}\int_0^{\pi /2}(cos^2\nu_+cos^2\nu_- 
+\Delta^{*2})^{1/2}d\nu_+d\nu_- \nonumber \\
\simeq{\cal E}_0-{16t_0\over\pi^2}\Delta^{*2}I^2;
\label{eq:O17}
\end{eqnarray}
for finite $t_1$, the divergence is cut off to $\sim |ln\delta |$, as in 
Eq.~\ref{eq:b30}.  Equation \ref{eq:O18} is essentially the same as that found 
in Ref.~\cite{RM8a}, the pseudogap calculation discussed in Section IX.A, and is
consistent with Eq. \ref{eq:rev2}.  However, this mode does {\it not} seem to be
directly involved in the structural instabilities of the cuprates; while the 
phonons do show signs of softening\cite{PiRe}, the actual soft modes found in 
the cuprates are instead associated with tilting modes -- of the CuO$_6$ 
octahedra, and (perhaps) of the CuO$_5$ pyramids in YBCO.  Analysis of the tilt 
mode instabilities led to the VHS Jahn-Teller model for the structural 
transitions. Once it was realized that short-range order and dynamic JT effects 
are important, the role of the in-plane stretch modes was reassessed, and it 
was found that they could still play an important role (Section VIII.D).
\par
The quadratically coupled modes can be treated similarly. Thus, the LTT 
distortion in Fig.~\ref{fig:21}a couples strongly to the in-plane shear mode,
Fig.~\ref{fig:c19}c.  In the one-band model, the energy dispersion becomes
\begin{equation}
E=-2t_0(\bar c_x(1+\delta)+\bar c_y(1-\delta)),
\label{eq:O19}
\end{equation}
with $\delta\propto\Phi^2$, and $\Phi$ the octahedral tilt angle.  Then
\begin{equation}
{\cal E}-{\cal E}_0\simeq -{4t_0\delta\over\pi^2}\int |{\bar c_x-\bar c_y|=-
{16t_0\delta\over\pi^2}\equiv -{\cal E}^*\Phi^2.}
\end{equation}
This does not have a logarithmic divergence, but can still lead to an 
instability.  The phonon energy has the form
\begin{equation}
E_{ph}={K\over 2}\Phi^2+{W\over 4}\Phi^4.
\end{equation}
Now $K\sim\hbar\omega_{ph}$, a phonon energy, and hence $K/2-{\cal E}^*$ is
likely to be negative; this is indeed the situation found in Ref. 
\cite{RM8b,RM8c}.
This quadratic phonon coupling has also been studied in the three-band 
model\cite{RM8a,RM8b}.  As discussed in Section VI.D.12, it may also play an
important role in generating d-wave superconductivity. 
Normand, et al.\cite{NKF} also studied this mode, but 
concentrated on the small, linear coupling which would result {\it below} the
LTO transition temperature, due to the presence of a finite static tilt.
\par
Equations \ref{eq:O18} and \ref{eq:O19} were used to calculate the energy
dispersions in Figs.~\ref{fig:c19}c and d.  The parameters were taken as: for 
Fig.~\ref{fig:c19}c: $t_0=125meV$, $\delta =0.48$; for Fig.~\ref{fig:c19}d:
$t_0=178meV$, $\tau =-0.3$, $\Delta_0^*=0.48$.  Note that it has been assumed 
that $t_1$ is unaffected by the structural distortion.  In this case, the gap 
falls at different energies in different parts of the Brillouin zone.  Clearly, 
opening a gap away from the Fermi level costs energy without contributing to the
electronic energy lowering.  Hence, for the figure, I have assumed an
extended-s-wave gap, $\Delta^*=\Delta_0^*|\bar c_x-\bar c_y|/2$, to minimize 
the gap away from the Fermi level.  

Many studies have employed a simple 2D Holstein model (one band or three 
band, with on-site phonon coupling), either at the RPA level\cite{BFal,RM8a} or
by Monte Carlo calculations\cite{Scal1,Mars1}.  The on-site phonon is usually 
interpreted as a vibration of the apical oxygen above a given Cu 
site\cite{Schu1}.  However, it is also possible to interpret the electronic band
as an `effective Cu atom', incorporating the coupling to a symmetric 
superposition of the wave function on the four neighboring O's (a Zhang-Rice 
singlet\cite{ZR}).  In this case, the phonon could also be a breathing mode of 
the planar O's\cite{Schu2,Yone}, and hence equivalent to the O-O stretch mode
discussed above.  However, the strong instability of Eq. \ref{eq:O18} requires
in addition that the Fermi level coincide with the VHS.
An inhomogeneous Hartree-Fock 
calculation\cite{Yone} finds that intersite phonons have a much stronger effect 
than intrasite phonons in the three-band model of the CuO$_2$ plane.

\subsection{Three-band Model, O-O Stretch Modes}

Even when analyzing only the O-O stretch modes, coupling to the VHS depends
sensitively on factors like the inter-row correlations of the stretch, and of
coupled $X$ and $Y$ stretch modes.  For instance, if the stretch is solely along
the $x$ axis, with the same alternation on each row, the associated phonon
is at $0.5(\pi /a,0)$, as in the lower mode in Fig.~\ref{fig:a22}a; if instead
there is a long-short-long alternation on alternate rows, Fig.~\ref{fig:20}b,
the phonon is at $0.5Q_0=0.5(\pi /a,\pi /a)$.  Only the latter mode causes a 
linear gap at the VHS, Fig.~\ref{fig:20}d, but it is the {\it former} mode which
is found to anomalously soften, possibly due to pseudogap formation. (Curiously,
this mode has a quadratic coupling to the same in-plane shear mode that couples 
quadratically to the LTT tilt.)  
\par
On the other hand, there can also be simultaneous stretches along $x$ and $y$.
There are two phonon modes of $0.5Q_0$ symmetry.  In a breathing mode 
(Fig.~\ref{fig:a22}a, upper frame), all four short O-lengths are associated 
with the same Cu, coupling the Cu to a Zhang-Rice combination of the 
four neighboring O's.  Alternatively, each Cu could have two short and two long 
O-bonds, with the short bonds alternating along $x$ and $y$ -- a quadrupole
mode, which is experimentally found to soften, Fig.~\ref{fig:b22}.  
\par
In the three-band model, one can compare the coupling of these different modes
to the VHS.  If the coupling is assumed to be due to modulation of the
magnitude of the hopping parameter $t$, then the dispersion for all three modes 
at $0.5Q_0$ can be written in a similar fashion (to lowest order in $\delta$,
Eq.~\ref{eq:CC1}):
\begin{equation}
E={\Delta\over 2}+\sqrt{({\Delta\over 2})^2+W^2},
\label{eq:CC2}
\end{equation}
\begin{equation}
W^2=4t^2\pm\sqrt{(2t^2[\bar c_x+\bar c_y])^2+(2nt\delta )^2},
\label{eq:CC3}
\end{equation}
where $n$ is the excess number of short O-bonds between the two inequivalent
Cu's in the distorted unit cell -- i.e., $n$ = 2 for the $x$-only tilts, 4 for
the breathing modes, and 0 for the quadrupole modes.  This would say that the
largest gap, and hence the strongest JT coupling, is associated with the
breathing modes, whereas these modes experimentally show the least anomalous
behavior.  The breathing mode introduces a unit cell doubling, with a modulation
of the occupancy of the Cu $d$ orbitals on the two sublattices, $\psi_{Cu}
\rightarrow\psi_{Cu}(1\pm\alpha )$, with $\alpha =-4t\delta /[E(E-\Delta )-4t^2
]$, to lowest order in $\delta$.  Hence, correlation effects near half filling
oppose a breathing-type distortion.
\par
To find a strong coupling to the quadrupole mode, it is necessary to examine
different forms of electron-phonon coupling -- perhaps a coupling to the Cu JT 
distortion.  This would then be similar to the proposed dynamic JT in 
La$_{1-x}$Sr$_x$MnO$_3$\cite{AJM}, Section XI.H.5.  Note that the relation
between the quadrupole mode and the 1D stretch modes is similar to that
between the LTO and LTT type distortions, with only the latter coupled to the
VHS.

\subsection{JT Parameters}
\par
Using the above results, one can estimate what parameters of the breathing mode
gap would be required to describe the photoemission studies on underdoped
Bi-2212, Section IX.A.2.  Thus, using the one-band model, Eq.~\ref{eq:O18}
($t_1=0$), Figure~\ref{fig:c19}a requires $t_0=(0.5/4)eV=125meV$, and for the
underdoped sample, the shift of the VHS is $2t_0\Delta^*\simeq 250meV$. Thus, 
for the breathing mode, $\delta=\Delta^*/4\simeq 0.25$.  Now $\delta =\delta t_0
/t_0=\beta\delta a/a$, with $a$ the lattice constant and $\beta\simeq 6-7
$\cite{Alig,NKF} (recall that $t_0\simeq t^2/\Delta$, where $t$ and $\Delta$ are
the three-band parameters).  Assuming $\beta =7$, $\delta a/a\simeq 1/30$.  This
should be compared with the expected lattice contraction associated with a
localized hole, $O^{2-}\rightarrow O^-$, $\delta a/a\simeq 0.1$\cite{RMXB}.  A
better estimate would be that the hole is delocalized over the four O's making
up the Zhang-Rice singlet, so $\delta a/a\simeq 1/40$.  [As discussed in
Section VIII.E, the electron-phonon interaction near a VHS may {\it not} be
reduced by correlation effects.  In that case, the above estimates should be
modified: $t_0=t^2/\Delta\simeq 1.3^2/6=0.28$, in which case $\delta$ and 
$\delta a/a$ would both be reduced by this factor of $\sim 2$.]
\par
Given the above estimate for the gap $\Delta^*$, the net energy lowering is 
(Eq.~\ref{eq:O17}) $\Delta{\cal E}\simeq 10-35meV$, where the uncertainty is
due to the different estimates for $t_0$ mentioned above, and $\phi_m$ was
assumed to be $\pi /2$.  This should be compared to the experimental value
$2E_{JT}\simeq$70meV, found in lightly doped cuprates\cite{Falc}, Section XI.B.
The underestimate is expected: the above calculation included only the linear
effects of the distortion, and hence ignored the important polaronic effects
(e.g., band narrowing).  Moreover, the molecular and band JT effects differ in
form, due to the electronic kinetic energy in the latter case; compare Eqs.~
\ref{eq:99d} and \ref{eq:O17}.  Note, however, that Fehrenbacher\cite{Fehr3} has
estimated a theoretical $E_{JT}\simeq$100-400meV for the (1D) breathing mode.

\section{RG Calculations}

Following Schulz\cite{Sch}, there are four types of electron-electron 
interaction constants, $G_1$, for scattering from, e.g., $X,Y\rightarrow
Y,X$ (labelling the electrons by the VHS they are associated with), $G_2:
(X,-X\rightarrow -X,X)$, $G_3:(X,-X\rightarrow Y,-Y)$, and $G_4:(X,Y
\rightarrow X,Y)$.  These coupling constants are renormalized by interaction,
and can diverge as the renormalization proceeds.  The various density wave and
superconducting instabilities are signalled by the divergence of certain
linear combinations of the $G$'s: $G_{CDW}=-2(2G_1+G_3-G_4)$, $G_{SDW}=2(G_3+
G_4)$, $G_{sSC}=-2(G_2+G_3)$, and $G_{dSC}=-2(G_2-G_3)$ (the latter for s- or 
d- wave superconductivity).  Additional divergences are possible\cite{Sch1},
related to spin currents and orbital antiferromagnets: $G_{JS}=-2(G_3-G_4)$ and
$G_{OAF}=-2(2G_1-G_3-G_4)$.
\par
In lowest-order perturbation theory, there are divergent corrections to the
electron-electron vertices, $\Gamma_i$, e.g.:
\begin{equation}
\Gamma_3=G_3-G_3[G_1+G_2-2G_4]I(\epsilon_c)
\label{eq:C1}
\end{equation}
with
\begin{equation}
I(\epsilon_c)={1\over\pi}[ln^2({\pi^2t\over\omega})-ln^2({\pi^2t\over\epsilon_c
})],
\label{eq:C2}
\end{equation}
and $\epsilon_c$ an ultraviolet cutoff.  The renormalization scheme is to
reduce the cutoff closer to the Fermi level, and require that the low energy
properties remain unchanged.  With scaling variable $l=ln^2(\pi^2t/\epsilon_c)/
2\pi$, the renormalization equations become
\begin{equation}
G_1^{\prime}=-2G_1[G_1-G_4]\delta ,
\label{eq:C3}
\end{equation}
\begin{equation}
G_2^{\prime}=-G_2^2-G_3^2,
\label{eq:C4}
\end{equation}
\begin{equation}
G_3^{\prime}=-2G_3[G_2+(G_1-2G_4)\delta ],
\label{eq:C5}
\end{equation}
\begin{equation}
G_4^{\prime}=[G_3^2+G_4^2]\delta ,
\label{eq:C6}
\end{equation}
where the prime means derivative with respect to $l$, and for now $\delta =1$.  
These equations are integrated numerically, given the initial values $G_i=U/4\pi
t\equiv G_0$ for the Hubbard model, until one diverges.  
The competition between various instabilities can be understood by looking at
the various $G$'s in terms of their initial values: $G_{SDW}\sim 4G_0$,
$G_{CDW}\sim G_{sSC}\sim -4G_0$, and the other three are $\sim G_0^0$.  Thus,
for positive $U$ (positive $G_0$) $G_{SDW}$ is dominant, and if it can be
suppressed then d-wave superconductivity (or orbital antiferromagnetism or spin
currents) becomes possible.  On the other hand, if the sign of $U$ can be
changed (by, e.g., boson exchange), then the competition between CDW's and
s-wave superconductivity becomes more favorable, since both terms are of the
same order in $G_0$.  In this Appendix, I will analyze the magnetic case $U>0$
for variety.  From Fig.~\ref{fig:33}a, it can be seen that the SDW instability
wins out at half filling, but the dSC and OAF are nearly as singular (they are
essentially degenerate with one another), while the other three responses are
suppressed by the renormalization.  Schulz found that by doping away from
half filling (moving away from the VHS), the divergences are reduced or cut off 
in such a manner that d-wave superconductivity can arise above a critical 
doping.
\par
However, in the real cuprates, the VHS falls away from half filling, so
doping actually moves the system closer to the VHS.  Hence, Schulz's scheme
must be modified.  I proposed a `floating VHS' model of doping, wherein 
correlation effects pin the VHS at the Fermi level for all dopings, with the 
shape of the Fermi surface continually changing to preserve the 
pinning\cite{RM8a} (see the discussion in Section VII.B).  In the present RG 
scheme, a deviation of the Fermi surface 
from square means that the inter-VHS scattering ($G_1$, $G_3$) gets cut off from
$ln^2$ to $ln^1$, with the cutoff depending on the curvature of the Fermi 
surface at the VHS, and hence on doping.  The terms in Eqs.~
\ref{eq:C3}-\ref{eq:C6} which are cut off are those proportional to $\delta$.
\par
To modify the above equations, let $y=\sqrt{l}$.  Then, if $y>y_c$, the cutoff 
value, Eqs. \ref{eq:C3}-\ref{eq:C6} hold with the prime reinterpreted as
derivative with respect to $y$, if the right hand sides of all four equations
are multiplied by $2y$.  This same perscription will also work for $y<y_c$, if
now $\delta\rightarrow y_c/2y$.  Figure \ref{fig:33}b shows that, as $y_c$
decreases, the SDW divergence is cut off, and ultimately the dSC instability
wins out.  The OAF is now no longer degenerate with the dSC state, and 
diverges somewhat more weakly.  
These results can be converted into a phase diagram of $T_c$ vs 
doping by identifying $\epsilon_c\rightarrow 2T$.  From Eq.~\ref{eq:b30}, $y_c=
ln[(1+A)/(1-A)]$.  For the one band model, 
\begin{equation}
A=\sqrt{1+2\tau\over 1-2\tau},
\label{eq:C7}
\end{equation}
so the doping dependence of $y_c\simeq ln(1/-\tau )$ can be found from the 
results of Appendix B.  Since $y_c=ln(E/T_c)$, the phase diagram can be
calculated; Fig.~\ref{fig:33}c shows the resulting phases, when $E$ is adjusted
to reproduce the AFM transition at zero doping.  The results are very similar to
those found by Schulz\cite{Sch}, and are qualitatively similar to the 2D RPA
calculations, Fig.~\ref{fig:23}b, although differing in detail.
\par
{\bf Note:} After this Appendix was written, I obtained a preprint of Ref.
\cite{IoMi}, which contains a similar RG calculation.

\newpage
\ 
\vskip 0.3in
\begin{tabular}{||c||c|c|c|c|c||}        \hline
\multicolumn{6}{c}{{\bf Table I: One Band Model Parameters}} \\ 
            \hline\hline
Material & $t_0$ (eV)&$\tau$&$t_z/t_0$&$x_c$&Ref.
     \\   
    \hline\hline
LSCO & 0.24 & -0.19 & ---  & 0.16 & \cite{Xu}     \\     \hline       
YBCO & 0.24 & -0.45 & 0.45 & 0.53 & \cite{YFree}   \\     
     & 0.35 & -0.28 & 0.40 & 0.25  & \cite{OKA2}     \\   \hline
\end{tabular}
\vskip 0.3in
\begin{tabular}{||c||c|c|c||}        \hline
\multicolumn{4}{c}{{\bf Table II: Three Band Model Parameters}} \\ 
            \hline\hline
Material & $t$ (eV)&$\Delta$ (eV)&$t_{OO}$ (eV)      \\   
    \hline\hline
LSCO & 0.60 & 0.0 & 0.15   \\     \hline       
YBCO & 0.10 & 0.0 & 0.16  \\     \hline       
Bi-2212 & 0.35 & 0.24 & 0.25  \\     \hline       
\end{tabular}
\vskip 0.3in
\begin{tabular}{||c||r|r|r||r||}        \hline
\multicolumn{5}{c}{{\bf Table III: Bare Parameters (eV)}} \\ 
            \hline\hline
\multicolumn{1}{c}{Parameter} &
  \multicolumn{3}{c}{LSCO} & \multicolumn{1}{c}{YBCO} \\   \hline
\multicolumn{1}{c}{Reference} & \multicolumn{1}{r}{\cite{Hyb}} &
  \multicolumn{1}{r}{\cite{GraM}} & \multicolumn{1}{r}{\cite{Esk}} 
& \multicolumn{1}{r}{\cite{OKA2}} \\
    \hline\hline
$\Delta$ & 3.6 & 3.51 & 3.5 & 3.0    \\     \hline       
$t_{CuO}$ & 1.3 &  1.47 & 1.3 & 1.6  \\     \hline
$t_{OO}$ & 0.65 &  0.61 & 0.65 & 1.1  \\     \hline
$V$ & 1.2 & 0.52   & $<1$  & --     \\     \hline
\end{tabular}
\vskip 0.3in
\begin{tabular}{||c||c|c||}        \hline
\multicolumn{3}{c}{{\bf Table IV: Effective Parameter Values of LSCO}} \\ 
    \hline\hline
{Parameter} & Scaling & Corrected Bare Value
     \\   
  & & (eV) \\
    \hline\hline
$t_{CuO}$ &  & 1.3 \\     \hline
$\Delta$ & $\Delta_0+2V$ & 5-6  \\   \hline       
$t_{OO}$ & $t_{OO}^1-X/\Delta$  & 0.14 (LSCO)  
    \\     
  &  &  0.3  (YBCO)    \\     \hline
\end{tabular}

\newpage
\ 
\newpage
\vskip 0.3in
\begin{tabular}{||c||c|c|c|c||}        \hline
\multicolumn{5}{c}{{\bf Table V: Experiments on Phase Separation}} \\ 
            \hline\hline
Material & Doping Range & Property & Section & Refs.
     \\   
    \hline\hline
\multicolumn{5}{c}{{\bf Underdoped Regime}} \\ 
             \hline
La$_2$CuO$_{4+\delta}$ & 0.06$\le\delta\le$0.11 & large period superlattices & 
    XI.A.1 & --- \\ \hline
LSCO & $x\le 0.14$ & incommensurate diffraction peaks & VI.E.2 & \cite{BPT} \\ 
    \hline
LSCO & $x\simeq 0$ & 1 hole = dielectric polaron & XI.B & --- \\ \hline
LSCO & $x\le 0.04$ & magnetic domain size from neutron diffraction & XI.C & 
   \cite{Keim}  \\ \hline
LSCO & $x\le 0.02$ & AFM stripes & XI.C & \cite{Cho} \\ \hline
LSCO & $x\le 0.02$ & AFM islands & XI.C & \cite{DJ4} \\ \hline
LSCO & $0.02\le x\le 0.08$ & spin glass & XI.C & \cite{DJ5} \\ \hline
LSCO & $x\le 0.16$ & magnetic fraction from M\"ossbauer & XI.D & \cite{Imb}
    \\ \hline
LSCO & $x\le 0.15$ & magnetic fraction from $\mu$SR & XI.D & 
    \cite{Weid,Kief,Nied}  \\ \hline
YBCO & 0.0$\le\delta\le$0.7 & photoemission doping dependence & VII.D.3 & --- 
   \\     \hline       
\multicolumn{5}{c}{{\bf Paramagnetic Centers (?)}} \\ 
             \hline
LSCO & $x\le 0.16$ & two populations of Cu NQR & XI.F.2 & --- \\ \hline
YBCO & $\delta\approx 0.6-0.7$ & AF polarons (?) & XI.B & \cite{Burl,AFp} \\ 
   \hline
LSCO,YBCO & --- & EPR (?) & XI.F.3 & --- \\  \hline
YBCO & --- & linear term in specific heat & XI.F.3 & \cite{Phil,Phil2} \\ 
   \hline
\multicolumn{5}{c}{{\bf Crossover at VHS}} \\ 
             \hline
LSCO & $x\le 0.3$ & $T_c(x)$ & VI.D.5 & --- \\ \hline
LSCO & $x\le 0.3$ & Meissner fraction & XI.E.1 & --- \\ \hline
LSCO & $x\le 0.3$ & susceptibility & XI.E.2 & --- \\ \hline
LSCO & $x\le 0.3$ & $T_1^{-1}$ & XI.E.3 & \cite{NQR1} \\ \hline
LSCO,YBCO & --- & intrinsic weak links & XI.E.4 & --- \\ \hline
\multicolumn{5}{c}{{\bf Overdoped Regime}} \\ 
             \hline
LSCO & $x\ge 0.15$ & `fishtail' magnetization & XI.E.4 & \cite{fish} \\ \hline
La$_2$CuO$_{4+\delta}$ &$\delta\ge$0.11 & `extra' superconductivity (60K) & 
    XI.F.4 & --- \\ \hline
YBCO & $\delta\le$0.1 & two T$_c$'s & XI.F.4 & --- \\ \hline
\multicolumn{5}{c}{{\bf Macroscopic}} \\ 
             \hline
La$_2$CuO$_{4+\delta}$ & 0$\le\delta\le$0.06 & mobile O & XI.A.1 & --- \\ 
  \hline
La$_2$CuO$_{4+\delta}$ & 0$\le\delta\le$0.06 & staging & XI.A.1 & 
  \cite{Custag} \\  \hline
YBCO & 0.1$\le\delta\le$1.0 & mobile O & XI.A.2 & --- \\ \hline
Bi-2212 & -- & polaronic phase separation O & XI.A.4 & \cite{Bian2} \\ \hline
LBCO & 0.125 & LTT = commens. charge-spin domain & XI.A.5 & \cite{Tran}
  \\ \hline
LSCO & 0.15$\le x\le$0.4 & miscibility gap & XI.F.1 & --- \\ \hline
La$_2$CuO$_4$ & 0& photodope & XI.A.6 & --- \\ \hline
YBCO & $\delta\ge 0.7$ & photodope & XI.A.6 & --- \\ \hline
YBCO & $\delta\ge 0.7$ & H-dope (?) & XI.A.7 & --- \\ \hline
NCCO & 0.1-0.2 & sc near miscibility gap & XV & \cite{JDJ} \\  \hline       
\multicolumn{5}{c}{{\bf Nickelates}} \\ 
             \hline
LNO & --- & miscibility gaps  & XI.H.1 & --- \\ \hline
LNO & --- & staging & XI.H.1 & --- \\ \hline
LSNO & --- & spin glass phase & XI.H.2 & \cite{LSNO2,LSNO6} \\ \hline
LSNO & --- & incommensurate diffraction peaks & XI.H.2 & --- \\ \hline
LSNO & --- & spin-charge superlattices  & XI.H.2 & \cite{LSNO,LSNO8} \\ \hline
LSNO & --- & incommensurate diffuse scattering & XI.H.2 & \cite{LSNO9} \\ \hline
\end{tabular}
\vskip 0.3in
\begin{tabular}{||c||c|c|c|c|c||}        \hline
\multicolumn{4}{c}{{\bf Table VI: Thermodynamic Functions of Eq. 
\ref{eq:A14b}}}   \\ 
            \hline\hline
$X$ & $X_0$ &$\gamma_X$&$\alpha_X$ &     \\   
    \hline\hline
$\tilde N(\mu )$ & $N_1$ & 1 & 1& \\     \hline       
$\Delta U(T)$ & ${N_1(\pi k_BT)^2\over 6}$ & $e$ & $e^{(z-1)}=0.635$ & \\ \hline
$C_v$ & ${N_1(\pi k_B)^2T\over 3}$ & $e$ & $0.635\sqrt{e}=1.047$ & \\ \hline
$S$ & ${N_1(\pi k_B)^2T\over 3}$ & $e$ & ${1.047\over e}=0.385$ & \\ \hline
\end{tabular}
\newpage
\ 
\newpage

\begin{figure}
\caption{(a) Energy dispersion of a 2D VHS.  Calculations are based on the 
antibonding band of the 3 band model for the CuO$_2$ plane, 
Eqs.~\protect\ref{eq:1} and \protect\ref{eq:2a}, with (renormalized) 
parameters $t_{CuO}=0.347eV$, $\Delta =0.244eV$, and $t_{OO}$ = 0.25eV (dashed
line) or 0 (solid line). (b) Selected Fermi surfaces, corresponding to the two
choices for $t_{OO}$: $E_F$ = 0.7eV (squares), 0.825eV$\equiv E_{VHS}
$(triangles), or 0.9eV (circles). (c) Corresponding density of states (dos), and
(d) total number of electrons per unit cell. }
\label{fig:1}
\end{figure}
\begin{figure}
\caption{Model Fermi surface of Ref. \protect\cite{Pin}, showing location of hot
spots (heavy black lines), in the immediate vicinity of the VHS's.}
\label{fig:a1}
\end{figure}
\begin{figure}
\caption{(a) Schematic Fermi surface of LSCO, $x=0.15$, illustrating both
conventional ($Q_1,Q_2$) and VHS ($Q_0$) nesting. (b) 
Calculated\protect\cite{Pick1} nesting function, showing the peaks in the joint 
dos associated with the various nesting vectors [Ref.~\protect\cite{Recon}].}
\label{fig:2}
\end{figure}
\begin{figure}
\caption{Modification of dos at VHS, due to interlayer 
coupling\protect\cite{RM6}: (a) single
layer, (b) four coupled layers, (c) infinite layers (dashed lines:  approximate
calculations).  Inset in (c): calculated dos for Ni, Ref.~\protect\cite{Muel}.}
\label{fig:a2}
\end{figure}
\begin{figure}
\caption{Photoemission-derived CuO$_2$ antibonding band energy dispersion, for 
(a) Bi-2212\protect\cite{PE1}, (b) Bi-2201\protect\cite{PE4}, (c) 
YBCO\protect\cite{Liu,PE3a}, (d) Y-124\protect\cite{PE3a}, (e) 
NCCO\protect\cite{PEn}. [Ref. \protect\cite{PE0}] }
\label{fig:4}
\end{figure}
\begin{figure}
\caption{Experimental Fermi surfaces of Bi-2212: (a) Ref. \protect\cite{PE1}; 
(b) Ref. \protect\cite{Aeb2} -- top view: $\vec k_{\parallel}$ map of 
photoelectron intensity (logarithmic scale), bottom: sketch of principal
features; (c) Ref. \protect\cite{Pgap9} -- top: Fermi surfaces, including 
replicas due to incommensurate superlattice (light solid lines), bottom: energy
dispersion.}
\label{fig:5}
\end{figure}
\begin{figure}
\caption{Theoretical Fermi surfaces of YBCO: (a) Ref. \protect\cite{Pick2}, (b) 
Ref. \protect\cite{OKA}, (c) Ref. \protect\cite{Mass}. The lower (upper) frames
are the $k_a-k_b$ plane for $k_c$ = 0 ($\pi /c$). Dashed lines suggest
chain-dominated sections of Fermi surfaces.}
\label{fig:a5}
\end{figure}
\begin{figure}
\caption{Experimental Fermi surfaces of (twinned) YBCO: (a) Ref. 
\protect\cite{PEY2}, (b) Ref. \protect\cite{Liu}; and energy dispersions in
untwinned (c) YBCO\protect\cite{PE3a} and (d) Y-124\protect\cite{PE3} near the
extended VHS's.  The theoretical curves are from (a) Ref. \protect\cite{YMFK}, 
(b) Ref. \protect\cite{Pick2}, and (c) Ref. \protect\cite{sen}.  The arrow in 
(c) illustrates a special inter-VHS scattering between bifurcated VHS's.}
\label{fig:b5}
\end{figure}
\begin{figure}
\caption{Contour map\protect\cite{PE3} of the energy dispersion near an 
extended VHS in Y-124 of Fig.~\protect\ref{fig:b5}(d).}
\label{fig:c5}
\end{figure}
\begin{figure}
\caption{Fermi surfaces of NCCO, $x$ = 0.15 (a), 0.22 (b), compared to LDA 
calculations, after Ref. \protect\cite{PEn} [Ref. \protect\cite{PE0}].}
\label{fig:6}
\end{figure}
\begin{figure}
\caption{Comparison of the Fermi surfaces of Bi-2201, Bi-2212, and NCCO [Ref.
\protect\cite{PE4}]. }
\label{fig:a6}
\end{figure}
\begin{figure}
\caption{Theoretical Fermi surfaces of LSCO\protect\cite{Xu} for $x$ = 0 (a),
0.17 -- the VHS (b), and 0.2 (c).  The Fermi surfaces are shown both in a 
flattened irreducible wedge (top) and in the extended zone scheme (bottom).}
\label{fig:7}
\end{figure}
\begin{figure}
\caption{Theoretical Fermi surfaces of Hg-1223\protect\cite{NoFree}.}
\label{fig:8}
\end{figure}
\begin{figure}
\caption{(a) Experimental (circles)\protect\cite{Well} and calculated 
(lines)\protect\cite{RM11} energy dispersion of SCOC.
(b) Comparison of slave boson\protect\cite{RM11} and cluster Monte 
Carlo\protect\cite{LMan,DagN} calculations of
the energy dispersion of a Mott insulator at half filling. }
\label{fig:a8}
\end{figure}
\begin{figure}
\caption{(a) Calculated energy dispersion of BaBiO$_3$\protect\cite{MaHa}. (b) 
Brillouin zone for simple cubic lattice. (c) Calculated dos, showing Fermi level
for $x$=0,0.29 (assuming uniform doping)\protect\cite{Hama}}
\label{fig:c8}
\end{figure}
\begin{figure}
\caption{(a) Calculated Fermi surface of BaBiO$_3$, at the 
VHS\protect\cite{ABan}. (b) Cross sections of the Fermi surface, for a series of
energies spanning the VHS.}
\label{fig:AB}
\end{figure}
\begin{figure}
\caption{Calculated Fermi surface of disorder-averaged M$_3$C$_{60}.
$\protect\cite{ME2}}
\label{fig:d8}
\end{figure}
\begin{figure}
\caption{Calculated plasma frequency, $\omega_{pl}$ (solid line) and dos (dashed
line) near a VHS.}
\label{fig:b8}
\end{figure}
\begin{figure}
\caption{Linearizing the Fermi surface near a VHS, to calculate inter-VHS
susceptibility. (a) Perfect nesting, $\alpha =45^o$; (b) imperfect nesting, 
$\alpha <45^o$.}
\label{fig:a9}
\end{figure}
\begin{figure}
\caption{(a) Discontinuity of heat capacity $\Delta C/T_c$, plotted vs 
fractional distance in energy from a VHS, $\delta/W$\protect\cite{NewC}. The two
lines are theoretical, with different values assumed for the prefactor of the 
logarithmic dos, the symbols are data from Refs. \protect\cite{C1,C2,C3,C4}. (b)
Measured heat capacity, including discontinuity at T$_c$ (diamonds), residual 
part (circles and dashed line), and sum (solid line)\protect\cite{Kuma}.}
\label{fig:10}
\end{figure}
\begin{figure}
\caption{Susceptibility of LSCO (open squares)\protect\cite{Gp3} and YBCO
(open circles)\protect\cite{Allo}, with fits to broadened VHS's, with $t_z$ =
2 (dotdashed line), 4 (solid line), 6 (dotted line), and 20 meV (dashed line).
The amplitudes of the theoretical curves were taken as adjustible parameters.
Filled circles: YBCO data, with a constant term subtracted.}
\label{fig:b10}
\end{figure}
\begin{figure}
\caption{Hole-hole scattering rate near (but slightly off of) a VHS.  Solid 
lines: theory\protect\cite{Patt}, dashed lines: infrared\protect\cite{Schl} 
and photoemission\protect\cite{Ols1,PEY2} experimental results.}
\label{fig:a10}
\end{figure}
\begin{figure}
\caption{(a) `Universal' form of thermopower $S$ vs T for different 
cuprates\protect\cite{TEP}. (b) `Universal' form of $S(290K)$ vs hole 
doping\protect\cite{TEP}. (c)Anomalous $S(T)$ for YBCO\protect\cite{TEPY}. 
(d) Calculated $S(T)$ near VHS\protect\cite{NewS} }
\label{fig:11}
\end{figure}
\begin{figure}
\caption{Magnetic breakdown vs switching orbit near a VHS. Frames a-d show the
Fermi surfaces (extended zone scheme) of a 2D metal as the Fermi level gradually
increases.  Frame (d) illustrates a typical situation for magnetic breakdown,
from the lens and diamond orbits to the underlying circular orbit.  On the other
hand, in (b) the metal is at the VHS, and an electron approaching the VHS has a
choice of staying on the same orbit (as in (a)) or {\it switching} to a hole
orbit, as in (c)\protect\cite{magsw}. }
\label{fig:12}
\end{figure}
\begin{figure}
\caption{(a) Effective electron-electron interaction potential vs excitation
energy $\hbar\omega$ in BCS theory. (b) Potential vs energy when the 
Fermi level coincides with a VHS.  (c) Coulomb potential vs imaginary frequency 
$\nu$ for several values of $q_x/k_c$ at fixed $q_y=0.05k_c$ (where $k_c$ is a 
momentum cutoff).  Solid curves: $E_F=0$ (i.e., at the VHS); dotted curves:
$E_F=0.05B$ (with $B$ the bandwidth)\protect\cite{NewP}.}
\label{fig:14}
\end{figure}
\begin{figure}
\caption{Isotope effect near a VHS. Dotted lines: VHS excitonic 
superconductivity model\protect\cite{NewP}, dashed line: phase separation 
model\protect\cite{RM7}, solid lines: guides to the eye; circles: data of Ref.
\protect\cite{isot1}, triangles: data of Ref. \protect\cite{isot2}, squares: 
data of Ref. \protect\cite{isot3}.}
\label{fig:a14}
\end{figure}
\begin{figure}
\caption{Uemura plot of T$_c$ vs $n_s/m$. (a) For cuprates and other high-T$_c$
superconductors\protect\cite{Uem1,Uem4}. (b) Showing modifications due to 
overdoping\protect\cite{Uem2}. (c) Expected behavior near a VHS (replotted from 
Fig.~\protect\ref{fig:b8}, as $N(E_F)$ vs $\omega_{pl}^2$). }
\label{fig:c14}
\end{figure}
\begin{figure}
\caption{(a) T$_c$ vs hole-doping in the cuprates. Lines are calculated from VHS
model\protect\cite{MG}, symbols are experimental data\protect\cite{Uem1}. (b) 
Universal parabolic dependence of T$_c$ on doping\protect\cite{Univ}.}
\label{fig:b14}
\end{figure}
\begin{figure}
\caption{Universal plots of T$_c$ vs pressure for the cuprates: (a),(b): 
comparison of the pressure coefficient (a) with the isotope effect coefficient
(b)\protect\cite{pres1}; (c) illustrating effect of overdoping, and the large
pressure coefficients which can be obtained if $T_c$ is sufficiently 
reduced\protect\cite{pres2}.}
\label{fig:d14}
\end{figure}
\begin{figure}
\caption{Pressure dependence of T$_c$ in Hg-12(n-1)n, 
n=1-3\protect\cite{HgChu1}.}
\label{fig:e14}
\end{figure}
\begin{figure}
\caption{Incommensurate diffraction peaks in (a) LSCO\protect\cite{inco}, (b)
YBCO$_{6.93}$\protect\cite{Mook}, and (c) LSNO\protect\cite{LSNO2}.}
\label{fig:15}
\end{figure}
\begin{figure}
\caption{Magnetic energy dispersion in the charge-transfer insulator phase of
the 3-band tJ model at half filling: (a) uniform phase, 
Eq.~\protect\ref{eq:13e}, (b) flat-band nesting flux phase, 
Eq.~\protect\ref{eq:13g}, (c) VHS-nesting flux phase,
Eq.~\protect\ref{eq:13h}, (d-f) are similar to (a-c), with added N\'eel order, 
Eq.~\protect\ref{eq:13f}, after \protect\cite{RM11}.}
\label{fig:a15}
\end{figure}
\begin{figure}
\caption{Slave boson calculations for the cuprates: (a) $t_R$, (b) $E_F$, 
(c) $n_{VHS}=1-x_{VHS}$, and (d) $\Delta E$, assuming $\Delta_0$ = 4eV (solid
lines), 5eV (dashed lines), or 6 eV (dot-dashed lines)\protect\cite{RMXA}.}
\label{fig:16}
\end{figure}
\begin{figure}
\caption{Dos vs energy, at several different dopings, for 
LSCO\protect\cite{New}. Fermi levels indicated by short horizontal lines.}
\label{fig:a16}
\end{figure}
\begin{figure}
\caption{Slave boson calculation (3-band tJ model) of YBCO dispersion as a 
function of hole doping (solid lines) or electron doping (dashed lines).  From
bottom to top, curves are for $x$ = -0.6, -0.45, -0.3, -0.1 (dashed lines), 
0, .02, .1, .15, .3, .5 (solid lines).}
\label{fig:19}
\end{figure}
\begin{figure}
\caption{$E_F$ vs doping, comparing slave boson and cluster calculations, for
$\Delta_0/t$ = 4 ($\times$)\protect\cite{Dopf}, 3 (open 
squares)\protect\cite{Scal}, 2 (open diamonds)\protect\cite{Scal}, and 
1 ($+$)\protect\cite{Dopf}.  In terms of the parameters of Scalettar, et 
al.\protect\cite{Scal}, $\Delta_0/t=\epsilon +2V$.}
\label{fig:17}
\end{figure}
\begin{figure}
\caption{(a) Comparison of Bi-2212 dispersion to tight binding model, using
either bare LDA-derived parameters (dotted lines), or slave boson calculations
(solid lines)  Dashed lines denote ghost Fermi surfaces. (b) Additional
narrowing of VHS peak associated with polaronic correction. [Ref. 
\protect\cite{RMXB}]}
\label{fig:18}
\end{figure}
\begin{figure}
\caption{(a) Angle-resolved photoemission spectra showing the `1eV' peak in 
YBCO\protect\cite{Liu2}.  (b) Model dispersion in phase separation model. (c) 
Shift of the `1eV' peak with doping.}
\label{fig:d19}
\end{figure}
\begin{figure}
\caption{Angle-resolved photoemission spectra along $S-Y$ or $\Gamma -Y$ of (a) 
Y-124, (b) YBCO$_{6.9}$\protect\cite{Camp}, (c) YBCO$_{6.5}$, (d) 
YBCO$_{6.3}$\protect\cite{Liu2}.}
\label{fig:a19}
\end{figure}
\begin{figure}
\caption{Angle-integrated photoemission spectra of (a) Ca$_{1-x}$Sr$_x$VO$_3$ at
several dopings\protect\cite{Mottran} and (b) a number of Mott-Hubbard
insulators, comparing experiment (symbols) with theory (lines) for the
conducting band\protect\cite{Mottran1}.}
\label{fig:b19}
\end{figure}
\begin{figure}
\caption{Pseudogaps and `shadow bands,' or `ghost' Fermi surfaces: (a) 
period-doubling Umklapp scattering leading to (b) hole pockets centered on
$\bar M=(\pi /2a,\pi /2a)$\protect\cite{Halla}; (c-e) associated pseudogap 
developing in dos, for (c) 1D CDW at several temperatures\protect\cite{LRA} 
(d) 2D CDW in cuprates, at several temperatures\protect\cite{RM5}, (e) 2D SDW, 
in Hubbard model, for several $U$ values\protect\cite{KaSch}.}
\label{fig:c22}
\end{figure}
\begin{figure}
\caption{Electronic dispersion in the presence of structural distortion or flux
phase: (a),(b) experimental dispersion in underdoped Bi-2212\protect\cite{Gp0};
(c) shear-strain-induced splitting of the VHS degeneracy\protect\cite{RMXB}; 
(d) breathing-mode-induced splitting of the VHS degeneracy; 
(e) VHS splitting in the flux phase\protect\cite{WeL}.  In c-e, the dashed lines
show the dispersion from $X$ ($Y$) to $\bar S\equiv (\pi /2a,\pi /2a)$, while 
the dot-dashed lines show the dispersion near $Y$, where that differs from near 
$X$.  In (e), the heights of the vertical bars indicate the local spectral
weights.  Parameters used in the calculations are discussed in Appendix C.1}
\label{fig:c19}
\end{figure}
\begin{figure}
\caption{Octahedral tilt distortions associated with the (a) LTT and (b) LTO
phases.}
\label{fig:21}
\end{figure}
\begin{figure}
\caption{In-plane phonon stretch modes which affect the Cu-O bond length:
(a) Cu-Cu stretch; (b) O-O stretch; (c,d) splitting of electronic band 
dispersion due to a static distortion of the corresponding 
symmetry.  Dot-dashed lines: dispersion along lines $X-\bar 
M$.\protect\cite{RMXB}.}
\label{fig:20}
\end{figure}
\begin{figure}
\caption{(a) Calculated energy vs. displacement for a number of $X$-point phonon
modes.  Inset: contour plot of same data.\protect\cite{LTTP}. (b) 
Self-consistently calculated phonon double well derived from the experimental 
T-dependence of the soft mode in LSCO\protect\cite{BuH}.}
\label{fig:a20}
\end{figure}
\begin{figure}
\caption{Comparison of anharmonic vs electron-phonon effects in the LTO
transition, illustrating phonon softening (a-c)\protect\cite{RM8b} and doping
dependence of La$_{2-x-y}$Nd$_y$Sr$_x$CuO$_4$ (d-f)\protect\cite{RMXB}.  While
all models have substantial anharmonicity, the top frames (a,d) have no
electron-phonon coupling, middle frames (b,e) have moderate, and bottom frames
(c,f) large electron-phonon coupling.  In frames (a-c), the dotted lines are 
guides to the eye; in frames (d-f), the solid lines and open circles are for 
$y=0$, the dashed lines and filled circles for $x=0.15$.}
\label{fig:bb21}
\end{figure}
\begin{figure}
\caption{Tilting mode distortions in LBCO, interpreted as a dynamic JT effect.}
\label{fig:b21}
\end{figure}
\begin{figure}
\caption{Ferroelectric mode distortions in BaTiO$_3$.}
\label{fig:a21}
\end{figure}
\begin{figure}
\caption{(a) Soliton model of dynamic JT phase\protect\cite{RM8d}. (b) 
Antiphase boundary of the LTT phase\protect\cite{APB}.  (c) Model of LTT phase, 
derived from electron microscopic examination\protect\cite{Zhu}. (d) Model of 
incommensurate Bi-2212, derived from EXAFS data\protect\cite{Bian2}.}
\label{fig:c21}
\end{figure}
\begin{figure}
\caption{Phonon dispersion curves showing anomalous softening of the O-O stretch
modes: (b) LSCO; (c) YBCO; (d) LNO\protect\cite{PiRe,PiRe2}. (a) Related phonon
distortions: top: breathing mode, bottom: 1D stretch mode.}
\label{fig:a22}
\end{figure}
\begin{figure}
\caption{Phonon dispersion curves showing anomalies of the planar quadrupolar
mode in several cuprates\protect\cite{PiRe}.}
\label{fig:b22}
\end{figure}
\begin{figure}
\caption{Pseudogap phase diagram for (a) LSCO\protect\cite{Hwa}, (b),(c) 
YBCO\protect\cite{RMPRL,Coop}, and (d) a combined diagram for both YBCO and 
LSCO\protect\cite{Bat1}.}
\label{fig:23}
\end{figure}
\begin{figure}
\caption{Pseudogap in the RVB model\protect\cite{NagL}.}
\label{fig:a23}
\end{figure}
\begin{figure}
\caption{NMR in LCO\protect\cite{Hamm}. (a,b) Full spectra at several 
temperatures. In (b), the three data sets were taken at different frequencies.
(c) Distribution of tilt angles, $P(\mu )$, with $\mu =cos\Phi$, required to
describe the NMR results.}
\label{fig:c23}
\end{figure}
\begin{figure}
\caption{Phonon anomaly in YBCO: optical conductivity in phonon region of
YBCO$_{6.6}$\protect\cite{Tim1}. Inset: T-dependence of conductivity at 410$
cm^{-1}$.}
\label{fig:e23}
\end{figure}
\begin{figure}
\caption{PDF's of LBCO\protect\cite{Bill}: theory (a) and experiment (b) for
the LTT (solid lines) and LTO (dashed lines) phases.}
\label{fig:d23}
\end{figure}
\begin{figure}
\caption{Phonon softening in LSCO\protect\cite{Noha}. Solid line shows the
result of the VHS theory.}
\label{fig:b23}
\end{figure}
\begin{figure}
\caption{Schematic phase diagram of nanoscale phase separation, illustrating the
free energy vs doping for the bulk phases (a) and the domain phases (b), with
the resulting phase diagram (c).  }
\label{fig:j23}
\end{figure}
\begin{figure}
\caption{Comparison between conventional CDW (a), CDW with discommensurations
(b), and phase separated CDW (c).}
\label{fig:f23}
\end{figure}
\begin{figure}
\caption{Phase separation due to combined magnetic and charge transfer
mechanisms\protect\cite{CDG}.  Shaded area: 2-phase regime.}
\label{fig:i23}
\end{figure}
\begin{figure}
\caption{(a) Phase diagram of LCO. Phases are labelled O(T) for orthorhombic 
(tetragonal).  (b) Modified phase diagram of LCO, found for 
annealed samples\protect\cite{Hor2}.  $Fmmm$ phases have recently been 
identified\protect\cite{Custag} as regions of stage ordering, as in LNO, 
Fig.~\protect\ref{fig:30}.}
\label{fig:g23}
\end{figure}
\begin{figure}
\caption{Theoretical\protect\cite{Ced} phase diagram for YBCO, with experimental
data of Ref.~\protect\cite{Yang}.}
\label{fig:24}
\end{figure}
\begin{figure}
\caption{$T_c$ vs hole doping $P$ in Ca and La substituted 
YBCO\protect\cite{Tal10}.  Note that the 60-70K plateau falls at a fixed hole
doping, and not at a fixed $\delta$.}
\label{fig:T10}
\end{figure}
\begin{figure}
\caption{ Phase diagram for LSCO (solid and dotted lines) and LBCO (dashed 
lines). Note change of scale at $T=100K$.  The AFM phase for LSCO is
controversial, having been detected by microscopic probes ($\mu$SR, NMR) only --
see discussion in Section XIII (dotted curve after Ref.~\protect\cite{Got}).
Inset defines $T_f$, $T_{gl}$.}
\label{fig:25}
\end{figure}
\begin{figure}
\caption{Roughening transition\protect\cite{THwa}. While the calculation can be
applied in many different situations, for present purposes, it represents a
transition between a striped (a) and an island (b) phase, as a function of 
increasing disorder.}
\label{fig:26}
\end{figure}
\begin{figure}
\caption{Magnetic domain fraction $P_{SG}$ in LSCO (a) and YBCO (b), found from 
M\"ossbauer measurements\protect\cite{Imb}. 
}
\label{fig:27}
\end{figure}
\begin{figure}
\caption{(a) Meissner fraction (squares) and diamagnetic phase fraction 
(circles) vs doping, for LSCO\protect\cite{Rad} and (b) Meissner fraction for a 
variety of cuprates (b)\protect\cite{AlSc}. Solid lines are indicative of 
expected behavior for phase separated material.  Long dashed lines in (a): 
guide to the eyes; short dashed line: LTO-HTT transition.}
\label{fig:28}
\end{figure}
\begin{figure}
\caption{Phase diagram of LNO\protect\cite{stag}. Filled circles (squares):
N\'eel temperatures of primary (secondary) phases.  The magnetic order crosses 
over from commensurate (C) to incommensurate (I) near $\delta =0.11$.}
\label{fig:30}
\end{figure}
\begin{figure}
\caption{Phase diagram of LSNO. Note change of scale at $T=100K$.  Dashed line
shows extent where $x=1/3$ phase is present. Open squares:
Ref.~\protect\cite{Chow}; filled squares: Ref.~\protect\cite{LSNO8}; remaining 
data: Ref.~\protect\cite{LSNO3} (triangles are actually taken from O-doped 
LNO).  }
\label{fig:31}
\end{figure}
\begin{figure}
\caption{Calculation of RG flow of various susceptibilities, near half filling, 
showing crossover from SDW to d-wave superconductivity, with (a) $y_c=5$, (b) 
$y_c=2.5$. (c) Resulting phase diagram.}
\label{fig:33}
\end{figure}
\begin{figure}
\caption{Phase diagram for BKBO\protect\cite{Slei}. }
\label{fig:34}
\end{figure}
\begin{figure}
\caption{Calculated thermodynamic functions: (a) $F(N)$, Eq. 
\protect\ref{eq:A7b} (solid line), with two approximate forms described in the 
text; (b) $\mu(x)$ near a VHS, Eq.~\protect\ref{eq:A8}, using either the exact 
or approximate form of $F(N)$; (c) temperature-dependent dos 
Eq.~\protect\ref{eq:A8b}, which enters the expressions for the compressibility 
$\kappa$ and the susceptibility $\chi$, at T=0 (dotted line) or 10meV (solid 
line), with approximate expression, using small-y limit of $F(y)$ (dashed line);
(d) comparing the exact internal energy $\Delta U^*\equiv\Delta U(T)/N_1
(k_BT)^2$, Eqs.~\protect\ref{eq:A2} and \protect\ref{eq:A9a} (solid line) with 
the approximate form, Eq.~\protect\ref{eq:A9c} (dashed line).}
\label{fig:A1}
\end{figure}
\begin{figure}
\caption{Calculated values for $\tau (x_c)$.}
\label{fig:B1}
\end{figure}

\end{document}